\documentclass[a4paper]{book}
\usepackage{graphicx}
\usepackage{amsmath}
\usepackage{axodraw2}
\usepackage{slashed}
\numberwithin{equation}{chapter}
\usepackage{amsfonts}
\usepackage{amssymb}
\usepackage{slashed}
\usepackage{adjustbox}
\usepackage[sort&compress]{natbib}
\setcitestyle{numbers,open={[},close={]},comma}
\usepackage[hidelinks]{hyperref}
\usepackage{bm}
\usepackage{MnSymbol}
\usepackage{pdfpages}
\usepackage{color}
\usepackage{physics}
\usepackage{afterpage}
\usepackage[normalem]{ulem}
\usepackage{tikz}
\usepackage{fancyhdr}
\fancypagestyle{plain}{%
	\fancyhead{} 
	 
}

\title{Thesis Draft}
\date{\today}
\author{Eamonn Weitz}
\begin{document}
\begin{titlepage} 
	\includepdf[pages={1}]{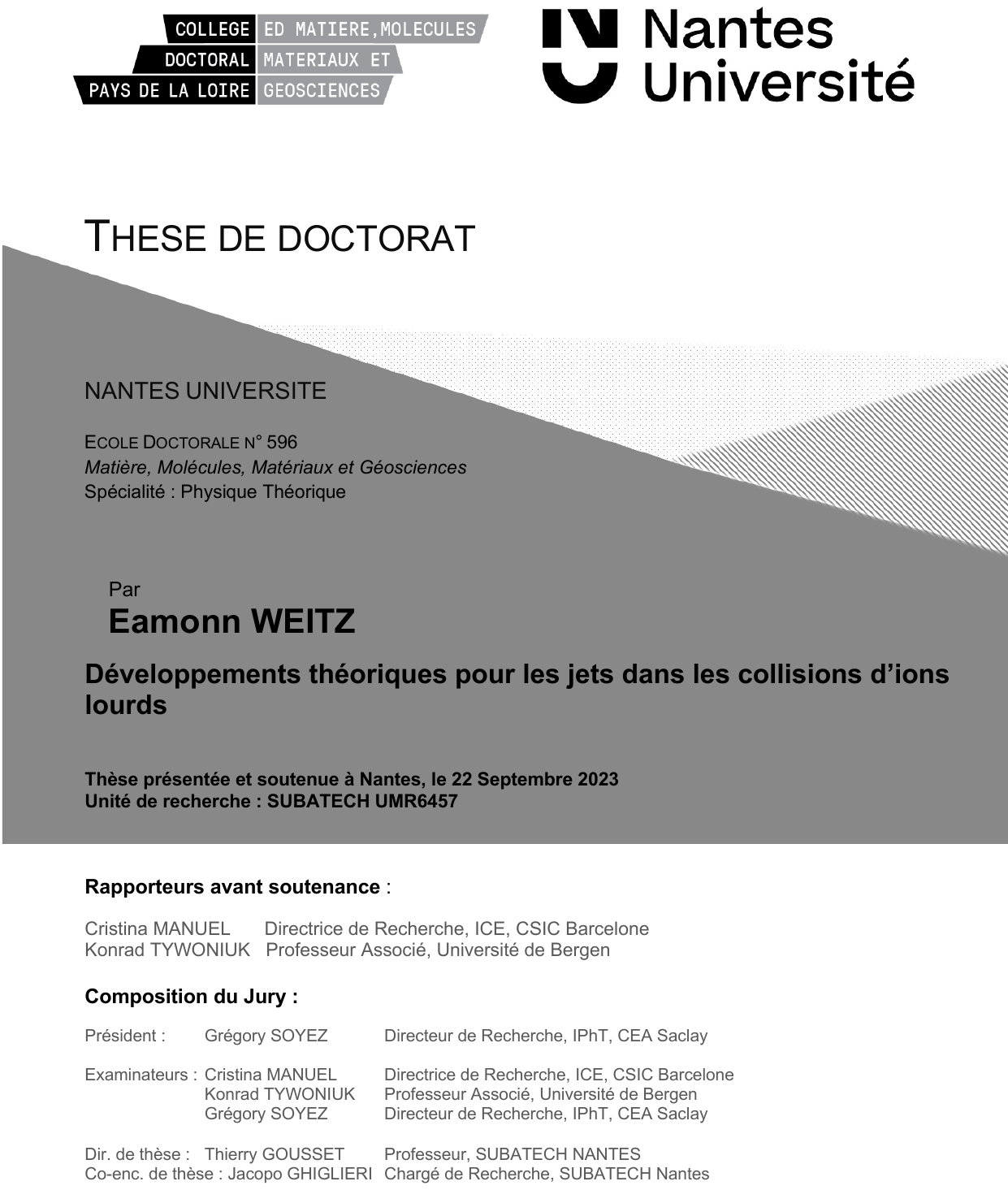}
\end{titlepage}
\pagenumbering{gobble}
\thispagestyle{empty}\null\newpage
\newpage
\clearpage
\begin{center}
    \thispagestyle{empty}
    \vspace*{\fill}
    \emph{To my grandparents}
    \vspace*{\fill}
\end{center}
\clearpage
\newpage
\section*{Acknowledgements/Reflections}

Completing this PhD has been an arduous, demanding 
yet profoundly gratifying journey. With that being 
said, I found myself contemplating on numerous occasions along the way, 
how much more difficult it would have been, had I not been 
fortunate enough to be surrounded by such tremendous people.

I arrived at Subatech in September 2020 in less than 
ideal conditions, amid the backdrop of a global pandemic, 
with a very basic level of French. For those first few months, 
during which time I was largely confined to my room because of 
lockdowns, there was no shortage of colleagues who realised the 
vulnerability of my situation and regularly checked up on me. To this, I am eternally grateful.

To my various officemates, from the basement with Nadiya and 
Hanna to H221 with Grégoire, Stéphane, Nathan, Jakub, Alexandre, 
Jean-Baptiste, Julie, Tobie and Jakub, I will always look back 
fondly on our interactions, be it through stimulating discussions 
or unapologetic procrastination. As well as being wonderful company, 
I owe Johannès, Michael, Victor Lebrin, Maxime Pierre and Arthur all
 much thanks for your patience and willingness to help me in situations 
 where I needed an english translator, particularly during my initial days. 
 I extend my thanks to many other PhD’s and post-docs: Vincent, 
 Marten, Claudia, Johan, Nicolas, Felix, Yohannes, Mahbobeh, Victor 
 Valencia, Keerthana, Pinar, Léonard, Quentin, Kazu, Jiaxing, Jakub; 
 it has been an absolute joy to share the lab with each of you over 
 these past three years.

I owe many thanks to the 
adminstrative and IT staff at Subatech, 
for their everpresent and impressive 
patience. Along with Pierrick, Farah, Sophie, Tanja and 
Nadège, I want to give a special thanks to Véronique 
and Séverine from the missions department; you have 
never made me feel like a burden or a chore, despite 
my inability to understand certain aspects of the reimbursement 
procedures, which I must admit persist to the current day.

Jacopo, your brilliance as a scientist coupled with your 
unyielding patience as a teacher has made it a pleasure 
to work under your supervision during the past three years. 
Pol-Bernard and Thierry, while we did not get the opportunity
 to work together directly as was originally envisioned, 
 you were reliable and helpful when I came to you for information 
 and I will always be thankful for that. I would also like to 
 thank Stéphane and François for organising the QCD Masterclass 
 in 2021 and Marlene, Marcus and Maxime Guillbaud for orchestrating 
 the heavy-ion school in 2022. 

In addition to Jacopo and Thierry, I would also like to thank Grégory, Cristina and Konrad for agreeing to 
be on my jury. Despite the inevitable stress surrounding my defense, I nevertheless found it to be 
a great learning opportunity; your questions and comments allowed me to understand my own work from a slightly 
different angle.

Away from the lab, I have my family to thank, 
who have provided their unrelenting support for as 
long as I can remember, always trying to do whatever 
they can to make my life easier. From my friends in Dublin
 to those scattered around the globe, you have all been sorely 
 missed over these last three years.

Last but certainly not least, I must thank my wife, Emerald. 
Besides being a such wonderful life companion, your constant 
encouragement and caring nature has allowed me to grow significantly 
as a person during our time together — I could not appreciate it more.

\newpage
\tableofcontents
\newpage
\pagenumbering{arabic}
\chapter{Introduction}\label{ch:into}

In the early 1980s, heavy-ion collisions were proposed 
as a means to provide a window into the 
inner workings of the early universe. In particular, it was hoped that 
these kinds of experiments would be able to produce 
a deconfined state of matter with quarks and gluons as the fundamental 
degrees of freedom. Around that time, Bjorken posited that objects known as jets, high-energy 
sprays of collimated hadrons would interact with such a state of matter if it were produced in these collisions 
and in doing so, would become extinguished or quenched \cite{Bjorken:1982tu}. Indeed, it has been over 20 years since 
the phenomenon of jet quenching was first observed at RHIC, providing strong 
evidence that for the existence of a \emph{quark-gluon plasma} (QGP) in heavy-ion collisions \cite{PHENIX:2001hpc,STAR:2002svs}.

In recent years, the arrival of more powerful experiments at the LHC 
and RHIC has put us in a position where we can reasonably ask sharp questions concerning the make-up and the nature
of the QGP, thereby definitively moving on from solely pondering its existence. Nevertheless, it goes without saying that with
such progress comes the need to improve the accuracy of our theoretical models to the stage where they 
can make pointed, meaningful predictions.

As an example, consider the phenomenon of transverse momentum broadening: in traversing the 
QGP, a jet receives kicks in transverse momentum space, giving rise to a diffusive process, characterised
 by the \emph{transverse momentum broadening coefficient}, $\hat{q}$. It recently has been suggested that such momentum 
 broadening effects may be imprinted on the jet acoplanarity observable \cite{Mueller:2016gko,Chen:2016vem,CMS:2017eqd,Ringer:2019rfk,Chien:2022wiq}. 
Moreover, $\hat{q}$ has an indirect but strong influence on the production of \emph{medium-induced radiation}. 
Quantum corrections to $\hat{q}$ (at $\mathcal{O}(g^2)$) have drawn a significant amount of attention over the past ten years or so since it 
was realised that they have the potential to be \emph{logarithmically enhanced} \cite{Liou:2013qya,Blaizot:2013vha}. We
follow this line of enquiry here, computing these logarithmic corrections using thermal field theory. In 
doing so, we rigorously analyse how the thermal scale affects the region of phase space from which these 
corrections are borne. Another important advancement in improving the quantification of jet quenching phenomena 
has come with the development of a resummation method, which provides a non-perturbative
evaluation of the \emph{classical contributions} to transverse momentum broadening \cite{CaronHuot:2008ni,Moore:2019lgw,Moore:2021jwe,Schlichting:2021idr}.

The \emph{asymptotic mass}, $m_{\infty}$ is a correction to a hard parton's dispersion relation as it forward 
scatters with plasma constituents and is also known to control in part the mechanism of medium-induced radiation.
The computation of non-perturbative corrections to $m_{\infty}$ is firmly underway \cite{Moore:2020wvy,Ghiglieri:2021bom}. In 
computing quantum corrections to $m_{\infty}$ (also at $\mathcal{O}(g^2)$) in the thermal 
field theory framework, we concretely further this program. We complete a matching 
calculation, which eliminates unphysical infinities from the previously computed non-perturbative evaluation while also beginning 
the full computation of the $\mathcal{O}(g^2)$ corrections.

This thesis is organised in such a way to slowly but surely build up the intuition and introduce the tools that 
are needed to understand the reasonably technical computations of quantum corrections to the aforementioned
quantities that come in the later chapters. 
Specifically, the thesis is laid out as follows:
\begin{itemize}
    \item Starting from the basics of quantum field theory, Ch.~\ref{ch:hic} is intended to convince 
        the reader of the existence of a QGP in heavy-ion collisions through the provision of both theoretical 
        arguments and experimental evidence. Along the way, we introduce the concept of jets and describe the 
        different stages of a heavy-ion collision.
    \item In Ch.~\ref{ch:eloss}, we set up the physical picture, which is helpful to keep in mind throughout the 
        rest of the thesis. Namely, a jet, which for our purposes can be taken as a hard parton, propagates through a quark-gluon 
        plasma, interacting with 
        the medium constituents and in doing so, is forced to radiate. We make a point to emphasise how the triggering mechanism for this radiation
        is controlled by the associated formation time and we identify the conditions under which Landau-Pomeranchuk-Migdal (LPM) interference must 
        be accounted for. We furthermore compute the radiation rate in the multiple scattering regime, where LPM resummation is necessary 
        and in the single scattering regime, where one can instead use the opacity expansion. The chapter closes with a brief overview 
        of the literature and a short discussion of some recent developments.
    \item Ch~\ref{chap:tft_eft} introduces some of the modern theoretical tools, which support the calculations 
        to follow in the later chapters. In particular, we give some general ideas regarding the basics of effective field theories 
        and thermal field theory before moving on to discuss the finite-temperature effective theories EQCD and HTL theory. In Sec.~\ref{sec:chtrick},
        we examine how a seminal idea from Caron-Huot \cite{Caron-Huot:2007zhp} can, in certain situations, provide 
        an extremely useful connection between these two theories, which 
        furthermore allows for the non-perturbative computation of classical corrections to the transverse momentum broadening kernel and the asymptotic mass.
        The chapter then continues with a demonstration of how this idea can also be utilised to efficiently extract the leading order soft 
        contributions to the transverse momentum broadening kernel and the asymptotic mass.
    \item Ch~\ref{chap:qhat_chap} is largely based on the work \cite{Ghiglieri:2022gyv}. There, we carefully 
        scrutinise the previous computations of double logarithmic corrections to $\hat{q}$ \cite{Liou:2013qya,Blaizot:2013vha}
        before showing how the double logarithmic phase space is deformed for a computation done in the setting of a weakly coupled 
        quark-gluon plasma. In addition, we clearly demonstrate how this region of phase space is situated with respect to 
        the one from which the classical corrections to $\hat{q}$ emerge. Most of our computation is performed in the harmonic oscillator 
        approximation. However, a pathway, which could allow us to go beyond this approximation is also presented. Technical 
        details are contained in App.~\ref{ch:qhat_app}.
    \item Ch~\ref{ch:asym_mass} contains our yet to be published work on the $\mathcal{O}(g^2)$ asymptotic mass corrections. We list all the 
        diagrams that can contribute at this order of perturbation theory, before arguing that only one of them, what we label 
        as diagram $(c)$ contributes. We show how some of the IR divergences associated with this diagram 
        cancel with corresponding UV divergences from the EQCD calculation, completed in \cite{Ghiglieri:2021bom}. We then proceed 
        to a computation of the full fermionic and gluonic parts of diagram $(c)$, which turn out to contain some unforeseen UV
        and collinear divergences. Technical details are contained in App.~\ref{app:amass_app}.
    \item A broad summary and conclusion are given in Ch.~\ref{ch:conc} in addition to discussions of how we plan to push the works contained 
        in Chs.~\ref{chap:qhat_chap} and ~\ref{ch:asym_mass} further.
    \item Our conventions are specified in App.~\ref{sec:conventions} and some details on the basics of Wilson loops 
        are located in App.~\ref{app:wloop}.
\end{itemize}
\newpage
\chapter{QCD and Heavy-Ion Collisions}\label{ch:hic}
The \emph{Standard Model of Particle Physics} is arguably one of the greatest intellectual achievements to have 
come out of the $20\textsuperscript{th}$ century. While it is not expected to be the end of the story 
as far as elementary physics is concerned, the well-tested Standard Model provides a remarkably clean and coherent description 
of three out of the four fundamental forces: the electromagnetic and weak 
forces, which are unified into the \emph{Electroweak Theory} as well as the strong force, described by
\emph{Quantum Chromodynamics} (QCD). This thesis is concerned with the study of the latter.

In this first chapter, we introduce QCD in Sec.~\ref{sec:qcd} through a somewhat formal lens before 
discussing some of its properties. In Secs.~\ref{sec:jets} and \ref{sec:qcdpd} we highlight 
how some of these properties manifest themselves in systems with a large number of particles. Sec.~\ref{sec:hic_qgp} 
is then devoted to a discussion of heavy-ion collisions, which 
have proved extremely successful in aiding the experimental exploration of the QCD phase diagram.
This is intended 
to appropriately set up and motivate what is to follow in the rest of the thesis, where we study in great detail 
how jets, high energy showers of radiation interact with the quark-gluon plasma.

\section{Introduction to QCD}\label{sec:qcd}
Locality is generally considered to be an essential feature of any (fundamental) quantum field theory (QFT). However, 
by insisting on a theory to be local, a certain redundancy is necessarily introduced. This redundancy manifests itself 
in the Lagrangian description as what is known as gauge freedom. To understand this in a more concrete way,
consider what happens when we demand that a Lagrangian describing $N_f$ fermionic fields, $\psi_i(X)$ ($i=1,2,...,N_f$) remain invariant under the local transformation
\begin{equation}
    \psi_i(X)\rightarrow \psi'_i(X)=\mathcal{U}(X)\psi_i(X),\quad\quad \mathcal{U}=\exp(i\xi^a(X)T_F^{a}),\quad\quad a=1,2,..,d_{A},\label{eq:quark_transform}
\end{equation}
with a sum over $a$ implied. The transformation, $\mathcal{U}$ is said to be in the \emph{fundamental representation} of the Lie group SU($N_c$), meaning 
in practice that $\mathcal{U}(X)$ is an $N_c\times N_c$ unitary matrix with $\det\mathcal{U}=1$. Since SU($N_c$) is a 
topologically trivial group, the matrices $\mathcal{U}$ are given in terms of the so-called generators, $T_{F}^a$, 
which are in the fundamental representation of the $\mathfrak{su}(N_c)$ Lie algebra. The $\xi_a(X)$ are then arbitrary 
functions of spacetime coordinates. The fact that the $T_{R}^a$ generate a Lie algebra means that they obey, for a general 
representation, $R$ 
\begin{align}
    [T_{R}^a,T_{R}^b]&=if^{abc}T_{R}^c.
    \\f^{abd}f^{dce}&+f^{bcd}f^{dae}+f^{cad}f^{dbe}=0
\end{align}
If the structure constants $f^{abc}$ are zero, then the matrices, $\mathcal{U}(X)$ are in a representation of an Abelian group. Otherwise, 
the group is said to be non-Abelian, which is the case that we are presently interested in.
Moreover, we have that 
\begin{equation}
    T_{R}^a T_{R}^a=C_{R}\mathbb{I},\quad\quad\Tr T_{R}^a T_{R}^b=T_{R}\delta^{ab},
\end{equation}
where $C_{R}, T_R$ are respectively the Casimir and index of the representation, $R$.

At this point, it is easy to see that a mass term for the quark fields, i.e $m\bar{\psi}_i\psi_i$ is invariant under the transformation Eq.~\eqref{eq:quark_transform}. 
Be that as it may, because of the spacetime dependence in $\mathcal{U}(X)$, any term with a derivative acting on $\psi_i$ 
will not be so fortunate since
\begin{equation}
\partial_{\mu}\psi_i(X)\rightarrow\Big[\partial_{\mu}\psi_i(X)\Big]'=\mathcal{U}(X)\partial_{\mu}\psi_i(X)+\Big(\partial_{\mu}\mathcal{U}(X)\Big)\psi_i(X).
\end{equation}
To fix the kinetic term, we must define a \emph{covariant derivative} operator, $D_{\mu}$ such that
\begin{equation}
    D_{\mu}\psi_i(X)\rightarrow\Big[\partial_{\mu}\psi_i(X)\Big]'=\mathcal{U}(X)D_{\mu}\psi_i(X).\label{eq:cov_tran}
\end{equation}
It turns out that the covariant derivative should take the form 
\begin{equation}
    D_{\mu}=\partial_{i}-ig A_{\mu},
\end{equation}
where $A_{\mu}=A_{\mu}^a T_F^a$ is a Lie-algebra-valued \emph{gauge field} and $g$ is the (bare) \emph{coupling constant} that 
characterises the strength of the interaction between this field and $\psi$. It transforms in such a way
\begin{equation}
    A_{\mu}(X)\rightarrow A'_{\mu}(X)=\mathcal{U}(X)A_{\mu}(X)\mathcal{U}^{-1}(X)-\frac{i}{g}\Big(\partial_{\mu}\mathcal{U}(X)\Big)\mathcal{U}^{-1}(X)\label{eq:adj_tran}
\end{equation}
to ensure that Eq.~\eqref{eq:cov_tran} holds true. The gauge field is said to transform according to the \emph{adjoint representation} 
of SU($N_c$), modified by an inhomogeneous term (the second term above). The adjoint representation of a group is defined such that the generators are given
in terms of the structure constants themselves
\begin{equation}
    (T^a_A)^{bc}=-if^{abc},
\end{equation}
where these are now $d_{A}\times d_{A}$ matrices. 
\begin{table}[ht]
    \centering
    \begin{tabular}{|c|c|c|c|}
    \hline
      \text{Quantity}& \text{Symbol} & \text{Value for QCD } \\
      \hline
        \text{Number of matter fields} & $N_f$ & $6$ \\
        \text{Number of charges} & $N_c$ & 3 \\
        \text{Fundamental Casimir} & $C_{F}=\frac{N^2_c-1}{2N_c}$ & $\frac{4}{3}$\\
        \text{Adjoint Casimir}  & $C_{A}=N_c$ & $3$\\
        \text{Fundamental index} & $T_{F}=\frac{1}{2}$ & $\frac{1}{2}$\\
        \text{Adjoint index}& $T_{A}=N_c$ & $3$\\
        \text{Fundamental dimension}& $d_{F}=N_c$ & $3$\\
        \text{Adjoint dimension}& $d_{A}=N_c^2-1$ & $8$\\
        \hline
    \end{tabular}
    \caption{Collection of group theory factors, specific to QCD, intended to be used for future reference. Note that while $6$ flavours 
    of quarks are included in the Standard Model, oftentimes perturbative calculations are performed with $N_f<6$ if the 
    energy scale of interest is much smaller than the heavier quarks.}
    \label{tab:group}
\end{table}

To briefly recap, by insisting on a Lagrangian description of a field $\psi$ that is blind to local, non-Abelian phases, the Lagrangian 
is forced to be furthermore invariant under the transformation Eq.~\eqref{eq:adj_tran} --  it is forced to be \emph{gauge-invariant}. Under these 
restrictions, the resultant theory that one derives, describing the interaction between dynamical gauge and matter fields goes by the name of \emph{Yang-Mills Theory} \cite{Yang:1954ek}. 

It turns out that Yang-Mills theory with $N_c=3$ \emph{colours} and
$N_f=6$ \emph{flavours} yields a theory that describes well the strong interaction, which is responsible for the binding of nucleons 
inside atomic nuclei. The colours: red, green and blue are non-Abelian charges and are mediated by the \emph{gluon field}, $A_{\mu}$.
Each flavour corresponds to a different \emph{quark field}: up, down, strange, charm, bottom, top, listed from least to most massive. The resultant 
theory is known as Quantum Chromodynamics and is given by the Lagrangian \cite{Fritzsch:1973pi}\footnote{An 
extra term $\mathcal{L}_{\theta}=\theta\varepsilon^{\mu\nu\rho\sigma}F_{\mu\nu}F_{\rho\sigma}$ is sometimes included. 
Despite the fact that it can be written as a total derivative, implying that it does not contribute at any order 
of perturbation theory, this term can be relevant in terms of the non-trivial vacuum structure of QCD. We will not be 
concerned with such a term in what follows.} constructed to be suitably gauge-invariant 
and renormalisable
\begin{equation}    
    \mathcal{L}_{\text{QCD}}=-\frac{1}{4}F^{a}_{\mu\nu}F^{\mu\nu\,a}+\bar{\psi}_i(i\slashed{D}-m_i)\psi_i\label{eq:qcd_lag},
\end{equation}
with a sum over $i$ implied. The field strength tensor is
\begin{equation}
    F_{\mu\nu}^a=\partial_{\mu}A_{\nu}-\partial_{\nu}A_{\mu}+g f^{abc}A^b_{\mu}A^c_{\nu}
\end{equation}
and $g$ is the strong coupling constant. An immediately remarkable aspect of Eq.~\eqref{eq:qcd_lag} is that of the self-interactions between 
gluons, a feature, which is not present in QCD's Abelian counterpart, \emph{Quantum Electrodynamics} (QED).
 
Eq.~\eqref{eq:qcd_lag} should be considered the Lagrangian for classical QCD. The quantisation of non-Abelian gauge theories 
is usually done using the functional integral formalism. Along the way, one runs into the issue of gauge-fixing at the level 
of the path integral, which is dealt with by introducing so-called \emph{ghost fields} in the Fadeev-Popov procedure. We refer the reader
to any of the textbooks \cite{Peskin:1995ev,Srednicki:2007qs,Schwartz:2014sze} for the details pertaining to the quantisation procedure.

In the form of Eq.~\eqref{eq:qcd_lag}, QCD does not possess any predictive power beyond tree-level -- it must be renormalised. Divergences 
that emerge at one-loop level and beyond can be absorbed by replacing the bare coupling constants of the theory, $g_i$ renormalised couplings, $g_i(\mu)$, which 
now depend on an additional scale, $\mu$. The idea is then to demand that physical observables be independent of this scale, which 
gives rise to a set of coupled differential equations known as \emph{renormalisation group equations} (RGE). The RGE for the strong coupling reads
\begin{equation}
    \frac{d g(\mu)}{d\mu}=\beta(g(\mu))\label{eq:beta_function},
\end{equation}
where the $\beta$-function, defined above, can be calculated order by order in perturbation theory. At one-loop level, one finds \cite{Politzer:1973fx,Gross:1973id}
\begin{equation}
    \beta(g(\mu))_{\text{one-loop}}=-\frac{g^3(\mu)}{16\pi^2}\beta_0,\quad\quad\beta_0\equiv\bigg(\frac{11}{3}C_{A}-\frac{4}{3}N_f T_F\bigg),\label{eq:beta_one}
\end{equation}
from which one can proceed to solve Eq.~\eqref{eq:beta_function}. The solution can be written as
\begin{equation}
    \alpha_s(\mu)=\frac{4\pi}{\beta_0}\frac{1}{\ln\frac{\mu^2}{\Lambda^2_{\text{QCD}}}},\quad\quad \alpha_s\equiv \frac{g^2}{4\pi},\label{eq:alpha_running}
\end{equation}
and is understood to only be valid for $\mu >\Lambda_{\text{QCD}}$. Two very 
striking features of QCD are brought to light by this equation:
\begin{itemize}
    \item As $\mu$ increases, the coupling gets weaker, signifying that QCD is \emph{asymptotically free}.
    \item As $\mu$ approaches the scale $\Lambda_{\text{QCD}}$, the coupling diverges, giving rise to what is known as a \emph{Landau Pole}. This 
        suggests that perturbative QCD should not be trusted at these energies and moreover, it hints at the \emph{confining nature} 
        of QCD, why free quarks and gluons\footnote{It is not correct to say that the presence of the 
        Landau pole proves confinement as this kind of RGE technology is strictly only applicable in the perturbative regime. An analytical explanation of confinement from
         ``first principles'' remains elusive up to the present day. Be that as it may, Lattice QCD 
         calculations have been able to reproduce observed hadron spectra.} and hence, coloured particles are never observed in nature.
\end{itemize}
The QCD $\beta$-function is currently known up to five-loop \cite{Baikov:2016tgj,Luthe:2016ima,Luthe:2017ttg,Chetyrkin:2017bjc}. With this value in hand and the 
2018 ``world-average'' value $\alpha_{s}(M_Z^2)=0.1181\pm0.0011$ \cite{ParticleDataGroup:2018ovx}, given in the $\overline{\text{MS}}$ subtraction scheme, one finds 
$\Lambda^{(6)}_{\text{QCD}}=(89\pm6)\text{MeV}$, where the apex denotes that such a calculation takes $N_f=6$.

If one takes the colour confinement hypothesis seriously, then it follows that only colour singlets -- states that are invariant under rotations
in colour space should be observed in nature. Such states are known as \emph{hadrons}. Hadrons are divided up into 
two categories: baryons, which are bound states of three quarks and mesons, which are bound states of a quark and an anti-quark\footnote{There is also of 
course an anti-particle for each baryon and charged meson.}.

Before moving on, let us go back to Eq.~\eqref{eq:qcd_lag} and consider its global symmetries, i.e, transformations under which it is invariant 
that are not spacetime dependent. Moreover, let us neglect for the moment the quark masses, which is not such a bad approximation 
if we consider energies much larger than $\Lambda_{\text{QCD}}$\footnote{It 
would indeed be a poor approximation to neglect all 6 of the quarks' masses. However, given
some energy scale, it is reasonable to take the quarks whose masses are below that scale to be \emph{light}
and neglect their masses. The \emph{heavy} quarks with masses above that scale then \emph{decouple} and their 
dynamics do not need to be considered.}. Then we can rewrite it as 
\begin{equation}
    \mathcal{L}_{\text{QCD}}=-\frac{1}{4}F^{a}_{\mu\nu}F^{\mu\nu\,a}+i\Big[\bar{\psi}_{i\,R}\slashed{D}\psi_{i\,R}+\bar{\psi}_{i\,L}\slashed{D}\psi_{i\,L}\Big]\label{eq:chiral_lag},
\end{equation}
where the chiral spinors are defined using the projectors $\psi_{i\,R/L}=\frac{1}{2}(1\pm\gamma_5)\psi_i$.
As well as Poincaré invariance, the quark sector is invariant under transformations belonging to the group 
\begin{equation}
    \mathcal{G}=\text{U}(1)_V\times\text{U}(1)_A\times\text{SU}(N_f)_L\times\text{SU}(N_f)_R.\nonumber
\end{equation}
$\mathcal{G}$ acts on the quark fields such that
\begin{align}
    \text{U}(1)_V&:\quad\quad\psi_{L}\to e^{i\alpha}\psi_L\quad,\quad\psi_{R}\to e^{i\alpha}\psi_R\nonumber
   \\ \text{U}(1)_A&:\quad\quad\psi_{L}\to e^{i\beta}\psi_L\quad,\quad\psi_{R}\to e^{-i\beta}\psi_R\nonumber
   \\ \text{SU}(N_f)_L&:\quad\quad\psi_{L}\to L\psi_L\quad,\quad\psi_{R}\to \psi_R\nonumber
   \\ \text{SU}(N_f)_R&:\quad\quad\psi_{L}\to \psi_L\quad,\quad\psi_{R}\to R\psi_R\nonumber
\end{align}
where $L,R$ are $N_f\times N_f$ matrices. Through Noether's Theorem, the abelian vector symmetry is associated with baryon number conservation\footnote{In the 
full Standard Model with the weak interaction included, baryon number is anomalous. However, the difference between baryon number 
and lepton number is conserved.}. In contrast, when quantising the theory, one finds that the path integral measure is not invariant 
under the axial abelian symmetry, resulting in what is known as a \emph{chiral anomaly} \cite{PhysRev.177.2426,Fujikawa:1979ay}. As for the 
non-abelian symmetries, consider the vacuum expectation value $\Delta=\langle\bar{\psi}_{i\,L}\psi_{j\,R}\rangle$, also known as the 
\emph{chiral condensate}, which has been shown to be non-zero by Lattice QCD calculations \cite{Giusti:1998wy}. It is not hard to see that this quantity is only 
invariant under transformations such that $L^{\dagger}=R$, implying that the chiral symmetry is \emph{spontaneously broken} into the 
diagonal subgroup:
\begin{equation}
    \text{SU}(N_f)_L\times\text{SU}(N_f)_R\to \text{SU}(N_f)_V\nonumber,
\end{equation}
where the pions are the Nambu-Goldstone bosons tied to the symmetry breaking\footnote{Technically, pions are the only Nambu-Goldstone 
bosons when $N_f=2$, where the non-abelian chiral symmetry is known as \emph{isospin symmetry}. For $N_f=3$, 
the three pions fit into an octet with the four kaons along with the $\eta$ meson.}.

We have up to this point discussed two kinds of transitions associated with the theory of QCD: confinement and chiral symmetry breaking, either 
side of which the fundamental degrees of freedom of the theory are clearly very different. Indeed, this 
points to the fact that QCD has a rich phase structure. We 
will come back to this point in Sec.~\ref{sec:qcdpd}, after we discuss what has been a historically successful testbed for perturbative QCD.

\section{Jets}\label{sec:jets}
Confinement greatly obstructs the prospect of ever being able to directly observe free quarks and gluons, which we collectively refer to 
as \emph{partons}\footnote{The term parton comes from Feynman's Parton Model \cite{Feynman:1969wa}. The Parton Model 
posits that at 
very high energies, hadrons can be thought of consisting of a collection of free partons, with each parton species, 
$i$ carrying a fraction, $x$ of the hadron's total momentum. In this sense, partons technically refer to any particle 
from the Standard Model. In this thesis, however, parton is restricted to mean either quark or gluon, without 
any reference to hadrons.}. Even though free partons are briefly produced in collider experiments, once their virtualities
decrease to the scale of $\Lambda_{\text{QCD}}$, they are expected to hadronise. So how can we ever expect to test 
perturbative QCD (pQCD) experimentally? Fortunately, the idea of quark-hadron duality, originally formulated 
by Poggio, Quinn and Weinberg \cite{PhysRevD.13.1958} suggests that certain inclusive hadronic cross-sections at high 
energies, being appropriately averaged over an energy range should coincide with what one calculates in pQCD. Because of this, one 
can use objects known as \emph{jets} to test the perturbative sector of QCD, without having a precise, fundamental description of the 
hadronisation process.

In, for instance, $e^+e^-$ collisions or proton-proton (p-p) collisions, one can have rare, \emph{hard} processes take place 
where large amounts of momentum are exchanged between participating constituents\footnote{At the Relativistic Heavy-Ion Collider (RHIC)
at the Brookhaven National Laboratory (BNL) in Long Island, New York, the initial momenta of these objects are on the order of  $10\,\text{GeV}$ \cite{Blaizot:2015lma}. At the Large Hadron Collider (LHC) in
CERN, Geneva, the momenta are an order of magnitude larger.}. The output of such a process, which is 
calculable in perturbative QCD, is a high-energy parton with large virtuality, $Q$. The way in which the parton 
radiates to shed its virtuality can in principle be calculated order by order in $\alpha_s$. However, it is more 
practical and furthermore accurate to seek a result, which instead focuses on including the contribution from certain regions of phase space, 
which greatly enhance higher order terms in the $\alpha_s$ expansion. The regions of phase space that we are talking about here are associated 
with \emph{collinear} and \emph{soft} emissions. By following such an approach, which is well-explained in the 
textbook \cite{Ellis:1996mzs}, the following picture emerges: the highly-virtual parton splits repeatedly, giving rise to
what is known as a \emph{parton shower}. The splitting, which generates the parton shower can be modelled as a Markovian process and 
can be readily implemented in Monte-Carlo \emph{event generator}.

The shower does not produce an incoherent splatter of particles but rather, because of the 
soft, collinear enhancements, a cone-like object. It is this self-collimated structure, which defines a jet. This is often referred to as \emph{angular-ordered} showering and is in 
fact a general feature of gauge theories. Therefore, this phenomenon also takes place in QED showers. However, the showering 
is accentuated for the case of QCD, due to the self-interacting gluon field, which is a novel feature of Eq.~\eqref{eq:qcd_lag}. Indeed, three jet 
events in $e^+e^-$ collisions provided the first direct evidence of the existence of the gluon and the quark-gluon vertex \cite{Ellis:1976uc,Wiik:1979cq}.
\begin{figure}[t]
    \begin{center}
        \includegraphics[width=0.8\textwidth]{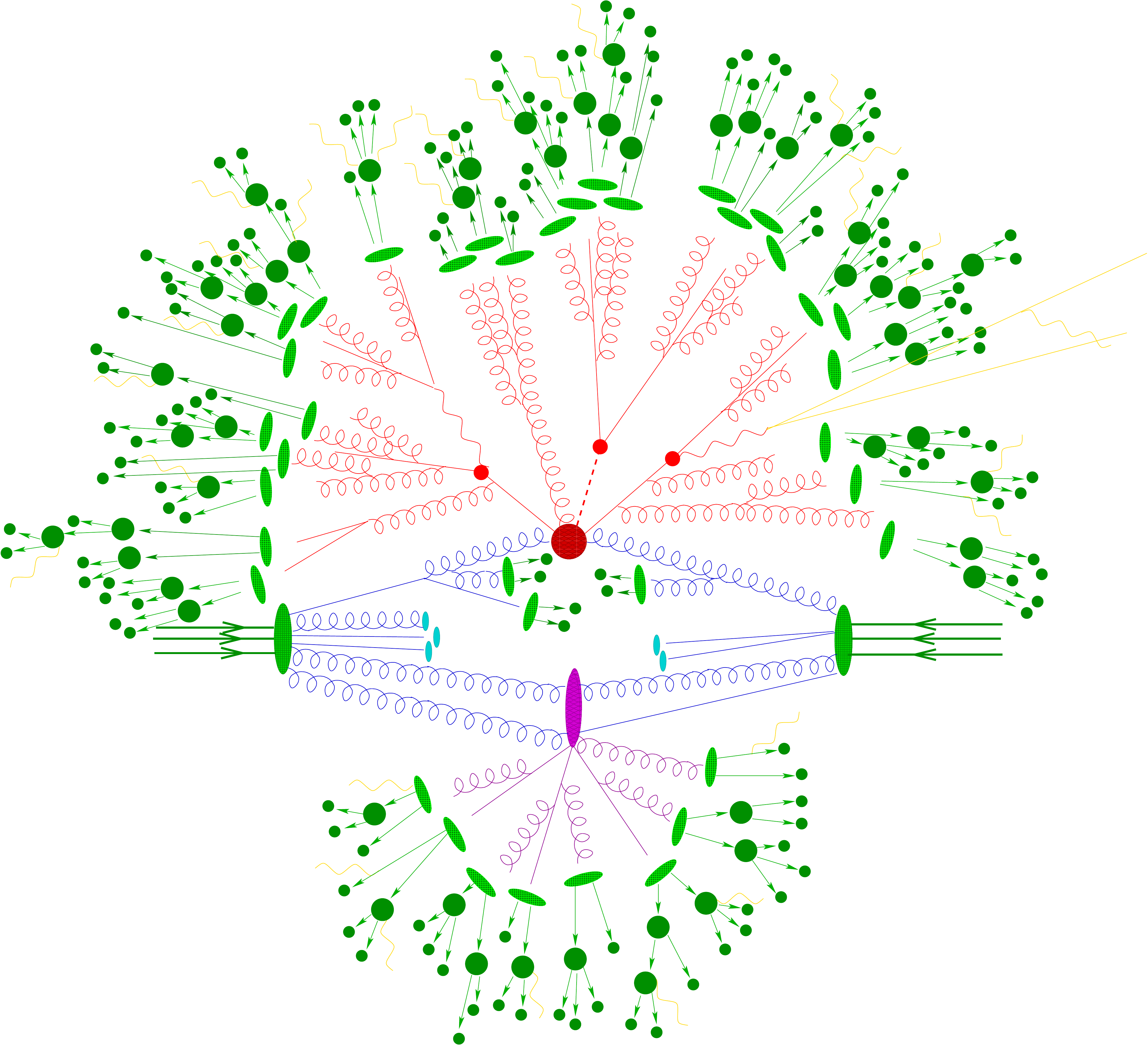}
    \end{center}
    \caption{Illustration of a hadron-hadron collision, taken from \cite{Hoche:2014rga}. The incoming green blobs represent the initial hadronic states. Blue lines denote initial-state radiation, 
    which is not discussed in the main text. The red and purple blobs are the sites at which the hard interactions take place and the 
    lines emanating from them represent the parton shower. Hadronisation processes are marked by the light-green blobs whereas dark-green 
    blobs mark hadron decays. Yellow lines signal soft photon radiation. Note that the angles associated with the parton splittings are greatly 
    exaggerated above.}
    \label{fig:event}
\end{figure}

There are a number of Monte-Carlo event generators\footnote{PYTHIA \cite{Bierlich:2022pfr}, HERWIG \cite{Bellm:2019zci} and SHERPA \cite{Sherpa:2019gpd}
are all examples of event generators. See \cite{Campbell:2022qmc} for a recent review.} on the market built to simulate parton showers in addition to the underlying 
hadron-hadron collision. Nevertheless, since hadrons are the objects that are measured in detectors, how can one be sure 
that the description provided by parton showers is accurate? To bridge the connection between theoretical, pQCD 
predictions and experimental observations, one needs to invoke a \emph{jet definition}. Jet definitions provide 
an algorithm through which experimentalists are able to cluster final-state particles together. The act of clustering leads to the reconstruction of the jet, all the way back to the initial hard 
interaction point. The method through which the final state particles are clustered should moreover be defined in such a way that is robust to the 
addition of an arbitrary number of soft particles --  it should be IR safe. Sequential recombination algorithms cluster particles in an iterative manner, according 
to some protocol until no particle remains. Examples include the Cambridge/Aachen \cite{Dokshitzer:1997in}, $k_t$ \cite{Catani:1993hr}
 and anti-$k_t$ \cite{Cacciari:2008gp} algorithms. Alternatively, there exist cone algorithms \cite{Salam:2007xv}, where one searches for stable cones where the jet axis and jet momentum
 align. See \cite{Salam:2010nqg} for a review.

Nowadays, the study of jets in $e^+e^-$ and $p-p$ collisions is considered to be a mature field, leaning on the side 
of what can be regarded as precision physics. Indeed, the ATLAS and CMS experiments at the LHC even try to search 
for physics beyond the Standard Model (BSM) by taking advantage of this level of precision 
\cite{ATLAS:2017bfj,CMS:2017zts,ATLAS:2018nda,CMS:2021snz,ATLAS:2023swa}. While we are not concerned with such 
a line of enquiry here, it is important to emphasise that the behaviour of jets in $p-p$ collisions or rather, 
\emph{jets in vacuum}\footnote{In some sense, this terminology can be considered 
outdated due to the recent evidence of medium formation in $p-p$ collisions. See Sec.~\ref{sec:collectivity}.} can be used 
as a reference point when comparing to how jets behave in \emph{heavy-ion collisions}, a topic, which 
we will move on to discuss shortly.

\begin{figure}[t]
    \begin{center}
        \includegraphics[width=\textwidth]{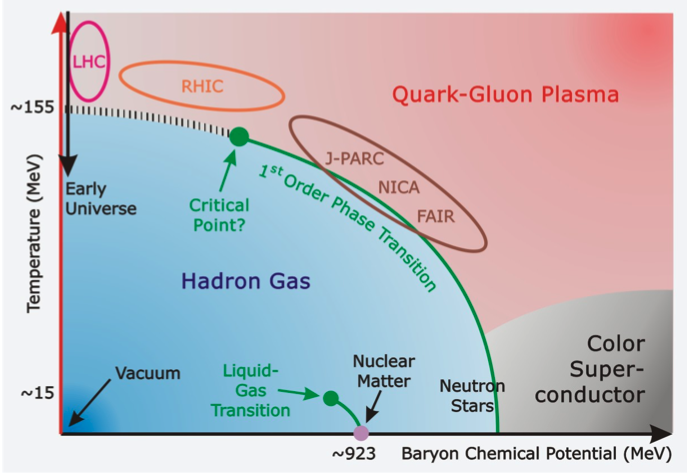}
    \end{center}
    \caption{Sketch of the QCD phase diagram, taken from \cite{Elfner:2022iae}. Close to the $\mu_{B}=0$ axis, the 
    transition from confined to deconfined matter, known as the quark-gluon plasma, is expected to be a smooth crossover. A hypothesised critical point then 
    separates this region from the one where a first-order transition is expected. Moving further to the right, there is the 
    possibility of a colour superconducting phase. The transition from the nuclear matter we observe at room temperature, well-described 
    by the Liquid-Drop Model \cite{Gamow1930MassDC,Weizsacker:1935bkz} to the so-called Hadron Resonance Gas, composed of free nucleons 
    is located at around $T=15\text{MeV}$ \cite{Karnaukhov:2003vp}.}
    \label{fig:qcdphase}
\end{figure}

\section{QCD Phase Diagram}\label{sec:qcdpd}

During Sec.~\ref{sec:qcd}, we discussed some of the basic features of QCD.
 Here, we progress to explore the \emph{QCD phase diagram}, first considered in \cite{Hagedorn:1965st,Cabibbo:1975ig}. A sketch is given in Fig.~\ref{fig:qcdphase}. The $y$-axis is temperature, $T$ whereas the $x$-axis denotes
baryon chemical potential, $\mu_{B}$, conjugate to the baryon number 
\begin{equation}
    \text{B}=\frac{1}{3}(N_q-N_{\bar{q}})
\end{equation}
and can be thought of as a proxy for density. Our aim here is not to provide a comprehensive review of the phase diagram but 
to explain some of its general features. We follow the reviews \cite{Fukushima:2013rx,Fischer:2018sdj,Gelis:2021zmx} closely.

In general, to obtain a quantitative understanding of a phase transition, one needs to identify an order parameter. In Sec.~\ref{sec:qcd}, the chiral condensate was mentioned, which can be thought of as an order parameter characterising the chiral 
phase transition. For the confinement-deconfinement transition, in the approximation where quarks are not dynamical 
(or rather infinitely heavy), a parameter often used 
is the \emph{Polyakov Loop} \cite{Polyakov:1975rs}, derived in the setting of Euclidean spacetime. 
It is given as (see App.~\ref{app:wloop} for a short introduction to Wilson lines)
\begin{equation}
    L(\vec{\mathbf{x}})=\mathcal{P}\exp\Big(ig\int_0^{\beta}d\tau A_4(\tau,\vec{\mathbf{x}})\Big),\quad\quad l(\vec{\mathbf{x}})=\frac{1}{N_c}\Tr L(\vec{\mathbf{x}}),\quad\quad\Phi\equiv\langle l\rangle
\end{equation}
where $\tau=it$ and $A_4$ is the Euclidean temporal component of the gauge field. It has been argued \cite{Kuti:1980gh,McLerran:1981pb} that this temporal Wilson line 
can be related to the heavy quark potential, $f_{\bar{q}q}$
\begin{equation}
    \langle l^{\dagger}(\vec{\mathbf{x}}') l(\vec{\mathbf{x}})\rangle=C\exp\Big(-\frac{1}{T}f_{\bar{q}q}(|\vec{\mathbf{x}}'-\vec{\mathbf{x}}|)\Big)
\end{equation}
through what is known as the Polyakov loop correlation function. In the confining phase, the potential should have a linearly rising potential, i.e $f_{\bar{q}q}(r)=\sigma r$ so that 
the correlation function decays exponentially at large separation. On the other hand, in the deconfined phase, the inter-quark 
potential is thermally screened so that $f_{\bar{q}q}(r\to\infty)=\text{const.}$, which yields a non-zero value for the 
correlation function. We can then state schematically that
\begin{align}
        \text{Confined Phase}\quad\quad\langle l^{\dagger}(\infty) l(0)\rangle&=0\quad\implies\quad\Phi=0\nonumber
        \\\text{Deconfined Phase}\quad\quad\langle l^{\dagger}(\infty) l(0)\rangle&\neq 0\quad\implies\quad\Phi\neq0\nonumber
\end{align}
where we have inferred the value of $\Phi$ from cluster decomposition, i.e $\langle l^{\dagger}(\infty) l(0)\rangle\to|\langle l\rangle|^2$. 
We pause here for a moment to note that the gauge group of (pure glue) QCD is not SU($N_c$) but 
rather SU($N_c$)/$\text{Z}_{N_c}$, the centre of SU($N_c$)\footnote{The centre of a group is a subgroup, whose 
elements commute with all of the other group elements.}. One can however check that 
$\Phi$ is not invariant under the action of $\mathcal{U}\in\text{SU}(N_c)/\text{Z}_{N_c}$. Thus, 
in analogy with the chiral condensate, $\Phi$ is an order parameter for the spontaneous breaking of centre symmetry.
\begin{figure}[t]
    \begin{center}
        \includegraphics[width=0.45\textwidth]{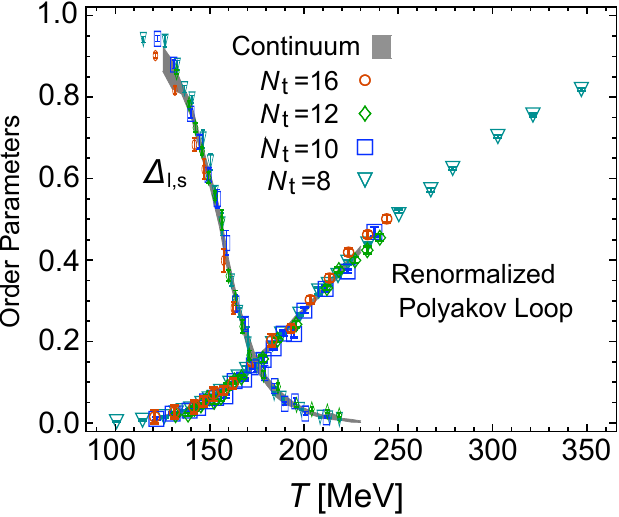}
        \includegraphics[width=0.45\textwidth]{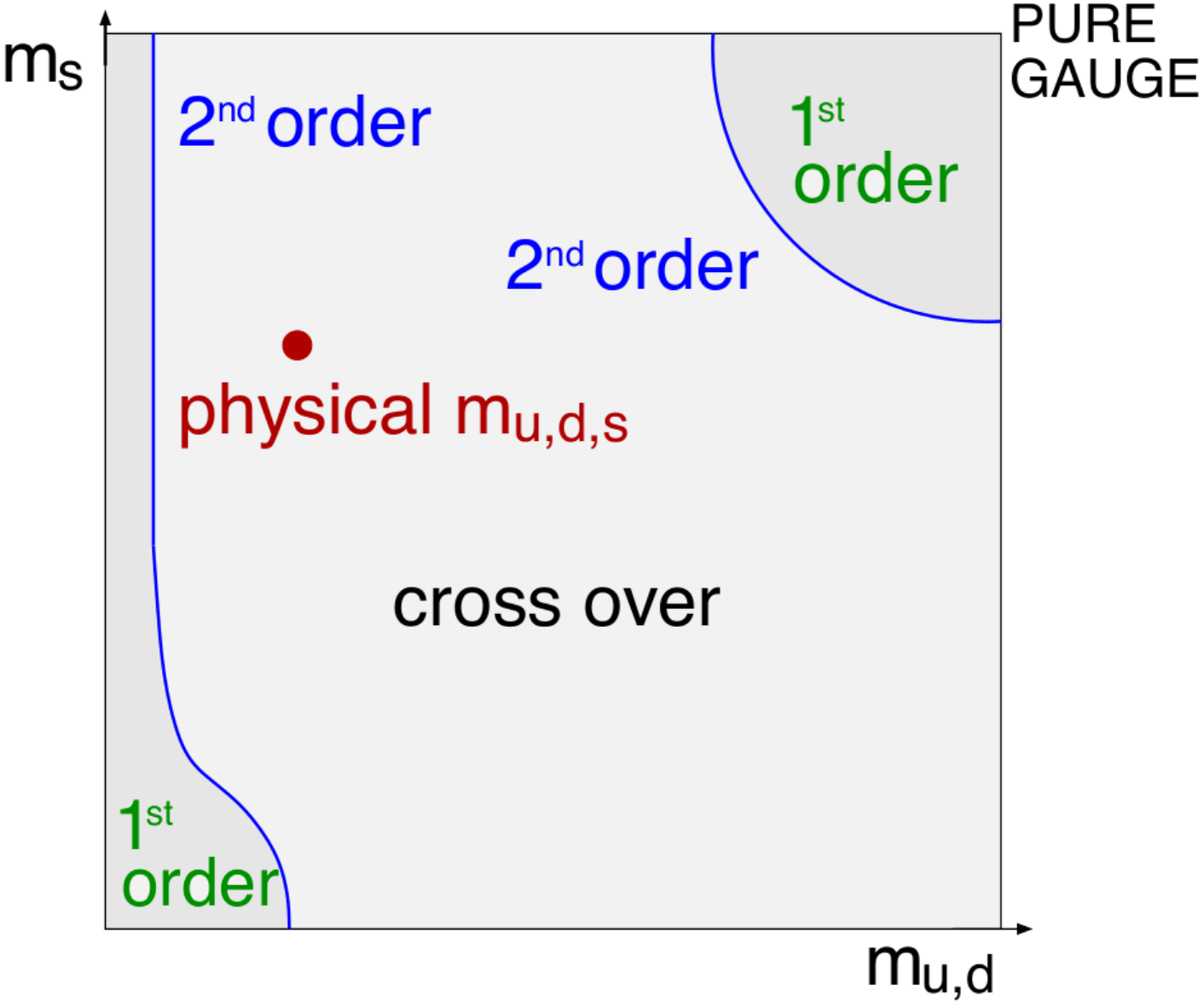}
    \end{center}
    \caption{(Left) Subtracted and normalised chiral condensate, along with renormalised Polyakov loop (see \cite{Fukushima:2013rx} for precise definitions), 
    in $2+1$ flavour LQCD simulation. Figure taken from \cite{Fukushima:2013rx}, which was in turn 
    adapted from \cite{Borsanyi:2010bp}. $N_t$ is the lattice site number along the temporal direction. (Right) ``Columbia plot'', taken from \cite{Fischer:2018sdj}, generated
    from $2+1$ flavour LQCD simulation.}
    \label{fig:order}
\end{figure}

It is possible to formulate QCD non-perturbatively by discretising Euclidean spacetime on a lattice, giving rise 
to what is known as \emph{Lattice QCD} (LQCD). Fig.~\ref{fig:order} shows two plots with data from LQCD calculations, which 
are intended to shed some light on the phase transitions that we have been discussing. On the left, we have the two order parameters, 
corresponding to the chiral condensate and the Polyakov loop. The main feature of this plot that we wish to draw attention to is 
the fact that the curves intersect at a point where the chiral condensate (Polyakov loop) is steeply decreasing (increasing), implying 
that both \emph{crossovers} (see below) occur at roughly the same temperature. This certainly does 
not provide a proof that the two crossovers occur at the same temperature in nature, since in reality, quarks 
cannot be considered infinitely heavy. Nevertheless, as we are not so concerned with the details of the crossover 
in what follows, we refer to both crossovers collectively in what follows as the QCD transition.

On the right of Fig.~\ref{fig:order}, we have what is known
as a ``Columbia plot'', with the name coming from the birthplace of the original paper \cite{PhysRevLett.65.2491}. These 
kinds of plots come from LQCD calculations where one can vary the number of flavours as well as the quark masses in 
order to investigate the role they play in observed phenomena. As is clearly visible, for physical values of the up, down 
and strange quarks, where the latter is much heavier than the former two (which are approximately equal)
\footnote{This is commonly known  as a $2+1$ flavour setup.}, the QCD transition
exhibits a crossover. As opposed to a first or second-order phase transition, this implies that there is no order parameter, 
which diverges or jumps at the critical temperature. Looking to Fig.~\ref{fig:qcdphase}, we note the depiction 
of this crossover near the $\mu_{B}=0$ axis. Recent results produced from the HotQCD LQCD collaboration \cite{HotQCD:2018pds} with two degenerate up and down 
dynamical quarks and a dynamical strange quark give a \emph{pseudo-critical} temperature of $T_{pc}=156.5\pm1.5\text{MeV}$. One can 
see by eye that this is where the dotted curve intersects the temperature axis in Fig.~\ref{fig:qcdphase}.

While LQCD is an extremely useful tool for understanding the strong coupling regime of QCD, its use is somewhat limited to 
the $\mu_{B}=0$ axis due to the infamous \emph{sign problem} \cite{Nagata:2021ugx}. The sign problem stems from the fact that, upon 
adding a chemical potential term to the discretised Euclidean path integral, one runs into the problem of complex probability measures, 
which are difficult to handle with numerical integration. At very small $\mu_{B}/T$, one can use various techniques 
\cite{Hasenfratz:1983ba,Alford:1998sd,Hands:1999md,Fodor:2001au,Gupta:2004pk} to track the QCD transition for non-zero values 
of $\mu_{B}$ but the accuracy of these methods deteriorates once $\mu_{B}/T$ becomes $\mathcal{O}(1)$.

Along with the extension of LQCD to non-zero $\mu_{B}$ mentioned above, one has to rely on holographic models (see
footnote~\ref{foot:holo}) or the 
Polyakov-extended Nambu-Jona-Lasino (PNJL) \cite{Fukushima:2003fw, Ratti:2005jh} model to study the regime of larger $\mu_{B}$, which suggest that the nature of phase transition could 
be first-order in this region. If such a first order transition exists, one would expect this region to be connected to the crossover region 
by a \emph{critical point}. A huge amount of experimental resources are currently being spent on trying to identify this critical point, in 
what are known as \emph{heavy-ion collisions}. We will come back to this topic in the next section.

\begin{figure}[t]
    \begin{center}
        \includegraphics[width=\textwidth]{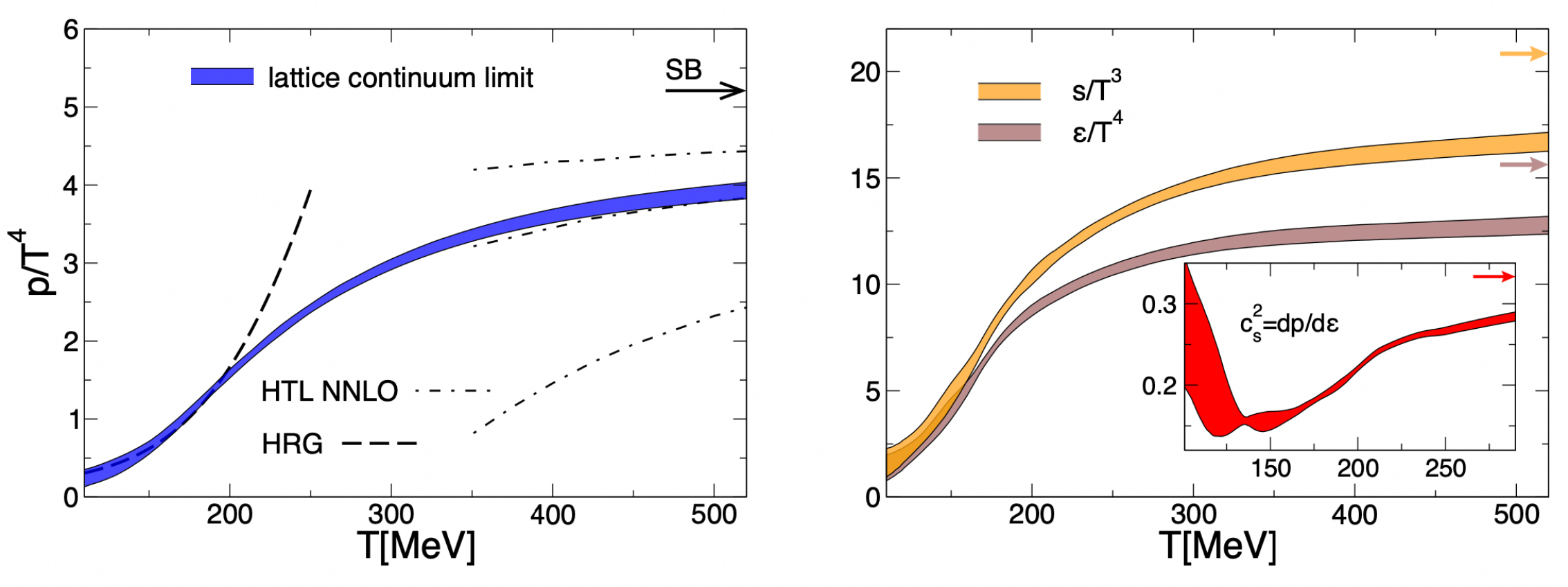}
    \end{center}
    \caption{Continuum-extrapolated equation of state from $2+1$ flavour 
    lattice simulations \cite{Borsanyi:2013bia}. The left panel 
    shows the pressure whereas the right shows the energy density, both suitably normalised. The 
    insert on the right panel shows the speed of sounds. The arrows indicate the 
    Stefan-Boltzmann limit for a non-interacting gas of massless quanta. HRG stands for 
    hadron resonance gas and HTL for hard thermal loop. There is a large 
    uncertainty associated with the HTL calculation, derived from the choice 
    of renormalisation scale.}
    \label{fig:eos}
\end{figure}

For large values of $T$ or $\mu_{B}$ (in the deconfined regime), one can take an analytic perturbative approach (to be introduced in Ch.~\ref{chap:tft_eft}). 
For large $T$ and small $\mu_{B}$ calculations from such an approach turn out to be consistent with LQCD results. In 
particular, looking to Fig.~\ref{fig:eos}-left one can see a good agreement between the LQCD equation of state at high
temperature and the perturbative (HTL) result. The large uncertainty there is associated with the choice 
of renormalisation scale, $\mu$, which agrees well with the lattice result for $\mu=2\pi T$. We 
furthermore note that the lattice result agrees well with the prediction from the hadron resonance gas model 
for lower temperatures, before it begins to diverge at $T\sim 150\text{MeV}$.

In contrast, for large $\mu_{B}$ and small $T$, a colour super-conducting phase is implied \cite{Alford:1997zt,Berges:1998rc,Son:1998uk,Alford:2007xm}. There is 
an intense ongoing effort to understand whether this deconfined phase of quark matter is found in neutron star cores \cite{Annala:2019puf}. Such
an effort has been boosted on the experimental side by the detection of gravitational waves from neutron 
star mergers by the LIGO and VIRGO collaborations \cite{PhysRevLett.119.161101,PhysRevLett.121.161101}.

\section{Heavy-Ion Collisions and the Quark-Gluon Plasma}\label{sec:hic_qgp}
In the previous section, we tried to give a relatively brief overview of the QCD phase diagram, picking out and 
discussing some of its defining features. If we keep increasing the temperature indefinitely,
 one expects from asymptotic freedom to find a weakly coupled plasma, where the fundamental degrees of freedom 
are quarks and gluons --  the quark-gluon plasma (QGP). The term QGP was introduced in the early 1980s by Shuryak \cite{Shuryak:1980tp}, by which point it 
had been realised that this state of matter should have existed in the very early universe, a few microseconds after the Big Bang. 
This realisation set off a coordinated effort to try and recreate the QGP in a laboratory setting, which 
in turn, prompted the colliding of heavy ions at several colliders\footnote{Namely, the Bevatron at LBNL, the Alternating 
Gradient Synchrotron at BNL and the Super Proton Synchrotron at CERN.}, thereby giving birth to the notion of creating and probing the QGP
in heavy-ion collisions (HIC). These days, the most energetic HIC take place at the LHC, where mainly Pb-Pb collisions are performed with centre of mass energy per 
nucleon, $\sqrt{s_{\text{NN}}}$ on the order of a few TeV. 
At RHIC, Au-Au and U-U collisions are instead performed going up to $\sqrt{s_{\text{NN}}}\approx 200\text{GeV}$. Extensions 
of this program to even lower energies are planned at the Japan Proton Accelerator Research Complex (J-PARC), the Facility for Antiproton and Ion Research (FAIR) in Darmstadt, Germany as 
well as the Nuclotron-based Ion Collider fAcility (NICA) in Dubna, Russia\footnote{The Beam Energy Scan (BES)
at RHIC also collided heavy-ions at lower energies ($\sim\mathcal{O}(\text{GeV}$)) to 
enlarge its search for the critical point.}. The motivation for running less energetic HIC experiments 
is that it allows one to scan the QCD phase diagram across larger values of $\mu_{B}$, thus providing a systematic 
strategy for the search of the QCD critical point\footnote{Indeed, 
it has recently been observed \cite{Gardim:2019xjs} that the mean transverse momentum 
of produced charged hadrons is proportional to the temperature of the plasma.}. A sketch of this is shown in Fig.~\ref{fig:qcdphase}.

\subsection{Stages of a Heavy-Ion Collision}\label{sub:stages}
Can we say with some level of confidence whether or not the QGP is created in HIC? The answer is yes, but 
we refrain from diving further into this topic until after we lay out the different stages of the HIC, following mainly \cite{Iancu:2012xa,Busza:2018rrf,Gelis:2021zmx,Elfner:2022iae}.

\begin{figure}[t]
    \begin{center}
        \includegraphics[width=0.8\textwidth]{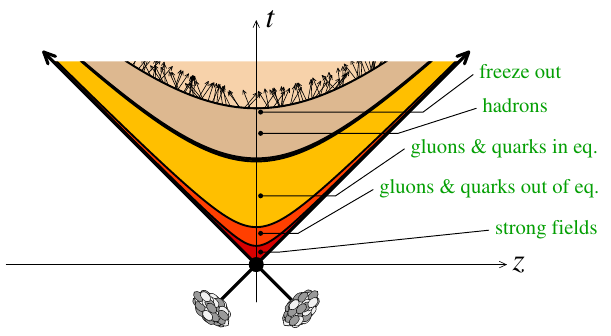}
    \end{center}
    \caption{The main stages associated with a heavy-ion collision, as described in the main text. Figure taken from \cite{Gelis:2021zmx}.}
    \label{fig:stages}
\end{figure}

In HIC, the QGP is created and expands, primarily along the beam axis, which 
we identify as the $x^1$ direction. An idea originally proposed by Bjorken \cite{Bjorken:1982qr}, 
says that such collisions, consisting of greatly Lorentz-contracted nuclei, are approximately independent of the spatial rapidity variable, 
\begin{equation}
    y\equiv\frac{1}{2}\ln\frac{x^0 +x^1}{x^0-x^1}.
\end{equation}
 In this way, one is permitted to order the different stages of the collision 
 in terms of\\ $\tau=\sqrt{({x^0})^2-({x^1})^2}$, the proper time of an observer comoving with a fluid element of the QGP. These hyperbolas, which 
 can be identified as hypersurfaces of constant energy density are marked clearly in Fig.~\ref{fig:stages}
 \begin{enumerate}
    \item $\tau=0\,\text{fm}/\text{c}$: The incoming Lorentz-contracted nuclei (in the laboratory frame), give rise to the typical 
        picture of two colliding pancakes\footnote{Such a picture is expected to be 
        less accurate at FAIR and NICA, where the centre of mass energies are much lower. In particular, see Fig.~\ref{fig:qcdphase}.}. These pancakes are composed mostly of gluons, carrying only tiny 
        fractions $x\ll 1$ of the longitudinal momenta of their parent nucleons, albeit with a density that increases with $1/x$. By the 
        uncertainty principle, such a high-density system can carry large amounts of transverse momentum.
        It turns out that this high-density gluonic form of matter dominates the hadron wavefunction at mid-rapidity and can be described using 
        the Colour Glass Condensate (CGC) effective theory \cite{McLerran:1993ni,McLerran:1993ka}. See below Fig.~\ref{fig:CGC} 
        for a short elaboration.

        As the actual collision takes place, \emph{hard} processes occur, i.e those where the outgoing 
        particles possess large amount of transverse momenta, $Q$. Taking place over a timescale of $1/Q$, particles produced
        in hard collisions: jets (of light quarks), heavy quarks, vector bosons, photons and dileptons, bear witness to the 
        entire HIC event and are often used to characterise the topology of the final state. We will come back to discuss some 
        of these \emph{hard probes} shortly.
        
        \begin{figure}[t]
            \begin{center}
                \includegraphics[width=0.6\textwidth]{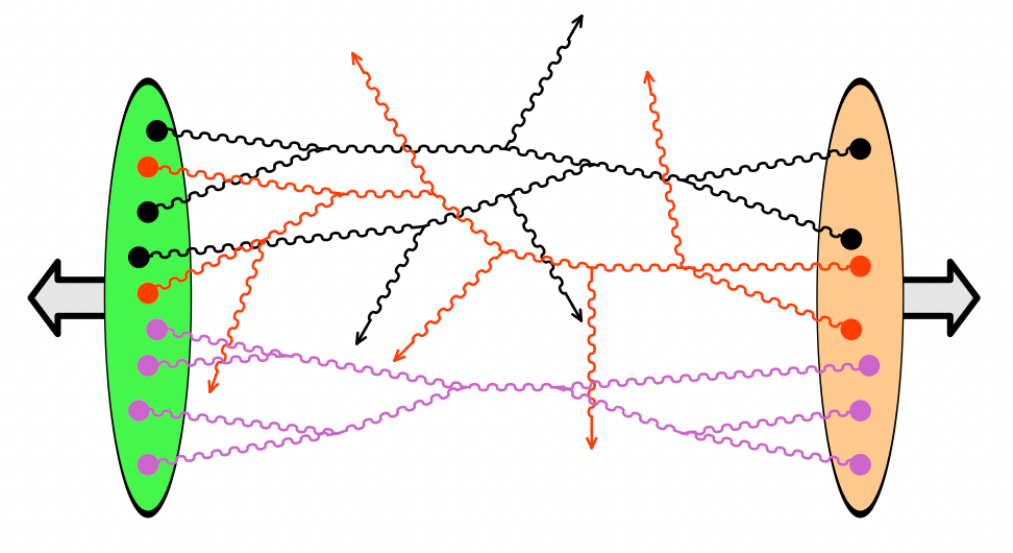}
            \end{center}
            \caption{Out-of-equilibrium gluons liberated by the nucleus-nucleus collision. Figure taken from \cite{Gelis:2010nm}. 
            In the Colour Glass Condensate (CGC) model, 
            one assumes that the energy deposition is dominated by the liberation of gluons from the colliding nuclei and that 
            the relevant gluon content can be described by semiclassical gauge fields. In such a description, the relevant information 
            about an incoming nucleus is contained in the colour current it carries, which acts as a source for the colour field.}
            \label{fig:CGC}
        \end{figure}

    \item $\tau\sim 0.2\,\text{fm}/\text{c}$: The bulk of the partonic constituents of the colliding nuclei (the gluons making up the CGC) is 
    liberated. Most of the hadronic content observed in the detectors stems from this liberation. Before hitting the detectors, 
    these particles undergo a complex evolution, forming an out-of-equilibrium, dense medium, known as the \emph{glasma}.

    \item $\tau\sim 1\,\text{fm}/\text{c}$: If the produced partons did not interact with each other, or if they interacted weakly, 
        they would independently evolve, racing towards the detectors. However, there is evidence that they in fact do interact strongly, exhibiting
        collective phenomena, which we come back to in Sec.~\ref{sec:collectivity}. In particular, by $\tau\sim 1\,\text{fm}/\text{c}$ the partons are 
        thought to have thermalised, giving rise to a (locally) equilibrated state of matter, the quark-gluon plasma. A microscopic description of this 
        thermalisation is provided by \emph{bottom-up thermalisation} \cite{Baier:2000sb,Kurkela:2018vqr}, where the harder gluons contained in the colliding nuclei radiate 
        soft gluons. These soft gluons form a thermal bath that drains away energy from the harder gluons through elastic scattering (see Fig.~\ref{fig:el_and_inel_coll}-left), 
        and radiation (see Fig.~\ref{fig:el_and_inel_coll}-right) thereby thermalising them.

    \item $\tau\sim 10\,\text{fm}/\text{c}$: Upon its formation the QGP, continues to expand and cool. By this time,
        the local temperature decreases to that where the transition to a confined state lies ($T_c\sim 150\,\text{MeV}$) and 
        the medium's constituents begin to hadronise. This hadronic system is still relatively dense, so it preserves thermal 
        equilibrium while expanding, inhabiting the hadron gas form of matter mentioned previously. At some point, the hadrons continue 
        to interact with each other but only elastically, which is known as the \emph{chemical freeze-out}.

    \item $\tau\sim 20\,\text{fm}/\text{c}$: By this time, the density becomes so low that the hadrons no longer interact 
        with each other at all, indicating the \emph{kinetic freeze-out}. They then proceed to free-stream all the way to the detector.
\end{enumerate}
In this thesis, we are primarily interested in the intermediate stage, where the QGP has equilibrated. Still, it 
is remarkable that all of the complexity, arising throughout the various stages above stems from the QCD Lagrangian, Eq.~\eqref{eq:qcd_lag}. 
With this picture in mind, we proceed to explore in more detail some of the various probes of the QGP, and how they can 
inform us further about its properties.

\subsection{Jet Quenching}\label{sec:jqq}
Note that up to this point, we have essentially assumed the presence of the QGP 
without providing any direct evidence for its existence. 
It was first proposed by Bjorken in the early 1980's \cite{Bjorken:1982tu} that the presence of a QGP in high multiplicity hadron-hadron collisions would result in the 
 suppression of high-$p_T$ jets. This sparked a great effort to try and
 observe this distorting or \emph{quenching} of jets in HIC. The realisation of 
 this goal arrived at RHIC 
 in the early 2000's \cite{PHENIX:2001hpc, STAR:2002svs}, where a suppression of the high $p_{T}$ spectra in central collisions 
 was observed. A few years later, evidence towards the existence of jet quenching was further solidified at the
 LHC \cite{ATLAS:2010isq,CMS:2011iwn} with fully reconstructed jets.

 Based on our discussion in Sec.~\ref{sec:jets}, it should not come as a surprise 
 that jets serve as an ideal hard probe of the QGP. As objects created at the beginning of the collision in
 hard processes, they persist through all phases of the HIC and can be used as a differential 
 tool to study the QGP across a multitude of scales.

 \begin{figure}[ht]
	\centering
	\includegraphics[width=\textwidth]{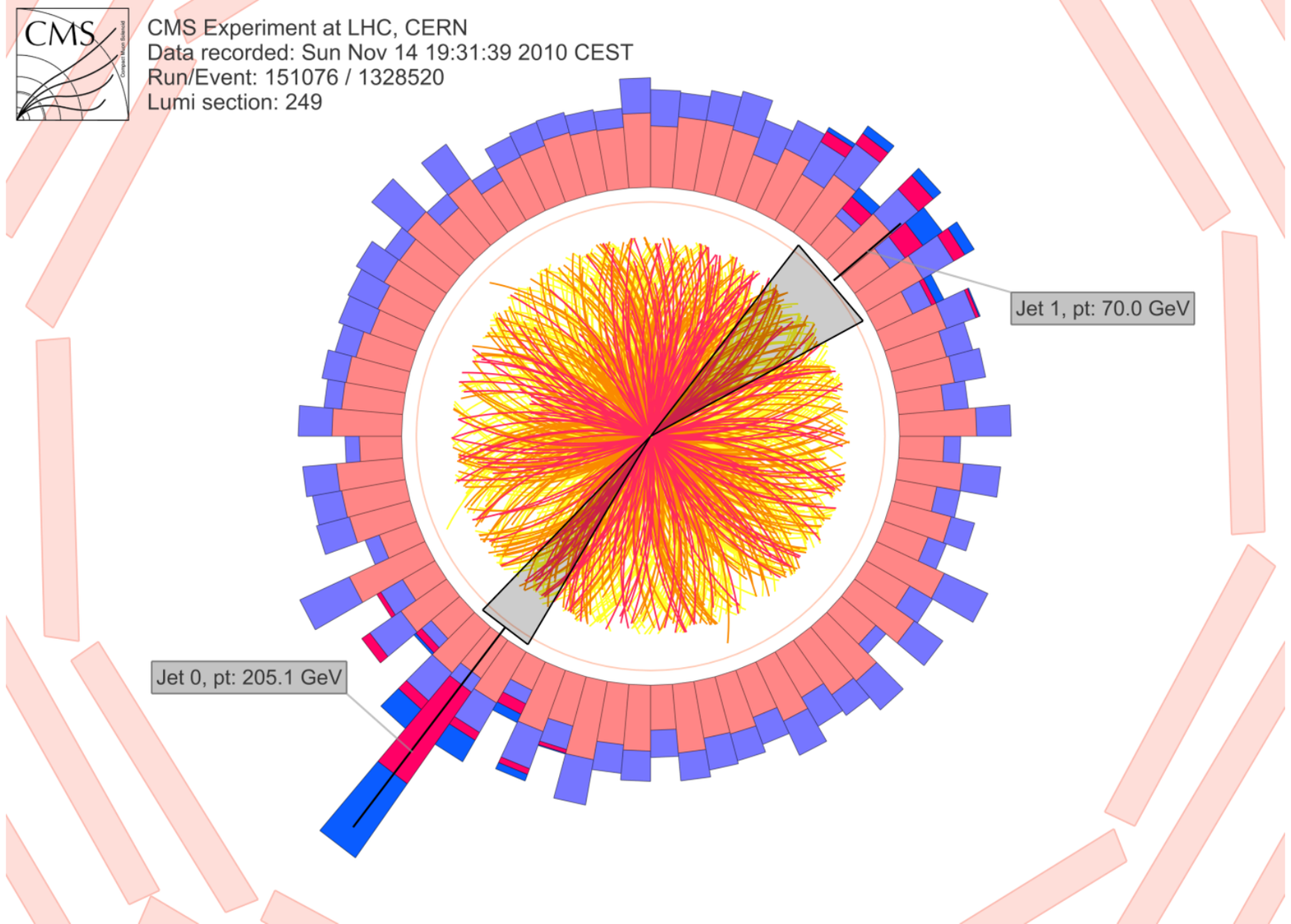}
	\caption{(Left) Event display showing dijet asymmetry in a $\text{Pb}+\text{Pb}$ collision at $\sqrt{s_{\text{NN}}}=2.76\,\,\text{TeV}$(CMS).}
    \label{fig:dijetsupp}
\end{figure}

In the absence of a QGP, dijets, produced back-to-back are expected to propagate
all the way to the detectors, registering (approximately) equal amounts of energy in the opposing calorimeters. 
Yet in HIC, it is possible that one of the dijets will pass through more of the QGP than the other and become quenched in comparison. This
expectation seems to be realised in nature, based on what is shown in Fig.~\ref{fig:dijetsupp}-left. In this context, 
the quenching can be quantified by the dijet asymmetry, $A_J=(p_{T\,1}-p_{T\,2})/(p_{T\,1}+p_{T\,2})$, where $p_{T\,1}$ ($p_{T\,2}$) is the 
transverse momentum of the reconstructed leading (sub-leading) jet.

In an ideal scenario, one would like to be able to compare what happens to a jet when it passes through the QGP to what happens when there
is no QGP. This is what is roughly expected p-p collisions\footnote{There has been 
recent evidence of medium formation in p-p collisions as well. We come back to discuss this briefly in Sec.~\ref{sec:collectivity}.}, which, as we mentioned in Sec.~\ref{sec:jets} are relatively 
well-understood. Indeed, to try and quantify the quenching in this way, one can use the \emph{nuclear modification factor} \cite{Wang:1998bha}, 
related to the ratio of the jet yield in HIC to the cross-section in $p-p$ collisions
\begin{equation}
    R_{\text{AA}}=\frac{\frac{1}{N_{\text{evt}}}\frac{d^2 N_{\text{jet}}}{dp_{T}dy}}{\langle T_{\text{AA}}\rangle\frac{d^2\sigma_{\text{jet}}}{dp_{T}dy}}
\end{equation}
where $N_{\text{evt}}$ is the number of events, $T_{\text{AA}}$ the nuclear thickness function, which can be calculated using the Glauber model \cite{Shukla:2001mb}\footnote{
    Note that the Glauber model can also be implemented to model the initial stage of the collision \cite{Stock:2020blh} instead of using the CGC effective theory.}. By taking 
    such a ratio, one hopes that hadronisation effects in both kinds of collisions ``cancel out''. In Fig.~\ref{fig:plots}-left, $R_{\text{AA}}$
is plotted for several centrality classes\footnote{Centrality corresponds to the size of the impact parameter vector, which stretches
between colliding participants. Small (large) centrality classes then correspond to selecting 
events with a small (large) 
impact parameter. In practice, experimentalists cannot measure the impact parameter directly and thus to resort 
defining the centrality through the total particle multiplicity or the total energy deposited in the detectors. Another 
option is to define the centrality given the amount of energy deposited in the zero-degree calorimeter, 
located close to the beam direction.} in Pb-Pb collisions. $R_{\text{AA}}$ is less than unity for the entirety 
of the kinematic range and is even smaller in more central collisions, which is again consistent 
with Bjorken's ``jet extinction'' hypothesis.

Indeed, $R_{\text{AA}}$ is not restricted to quantifying the relative suppression of jets -- it can be defined analogously for
 other hard probes, a few of which will be mentioned in the next section. One issue with $R_{\text{AA}}$ is that depends 
 somewhat on the underlying $p_T$ spectrum of the objects of interest; for fixed energy loss, a spectrum with 
 steeper $p_T$ dependence will have a smaller $R_{\text{AA}}$ value. The recent reviews \cite{Cunqueiro:2021wls,Apolinario:2022vzg} 
 should be consulted for more information on this issue as well as the more general 
 field of jet quenching phenomenology.

 The message that we have tried to impart here is simply 
 that experimental measurements have shown that jet quenching is realised at the LHC and RHIC. In Ch.~\ref{ch:eloss}, 
 we will dive further into the theoretical side of jet energy loss, showing how one can 
 quantify the quenching of a jet with some (relatively) simple analytical calculations.

\subsection{Other Hard Probes}\label{sec:other}

In Sec.~\ref{sec:qcdpd}, we mentioned that because of thermal screening, heavy quarks should 
not be able to form bound states, known as \emph{quarkonia}, in the deconfined phase. Historically, this 
lead to the idea by Matsui and Satz \cite{Matsui:1986dk} that quarkonia 
can be used as a ``thermometer'' for the medium: $\bar{Q}Q$ bound states formed 
in the initial stages of the HIC will begin to dissociate once the Debye screening length (to be discussed 
in Ch.~\ref{chap:tft_eft}) becomes shorter than the size associated with the bound state. Afterwards,
 the quark and antiquark should evolve independently in the medium until the plasma 
 cools down to $\Lambda_{\text{QCD}}$, at which they bind with light quarks 
 or antiquarks to form heavy-light mesons, leading to an overall 
 suppression of quarkonia. There is an additional complication, namely that, charm quarks which 
 are produceed in abundance in HIC may have a high enough density to then 
 recombine at a later stage of the collision. See the recent reviews \cite{Tang:2020ame,Apolinario:2022vzg} for more information.

 \begin{figure}[ht]
	\centering
    \includegraphics[width=0.5\textwidth]{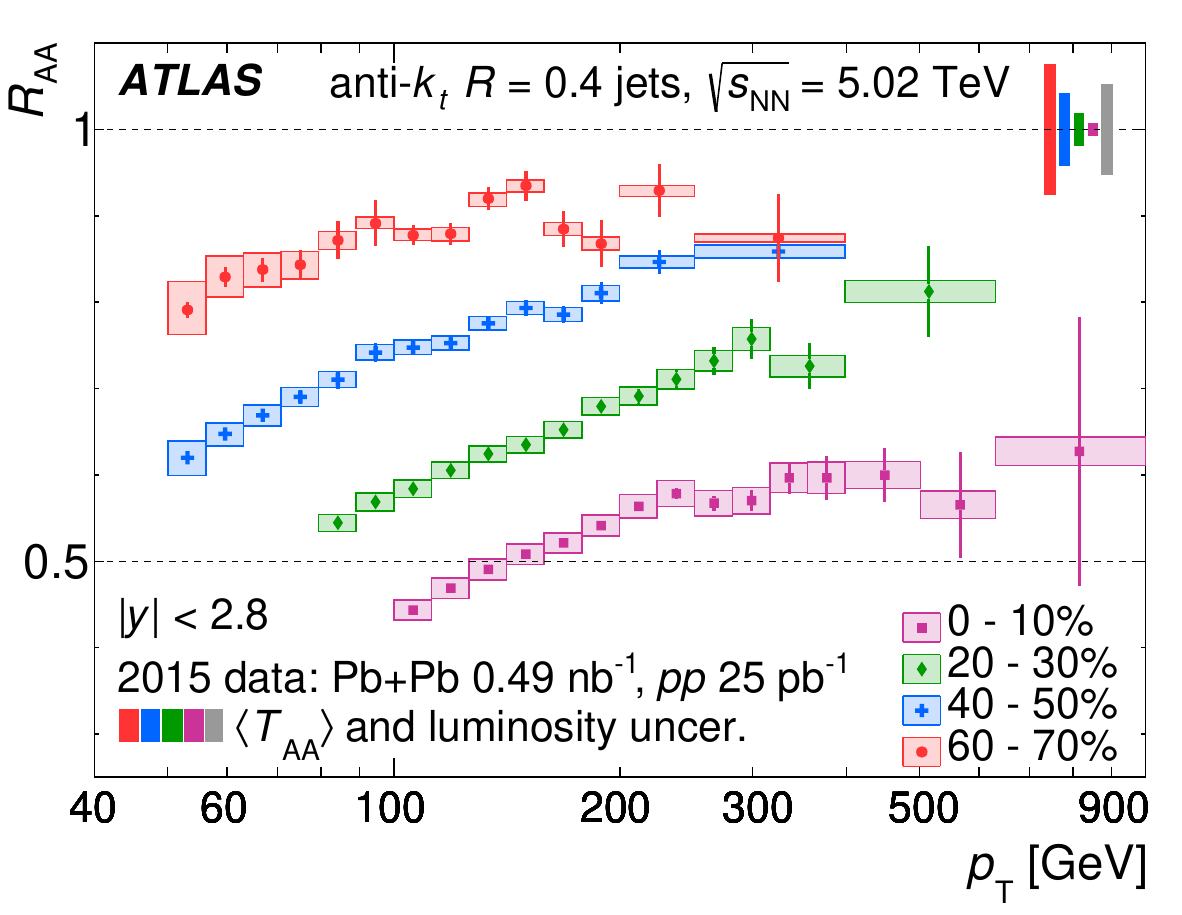}
    \includegraphics[width=0.45\textwidth]{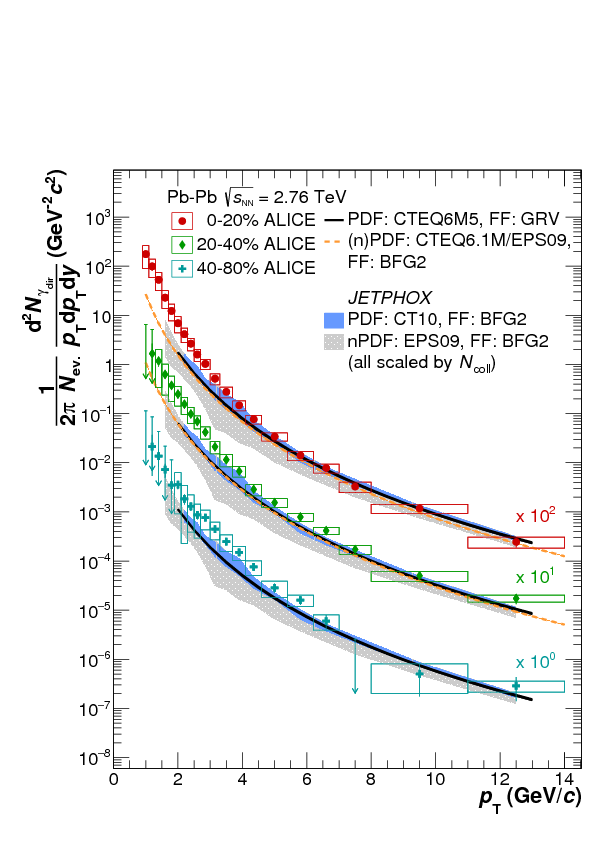}
	\caption{(Left) Nuclear modification factor plotted for a range of different centrality classes, taken from \cite{ATLAS:2018gwx}.
    (Right) pQCD prediction of direct photon production compared with experimental results from the ALICE collaboration \cite{ALICE:2015xmh}.}
    \label{fig:plots}
\end{figure}

Electromagnetic radiation can also provide a window into the inner workings of the QGP, but 
in a somewhat different way to that of jets and heavy quarks. Since the mean free path 
of photons produced in the HIC is considerably larger than the size of the QGP, they never 
equilibrate with the medium. Coupled with the fact that neither photons nor the dilepton 
pairs they decay into hadronise, it is reasonable to assert that 
perturbative calculations may do a decent job of predicting photon and dilepton production. 
On the experimental side, an issue is that one has to work very hard 
to understand the origins of the electromagnetic radiation: there are \emph{direct}
photons, produced as a result of partonic interactions in addition 
to \emph{decay} photons, which are produced from the decay of light hadrons.
At the same time, the various forms of production allow one to use 
the observed spectrum to report on local conditions at the radiation's creation point \cite{Gale:2009gc}. Indeed, Fig.~\ref{fig:plots}-right 
shows very good agreement of an NLO pQCD calculation of the direct photon spectrum with experimental results from the ALICE collaboration. 
Moreover, the exponential shape of the spectrum suggests that these photons were emitted from a \emph{thermalised medium}.

\begin{figure}[ht]
	\centering
    \includegraphics[width=\textwidth]{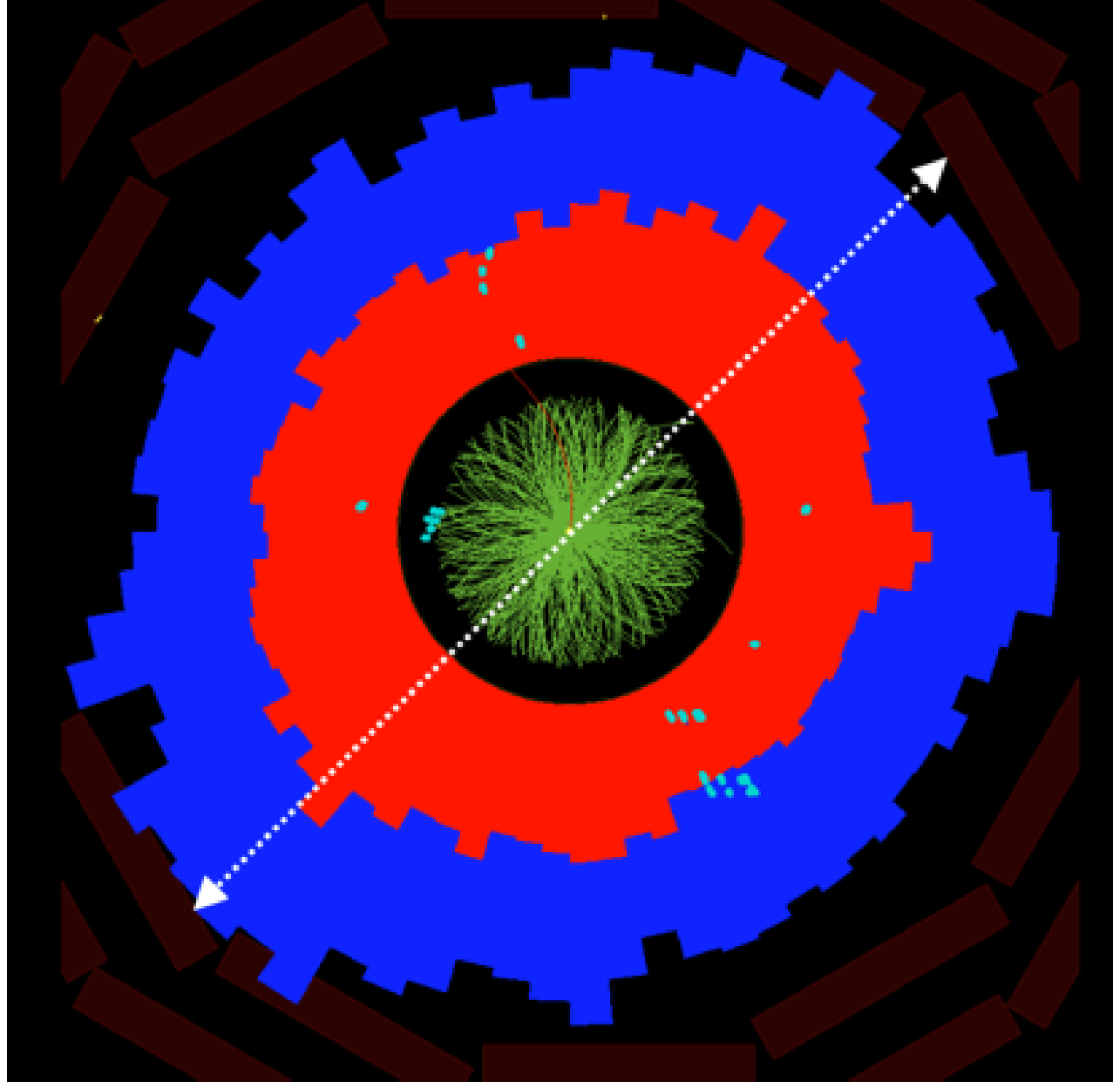}
	\caption{CMS event dispay, taken from \cite{Wiedemann:2021bwz}. The green 
    lines are charged tracks and energy in the electromagnetic and hadronic 
    are denoted by red and blue respectively. The dotted white line traversing the display highlights 
    the final state anisotropies in the azimuthal direction.}
    \label{fig:anisotropy}
\end{figure}

\subsection{Collective Flow}\label{sec:collectivity}
Up to this point, we have concentrated solely on hard probes of the QGP. That is not 
to say that the soft content produced in HIC cannot tell us anything about the medium. From Fig.~\ref{fig:anisotropy}, which shows 
the event display in the azimuthal plane, we note that the soft particles
are not deposited uniformly throughout the detector: there is increased 
production along the dotted line. This conclusion can be reached in a more quantitative manner 
by looking at the distribution of particle pairs, differential in the azimuthal angle and rapidity between the two particles \cite{ATLAS:2012at}.
There, one observes a cosine modulation in the azimuthal direction, which 
 implies a \emph{momentum anisotropy} in the final state. This anisotropy is quantified
by the so-called \emph{flow coefficients}, $v_n$, defined as the Fourier coefficients of the 
azimuthal distribution of particles.

This phenomenon is particularly prominent in non-central collisions, where a relatively large 
impact parameter implies an almond-shaped overlap zone between colliding nuclei, generating 
increased pressure in the azimuthal direction, which gives rise 
to a \emph{spatial anisotopy} in the initial state. Modelling the QGP as a fluid through the use of \emph{viscous hydrodynamics}
has proven extremely successful in showing 
how an initial state spatial anisotropy can give 
rise to final state momentum anisotropies \cite{Ollitrault:1992bk,Dusling:2007gi,Song:2007ux,Alver:2010dn}. 
Moreover, in the hydrodynamical framework, where the medium's properties 
are characterised by \emph{transport coefficients}, one can 
reproduce values for the shear viscosity to entropy ratio, $\eta/s\sim0.1-0.3$ that are 
consistent with those measured at the LHC and RHIC \cite{phenix:2003qra,star:2004jwm,ALICE:2011ab,cms:2013jlh,ALICE:2016kpq}. These values 
also happen to be reasonably close to $\eta/s=1/4\pi$, calculated using the strong-coupling 
technique, holography \cite{Policastro:2001yc,Kovtun:2003wp,Kovtun:2004de}\footnote{
    Specifically, the AdS-CFT correspondence \cite{Maldacena:1997re} is a particular realisation of the holographic principle, where a
    strongly coupled gauge theory is dual to a weakly coupled gravitational theory in one higher dimension. The duality 
    is made manifest through the \emph{field-operator map}, which allows one to compute correlation 
    functions in the strongly coupled theory.}.
Strikingly, a small value for $\eta/s$ would imply that the QGP is so \emph{strongly coupled} 
that very little (net) momentum can be transferred to nearby fluid elements. Moreover, such a 
value would imply that fluid cannot be described in terms of quasiparticles with 
mean free path: to do so, one would have to require mean free paths smaller than $1/T$. 
Perhaps unsurprisingly, calculations using finite-temperature perturbation
theory \cite{Arnold:2000dr,Arnold:2003zc,Ghiglieri:2018dib} are not able to reproduce such 
a small value of $\eta/s$.

As an aside, we note that collective effects have recently been observed in ``small systems'', namely p-p and p-Pb collisions \cite{CMS:2010ifv}, implying
that a QGP may be formed in that case as well. So far however, there has been no signature of jet quenching in these 
collisions, implying that the pursuit of quantifying jet quenching by comparing to p-p collisions 
may not be such a hopeless one after all. See \cite{Nagle:2018nvi} for a recent review.

Getting back to the strong coupling discussion, it is fair to say that the 
success of viscous hydrodynamics to describe the QGP's strongly coupled, collective
behaviour sparked the increased employment of strong coupling techniques. The rest 
of this thesis will nevertheless be based on the utilisation finite-temperature perturbation theory, applied to jet quenching. 
We argue that such an endeavour is not a futile one for the following reasons:
\begin{itemize}
    \item QCD perturbation theory, by definition maintains a strong connection with first principles and 
        can sometimes provide analytical results where LQCD cannot. In the holographic setup, while 
        analytical results can undoubtedly be obtained, the strongly coupled theory that one is studying 
        is not QCD but rather $\mathcal{N}=4$ super Yang-Mills theory, a supersymmetric
         conformal theory\footnote{\label{foot:holo}There do exist deformations from Maldacena's original
         correspondance \cite{Polchinski:2001tt,Erlich:2005qh,DaRold:2005mxj,Casero:2007ae,Gursoy:2007cb,Gursoy:2007er,Jarvinen:2011qe,Alho:2012mh}, 
         although they are still only intended to produce an effective theory that 
         resembles QCD in the IR.}. Therefore, holography 
        is considered more as a guiding light, as opposed to a tool with which one can make concrete, observable 
        predictions.
    \item Even if the QGP is strongly coupled, the jet itself is still weakly coupled. More precisely, 
        if one wants to calculate the probability of a quark jet radiating a gluon, the factor 
        of $\alpha_s$ associated with this probability should be much smaller than one, given the jet's extremely large energy. 
    \item In a strong coupling picture, interactions between the jet and QGP constituents will be characterised by a value 
        of $\alpha_s$ which is not smaller than one. However, as we will see in Sec.~\ref{sec:np_classical} and 
        Ch.~\ref{ch:asym_mass}, it is sometimes possible to evaluate thermal correlators on the lightcone using
        a certain resummation procedure, which thus provides a non-perturbative evaluation. 
\end{itemize}
With these points in mind, we conclude this chapter and proceed 
to the next one, where we explore some parts of the jet energy loss literature.

\newpage
\chapter{Jet Energy Loss}\label{ch:eloss}

Given the large 
background present in HIC (looking again to Fig.~\ref{fig:dijetsupp}), one immediately understands why it is extremely difficult 
to extract precise details of the QGP \cite{Connors:2017ptx} from experiment. This means that on the theory side, it is imperative 
to have a quantitatively precise understanding of the jet-medium interaction. One of the purposes of this chapter is
to review the progress made towards this pursuit. However, its primary goal is to provide 
the reader with a clear physical picture, which can be relied on throughout the rest of the 
thesis.

In Sec.~\ref{sec:kt} we discuss the different ways that the jet can interact with the medium through 
the lens of kinetic theory. In doing so, we understand that depending on the 
relevant region of phase space, bremsstrahlung can be triggered by \emph{multiple scatterings}
between the jet and medium constituents. To appropriate deal with such a region, one 
needs to take care of \emph{LPM interference} and we sketch how this is done in Secs.~\ref{sec:rrate} and ~\ref{sec:hoa}.
In Sec.~\ref{sec:scatt_kernel}, we diverge momentarily to discuss the transverse scattering kernel, an object, 
which controls how the jet diffuses in transverse momentum space before 
exploring another dominant region of phase space, the \emph{single-scattering regime} in Sec.~\ref{sec:single}. Secs.~\ref{sec:formalisms} and ~\ref{sec:recent}
are then respectively devoted to giving a bird's-eye view of the jet energy loss 
literature and a brief summary of some recent developments.

A word of caution is in order: in the previous chapter, 
it has been highlighted that jets are complicated objects and that Monte-Carlo 
methods are needed to simulate their full evolution. In what 
is to follow, however, the jet is interchangeably referred to as a single hard parton with initial four momentum $P=(E,0,0)$\footnote{Immediately after the hard collision, which originally seeds the jet, the parton will of course have a very 
large virtuality. Thus, initial in this context refers
to the state of the hard parton after it has radiated away most of its virtuality through a vacuum-like shower. Formalisms such as \cite{Wang:2001ifa,Majumder:2009zu,Caucal:2018dla,Caucal:2019uvr,Caucal:2020xad} 
include the interplay between vacuum-like and medium-induced emissions.}. By hard, we mean that $E\gg T$ with $T$ the temperature of the plasma. 
In this chapter, both the jet and its radiation (which is strictly speaking, also part of the jet) is 
assumed to be hard. We relax this latter assumption in Ch.~\ref{chap:qhat_chap}. See App.~\ref{sec:conventions} 
for conventions and notation.

\section{Jet Quenching in Kinetic Theory}\label{sec:kt}
\begin{figure}[ht]
	\centering
	\includegraphics[width=1\textwidth]{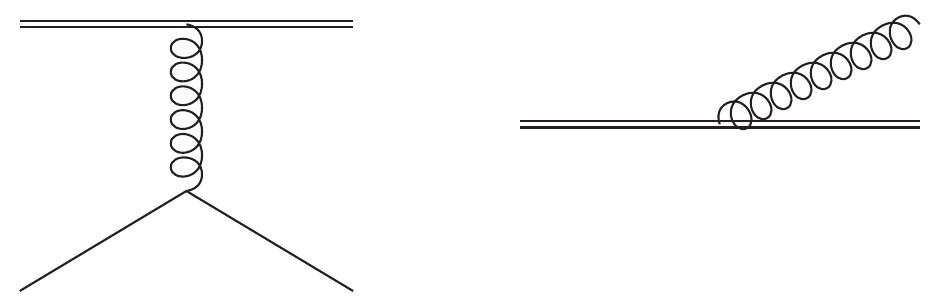}
	\caption{On the left, we have $2\rightarrow 2$ scattering, 
    also known an elastic collision, with the bottom single line representing a medium quark or gluon. 
    On the right, we have $1\rightarrow 2$ scattering or an inelastic collision.
    In both cases, the double line can be either a hard quark or a hard gluon. The hard 
    partons are deliberately denoted by straight lines to emphasise their eikonal nature. As is true for all diagrams 
    in this chapter, time flows from left to right.}
    \label{fig:el_and_inel_coll}
\end{figure}
Before embarking on the derivation of an actual energy loss calculation, we will classify
the different kinds of collisions between the jet and medium and mention their relative impact on energy loss.

In the very same paper where Bjorken first proposed using jet quenching as a way to detect the presence of the 
QGP \cite{Bjorken:1982tu}, he 
provided an estimate of collisional energy loss of a massless parton in colour representation $R$  
\begin{equation}
    \frac{dE}{dz}= C_{R}\pi\alpha_s^2 T^2\left(1+\frac{N_{f}}{6}\right)\ln\frac{ET}{m_{D}^2}
\end{equation}
per unit length, $z$ traversing a plasma of temperature, $T$. Collisional or elastic energy loss 
is classified as that which comes directly from $2\to 2$
scatterings between the jet and QGP constituents. A typical diagram contributing to collisional energy
loss is depicted in Fig.~\ref{fig:el_and_inel_coll}-left. For the case of a hard, light parton, this 
form of energy loss is dominated by Coulomb scatterings.
 The argument of the logarithm is a ratio of the maximum possible momentum exchange between the jet with energy $E$
  and medium constituent with energy $\sim T$ and the minimum possible momentum exchange, dictated by the Debye screening mass, $m_{D}$ \cite{Thoma:1990fm,Mrowczynski:1991da}.
This result has since been built upon, taking into account finite size effects \cite{Djordjevic:2006tw}, considering heavy quarks \cite{Braaten:1991jj}
and the effect of Compton scattering \cite{Peigne:2007sd}. See \cite{Peigne:2008wu} for a review. We mention that for RHIC and LHC conditions, collisional energy loss on its own is expected to be negligible in comparison to radiative energy loss
\cite{Baier:2000mf,Zakharov:2007pj}, which we move on to discuss now.

In the early 2000's, Arnold, Moore and Yaffe (AMY) developed an effective kinetic theory \cite{Arnold:2000dr,Arnold:2001ba,Arnold:2001ms,Arnold:2002ja,Arnold:2002zm,Arnold:2003zc}
\footnote{Their formalism was originally developed to study photon and dilepton production but was soon after  
adapted to the case of jet energy loss \cite{Jeon:2003gi}.} to study jet energy loss, centred around the Boltzmann equation
\begin{equation}
    \bigg(\frac{\partial}{\partial t}+\vec{\mathbf{v}}\cdot\vec{\nabla}_{\vec{\mathbf{x}}}\bigg)f^{a}(\vec{\mathbf{p}},\vec{\mathbf{x}},t)=-C_{a}^{2\to 2}[f]-C_{a}^{1\to 2}[f],
    \label{eq:amy_boltzmann}
\end{equation}
where $f^{a}(\vec{\mathbf{p}},\vec{\mathbf{x}},t)$ is the classical phase space distribution for a single colour and helicity 
state quasiparticle. Roughly speaking, the Boltzmann equation describes how changes in the hard parton's momentum, $\vec{\mathbf{p}}$ occur 
by loss or gain through scattering. The different kinds of scattering processes are then included through the collision operators on 
the right hand side; elastic collisions are incorporated through the operator $C_{a}^{2\to 2}[f]$. The second term $C_{a}^{1\to 2}[f]$, 
instead reflects the fact that the jet can also shed energy by splitting, be it through bremsstrahlung or pair production. A 
bremsstrahlung diagram is depicted in Fig.~\ref{fig:el_and_inel_coll}-right.

 For on-shell particles in the vacuum, processes such as the one in Fig.~\ref{fig:el_and_inel_coll}-right are kinematically disallowed. 
 However, when passing through the QGP, the hard parton may interact with the medium, exchanging a soft gluon (i.e through an 
 elastic collision with momentum exchange $\sim gT$) and in doing so be pushed slightly
 off-shell. Upon picking up this small virtuality, the parton will then have the ability to radiate; it is for this reason that 
 the operator $C_{a}^{1\to 2}[f]$ captures what is known as radiative or medium-induced energy loss.
 \begin{figure}[ht]
	\centering
	\includegraphics[width=1\textwidth]{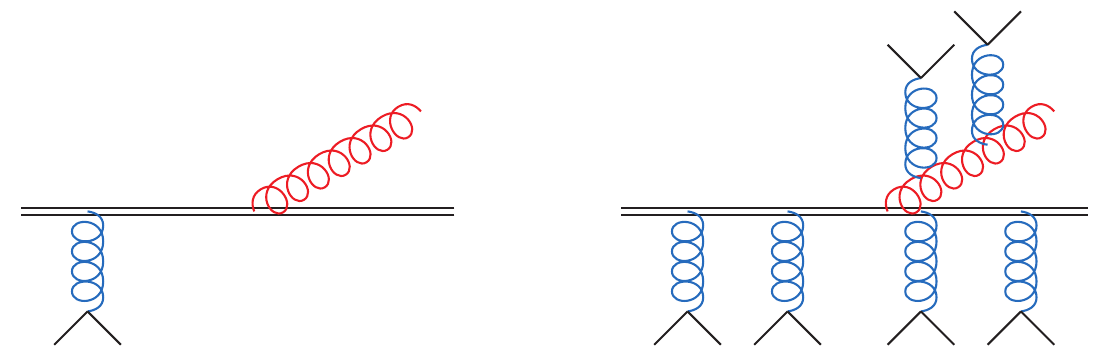}
	\caption{Single-scattering (left) and multiple scattering (right). Blue gluons are those that are exchanged in the jet-medium 
    interaction whereas red gluons are medium induced radiation. In the multiple scattering regime, bremsstrahlung is unable 
    to resolve individual elastic collisions, which is a manifestation of LPM interference.}
    \label{fig:rad_eloss}
\end{figure}

It is not obvious why diagrams such as the one in Fig.~\ref{fig:rad_eloss}-left
 contribute at the same order as the one in Fig.~\ref{fig:el_and_inel_coll}-left since the former possesses 
 an extra factor of $g$. Nevertheless, it turns out that the contribution from these kinds of diagrams is enhanced when,
 looking to Fig.~\ref{fig:rad_eloss}: the momentum exchanged with the medium, $K$ is spacelike and soft$\sim gT$\footnote{More
 precisely, $K\gtrsim gT$ with $k^+\sim k_{\perp}\gtrsim gT,\,k^-\lesssim g^2T$, coupled with the 
 condition that the outgoing hard partons are nearly on-shell and collinear to each other ($\theta\lesssim g$ with $\theta$ the angle
 between the outgoing hard partons). This process will be studied in detail in Sec.~\ref{sec:scatt_lo_soft}.}. We will not wade through the details here but one 
 can show \cite{Aurenche:1998nw,Arnold:2001ba} that if these conditions are met, the propagators from the internal lines
 in Fig.~\ref{fig:rad_eloss} compensate for the extra factor of $g$ in the numerator of the same diagram, implying that radiative energy loss in \emph{at least} as
 parametrically as important as collisional energy loss\footnote{In fact, for the case that we are considering here, where 
 the energy of the jet large compared to the temperature, the $C_{a}^{1\to 2}[f]$ term dominates.}. Be that as it may, it transpires that in order to 
 properly include these radiative scattering processes into the kinetic theory, one needs to overcome another major hurdle.

One of the assumptions tied to the validity of the Boltzmann equation Eq.~\eqref{eq:amy_boltzmann} is that the de Broglie wavelength 
of the hard partons must be small compared to their mean free path. This allows one to treat the partons as classical particles between 
scatterings, meaning that $\vec{\mathbf{x}}$ and $\vec{\mathbf{p}}$ can be treated as classical variables in the Boltzmann equation. 
For a jet particle with energy $E\gg T$, its de Broglie wavelength is of course $\lambda\sim 1/E$. To see if this assumption holds here, 
let us estimate the cross-section of an elastic collision such as that in Fig.~\ref{fig:el_and_inel_coll}-left, with the 
exchanged momentum soft, $K \sim gT$. After squaring the amplitude and integrating over the final state phase space, 
\begin{equation}
    \sigma_{\text{el}}\sim \int d(K^2)\bigg\vert\frac{g^2}{K^2}\bigg\vert^2\sim\frac{g^2}{T^2}.
\end{equation}
The mean free path between collisions is then
\begin{equation}
    l_{\text{el}}=\frac{1}{n\sigma_{\text{el}}}\sim\frac{1}{g^2T},
\end{equation}
where we have used that the density of scatterers is $n\sim T^3$.
Clearly, we have that $\lambda\ll l_{\text{el}}$ and we can conclude that this assumption is consistent with the framework of kinetic theory.
Another requirement is that the quantum mechanical duration of individual scattering 
events be much smaller than the mean free time between collisions. If this condition is not satisfied, there will 
be quantum interference between successive scatterings, implying that they cannot be treated independently.
The mean free time between elastic collisions is 
\begin{equation}
    t_{\text{el}}\sim\frac{1}{n\sigma_{\text{el}}v}\sim\frac{1}{g^2T}.
\end{equation}
The scatterers are at rest with respect to the plasma frame 
so that we can set the relative velocity, $v\sim 1$. The scattering duration associated with these elastic collisions is instead
\begin{equation}
    \tau_{\text{el}}\sim\frac{1}{K}\sim\frac{1}{gT},
\end{equation}
implying that the aforementioned condition is indeed satisfied. Hence, the energy loss coming from the elastic collisions can be safely 
implemented in Eq.~\eqref{eq:amy_boltzmann}. But what about radiative energy loss?

\begin{figure}[ht]
	\centering
	\includegraphics[width=1\textwidth]{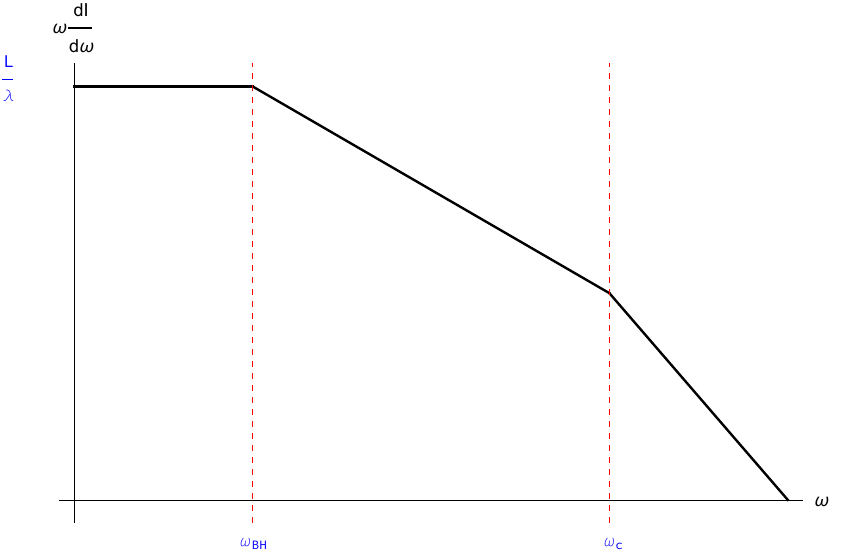}
	\caption{A log-log plot of the radiative energy loss spectrum across a range of different energies. As is described 
    in the text, the BDMPS-Z and AMY formalisms capture the middle region, for which $\omega_{BH}\leq\omega\leq\omega_c$. 
    The OE then correctly describes the BH region on the left, along with the single-scattering region on the right.}
    \label{fig:whole_spectrum}
\end{figure}

Let us now go back to Fig.~\ref{fig:rad_eloss}-left, specialising to the case of democratic $q\to qg$ splitting (so that we may
estimate the energy of the radiation as $\sim E$). It turns out 
that the internal quark line will be pushed off-shell in energy by an amount
\begin{equation}
        \delta E\sim \frac{Q_{\perp}^2}{E},
\end{equation}
where $Q_{\perp}$ is the size of momentum transfer (transverse to the hard parton's direction of motion) in the underlying scattering 
process\footnote{$Q_{\perp}$ is not necessarily 
identified with $k_{\perp}\sim gT$ from before, since in what follows it can denote the momentum transfer coming 
from either one collision or from multiple collisions.}. Fourier transforming, we can then identify the duration
associated with the bremsstrahlung process as the formation time,
\begin{equation}    
    \tau\sim\frac{1}{\delta E}\sim\frac{E}{Q_{\perp}^2}.
\end{equation}
$Q_{\perp}$ is bounded by below by $m_{D}\sim gT$ (see Sec.~\ref{sec:scatt_kernel} for a justification). Depending\footnote{Up until this point, in line 
with the original AMY formalism, we have essentially assumed the medium to extend infinitely in the longitudinal direction. 
However, the length of the medium can in practice dictate whether the radiative energy loss will be dominated 
by single or multiple scattering. See for example Fig.~\ref{fig:whole_spectrum} and the discussion preceding it.} on the values of $Q_{\perp}$ and $E$, 
one can have $\tau\ll t_{\text{el}}\sim 1/(g^2T)$, a region of phase space
known as the single-scattering regime. In this region, one can really consider a single collision with the medium to trigger 
bremsstrahlung as in Fig.~\ref{fig:rad_eloss}-left.

Be that as it may, as $E\to\infty$ we become sensitive to a region of phase space where the
formation time becomes comparable or larger
than the mean free time between elastic collisions. Consequently, successive elastic collisions can no longer be treated as quantum mechanically independent for the calculation of 
bremsstrahlung. In this region, $Q_{\perp}$ should be thought of as the size of the momentum transfer during the formation time, effectively coming 
from multiple collisions. This coherence effect was first investigated in the context of QED by Landau, Pomeranchuk and Migdal \cite{Landau:1953um,Landau:1953gr,Migdal:1956tc} and
is known as LPM interference.

 In practice, in order to correctly account for radiative energy loss in this region known as the multiple scattering 
 regime, one needs to sum an infinite number of interference terms, which are composed of diagrams such as Fig.~\ref{fig:rad_eloss}-right.

\section{Radiation Rate in the Multiple Scattering Regime}\label{sec:rrate}

\begin{figure}[ht]
	\centering
	\includegraphics[width=1\textwidth]{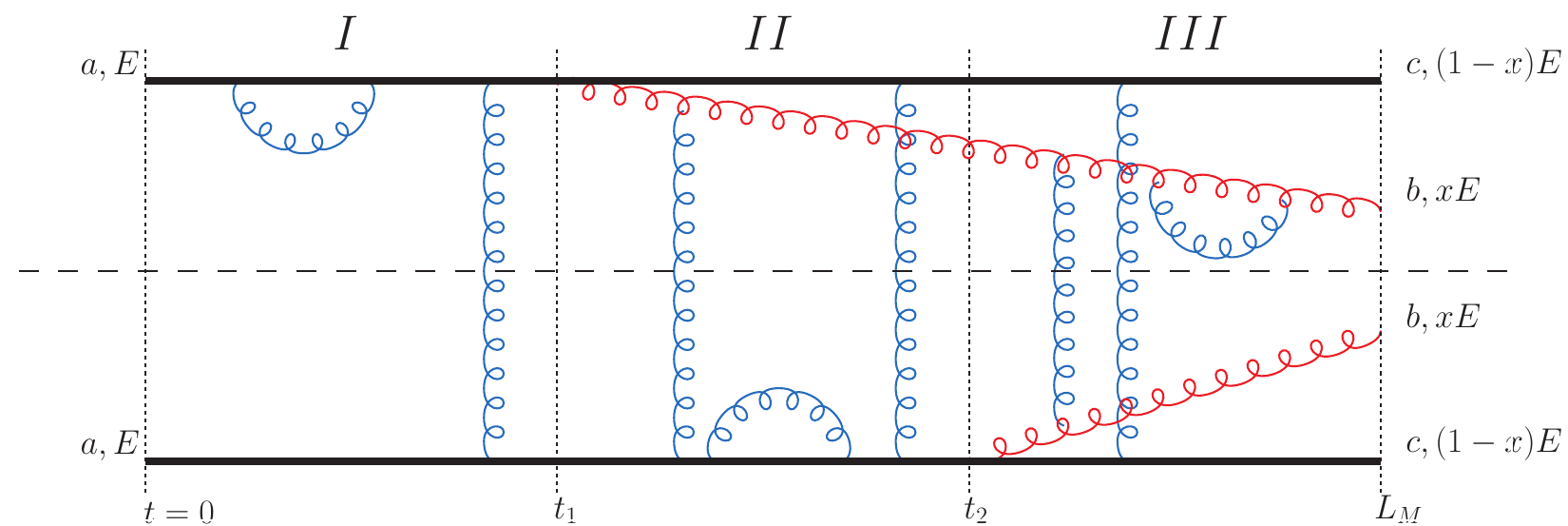}
	\caption{Interference terms contributing to the collinear splitting probability (see Eq~\eqref{eq:radprob_start})  of a 
    hard quark, $a$ created at $t=0$ with initial energy $E\gg T$ radiating a gluon, $b$ with energy $\omega=xE$, before exiting the medium at $t=L_{M}$.
    The horizontal dashed line represents the cut between the amplitude above and the conjugate amplitude below. The upper and lower
    red curly lines are the radiated gluons in the amplitude and conjugate amplitude respectively. Exchanges of 
    gluons with the medium are denoted by blue lines -- these include both forward scatterings (loops) and 
    true scatterings (vertical rungs), through which the initial or final state partons pick up transverse momentum kicks from the medium.
    It is assumed that all hard partons $a, b$ and $c$ propagate eikonally --  this allows for a clean partitioning 
    of the diagram into three regions, $I, II$ and $III$, separated by the hard, DGLAP vertices.
    \newline Figure reproduced from
    \cite{CaronHuot:2010bp,Blaizot:2015lma}.}
    \label{fig:gluon_emission_lpm}
\end{figure}

In this section, we will demonstrate how the expression for the collinear radiation rate is obtained. The collinear radiation rate is 
a central ingredient within AMY's kinetic theory formalism, originally produced through a strict diagrammatic derivation (within the 
context of photon production) in \cite{Arnold:2001ba}. For the sake of brevity however, we choose to present here a formulation which resembles 
closer the work of Zakharov, while attempting to make contact with the AMY formalism along the way. The reason for insisting to 
maintain a connection with the AMY formalism (see Sec.~\ref{sec:mult_sing}) is that it is derived using Thermal Field Theory, to be discussed in some 
detail during the next chapter.

Before starting, we give the assumptions along with their justifications, while also making a point to state 
the simplifications that arise due to these assumptions:
\begin{itemize}
    \item We require initial and final state partons propagate eikonally or in straight lines,
     with energies $\gg T$\footnote{We will make a point of relaxing this assumption at a later stage in the thesis.}. This 
     assumption allows us to neglect the presence of any statistical distribution functions (see Eqs.~\eqref{eq:nbedef}, ~\eqref{eq:nfedef}), which 
     are exponentially suppressed in this regime. Consequently, the partons
     probe the classical nature of the medium gauge field\footnote{The justification for this 
     statement is provided by the arguments given in Sec.~\ref{sec:chtrick}.}.
     \vspace{0.3cm}
     \item The opening angle of the bremsstrahlung is small so that the emission itself can be described by the leading order DGLAP expressions \cite{Altarelli:1977zs}. 
     Loop corrections to these vertices can then be neglected, provided that the factors of $g$ controlled by the transverse momentum picked up during 
     bremsstrahlung formation time are also small. See Section VI. C of \cite{Arnold:2008zu} for a discussion.
     \vspace{0.3cm}
    \item The hard partons receive transverse momentum kicks through interactions which 
    are instantaneous in comparison to the bremsstrahlung formation time. This is reasonable because the multiple scattering regime is dominated by soft collisions
    with a duration of order $\mathcal{O}(1/gT)$, whereas the mean free time between collisions is $\mathcal{O}(1/g^2T)$. 
    We are then permitted to interpret time as flowing monotonically from left to right in Fig.~\ref{fig:gluon_emission_lpm}, with the amplitude 
    squared being cleanly separated into $3$ regions, defined by the radiation in the amplitude (region $II$) and the 
    conjugate amplitude (region $III$). Since we are only interested in computing a radiation probability, which is 
    not differential in transverse momentum, the system can then just be propagated through region $II$, 
    since soft scatterings suffered by the partons in regions $I$ and $III$
    will not contribute to LPM interference and thus need not be resummed. If one wants to remain differential in transverse momentum, 
    a more sophisticated formalism \cite{Apolinario:2014csa,Isaksen:2023nlr} is needed to account for regions $I$ and $III$.
    \vspace{0.3cm}
    \item The medium is homogeneous in the transverse plane and static. These simplifications are of course difficult to justify making use of in a heavy-ion experiment. 
    Nevertheless, improvements on these approximations will not impact our goal here, which is to demonstrate the application of the 
    harmonic oscillator approximation.
\end{itemize}
Given these assumptions, we can then proceed, following \cite{Arnold:2008iy,CaronHuot:2010bp,Arnold:2015qya,Ghiglieri:2020dpq} to write down the differential 
splitting probability for a hard quark with initial energy $E$ to radiate a gluon with energy $\omega=xE$
 \begin{align}
    &\frac{dI}{d\omega}=\frac{2}{E}\text{Re}\bigg\{\frac{E}{2\pi V_{\perp}}\int_{t_{1}<t_{2}}dt_{1}dt_{2}\sum_{\text{pol.}}\int_{\mathbf{p}_{f},\mathbf{k}_{f}}
    \int_{\mathbf{p}_{1},\bar{\mathbf{p}}_{2},\mathbf{k}_{1}}\nonumber
    \\&\times\bigg\llangle\Big(\langle\mathbf{p}_{f}\mathbf{k}_{f},t_{2}|\mathbf{p}_{1}\mathbf{k}_{1},t_{1}\rangle
    \langle\mathbf{p}_{1}\mathbf{k}_{1}|-i\delta H|\mathbf{p}_{i}\rangle\Big)
    \Big(\langle\mathbf{p}_{i},t_{1}|\bar{\mathbf{p}}_{2},t_{2}\rangle
    \langle\bar{\mathbf{p}}_{2}|i\delta H|\mathbf{p}_{f}\mathbf{k}_{f}\rangle\Big)\bigg\rrangle\bigg\}.
    \label{eq:radprob_start}
\end{align}
Let us take a moment to unpack the notation: $\mathbf{p}$ is transverse momentum of the hard quark and $\mathbf{k}$ 
the transverse momentum of the emitted gluon that splits with momentum fraction $x$. $\mathbf{p}_{i}, \mathbf{p}_{f}$ denote the time 
momenta at time $t=1$ and $t=2$ respectively. The double-angled brackets
$\llangle...\rrangle$ define a thermal average over the medium, composed of classical gauge fields, 
through which the partons propagate\footnote{We ignore correlations just before $t_1$ and just after $t_2$ within
one correlation length of the plasma, which is consistent with modelling of the jet-medium interaction through an instantaneous potential. See footnote $21$ of \cite{Arnold:2015qya} for a further
 discussion.}. The first set of round brackets on the second line contains the amplitude associated with the
in-medium splitting process and the second set with the conjugate amplitude. Packed into these amplitudes are the collinear 
splitting matrix elements, i.e $\langle\mathbf{p}_{1}\mathbf{k}_{1},t_{1}|-i\delta H|\mathbf{p}_{i}\rangle$, 
where $\delta H$ is the part of the Hamiltonian containing the DGLAP vertices for the hard particles. 
In addition, there are the dressed propagators, such as $\langle\mathbf{p}_{f}\mathbf{k}_{f},t_{2}|\mathbf{p}_{1}\mathbf{k}_{1},t_{1}\rangle$, 
which resum multiple soft scatterings from the time of splitting in the amplitude, $t_{1}$ up until the time
of splitting in the conjugate amplitude $t_{2}$. The sum is over the polarisation 
of the final state partons and the factors of  $\frac{E}{2\pi}$ and $V_{\perp}$ are respectively artefacts of the normalisation of the initial state, 
$\langle\mathbf{p}_{i}|\mathbf{p}_{i}\rangle=V_{\perp}$ and having written $dk_{z}/2\pi=(E/2\pi)dx$. In
equilibrium, the integrand can only depend on the difference between times, $t\equiv t_{2}-t_{1}$. We will nevertheless hold off on 
making this dependence explicit until later in the section.

We now move on to show how this two-particle, four-dimensional problem can be reduced to that of a one-particle, two-dimensional
quantum mechanics problem. To start, let
\begin{align}
    \langle\mathbf{p}_{f}\mathbf{k}_{f},t_{2}|\mathbf{p}_{1}\mathbf{k}_{1},t_{1}\rangle&=
    \langle\mathbf{p}_{f}\mathbf{k}_{f}|e^{-iH(t_{2}-t_1)}|\mathbf{p}_{1}\mathbf{k}_{1}\rangle,
    \\\langle\mathbf{p}_{i},t_{1}|\bar{\mathbf{p}}_{2},t_{2}\rangle&=
    \langle\mathbf{p}_{i}|e^{iH_{*}(t_{2}-t_1)}|\bar{\mathbf{p}}_{2}\rangle.
\end{align}
$H$ propagates the quark and gluon through the medium in the amplitude, along with $H_{*}$, 
which propagates the quark through the medium in the conjugate amplitude. The medium average in Eq~\eqref{eq:radprob_start} 
need then only be restricted to the factors depending on this part of the Hamiltonian, i.e 
\begin{equation}
    \bigg\llangle e^{-iH(t_2 - t_1)}|\mathbf{p}_{1}\mathbf{k}_{1}\rangle\langle \mathbf{p}_{i}| e^{iH_{*}(t_{2}-t_{1})}\bigg\rrangle.
\end{equation}
Written like this, we can think of the object, $|\mathbf{p}_{1}\mathbf{k}_{1}\rangle\langle \mathbf{p}_{i}|$ 
 as living in the Hilbert space
\begin{equation}
    \bar{\mathbb{H}}_{\mathrm{q}}\otimes\mathbb{H}_{\mathrm{q,g}}=\bar{\mathbb{H}}_{\mathrm{q}}\otimes{\mathbb{H}}_{\mathrm{q}}\otimes\mathbb{H}_{\mathrm{g}},
\end{equation}
with $\mathbb{H}_{\mathrm{q}}$ the Hilbert space of states of hard quark and $\mathbb{H}_{\mathrm{q,g}}$ the Hilbert space of states
with a collinear quark-gluon pair. One then interprets this product rather as a Fock space of three particles: 
one quark, one gluon and one conjugate quark so that we may instead write
\begin{equation}
    \bigg\llangle e^{-i(H-H_*)(t_{2}-t_{1})}|\mathbf{p}_{a},\mathbf{p}_{b},\mathbf{p}_{c}\rangle\bigg\rrangle
    =e^{-i\mathcal{H}(t_{2}-t_{1})}|\mathbf{p}_{a},\mathbf{p}_{b},\mathbf{p}_{c}\rangle
    \label{eq:effprop}
\end{equation}
with
\begin{align}
    (\mathbf{p}_{a},\mathbf{p}_{b},\mathbf{p}_{c})&=(-\mathbf{p}_{i},\mathbf{k}_{1},\mathbf{p}_{1})
    \quad\quad\quad \mathrm{at}\quad t=t_1,
    \\\mathbf{p}_{a}+&\mathbf{p}_{b}+\mathbf{p}_{c}=0.
\end{align}
The effective light-cone Hamiltonian is given as 
\begin{align}
    \mathcal{H}=\delta E+&V(\mathbf{b}_{a}, \mathbf{b}_{b}, \mathbf{b}_{c})
    \\\mathcal{H}=\varepsilon_{P_{b}}+&\varepsilon_{P_{c}}-\varepsilon_{P_{a}}+V(\mathbf{b}_{a}, \mathbf{b}_{b}, \mathbf{b}_{c}),\nonumber
    \\&\varepsilon_{P_j}=\sqrt{p_j^2+m^2_{\infty j}},
\end{align}
with the $\mathbf{b}_i$'s transverse positions, conjugate to the $\mathbf{p}_i$'s. In the eikonal limit, where most of the particles' momentum is carried
by their $+$ components, the kinetic terms simplify to
\begin{align}
    \mathcal{H}=\frac{\mathbf{p}^2_{a}+m^2_{\infty a}}{2p^+_{a}}+&\frac{\mathbf{p}^2_{b}+m^2_{\infty b}}{2p^+_{b}}
    +\frac{\mathbf{p}^2_{c}+m^2_{\infty c}}{2p^+_{c}}+V(\mathbf{b}_{a}, \mathbf{b}_{b}, \mathbf{b}_{c}). \label{eq:lpm_ham_a}
    \\(p_a^{+},p_b^{+},p_c^{+})&=E(x_a,x_b,x_c)=E(-1,x,1-x)
\end{align}
where the $m_{\infty i}$'s represent the asymptotic or in-medium masses\footnote{By thermal mass, we mean the 
mass picked up by the parton due to forward scattering with plasma constituents. In general, such a mass approaches a constant
known as the asymptotic mass as the parton approaches the lightcone. See Secs.~\ref{sec:htl} and ~\ref{sec:soft_jq}. Since partons will always be assumed 
to propagate in this limit, we use the terms thermal mass and asymptotic mass interchangeably.}.

The potential, $V(\mathbf{b}_{a}, \mathbf{b}_{b}, \mathbf{b}_{c})$ for the case of $q\rightarrow qg$ splitting is given as
\begin{align}
    V(\mathbf{b}_{a}, \mathbf{b}_{b}, \mathbf{b}_{c})&=-\frac{i}{2C_{F}}\Big\{\left(2C_{F}-C_{A}\right)\mathcal{C}_{F}(\mathbf{b}_{b}-\mathbf{b}_{a})\nonumber
    \\&+C_{A}\mathcal{C}_{F}(\mathbf{b}_{c}-\mathbf{b}_{b})+C_{A}\mathcal{C}_{F}(\mathbf{b}_{a}-\mathbf{b}_{c})\Big\},\label{eq:dip_pot}
    \\\mathcal{C}_{R}(\mathbf{b})&\equiv\int_{\mathbf{q}}\mathcal{C}_{R}(q_{\perp})(1-e^{i\mathbf{b}\cdot\mathbf{q}}).\label{eq:scatt_ker_pos}
\end{align}
We pause here to make some comments on the potential above:
\begin{itemize}
    \item It is assumed that the potential takes this dipole-factorised form, with each term containing only depending on the difference of
        two transverse positions. This assumption holds up to at least NLO \cite{CaronHuot:2008ni}, where the
        threepole contributions have been shown to vanish.
    \item The potential is evidently imaginary as it describes the transverse momentum damping of a hard parton 
    through its interactions with the medium. This of course implies that $\mathcal{H}$ is not Hermitian.
    \vspace{0.3cm}
    \item The factor of $g^2$ appearing at the beginning of Eq.~\eqref{eq:dip_pot} characterises the strength of the interaction between
    partons and the medium and so the relevant scale for its running can be as low as that of the Debye mass \cite{Peshier:2008zz}.
    \vspace{0.3cm}
    \item $\mathcal{C}_{R}(q_{\perp})$ is the 
        transverse scattering kernel, related in the eikonal limit to the transverse scattering rate,\footnote{We purposely write
        $\mathcal{C}_{F}$ as a function of the scalar, $q_{\perp}$ in order to emphasise the assumption of an isotropic 
        medium in the transverse directions.}
        \begin{equation}
            \lim_{p^+\to\infty}\frac{d\Gamma_{R}(\mathbf{p},\mathbf{p}+\mathbf{q})}{d^2q_{\perp}}=\frac{\mathcal{C}_{R}(q_{\perp})}{(2\pi)^2}.
        \end{equation}
        for a high-energy parton in colour representation $R$.
        In an abuse of terminology, we refer to $\mathcal{C}_{R}(q_{\perp})_{R}$ interchangeably as the transverse scattering rate 
        and the transverse scattering kernel in what follows. In the multiple scattering regime, with $q_{\perp}\gtrsim gT$ the appropriate scattering kernel can be computed using Hard Thermal Loop (HTL) effective theory \cite{Aurenche:2002pd},
        \begin{equation}
            \mathcal{C}_{R}(q_{\perp})_{\text{HTL}}\equiv\frac{g^2 C_{R}T m_{D}^2}{q_{\perp}^2(q_{\perp}^2 + m_{D}^2)}.
            \label{eq:scattkerhtl}
        \end{equation}    
        We will come back to discuss the scattering kernel further at a later point (see Sec.~\ref{sec:scatt_kernel}).
    \vspace{0.3cm}
    \item Looking back to Fig.~\ref{fig:gluon_emission_lpm} can help us further understand how to interpret the different
    terms in $\mathcal{C}_{F}(\mathbf{b})$:
    the first one captures the correlations between the amplitude and itself
    or the conjugate amplitude and itself whereas the second term corresponds to background field correlations between the amplitude 
    and the conjugate amplitude, separated by the transverse vector $\mathbf{b}$.
\end{itemize}
A further reduction of the dimensionality comes from the observation that there is an inherent symmetry with 
respect to how we choose the axis of jet propagation; it is possible to choose 
the $z$ axis to point in a slightly different direction while still maintaining the collinear approximations that we have made. 
Consequently, the splitting rate that we are deriving should be invariant under the transformations
\begin{equation}
    (\mathbf{p}_{i},p_{i\,z})\rightarrow(\mathbf{p}_{i}+p_{i\,z}\mathbf{\xi},p_{i\,z}). 
\end{equation}
Since $p^{+}_{i}=x_{i}E$ and because of the zero sum of the $x_{i}$ and the $\mathbf{p}_i$,
a single independent combination emerges\footnote{Note that this differs from the AMY convention by a factor of $E$, i.e $\mathbf{h}_{\text{AMY}}=E\mathbf{h}$.}
\begin{equation}
    \mathbf{h}\equiv x_{b}^+\mathbf{p}_a-x_{a}^{+}\mathbf{p}_b=x_c^+\mathbf{p}_b-x_b^+\mathbf{p}_c=x_a^+\mathbf{p}_c-x_c^+\mathbf{p}_a,
\end{equation}
which is invariant under the aforementioned set of transformations. The kinetic part of the Hamiltonian, $
\delta E$ in Eq~\eqref{eq:lpm_ham_a} then reduces to 
\begin{equation}
    \delta E =-\frac{h^2}{2Ex_a^+ x_b^+ x_c^+}+\sum_{i=a}^{c}\frac{m_{\infty\,i}^2}{2Ex_i^{+}}
    \label{eq:deltaE_h}
\end{equation}
This reduction 
will also allow us to rewrite the potential in Eq.~\eqref{eq:lpm_ham_a} by introducing the corresponding
position variable 
\begin{equation}
    \mathbf{B}\equiv\frac{\mathbf{b}_a-\mathbf{b}_b}{x^+_a+x^+_b}=\frac{\mathbf{b}_b-\mathbf{b}_c}{x^+_b+x^+_c}=\frac{\mathbf{b}_c-\mathbf{b}_a}{x^+_c+x^+_a}
\end{equation}
All together, the effective Hamiltonian then reads
\begin{align}
    \mathcal{H}&=-\frac{h^2}{2Ex_a^+ x_b^+ x_c^+}+\sum_{i=a}^{c}\frac{m_{\infty\,i}^2}{2Ex^+_{i}}-\frac{i}{2C_{F}}\Big\{\left(2C_{F}-C_{A}\right)\mathcal{C}_{F}(-(x^+_a+x^+_b)\mathbf{B})\nonumber
    \\&+C_{A}\mathcal{C}_{F}(-(x_b^+ + x_c^+)\mathbf{B})+C_{A}\mathcal{C}_{F}(-(x_c^+ + x_a^+)\mathbf{B})\Big\}
    \label{eq:ham_ready}
\end{align}
We can now return to Eq.~\eqref{eq:radprob_start}, and rewrite everything in 
terms of the momentum, $\mathbf{h}$, while also unpacking the hard vertices
\begin{equation}
    \frac{dI}{d\omega}=\frac{\alpha_sP_{q\to qg}(x)}{x^2(1-x)^2E^3}2\text{Re}\int_{t_{1}<t_{2}}dt_{1}dt_{2}\int_{\mathbf{h}_{2},\mathbf{h}_{1}}
    \mathbf{h}_{2}\cdot\mathbf{h}_{1}\langle\mathbf{h}_{2},t_{2}|\mathbf{h}_{1},t_1\rangle.
    \label{eq:radprob_onepart_mom}
\end{equation}
$P_{q\to qg}(x)$ is the standard spin-averaged DGLAP splitting function for a $q\to qg$ splitting \cite{Altarelli:1977zs}. It reads
\begin{equation}
    P_{q\to qg}(x)=C_{F}\frac{1+(1-x)^2}{x}.
\end{equation}
Eq.~\eqref{eq:radprob_onepart_mom} makes manifest the reduction to a 1-particle quantum mechanics problem. We are now going to try and make contact with the AMY formalism. Fourier
transforming to $\mathbf{B}$ space
\begin{equation}
    \frac{dI}{d\omega}=\frac{\alpha_sP_{q\to qg}(x)}{x^2(1-x)^2E^3}\text{Re}\int_{t_{1}<t_{2}}dt_{1}dt_{2}
    \mathbf{\nabla}_{\mathbf{B}_{2}}\cdot\mathbf{\nabla}_{\mathbf{B}_{1}}\langle\mathbf{B}_{2},t_{2}|\mathbf{B}_{1},t_1\rangle\Big\vert_{\mathbf{B}_{2}=\mathbf{B}_{1}=0}.
    \label{eq:radprob_onepart_pos}
\end{equation}
In equilibrium, the propagators only depend on the difference of times. Upon changing integration variables 
from $\{t_1, t_{2}\} \to \{t_1, t\}$, we then see that the entire integrand only depends on $t$. By differentiating 
with respect to $t_1$, we can then obtain an expression for the rate 
\begin{equation}
    \frac{d\Gamma}{d\omega}=\frac{\alpha_sP_{q\to qg}(x)}{x^2(1-x)^2E^3}\text{Re}\int_{0}^{\infty}dt
    \mathbf{\nabla}_{\mathbf{B}_{2}}\cdot\mathbf{\nabla}_{\mathbf{B}_{1}}\langle\mathbf{B}_{2},t|\mathbf{B}_{1},0\rangle\Big\vert_{\mathbf{B}_{2}=\mathbf{B}_{1}=0},\label{eq:before_amy}
\end{equation}
where the $t$ is now the time between the emission vertices in the amplitude and the conjugate amplitude. In integrating 
$t$ up to infinity, we are assuming the medium to have infinite length, in line with the AMY formalism. We specify that $\langle\mathbf{B}_{2},t_{2}|\mathbf{B}_{1},t_1\rangle$ is a Green's function 
of the Schrödinger equation defined in Eq.~\eqref{eq:ham_ready}, i.e 
\begin{equation}    
    i\partial_{t}\Psi(\mathbf{B},t)=\mathcal{H}\Psi(\mathbf{B},t)
    \label{eq:schrod_pos}
\end{equation}
with initial condition
\begin{equation}
    \langle\mathbf{B}_{2},t_1|\mathbf{B}_1,t_1\rangle=\delta^2(\mathbf{B}_{1}-\mathbf{B}_1).
\end{equation}
Then, we define
\begin{equation}
    \mathbf{F}(\mathbf{B},t)\equiv2i\mathbf{\nabla}_{\mathbf{B}_{1}}\langle\mathbf{B},t|\mathbf{B}_1,0\rangle\Big\vert_{\mathbf{B}_1=0}.
\end{equation}
This quantity obeys the same Schrödinger equation Eq.~\eqref{eq:schrod_pos} as $\langle\mathbf{B}_{2},t|\mathbf{B}_1,0\rangle$, i.e 
\begin{equation}
    i\partial_{t}\mathbf{F}(\mathbf{B},t)=\mathcal{H}\mathbf{F}(\mathbf{B},t),
    \label{eq:schrod_mom}
\end{equation}
albeit with an altered initial condition
\begin{equation}
    \mathbf{F}(\mathbf{B},0)=-2i\mathbf{\nabla}_{\mathbf{B}}\delta^2(\mathbf{B}).
\end{equation}
We then integrate Eq.~\eqref{eq:schrod_mom} over time, using that $\mathbf{F}(\mathbf{B},t)$ decays as $t\to\infty$ because of the imaginary 
potential, which leaves us with
\begin{equation}
    -2\mathbf{\nabla}_{\mathbf{B}}\delta^2(\mathbf{B})=\mathcal{H}\mathbf{F}(\mathbf{B}),
    \label{eq:schrod_timeint}
\end{equation}
where we have defined the time-integrated amplitude
\begin{equation}
    \mathbf{F}(\mathbf{B})\equiv\int_{0}^{\infty}dt\mathbf{F}(\mathbf{B},t).
\end{equation}
Fourier transforming Eq.~\eqref{eq:schrod_timeint} back to $\mathbf{h}$ space, while using the explicit form of the effective Hamiltonian, 
Eq.~\eqref{eq:ham_ready} then yields
\begin{align}
    2\mathbf{h}&=i\delta E \mathbf{F}(\mathbf{h})+\frac{1}{2C_F}\int_{\mathbf{q}}
    \mathcal{C}_{F}(q_{\perp})
    \bigg\{(2C_{F}-C_{A})[\mathbf{F}(\mathbf{h})-\mathbf{F}(\mathbf{h}-x\mathbf{q_{\perp}})]\nonumber
    \\ &+C_{A}[\mathbf{F}(\mathbf{h})-\mathbf{F}(\mathbf{h}+\mathbf{q_{\perp}})]+C_{A}
    [\mathbf{F}(\mathbf{h})-\mathbf{F}(\mathbf{h}-(1-x)\mathbf{q_{\perp}})]\bigg\}.
    \label{eq:defimplfulla}
\end{align}
In addition, we can now rewrite Eq.~\eqref{eq:radprob_onepart_mom} in terms of 
$\mathbf{F}(\mathbf{h})$ which now reads\footnote{Eq.~\eqref{eq:jmcoll} matches with the ($q\to qg$ case of the) LO AMY collinear radiation rate \cite{Arnold:2002zm}, up to final state statistical factors. 
These statistical factors can nonetheless be safely neglected as long as the temperature of the medium is negligible compared
to the energy of the least energetic daughter.}
\begin{equation}
    \frac{d\Gamma}{d\omega}=\frac{\alpha_sP_{q\to qg}(x)}{4x^2(1-x)^2E^3}\int_{\mathbf{h}}\mathrm{Re}[2\mathbf{h}\cdot\mathbf{F}(\mathbf{h})].
    \label{eq:jmcoll}
\end{equation}

Let us pause for a moment to recap. We have set out to compute the radiation rate of a hard quark traversing a weakly coupled QGP,
while correctly taking into account LPM interference. The latter goal is explicitly realised through Eq.~\eqref{eq:defimplfulla}, 
which resums multiple scatterings between the hard partons and the medium. Its solution, $\mathbf{F}(\mathbf{h})$ 
is then fed into Eq.~\eqref{eq:radprob_onepart_mom} thereby determining the rate, $d\Gamma/d\omega$. 

\section{The Harmonic Oscillator Approximation}\label{sec:hoa}

 We now proceed to introduce the Harmonic Oscillator Approximation (HOA) within 
the setup that has been put forward in the previous section. Our motivation for doing so is two-fold: on one hand, it will allow
us to relatively easily obtain a well-known analytical result from the literature. Perhaps more importantly, it will give us an 
opportunity to clearly and precisely state the assumptions tied to the HOA, hence simultaneously determining the region of phase 
space in which it can be safely applied.

Before going on to solve Eq.~\eqref{eq:defimplfulla} in the HOA, it will be useful to first define 
$\hat{q}_{R}(\mu_{\perp})$, the transverse momentum broadening coefficient
\begin{equation}
    \hat{q}_{R}(\mu_{\perp})\equiv\int^{\mu_{\perp}} \frac{d^{2}q_{\perp}}{(2\pi)^2}q_{\perp}^2\mathcal{C}_{R}(q_{\perp}).
    \label{eq:qhatdef}
\end{equation}
While it will soon be clear how $\hat{q}_{R}(\mu_{\perp})$ is related
to the problem at hand, for the moment let us consider it as a transport coefficient, describing the hard parton's diffusion in transverse 
momentum space. Asymptotic freedom should render the integral above UV finite. However, we would then be including harder, so-called
Molière scattering \cite{Moliere:1947zza,Moliere:1948zz}, which gives rise to two hard partons in the final state and is of course not
a diffusive process. It is for this reason that $\hat{q}$ comes equipped with a UV cutoff, $\mu_{\perp}$.

$\hat{q}_{R}(\mu_{\perp})$ visibly depends on the transverse scattering kernel $\mathcal{C}_{F}(q_{\perp})$, which we take here to be the 
HTL kernel, defined in Eq.~\eqref{eq:scattkerhtl}.
Let us now go back to Eq.~\eqref{eq:defimplfulla} and note that the $\mathbf{q}$ integration is logarithmically enhanced 
 when:
\begin{itemize}
    \item $h\sim Q_{\perp}\gg q_{\perp}$ with $Q_{\perp}$ the total momentum exchanged between the medium and parton during
    gluon emission. In 
    other words, the momentum exchanged between the medium and parton during each collision should be much smaller than the total momentum
    exchanged during the formation time associated with the bremsstrahlung. We can then expand all of the $\mathbf{F}(\mathbf{h},x,\mathbf{q})$'s 
    in Eq.~\eqref{eq:defimplfulla} around $\mathbf{F}(\mathbf{h})$. 
    \vspace{0.2cm}
    \item $Q_{\perp}\gg m_{D}\sim gT$, so that the resulting logarithm can be considered large. One can then justify 
    solving Eq.~\eqref{eq:defimplfulla} to leading log (LL) accuracy, meaning that no effort is made to determine the constant under the logarithm. 
    The solution to Eq.~\eqref{eq:defimplfulla} to NLL accuracy has been obtained in \cite{Arnold:2008zu}.
\end{itemize}
If both of these criteria are satisfied, all of the mass terms will drop out of Eq.~\eqref{eq:deltaE_h}, 
since $m_{\infty}\sim gT$ at LO. The expansion then leaves us with a much-simplified form of Eq.~\eqref{eq:defimplfulla}
\begin{equation}
    2\mathbf{h}\approx i\frac{h^2}{2Ex(1-x)}\mathbf{F}(\mathbf{h})-\frac{C_{A}x^2+(2C_{F}-C_{A})+C_{A}(1-x)^2}{2}\int_{\mathbf{q}}
    \mathcal{C}_{F}(q_{\perp}) 
   \frac{ q_{\perp}^2}{4}\nabla^2\mathbf{F}(\mathbf{h}).
    \label{eq:defimplfullho}
    \end{equation}
At this point, we can then use Eq.~\eqref{eq:qhatdef} to write 
\begin{equation}
    2\mathbf{h}\approx i\frac{h^2}{2Ex(1-x)}\mathbf{F}(\mathbf{h})-\frac{C_{A}x^2+(2C_{F}-C_{A})+C_{A}(1-x)^2}{2C_{F}}
   \frac{\hat{q}_{F}(\mu_{\perp})}{4}\nabla^2\mathbf{F}(\mathbf{h}).  
\end{equation}
where 
\begin{equation}
    \hat{q}_{R}(\mu_{\perp})=\frac{g^2C_{R}Tm_{D}^2}{4\pi}\ln\frac{\mu_{\perp}^2}{m_{D}^2}\label{eq:qhat_soft_lo}
\end{equation}
to leading log (LL) accuracy. By LL accuracy, we mean that there has been no attempt made to determine the constant under the logarithm. The logarithmic sensitivity on $\mu_{\perp}$ signals a contribution at the same order and with an 
opposite IR log divergence from the region $h\sim  Q_{\perp}\sim  q_{\perp}$. In this region, a single, harder scattering between the medium 
and the parton is responsible for the momentum exchanged, $Q_{\perp}$ during the formation time. In taking the HOA, this region is ignored,
 treating $\hat{q}_{R}(\mu_{\perp})$ itself as a parameter as opposed to some UV-regulated quantity. We call this parameter $\hat{q}_{0R}$\footnote{Through taking the 
  HOA,
 the potential in Eq.~\eqref{eq:ham_ready} becomes $V(\mathbf{B})\sim\hat{q}_{0F}\mathbf{B}^2$. Evidently, this is where the term 
 ``Harmonic Oscillator Approximation'' originates.}.
Continuing in this manner, we get
\begin{equation}
    2\mathbf{h}=i\frac{h^2}{2Ex(1-x)}\mathbf{F}(\mathbf{h})-\frac{C_{A}x^2+(2C_{F}-C_{A})+C_{A}(1-x)^2}{2C_{F}}\frac{\hat{q}_{0F}}{4}
\nabla^2\mathbf{F}(\mathbf{h}).
\label{defimplfullhoqhat}
\end{equation}
By rotational invariance  $\mathbf{F}(\mathbf{h})\equiv\mathbf{h} \mathrm{F}(h^2)$  and thus
\begin{equation}
    2=i\frac{h^2}{2Ex(1-x)}\mathrm{F}(h)-\frac{C_{A}x^2+(2C_{F}-C_{A})+C_{A}(1-x)^2}{2C_{F}}\frac{\hat{q}_{0F}}{4}
     \left(\mathrm{F}''(h)+3\frac{\mathrm{F}'(h)}{h}\right).
    \label{defimplfullho}
    \end{equation}
This equation can be solved by applying the boundary conditions that $\mathbf{F}(\mathbf{h})$ remain finite as $h\to 0$
and as $h\to\infty$
\begin{align}
    \mathbf{F}(\mathbf{h})&=\frac{4iEx(1-x)}{h^2}\left(\exp\left(-\frac{(\frac{1}{2}+\frac{i}{2})h^2}{H}\right)-1\right)\mathbf{h},
    \\H&\equiv\Big(Ex(1-x)(1-\frac{C_{A}}{C_{F}}x(1-x))\hat{q}_{0R}\Big)^{\frac{1}{2}}.
\end{align}
In order to get the rate, we need to extract the real part of $\mathbf{F}(\mathbf{h})$ and compute
\begin{align}
    \int_{\mathbf{h}}2\mathbf{h}\cdot\mathrm{Re}\,\mathbf{F}(\mathbf{h})
    &=\int_{\mathbf{h}}2(\mathbf{h}\cdot\mathbf{h})\frac{4Ex(1-x)}{h^2}\exp\left(\frac{-h^2}{2H}\right)\sin\left(\frac{h^2}{2H}\right)
    \\&=\frac{2Ex(1-x)H}{\pi}.
\end{align}
Plugging this back into Eq.~\eqref{eq:jmcoll}, we get a result for the rate
\begin{align}
    \frac{d\Gamma}{d\omega}^{\text{LPM}}&=\frac{g^2 C_{F}}{8\pi E^3}
    \frac{1+(1-x)^2}{x^3(1-x)^2}
    \nonumber
    \frac{Ex(1-x)\Big(Ex(1-x)(q-\frac{C_{A}}{C_{F}}x(1-x))\hat{q}_{0F}\Big)^{\frac{1}{2}}}{\pi}\nonumber
    \\&=\frac{g^2C_{F}}{8\pi^2}\frac{1+(1-x)^2}{x^2(1-x)}\sqrt{\frac{x(1-x)(1+\frac{C_{A}}{C_{F}}x(1-x))\hat{q}_{0F}}{E^3}}.
    \label{eq:coll_rate_final}
\end{align}
Within the AMY framework this collinear rate is then fed into the term with the collision operator $C^{1\to 2}[f]$ on the right hand 
side of Eq.~\eqref{eq:amy_boltzmann} along with the $g\to gg$ and $g\to q\bar{q}$ rates, after which point one can go and solve 
the (linearised) Boltzmann equation. Comparison with experimental data can be performed through, for instance, 
the embedding of these rates in the MARTINI event generator \cite{Qin:2007rn,Schenke:2009gb}.

In deriving Eq.~\eqref{eq:coll_rate_final}, we have sketched a calculation of the 
 collinear radiation rate, while showing how to properly account for LPM interference, as is necessary in the multiple scattering regime. 
 Furthermore, we have showed how one can use the HOA in order to obtain an analytical solution in this regime. In what follows,
we will demonstrate how this result fits within the context of the rest of the radiative jet energy loss literature.
 Before doing so, however, we will take a moment to clarify some of the details regarding the transverse scattering kernel, $
\mathcal{C}_{F}(q_{\perp})$, which in principle depends on the model of the medium, as well as the scale of the transverse momentum exchanges 
between the jet and the QGP.

\section{Choice of \texorpdfstring{$\mathcal{C}_{F}(q_{\perp})$}{TEXT}}\label{sec:scatt_kernel}
As was already mentioned, the HTL kernel was used in the calculation of the quark collinear radiation rate,
 Eq.~\eqref{eq:coll_rate_final}. In the context of this thesis, where we are using finite temperature perturbation 
 to study what we call a weakly coupled QGP, it is the obvious choice. Yet, in the previous section, where 
 we computed the radiation rate to LL accuracy, we never actually used the explicit form of Eq.~\eqref{eq:scattkerhtl}. Rather, we 
 just made use of the fact that upon plugging the HTL kernel in the integral equation, Eq.~\eqref{eq:defimplfulla} the $q_{\perp}$
 integration (in the multiple scattering regime) gives rise to a (large) logarithm.

In terms of calculating the collinear radiation rate to LL accuracy, we could have just as well used another kernel,
originally proposed by Gyulassy and Wang (GW) \cite{Gyulassy:1993hr}
\begin{equation}
    \mathcal{C}_{R}(q_{\perp})_{\text{GW}}\equiv\frac{g^4 C_{R}n}{(q_{\perp}^2 + m_{D}^2)^2},
    \label{eq:scattkergw}
\end{equation}
where $n$ is the density of scattering centres. Instead of considering the medium to be weakly coupled, the GW
kernel comes from modelling the medium as an 
ensemble of static scattering centres with Yukawa-like potentials.

In the case of the GW model, only static colour-electric fields are screened, 
whereas for the HTL kernel, since the colour
charges are considered to be dynamical, both colour-electric and colour-magnetic fields need to be screened. However,
the details with regard to screening are irrelevant at LL accuracy, precisely because this is the limit 
in which $q_{\perp}\gg m_{D}$. At NLL accuracy, individual momentum
transfers of $q_{\perp}$ of order $m_{D}$ become important \cite{Arnold:2008zu}, in which case there is a difference between 
a full treatment of plasma screening and the static approximation.

Returning to the setting of a weakly coupled plasma, we remark that the HTL kernel should only be used for 
momenta $T\gg q_{\perp}$. For larger momentum exchanges $\sqrt{ET}\gg q_{\perp}\gtrsim T$, 
the appropriate kernel is
\begin{equation}
    \mathcal{C}_{R}(q_{\perp})_{\text{hard}}=\frac{2g^4C_{R}}{q_{\perp}^4}\int_l\frac{l-l_z}{l}[C_{A}n_{B}(l)(1+n_{B}(l'))+2N_{f}T_{F}n_{F}(l)(1-n_{F}(l'))],\label{eq:scat_hard}
\end{equation}
where $l'=l+\frac{l_{\perp}^2+2\mathbf{l}\cdot\mathbf{q}}{2(l-l_z)}$. For a
kernel that interpolates between $\mathcal{C}_{F}(q_{\perp})_{\text{HTL}}$ and $\mathcal{C}_{F}(q_{\perp})_{\text{hard}}$, 
one can use \cite{Arnold:2008vd,Ghiglieri:2022gyv}
\begin{equation}
    \mathcal{C}_{R}(q_{\perp})_{\text{smooth}}^\mathrm{LO}=
    \frac{2g^4C_R}{q_{\perp}^2(q_{\perp}^2+m_D^2)}\int_l\frac{l-l_z}{l}
    \Big[C_A\,n_{\text{B}}(l)(1+n_{\text{B}}(l'))
    +2 N_fT_F\,n_{\text{F}}(l)(1-n_{\text{F}}(l'))\Big].
    \label{hardqcdresum}
\end{equation}
When $\mu_{\perp}\gg T$, 
the value for $\hat{q}_{R}(\mu_{\perp})$ reads
\begin{equation}
    \hat{q}_{R}(\mu_{\perp}\gg T)=\frac{g^4 C_{R}\xi(3)T^3 m_{D}^2}{\pi }[2C_{A}+3N_{f}T_{F}]\ln\frac{\mu_{\perp}^2}{T^2},\label{eq:qhat_hlimit}
\end{equation}
which also includes a subleading $\ln T/m_{\text{D}}$ contribution not shown here. The ratio of the two leading log coefficients, (that is, comparing Eqs.~\ref{eq:qhat_soft_lo} and ~\eqref{eq:qhat_hlimit}) is
$4\alpha_sT^2\xi(3)[2C_{A}+3N_fT_{F}]/(\pi m_{D}^2)\sim 0.85$ for $N_c=N_f=3$ QCD.

We emphasise that everything discussed above applies to the restriction $q_{\perp}\gtrsim m_{D}$; the most frequent scatterings have momentum
 $m_{D}\gg q_{\perp}\gtrsim g^2T$ and are mediated by the exchange of low-frequency magnetic gluons. These interactions are not 
 cut off by Debye screening but instead by non-perturbative effects \cite{Arnold:2007pg}\footnote{See Sec.~\ref{sec:htl} for a 
 brief discussion.}. That being said, they will 
 not contribute to $\hat{q}_{R}(\mu_{\perp})$ and hence the radiation rate at leading order because of the extra factor 
 of $q_{\perp}^2$ in the numerator of Eq.~\eqref{eq:qhatdef}. As was mentioned in the previous chapter, 
 the contribution of these low-frequency exchanges has been incorporated through the Lattice EQCD calculation \cite{Moore:2020wvy,Moore:2021jwe}, 
 to be discussed in Sec.~\ref{sec:np_classical}.

 \section{Radiation Rate in the Single-Scattering Regime}\label{sec:single}
 As was already explained, the HOA needs to be invoked to make analytical calculations 
 of radiative energy loss in the multiple scattering regime possible. Yet in doing so, one neglects the contribution 
 from the single-scattering regime where only one scattering with the medium is responsible for triggering 
 bremsstrahlung. Such a contribution may be obtained using the \emph{opacity expansion}, where 
 one essentially expands in the number of collisions between the jet and the medium. In this section, we will show how to obtain this contribution, again for
 $q\to qg$ splitting in the soft limit, starting from Eq.~\eqref{eq:lpm_ham_a},
with $x\to 0$
\begin{equation}
    \mathcal{H}_{\omega\to 0}=\frac{-\mathbf{\nabla}_{\mathbf{B}}^2+m^2_{\infty g}}{2\omega}-i\mathcal{C}_{A}(\mathbf{B}).\label{eq:neqone_ham}
\end{equation}
We specify that while $x\to 0$, we still assume that $\omega\gg T$ so that we may neglect the presence of any statistical functions.
We moreover take the soft limit in Eq.~\eqref{eq:before_amy} and allow for a medium of finite length\footnote{The motivation 
for the relaxation of this assumption will become clear shortly.}, leaving us with 
\begin{equation}
\frac{d\Gamma}{d\omega}\bigg\vert_{\omega\to 0}=\frac{2\alpha_sC_{F}}{\omega^3}\mathrm{Re}\int_0^{L_M}dt\mathbf{\nabla}_{\mathbf{B}_{2}}\cdot\mathbf{\nabla}_{\mathbf{B}_{1}}
\bra{\mathbf{B}_2,t}\ket{\mathbf{B}_1,0}\Big\vert_{\mathbf{B}_2=\mathbf{B}_1=0}.\label{eq:neqone_rate}
\end{equation}
We define the Fourier transform of the propagator, which we henceforth refer to as
\begin{equation}
    G(\mathbf{B}_2,\mathbf{B}_1;t)=\int_{\mathbf{p},\mathbf{q}}e^{i\mathbf{B}_2\cdot\mathbf{p}}e^{-i\mathbf{B}_1\cdot\mathbf{q}}G(\mathbf{p},\mathbf{q};t).
\end{equation}
and note that the action of the potential term from Eq.~\eqref{eq:neqone_ham} reads, in momentum space
\begin{equation}
    [\mathcal{C}_F(\mathbf{B}_2)G(\mathbf{B}_2,\mathbf{B}_1;t)](\mathbf{p},\mathbf{q},t)=\int_{\mathbf{l}}\mathcal{C}_{A}(l_{\perp})\left(G\left(\mathbf{p},\mathbf{q};t\right)-G\left(\mathbf{p}+\mathbf{l},\mathbf{q};t\right)\right),
\end{equation}
which allows us to rewrite the Schrödinger equation 
\begin{equation}
    i\partial_tG(\mathbf{p},\mathbf{q},t)-\frac{p_{\perp}^2+m_{\infty g}^2}{2\omega}G(\mathbf{p},\mathbf{q},t)
    +i\int_{\mathbf{l}}\mathcal{C}_A(l_{\perp})\left(G\left(\mathbf{p},\mathbf{q};t\right)-G\left(\mathbf{p}+\mathbf{l},\mathbf{q};t\right)\right)=0.\label{eq:sch_mom}
\end{equation}
In the meantime, we can write Eq.~\eqref{eq:neqone_rate} 
\begin{equation}    
    \frac{d\Gamma}{d\omega}\bigg\vert_{\omega\to 0}=\frac{2\alpha_sC_{F}}{\omega^3}\mathrm{Re}\int_0^{L_M}dt\int_{\mathbf{p},\mathbf{q}}
    \mathbf{p}\cdot\mathbf{q}\Big[G(\mathbf{p},\mathbf{q};t)-\text{vac}\Big].
\end{equation}
One can then use the trick described around Eq. (5) of \cite{CaronHuot:2010bp} to rewrite the time integration
\begin{align}
    \frac{d\Gamma}{d\omega}\bigg\vert_{\omega\to 0}&=\frac{2\alpha_sC_{F}}{\omega^3}\mathrm{Re}\int_0^{L_M}dt\int_{\mathbf{p},\mathbf{q}}
    \mathbf{p}\cdot\mathbf{q}\frac{2\omega}{p_{\perp}^2+m_{\infty g}^2}\Big[-\frac{p_{\perp}^2+m_{\infty g}^2}{2\omega}G(\mathbf{p},\mathbf{q};t)
    +i\partial_t G(\mathbf{p},\mathbf{q};t)\Big]\nonumber
    \\&=-i\frac{4\alpha_sC_{F}}{\omega^2}\mathrm{Re}\int_0^{L_M}dt\int_{\mathbf{p},\mathbf{q},\mathbf{l}}\mathcal{C}_{A}(l_{\perp})\frac{\mathbf{p}\cdot\mathbf{q}}{p_{\perp}^2+m_{\infty g}^2}
    \Big[G(\mathbf{p},\mathbf{q};t)
    - G(\mathbf{p}+\mathbf{l},\mathbf{q};t)\Big].
\end{align}
Since we are interested in the single-scattering contribution, (also known as the $\mathcal{N}=1$ term in the opacity
expansion) we take
\begin{equation}
    G_{F}(\mathbf{p},\mathbf{q};t)=(2\pi)^2\delta^2(\mathbf{p}-\mathbf{q})e^{-i\frac{p_{\perp}^2+m_{\infty g}^2}{2\omega}t},
\end{equation}
which solves Eq.~\eqref{eq:sch_mom} in the absence of the potential. We can then do the $t$ integration 
\begin{align}
    \frac{d\Gamma}{d\omega}\bigg\vert^{\mathcal{N}=1}_{\omega\to 0}&=\frac{2g^4C_{F}C_{A}Tm_{D}^2}{\pi\omega}\int_{\mathbf{q},\mathbf{l}}
    \frac{1}{l_{\perp}^2(l_{\perp}^2+m_{D}^2)}\bigg(\cos\Big(\frac{q_{\perp}^2+m_{\infty g}^2}{2\omega}L_{M}\Big)-1\bigg)\nonumber
    \\&\times\bigg(\frac{q_{\perp}^2}{(q_{\perp}^2+m_{\infty g}^2)^2}-\frac{\mathbf{q}\cdot(\mathbf{q}+\mathbf{l})}{((q_{\perp}^2+l_{\perp}^2)^2+m_{\infty g}^2)^2}\bigg),\label{eq:op_ls}
\end{align}
where we have used the HTL form of the kernel, Eq.~\ref{eq:scattkerhtl} and used that it is invariant under flips in $l_{\perp}$.
The $\mathbf{q}$ integration cannot be done analytically in general but we will discuss a couple 
of limiting cases in Sec.~\ref{sec:opacity}.

\section{Energy Loss Formalisms}\label{sec:formalisms}
 We use this section to give a pedagogical, though not exhaustive summary of some of the formalisms used to compute radiative energy loss in 
 order to give some context to the calculations completed earlier in the chapter. We refer the reader to \cite{Armesto:2011ht} 
 for a comprehensive comparison of the formalisms mentioned subsequently in the case of a brick medium. In order to make 
 contact with Secs.~\ref{sec:hoa}, ~\ref{sec:single}, all of the 
 results quoted in this section are for $q\to qg$ processes. Results for other splittings can be obtained 
 by suitably exchanging $C_{F}\to C_{R}$\footnote{That is, with the exception of the large $L_M$ case of Eq.~\ref{eq:opacity_ss}, 
 where the identification $m_{D}=2m_{\infty g}$ is necessary in order to perform the integration analytically.}

 \subsection{Multiple Scatterings}
Formalisms introduced concurrently by Baier, Dokshitzer, Mueller, Peigné and Schiff (BDMPS)
 \cite{Baier:1994bd,Baier:1996vi,Baier:1996kr,Baier:1996sk}
 and Zakharov \cite{Zakharov:1996fv,Zakharov:1997uu,Zakharov:1998sv}\footnote{These two formalisms were later shown to be 
 equivalent \cite{Baier:1998kq}.}  were the first to 
  resum multiple soft scatterings between the jet and a dense medium composed of static scattering centres. Their result, 
  obtained using the HOA for gluon emission spectrum from a hard quark with energy $\omega$ reads 
\begin{equation}
    \omega\frac{dI}{d\omega}_{\text{BDMPS-Z}}=\frac{2\pi\alpha_sC_{F}}{\pi}\bigg\vert\cos\Big[(1+i)\sqrt{\frac{\omega_c}{2\omega}}\Big]\bigg\vert,
\end{equation}
yielding for the limiting case, where $\omega\ll\omega_c$
\begin{equation}
    \omega\frac{dI}{d\omega}_{\text{BDMPS-Z}}\approx\frac{2\pi\alpha_sC_{F}}{\pi}
        \sqrt{\frac{\omega_{c}}{2\omega}}\label{eq:bdmpsz_eloss}.
\end{equation}
Upon considering the maximum transverse momentum that a hard parton
can pick up through multiple soft scatterings\footnote{The scale $Q_s$ is commonly known in the jet energy 
loss literature as the \emph{saturation scale}.}, $Q_s^2\equiv\hat{q}_{0R} L_M$, one derives a corresponding maximum
allowed frequency in this region, given by $\omega_c\equiv\hat{q}_{0R}L_M^2$. The $1/\sqrt{\omega}$ tail is characteristic of the suppression stemming
from LPM interference. The work of Arnesto, Salgado and Wiedemann \cite{Salgado:2003gb,Armesto:2003jh} starts from the same 
set of assumptions and attempts to tackle the problem of multiple gluon emissions, treated independently with a 
Poisson distribution.

As was already mentioned, AMY developed an effective kinetic theory \cite{Arnold:2000dr,Arnold:2001ba,Arnold:2001ms,Arnold:2002ja,Arnold:2002zm,Arnold:2003zc},
 capable of calculating medium-induced energy loss (see Eq.~\eqref{eq:coll_rate_final}). Instead of modelling 
 the medium as an ensemble of static scattering centres, the AMY framework holds a stronger connection to first principles 
 through its modelling of the QGP with finite temperature QCD, where the medium constituents are dynamical. The framework was originally constructed to deal with a medium 
 of infinite length, although this obstacle was overcome in \cite{CaronHuot:2010bp}. The conclusions of AMY and BDMPS-Z are 
 essentially the same \cite{Jeon:2003gi}; one can recover Eq.~\eqref{eq:bdmpsz_eloss} by taking
 the soft limit ($x\ll 1$) in Eq.~\eqref{eq:coll_rate_final} and using that 
 \begin{equation}
    \omega\frac{dI}{d\omega}=\omega\int_0^{L_M}dt\frac{d\Gamma}{d\omega}.\label{eq:r_to_spec}
 \end{equation}
Before moving on, we seize this opportunity to parametrically estimate the total rate associated with Eq.~\eqref{eq:bdmpsz_eloss}
\begin{equation}
    \Gamma\sim g^2\sqrt{\frac{\hat{q}_{0F}}{\omega}}.\label{eq:rate_int}
\end{equation}
Neglecting final state factors (as we have done so far), the rate will be inversely proportional to the formation time introduced earlier,
$\tau$
\begin{equation}
\Gamma\sim\frac{g^2}{\tau},
\end{equation}
so that
\begin{equation}
    \tau\sim\sqrt{\frac{\omega}{\hat{q}_{0F}}}\equiv \tau_{\text{LPM}}\gg\frac{1}{g^2T}.\label{eq:lpm_ft}
\end{equation}
It is worth noting that Eq.~\eqref{eq:lpm_ft} and hence Eq.~\eqref{eq:rate_int} could have also been obtained from 
the fact that in the multiple scattering regime, the parton undergoes diffusion in transverse momentum space, i.e 
\begin{equation}
    Q_{\perp}^2\sim \hat{q}_{0F}\tau_{\text{LPM}},
\end{equation}
where $Q_{\perp}$ is the transverse momentum picked up from interactions with the medium during $\tau_{\text{LPM}}$. The formation time (in general) is also given as 
\begin{equation}
    \tau_{\text{LPM}}\sim\frac{\omega}{Q_{\perp}^2},
\end{equation}
with which can arrive at Eq.~\eqref{eq:lpm_ft}. For future reference, we label 
the region of phase space where this random walk takes place the \emph{deep LPM regime}. In Ch.~\ref{chap:qhat_chap}, these kinds of parametric 
estimations will play a very central role. 

\subsection{Single-Scattering}\label{sec:opacity}

The opacity expansion (OE)\cite{Gyulassy:2000er,Gyulassy:2001nm}, introduced briefly in Sec.~\ref{sec:single} was originally developed by Gyulassy, Levai, and Vitev (GLV)
and also independently by Wiedemann \cite{Wiedemann:2000za}. From Eqs.~\eqref{eq:op_ls} and ~\eqref{eq:r_to_spec}, one can easily 
obtain the $\mathcal{N}=1$ contribution (for $q\to qg$ splitting) to the spectrum in the following limits \cite{Mehtar-Tani:2019tvy}\footnote{The case for large $L_{M}$
in \cite{Mehtar-Tani:2019tvy} contains an unphysical logarithm. This logarithm 
is not present in Eq.~\eqref{eq:opacity_ss} as we have included the asymptotic mass in the calculation from Sec.~\ref{sec:single}. Moreover, 
in order to get a result analytically in this limit, one has to use that $m_{D}=\sqrt{2}m_{\infty g}$ at LO.}
\begin{equation}
    \omega\frac{dI}{d\omega}_{\mathcal{N}=1}\approx \alpha_s^2C_{F}C_{A}Tm_{D}^2L_{M}
        \begin{cases}
            \frac{\ln\frac{e^2}{2}}{\pi m_{\infty g}^2}&\text{for $\omega\ll m_D^2L_{M}$}
            \\\frac{L_{M}}{2\omega}&\text{for $\omega\gg m_D^2L_{M}$}\label{eq:opacity_ss}
        \end{cases}
        .
\end{equation}
The opacity expansion formalism has been extended beyond the soft gluon approximation
\cite{Ovanesyan:2011kn,Sievert:2018imd} and the case of a dynamical medium has been explored in \cite{Djordjevic:2008iz}.

There is also the higher-twist pioneered by Guo and Wang \cite{Wang:2001ifa,Majumder:2009zu}, which 
describes the medium in terms of expectation values of field correlation functions. In particular, properties 
of the medium enter energy loss calculations in terms of higher-twist matrix elements. It is similar to the OE in the 
sense it is formulated for media where the parton only scatters a few times per radiation (i.e media with low opacity), but with the added approximation 
that the transverse momentum carried away with the bremsstrahlung is much larger than the typical 
momentum exchanged with the medium.

\subsection{Comparison and Summary}\label{sec:mult_vs_single}
The traditional view is that the formalisms mentioned above are together capable of explaining the spectrum 
across a range of energies. We limit ourselves to the case of a dense medium, i.e $L_M\gg l_{\text{el}}$\footnote{For the case
of a dilute medium, i.e $l_{\text{el}}\gg L_M$, see \cite{Peigne:2008wu,Isaksen:2022pkj}. In this thesis, 
we will eventually send $L_{M}\to\infty$ and will therefore not be concerned with the case of a dilute medium.}. 
For this case, there is a range of $\omega$ for which the spectrum is captured by the BDMPS-Z and AMY formalisms\footnote{Note that the 
OE has a finite radius convergence \cite{Arnold:2008iy} and consequently cannot be relied upon in the multiple scattering regime.}. As was discussed previously, it is in this region where the spectrum of the
 hard parton suffering multiple (soft) scatterings is suppressed (compared to the harder, single-scattering region) due to LPM interference.

 The multiple scattering region is bounded from above by the 
 scale $\omega_c$, beyond which hard, single-scattering dominates. Perhaps unsurprisingly, this region 
 is well-described by the first term in the OE, i.e the second case of Eq.~\eqref{eq:opacity_ss}. The single 
 scattering region is neglected upon taking the HOA, as was demonstrated 
 in Eq.~\eqref{eq:defimplfullho}.

 From below, the LPM regime is bounded by $\omega_{\text{BH}}\equiv\hat{q}_{0F}l_{\text{el}}^2$\footnote{This scale can be obtained by 
 realising that in the BH regime, even though the parton will only have time to scatter once before it radiates, it will
 still undergo a random walk in transverse momentum space.}, the characteristic energy of the Bethe-Heitler (BH) regime. The BH spectrum can be thought of 
 as the incoherent limit of the LPM spectrum, where $l_{\text{el}}\gg \tau$ and individual elastic collisions can be resolved by radiation. The BH 
 region is also well-described by the OE. 

\section{Recent Developments}\label{sec:recent}
Jet energy loss is still an extremely active topic of research, with there being a number of exciting advancements over the past decade or so.
In particular, the Improved Opacity Expansion (IOE) is a powerful tool that has been developed \cite{Mehtar-Tani:2019tvy,Mehtar-Tani:2019ygg,Barata:2020rdn,Barata:2021wuf}
 to provide a unified description of the multiple, soft scattering and the single, hard scattering regimes for the case of a dense 
 medium. Whereas the usual OE involves expanding in the number of elastic collisions, which is equivalent to expanding in powers of the dipole potential, 
 $V(\mathbf{B})$ from Eq.~\eqref{eq:dip_pot}, the IOE shifts the expansion point to HOA solution, i.e for $V(\mathbf{B})\sim\hat{q}_{0R}\mathbf{B}^2$. 
 In doing so, the IOE resums multiple soft scatterings while also accounting for rare, harder scatterings. In fact, it has 
 been highlighted recently \cite{Isaksen:2022pkj} that for a dense medium, only the OE and IOE are needed to describe the entire 
 medium induced spectrum\footnote{The IOE is not capable of describing the spectrum in the BH regime \cite{Barata:2020rdn}.}. 
 Their conclusion is solidified by comparison with the full, numerically constructed in-medium spectrum \cite{Andres:2020vxs}.

There has also been much progress with respect to the understanding of multiple in-medium splitting. It was only understood 
relatively recently that the colour coherence phenomenon that parton showers possess in the vacuum does not carry over to the case of a dense medium \cite{Mehtar-Tani:2010ebp,Mehtar-Tani:2011hma,Mehtar-Tani:2011vlz}.
Instead, ``anti-angular ordering'' takes place, suppressing interference between successive gluon radiations. 
This has paved the way for the construction of an in-medium cascade \cite{Blaizot:2012fh,Blaizot:2013hx,Blaizot:2013vha,Blanco:2020uzy,Blanco:2021usa,Adhya:2022tcn}
 that is differential in transverse momentum. The validity of these approaches is predicated on the fact that interference effects between gluons with 
 overlapping formation times are not large. A series of papers by Arnold and collaborators \cite{Arnold:2015qya,Arnold:2016kek,Arnold:2016mth,Arnold:2018yjd,Arnold:2018fjr,Arnold:2019qqc,
 Arnold:2020uzm,Arnold:2021pin,Arnold:2022epx,Arnold:2022fku,Arnold:2022mby,Arnold:2023qwi} have been devoted to exploring whether or not these latter interference 
 effects are sizable. They claim, restricting the radiated gluons to have energies $\omega\gg T$ that most of the overlap effects can be absorbed into a redefinition of $\hat{q}_{R}(\mu_{\perp})$.

 A large part of this thesis will be devoted to the discussion of double logarithmic corrections to $\hat{q}_{R}(\mu_{\perp})$ \cite{Wu:2011kc,Liou:2013qya,Blaizot:2013vha,Blaizot:2019muz,Ghiglieri:2022gyv} at $\mathcal{O}(g^2)$. 
These are radiative, quantum corrections that arise from keeping track of the recoil during the medium-induced emission of a gluon. Determining the precise 
region of phase space from which the double log originates is of paramount importance, as it of course controls whether they are large in comparison to $g^2$.
While the calculations \cite{Liou:2013qya,Blaizot:2013vha} demonstrated the presence of these corrections, they did not strive to quantify them precisely, which, as we will see, 
turns out to be a difficult task. The double logarithmic corrections have been resummed \cite{Iancu:2014kga,Iancu:2014sha,Caucal:2021lgf,Caucal:2022fhc}, effectively
renormalising the leading order value of $\hat{q}_{R}(\mu_{\perp})$. It has also been shown that these double logs are universal \cite{Iancu:2014kga,Wu:2014nca,Blaizot:2014bha}; they also appear in calculations with two 
emitted gluons with overlapping formation times in the final state, i.e in precisely the context of multiple emissions mentioned above. 

\newpage
\chapter{Tools for Studying Jets at Finite Temperature}\label{chap:tft_eft}

Throughout the previous chapter, we have focused on motivating 
and discussing some of the basic ideas related 
to jet energy loss. Along the way, rigour is by and large sacrificed so as to 
make the physical picture more transparent. The theme of this thesis however 
is to provide and analyse corrections to two quantities, which in part control jet quenching, 
while maintaining a strong connection to first principles.

This connection 
is forged through the use of thermal field theory, which is to be 
introduced in Sec.~\ref{sec:tft}. Before diving into this somewhat intricate 
topic, we choose to prepend some information regarding the general principles and 
features of effective field theory in Sec.~\ref{sec:eft}. The rest of the chapter 
will then be focused on showing how two effective field theories: Electrostatic 
QCD and Hard Thermal Loop Effective Theory play a part 
in the higher order corrections to $\hat{q}$ and $m_{\infty}$ that 
we will ultimately be interested in computing.

\section{Effective Field Theory and The Method of Regions}\label{sec:eft}
A common approach, practiced across many facets of physics is to
 take advantage of inherent scale separation, present in the relevant physical 
 setup in order to drastically simplify calculations. Oftentimes, this 
 is done almost unconsciously: one does not need 
 to account for quantum gravity effects to accurately calculate 
 planets' orbits around the sun nor does one need worry 
 about the bottom quark when calculating hydrogen's energy 
 levels, to give a couple of examples. For systems, whose dynamics are
 completely determined through 
 their Lagrangian,\footnote{While it is true that quantum field theories are technically 
 defined through their associated generating functional, this will not affect any of the arguments that 
 follow.} the effective field theory (EFT) framework can provide a rigorous, yet extremely powerful 
 tool through which, one can make otherwise daunting calculations more tractable, while 
 at the same time rendering the physical picture more lucid. Below, we discuss
 some of the principles associated with the construction of EFTs, before briefly introducing a very helpful 
 technique used for calculating loop integrals with multiple scales. We follow 
 a combination of \cite{Pich:1998xt,Stewart:2014,Manohar:2018aog}.

 A Lagrangian 
density can generally be written as 
\begin{equation}
    \mathcal{L}=\sum_i C_i\mathcal{O}_i\label{eq:lagrangian_full}
\end{equation}
where the $\mathcal{O}_i$ are operators and the $C_i$ are their
respective couplings. For simplicity, we restrict ourselves to theories in $D=4$ spacetime dimensions. Since the action has to be dimensionless, a field's dimension is always determined by its kinetic term. 
Thereafter, for each non-kinetic term, one can deduce the mass dimension of associated $C_i$ since the dimensions of all $\mathcal{O}_i$'s, being 
products of (possibly different) fields have already been fixed. We can then distinguish 
three classes of operators, sorted according to their dimension $d_i$. Operators with:
\begin{itemize} 
    \item $d_i<4$ are known as relevant.
    \item $d_i=4$ are known as marginal.
    \item $d_i>4$ are known as irrelevant.
\end{itemize}
A simple counting argument (see for instance Chapter $18$ of \cite{Srednicki:2007qs}) shows that an infinite number of counterterms 
are required to cancel the UV divergences (to all orders in perturbation theory) coming from irrelevant operators, 
thereby stripping the overarching theory 
of any predictive power\footnote{In what follows, we aim to persuade the reader that this is an outdated perspective.}. Thus, if Eq.~\eqref{eq:lagrangian_full} is to describe a renormalisable QFT in the traditional sense, the sum should
 only include marginal and relevant operators. In contrast, as we will now see, irrelevant operators 
 turn out to be very important in effective field theories.

For simplicity, let us assume that we are interested in studying a QFT, describing two massive fields: $\phi_1$ with 
mass $m$ and $\phi_2$ with mass $M$. Moreover, we assume that the QFT is renormalisable in the traditional sense. 
We now want to write down an effective field theory (EFT), 
which only captures the dynamics below some scale, $\Lambda$ such that $M>\Lambda\gg m$. After implementing the cutoff, 
$\Lambda$ on the full theory's generating functional one can in principle integrate
out the heavy $\phi_2$ field from the path integral. Upon expanding the resultant Lagrangian powers of $1/M$, one recovers 
an infinite sum over local operators
\begin{equation}
    \mathcal{L}_{\text{EFT}}=\mathcal{L}_{d_i \leq 4}+\sum_i\frac{c_i}{M^{d_i}}\mathcal{O}_i\label{eq:lagrangian_eft},
\end{equation}
where the $c_i$ above are dimensionless and are called Wilson coefficients. Note that the first term above contains the part of the original 
Lagrangian that only depends on $\phi_1$. The second term instead contains a sum over irrelevant operators $\mathcal{O}_i$, built 
from products (and possibly derivatives) of the $\phi_1$ field. But how does this mesh with our previous statement 
that irrelevant operators leave theories unrenormalisable? Observe that, around the small mass scale $m$ these operators
should give a contribution to some measurable quantity that goes like $\sim c_i(m/M)^{d_i}$ with $d_i\geq 1$. 
It is then not difficult to see that one can truncate the sum in Eq.~\eqref{eq:lagrangian_eft} according to the desired 
accuracy for the perturbative computation -- here lies one of the great 
advantages of EFTs. Furthermore, even though the resultant EFT is not renormalisable in the traditional sense, it 
can be renormalised, order by order in the $(1/M)$ expansion, thereby maintaining its predictive capability.

The fact that terms with irrelevant operators in Eq.~\eqref{eq:lagrangian_eft} are suppressed for \\energies $\sim m\ll M$ is in some sense a formalising of the rather general comments 
 made at the beginning of the section. It is nevertheless worth pointing 
 out that the Wilson coefficients multiplying the irrelevant operators will inevitably contain 
 information about the field that has been integrated out. To extract the dependence of the Wilson coefficients
 on the parameters of the higher-level theory, one generally has to perform what is known
 as a matching procedure.

 In contrast to the notion of integrating out mentioned above, where one can actually do the functional integration, it is more 
 often than not the case that even trying to perform such a manoeuvre is impractical. Instead, what one generally does is to
 construct the EFT theory by writing down a Lagrangian, which respects all of the symmetries that one 
 wishes the low energy theory to respect, usually a subset of the symmetries associated with the higher-level 
 theory. Then, once the Lagrangian has been truncated to the desired precision, matching 
 conditions are imposed; by comparing diagrams from the two theories, $c_i's$ are chosen so that the 
 EFT and higher-level theory agree in the regime where they are both valid. Matching is then performed order by order
 in the expansion parameter, which in the previous example was $(m/M)$.

While it is not so clear from the discussion above, computations in an EFT are usually 
considerably more simple due to the fact that one has reduced the number of scales in the problem. Even 
if there are multiple scales in the EFT, as long as these scales possess some scale separation, one 
can iterate the above procedure, producing a family of EFTs, each of which is well-suited
to describe physics at a different scale. As an example, consider the reduction of QCD to 
non-relativistic QCD (NRQCD) \cite{Caswell:1985ui}, capable of describing heavy quarkonia. 
There, the hierarchy is $M\gg Mv$ where $M$ is the mass of the heavy quark, $v\ll1$ is the relative velocity
and $Mv$ is the typical momentum transfer. One can then integrate out the scale $Mv$, which is large
compared to the binding energy $Mv^2$ and in doing so obtain potential NRQCD (pNRQCD) \cite{Brambilla:1999xf}. 
Another example of a so-called \emph{top-down} EFT is soft collinear effective theory (SCET) \cite{Bauer:2000yr,Bauer:2001ct,Bauer:2001yt,Bauer:2002nz,Beneke:2002ph}, 
which capitalises on the separation between the so-called hard, collinear and soft scales in jet physics
to efficiently resum Sudakov logarithms. 

So far, we have restricted ourselves to EFTs where the underlying theory is known. There 
are also so-called bottom-up EFTs where the aforementioned method of matching is not applicable. Instead, one
constrains the Wilson coefficients using experimental data or data from LQCD. 
This is the case for Chiral Perturbation Theory \cite{Weinberg:1978kz,Gasser:1983yg}, which allows 
one to study the low-energy dynamics based on the underlying chiral symmetry\footnote{It 
goes without saying that the overarching theory, QCD, is indeed known in this context. Yet, the difficulty 
in matching stems from the difficulty of describing QCD in the non-perturbative regime.}. Indeed, 
a more modern take on QFT is that every theory is an EFT! The practice of adding 
irrelevant operators to the Standard Model Lagrangian goes under the name Standard Model Effective Theory or SMEFT 
\cite{Buchmuller:1985jz,Grzadkowski:2010es}.

Before moving on, by giving a simple example, we wish to elucidate the usefulness of a strategy that will be used 
throughout this thesis, known as the \emph{method of regions} \cite{Beneke:1997zp,Smirnov:2002pj}. To start, consider the 
following integral, where for the sake of simplicity we work in $2$ spatial dimensions
\begin{equation}
    I=\int\frac{d^2P}{(2\pi)^2}\frac{1}{(P^2+m^2)(P^2+M^2)}.\label{eq:regions_start}
\end{equation}
The integration is evidently finite so no regularisation is necessary. After Feynman parameterisation, we immediately get
\begin{equation}
    I=\frac{1}{4\pi(m^2 - M^2)}\ln\frac{m^2}{M^2}.
\end{equation}
Now let us consider the limit where there is a scale separation, i.e $m\ll M$ so that we can expand for large $M$
\begin{equation}
    I=\frac{1}{4\pi M^2}\ln\frac{m^2}{M^2}\Big(1+\frac{m^2}{M^2}+\mathcal{O}(m^4)\Big)\label{eq:regions_full}.
\end{equation}
So far, all we have done is calculated a simple integral. But let us now go back Eq.~\eqref{eq:regions_start} 
and see what happens if we perform the expansion before doing the integration. First, we consider the region 
where $P\sim m\ll M$. There,
\begin{align}
    I_{P\sim m}&=\int\frac{d^2P}{(2\pi)^2}\frac{1}{(P^2 +m^2)M^2}\left(1-\frac{P^2}{M^2}+\mathcal{O}(P^4)\right)\nonumber
    \\&=\frac{1}{4\pi M^2}\left(\ln\frac{\Lambda^2}{m^2}+\frac{m^2}{\Lambda^2}-\frac{\Lambda^2}{M^2}+\frac{m^2}{M^2}\ln\frac{\Lambda^2}{m^2}+\mathcal{O}(m^4)\right),\label{eq:region_small}
\end{align}
where we have cut off the integration in the UV with $\Lambda\gg m$. Next, let us 
consider the region $P\sim M\gg m$. Here, $\Lambda\ll M$ instead acts as an IR cutoff
\begin{align}
    I_{P\sim M}&=\int\frac{d^2P}{(2\pi)^2}\frac{1}{(P^2 +M^2)p^2}\left(1-\frac{m^2}{p^2}+\mathcal{O}(\frac{1}{P^4})\right)\nonumber
    \\&=\frac{1}{4\pi M^2}\left(\ln\frac{M^2}{\Lambda^2}+\frac{\Lambda^2}{M^2}-\frac{m^2}{\Lambda^2}+\frac{m^2}{M^2}\ln\frac{M^2}{\Lambda^2}+\mathcal{O}(m^4)\right).\label{eq:region_large}
\end{align}
If we add the contributions from the two regions together, we recover the full result Eq.~\eqref{eq:regions_full}. While this expected,
 note that for the 
combined result to be $\Lambda$ independent, all of the corrections
depending on a power of $\Lambda$ must cancel. This turns out to be a general result: given some loop integral with multiple scales (in our case, 
$m$ and $M$), we can make use of the strategy:
\begin{itemize}
    \item By power counting, identify the different regions that will give a finite contribution to the integral. In the example 
        above, these regions were $P\sim m$ and $P\sim M$.
    \item Expand appropriately in those regions and then perform the integration. While the integral above 
         was simple enough to calculate without resorting to this method, if there are more scales, or the integrand 
         is more complicated, this can be a formidable task.
    \item Add the contributions from the different regions together. Corrections depending 
     on a power law of the introduced cutoff will cancel, leaving only logarithms of ratios of the separated scales and/or finite 
     terms. Most of the time, we throw away the power law corrections prior to this step, since we know they will cancel out
     anyway. But in some cases, one can take advantage of this feature of power law corrections: if it is not \emph{a priori} clear 
     which regions contribute to the full integral, one can use power law correction cancellation as a proxy for checking 
     whether two regions are adjacent. This is essentially what is done in Sec.~\ref{sec:class}.
\end{itemize}
Such a protocol is followed for many of the calculations performed in this thesis.

Despite our example above being extremely simple, it easy to see how this strategy conveniently compliments 
the EFT philosophy. If some quantity of interest is given in terms of an integral, $I$, one can do a power counting, thereby 
identifying the different regions, over which integrating will contribute to $I$. For certain regions, the contribution 
may then be much easier to compute using some EFT. Consider $\hat{q}(\mu_{\perp})$, which receives a contribution from the region $P\sim T$ 
and also from the region $P\sim gT$ at LO in the strong coupling. The contribution from the former region is computed using finite temperature 
QCD, whereas the latter is computed either with Hard Thermal Loop effective theory or electrostatic QCD, both of 
which will be introduced in the next sections. The LO soft contribution is computed in Sec.~\ref{sec:scatt_lo_soft}. 
For $m_{\infty}$, the LO value only receives a hard contribution, whereas the NLO correction only receives 
a soft contribution. The computation of the NLO correction is contained in Sec.~\ref{sec:nlo_asym}.

Before moving on, we make a remark with respect to the choice of regulator used above. In that example, we regulated 
the integration in each region with a hard cutoff in order to explicitly show the cancellation of the terms depending 
on power laws of $\Lambda$. Nevertheless, whenever possible, we will choose to instead use dimensional regularisation (DR). 
This is well-justified as:
\begin{itemize}
    \item Unlike cutting off with a hard regulator, DR preserves Lorentz and gauge invariance.
    \item Scaleless integrals vanish. In thermal field theory calculations, where one has to do a vacuum subtraction, this 
        is an especially convenient choice since the vacuum terms are usually scaleless.
    \item Logarithmic divergences show up as $1/\epsilon$ poles. Conversely, power law divergences do not appear at all.
\end{itemize}
See App.~\ref{sec:conventions} for our conventions for DR.

 \section{Thermal Field Theory}\label{sec:tft}
The generalisation of perturbative quantum field theory to finite temperatures traditionally 
goes under the name of thermal field theory (TFT). This section will therefore be devoted 
to explaining how concepts from quantum statistical mechanics can be consistently implemented in a QFT 
context. In doing, so we will arrive at two different, but equivalent formulations of TFT:  the Imaginary Time Formalism (ITF)
 and the Real Time Formalism (RTF). Essentially all of the calculations in Ch.~\ref{chap:qhat_chap} and Ch.~\ref{ch:asym_mass}
will be done in the RTF. Much of Sec.~\ref{sec:rtf} will therefore be spent on introducing the propagators 
to be used in later chapters. We will nevertheless also 
make a point to explore the ITF, as it allows for a cleaner derivation of the effective theory, Electrostatic QCD, 
which plays a central role in the calculations that we are trying to build upon in this thesis. Throughout this 
section, we only work towards sketching a derivation of the finite temperature propagators for bosons. The fermion 
propagators will instead be stated, with the occasional remark made when necessary. Moreover, we 
set all chemical potentials to zero throughout, as the phenomena that we investigate in this 
thesis only take place near the $\mu=0$ axis of the QCD phase diagram. We follow 
a combination of \cite{Bellac:2011kqa,Arnold:2007pg,Ghiglieri:2020dpq}.

\subsection{The Imaginary Time Formalism}\label{sec:itf}
An important difference between zero and finite temperature QFT is that in the latter case, it is necessary 
to take into account statistical fluctuations (in addition to quantum fluctuations), which arise from only having limited, macroscopic information 
of the entire system at some moment $t_0$. The density operator
\begin{equation}
    \rho(t_0)=\sum_i p_i(t_0)\ket{i}\bra{i},
\end{equation}
is an important object in this context. In particular, the $p_i(t_0)$ represent the probabilities 
associated with the different pure states, whose linear combination define the mixed state, described by $\rho(t_0)$.
In equilibrium, it takes the form
\begin{equation}
    \rho_{\text{eq}}=\frac{1}{Z}e^{-\frac{H}{T}}\label{eq:rho_eq}
\end{equation}
where
\begin{equation}
    Z=\Tr e^{-\frac{H}{T}} \label{eq:partition}
\end{equation}
is the partition function and  $H$ is the Hamiltonian operator describing a field theory for a scalar field, $\varphi(X)$. The route 
to the derivation of the ITF starts by identifying  $\tau=-i/T$ above 
and then recognising that the operator in the trace is a time evolution operator in imaginary time, $e^{-iH\tau}$. One then proceeds to
use the standard techniques introduced by Feynman, evaluating the trace using the 
eigenstates of the field operator, $\varphi(X)$. Following \cite{Feynman:100771}, 
one can write $Z$ as a Euclidean path integral
\begin{equation}
    Z(\beta)=\int \mathcal{D}\varphi \exp\Big(-S_{\text{E}}(\beta)\Big)\label{eq:func_z}
\end{equation}
where 
\begin{equation}
    S_{\text{E}}(\beta)=\int_0^{\beta}d\tau\int d^3x\mathcal{L}(\varphi),
\end{equation}
and $\beta\equiv 1/T$. The cyclicity property of the trace in Eq.~\eqref{eq:partition} is moreover implemented by the 
condition that $\varphi(0,\vec{\mathbf{x}})=\pm\varphi(0,\vec{\mathbf{x}})$, where $+$ ($-$) is the correct choice 
for bosons (fermions). 

Of course, we would like to be able to compute thermal expectation values of composite operators
\begin{equation}
    \langle\mathcal{O}\rangle=\frac{\int\mathcal{D}\varphi\exp\Big(-S_{\text{E}}(\beta)\Big)\mathcal{O}}{Z(\beta)}.\label{eq:expec}
\end{equation}
Such a goal motivates the inclusion of a source term to the partition function
\begin{equation}
    Z(\beta)=\int \mathcal{D}\varphi\exp\Big(-S_{\text{E}}(\beta)+\int_0^\beta d^4X j(X)\varphi(X)\Big).
\end{equation}
so that the thermal expectation values can be obtained from the usual methods of functional 
differentiation.

After appropriately shifting the field $\varphi(X)$, one recovers
\begin{align}
    Z_{F}(\beta,\,j)=\int&\mathcal{D}\varphi\exp\Big(-\int_0^{\beta}d\tau\int d^3x\nonumber 
    \\&\times\varphi(X)\frac{1}{2}\Big[-\frac{\partial^2}{\partial\tau^2}-\nabla^2+m^2\Big]\varphi(X)-j(X)\varphi(X)\Big),
\end{align}
at which stage the Gaussian functional integration over $\varphi(X)$ can be performed, resulting in
\begin{equation}
      Z_{F}(\beta,\,j)=Z_{F}(\beta,\,0)\exp\Big(\frac{1}{2}\int_0^{\beta}d^4X d^4Y j(X)\Delta_F(X-Y)j(Y)\Big).
\end{equation}
As can be readily verified using Eq.~\eqref{eq:expec} and functional differentiation methods (see \cite{Bellac:2011kqa} for details), $\Delta_F(X-Y)$ is the time-ordered (in 
imaginary time) two-point function, 
 \begin{equation}
    \Delta_{F}(\tau,\vec{\mathbf{x}})=\langle\mathcal{T}\varphi(\tau,\vec{\mathbf{x}})\varphi(0,\vec{\mathbf{0}})\rangle.
 \end{equation}
$\mathcal{T}$ denotes time-ordering in imaginary time and the cyclicity of the trace in Eq.~\eqref{eq:partition} fixes
$\tau$ to lie in the interval $[0, \beta]$ with periodicity $\tau\to\tau-\beta$. Progressing onward, let us note that 
$\delta(\tau,\vec{\mathbf{x}})$ solves the partial differential equation
\begin{equation}
    \Big(-\frac{\partial^2}{\partial\tau^2}-\nabla^2+m^2\Big)\Delta_F(X-Y)=\delta(\tau_X-\tau_Y)\delta(\vec{\mathbf{x}}-\vec{\mathbf{y}}).
\end{equation}
It is then possible to find a solution by Fourier transforming to momentum space
\begin{equation}
    \Delta(\omega_n,\vec{\mathbf{k}})=\frac{1}{\omega_n^2+k^2+m^2},\label{eq:euc_prop}
\end{equation}
which leaves us with the Euclidean or Matsubara propagator \cite{Matsubara:1955ws}. The periodicity of imaginary time is 
reflected above through the fact that the so-called Matusbara frequencies
are forced to take on discrete values, $\omega_n=2\pi nT\,,\,n\in\mathbb{Z}$.

With the free propagator in hand, one can then deform the theory by adding a potential term and proceed 
to work out the Feynman rules, similarly to the $T=0$ case. The main difference here is however, that in loop computations, 
instead of integration over continuous frequencies, one is left having to perform an infinite sum over 
the discrete Matsubara frequencies. When computing time-dependent quantities in the ITF, the 
infinite sum must be evaluated before analytically continuing back to real time. For this reason, 
the ITF's main domain of applicability is largely restricted to the computation of thermodynamic quantities.

Before progressing to the case of QCD, we mention that analytical continuation to arbitrary (non-discrete) frequency values is provided by 
\begin{equation}    
    \Delta(z,\vec{\mathbf{k}})=\int_{-\infty}^{\infty}\frac{dk^0}{2\pi}\frac{\rho_{\text{B}}(k_0,\vec{\mathbf{k}})}{k^0-z}
\end{equation}
with $\rho_{\text{B}}(k_0,\vec{\mathbf{k}})$ the spectral function for bosons. Indeed, choosing $z=k^0\pm i\varepsilon$  with real $k^0$ allows us 
to recover the causal propagators. Specifically, 
\begin{equation}
    D_{R}(k^0,\vec{\mathbf{k}})=-i\Delta(z=k^0 + i\varepsilon)\quad,\quad D_{A}(k^0,\vec{\mathbf{k}})=-i\Delta(z=k^0 - i\varepsilon). \label{eq:eucl_causal}
\end{equation}
The retarded propagator, $D_{R}$ is thus analytic in the upper half complex $k^0$ plane whereas
the advanced propagator, $D_{A}$ is analytic in the lower half complex $k^0$ plane. The above relations 
will be used in Sec.~\ref{sec:chtrick}.

As in the $T=0$ case, much of what has been discussed above can be adapted without too much difficulty
to the case of a gauge theory, such as QCD. We point 
out that fermions in the partition function Eq.~\eqref{eq:func_z} should have anti-periodic boundary conditions in contrast 
to their bosonic counterparts. It then transpires that the Matsubara frequencies for the quarks are
instead $\omega_n =(2n+1)\pi T,\,n\in\mathbb{Z}$. Another issue that is familiar from zero temperature
QFT is that of the integration over unphysical gauge field configurations in the path integral. One 
can nevertheless bypass this issue through the inclusion of ghosts, 
through a suitably generalised Fadeev-Popov procedure. Since the details pertaining to such a procedure will 
not be relevant for what is to follow, we refer the reader to the textbooks \cite{Kapusta:1989tk,Bellac:2011kqa} for 
more information. Renormalisation, which is of course also relevant for the scalar field case, need not be worried 
about at finite temperature. This stems from the fact that the new scale, $T$ does not modify the theory
at short distances $\ll 1/T$, and thus it is safe to assume that UV divergences will carry over from 
the $T=0$ theory. When computing thermal contributions to quantities of interest in the ITF or RTF, 
one usually subtracts off the vacuum result anyway, which is thus equivalent to renormalisation\footnote{This statement only holds 
up to one-loop. In Ch.~\ref{ch:asym_mass}, we will see an explicit example where this fails at the two-loop level.}.
On the other hand, the scale $T$ significantly disturbs the IR structure of the theory (again, this is also 
true for the scalar field). We take this opportunity to write down the 
QCD Lagrangian (with massless quarks) in Euclidean space
\begin{equation}
    \mathcal{L}^{\text{QCD}}_{\text{E}}=\frac{1}{4}F_{\mu\nu}F^{\mu\nu}+\bar{\psi}_i\slashed{D}\psi_i,\label{eq:l_qcd_im}
\end{equation}
where the sum is over the quark flavours.

\subsection{Dimensional Reduction}\label{sec:dim_red}

At zero temperature, we can roughly say that perturbation theory is done in $g^2$, owing to the fact that loops 
are added at a cost of $g^2/4\pi$ (see Eq.~\eqref{eq:alpha_running}). As we will now sketch, things become much more complicated when $T>0$ because of the inclusion 
of additional scales. To show in a more concrete way how these complications arise, let us consider, 
as an example, the contribution to the pressure from the gluonic sector. It takes the following schematic form\footnote{This can 
be seen by computing the energy-momentum tensor, $T_{\mu\nu}$ associated with the gluonic part of
 Eq.~\eqref{eq:l_qcd_im} and then using that $P^G=1/3\langle T_{ii}\rangle$.}
\begin{equation}
    P^G \sim \int d^3 p \,p\, n_{\text{B}}(E_p) \label{eq:pressure_schematic}
\end{equation}
where $E_p$ is the energy and
\begin{equation}
    n_{\text{B}}(E_p)\equiv\frac{1}{e^{\frac{E_p}{T}}-1}\label{eq:nbedef}
\end{equation}
is the Bose-Einstein distribution function. $T$ is often denoted the hard scale: it represents the average 
energy of quarks and gluons in the plasma and is also the scale associated with the Matsubara frequencies\footnote{
For this reason, $\pi T$ is sometimes instead regarded as the hard scale.}. From Eq.~\eqref{eq:pressure_schematic}, its 
contribution to the pressure can be estimated as
\begin{equation}
    P^G_{\text{hard}}\sim T^4n_{\text{B}}(T)\sim T^4+\mathcal{O}(g^2).
\end{equation}
As long as $g^2\ll1$, the perturbative expansion is still evidently well-defined for contributions coming from this scale.
There is also the soft scale $gT$, which, as we will see in Sec.~\ref{sec:htl} is the momentum associated with
collective excitations in the plasma. Its contribution reads\footnote{The origin of the $\mathcal{O}(g)$ correction 
in Eq.~\eqref{eq:soft_pressure} and the $\mathcal{O}(1)$ term in Eq.~\eqref{eq:ultrasoft_pressure} can 
be understood from the following argument: in the ITF, as well as the usual factor of $g^2/4\pi$, 
additional loops come with an additional 
sum over Matsubara frequencies. Once such a sum has been carried out, it will give 
rise to several terms, with at least one containing an $n_{\text{B}}(E_p)$. For $E_p\ll T$, 
the $n_{\text{B}}(E_p)$ term dominates, resulting in the enhancements seen in Eqs.~\eqref{eq:soft_pressure}, ~\eqref{eq:ultrasoft_pressure}.}
\begin{equation}
    P^G_{\text{soft}}\sim (gT)^4n_{\text{B}}(gT)\sim g^3T^4+\mathcal{O}(g),\label{eq:soft_pressure}
\end{equation}
where the perturbative expansion still converges, albeit in a slower manner, due to the fact that 
$n_{\text{B}}(E_p)\sim T/E_p$ if $E_p\ll T$. The diverging of the statistical function in this regime is known 
as Bose enhancement. Finally, the 
ultrasoft scale $\sim g^2 T$ gives 
\begin{equation}
    P^G_{\text{ultra soft}}\sim (g^2T)^4n_{\text{B}}(g^2T)\sim g^6T^4+\mathcal{O}(1).\label{eq:ultrasoft_pressure}
\end{equation}
In other words, higher loop corrections to the contribution from the ultrasoft scale 
come at no cost! For this reason, the ultrasoft scale is sometimes called the 
non-perturbative scale. In this case, the contribution from the ultrasoft scale only enters at 
$\mathcal{O}(g^6)$, which turns out to be at four-loop order. Yet, 
there are other quantities of interest, such as the transverse scattering rate $\mathcal{C}(k_{\perp})$, discussed 
in Sec.~\ref{sec:scatt_kernel}, that already receive these kinds of non-perturbative contributions at leading order\footnote{To remind the 
reader, this was not apparent nor relevant in any of the calculations in the preceding chapter because we were only ever calculating 
the second moment of this object $\mathcal{C}(k_{\perp})$. See the paragraph below Eq.~\eqref{eq:qhat_hlimit}.}. Such 
a seemingly ubiquitous feature was identified by Linde in the 1980's \cite{Linde:1980ts}
and at first glance, seems to cast much doubt on the validity of perturbation theory at finite temperature. 
Note that this issue is unique 
to the gauge sector of QCD; the fermionic analogue of Eq.~\eqref{eq:pressure_schematic} would instead possess 
a factor of 
\begin{equation}
    n_{\text{F}}(E_p)\equiv\frac{1}{e^{\frac{E_p}{T}}+1}\label{eq:nfedef},
\end{equation}
the Fermi-Dirac distribution function, which instead has the property that 
$n_{F}(E_p)\sim 1/2$ for $E_p\ll T$.

Fortunately, much progress has been made in circumventing these issues related to 
the gauge sector's behaviour in the IR. Two approaches have proven successful in this 
pursuit: dimensional reduction and Hard Thermal Loop effective theory. The latter EFT will be introduced
subsequent to the discussion about the RTF.  

In the mid-1990s, Braaten and Nieto popularised the use of Electrostatic QCD (EQCD), an effective field theory derived by 
integrating out the hard scale, $\pi T$ or said differently, all but the zero Matsubara mode from finite temperature QCD \cite{Braaten:1995cm,Braaten:1995jr}. 
Since there is only one Matsubara mode left, the Euclidean time direction is effectively eliminated -- this 
is where the term dimensional reduction \cite{Kajantie:1995dw,Kajantie:1997tt} comes from. EQCD 
is thus also said to be a theory of \emph{static modes}. 
A novel consequence of this reduction is that the $A^0$ field, no longer possessing a gauge symmetry 
is allowed to acquire a mass, $m_D\sim  gT$, known as the \emph{Debye mass}. This component, which 
we denote as $\Phi=iA_0$, now transforms as a scalar in the adjoint representation of SU(N). EQCD's Lagrangian can be derived, following the procedure roughly described in the previous section\footnote{The normalisation
is fixed in order to keep with the convention of \cite{Moore:2020wvy,Ghiglieri:2021bom}. Building on these 
works will be the subject of Ch.~\ref{ch:asym_mass}.}
\begin{equation}
    \mathcal{L}_{\text{EQCD}}=\frac{1}{2}\Tr F_{ij}F_{ij}+\Tr D_i\Phi D_i\Phi+m_D^2\Tr\Phi^2+
    \lambda_{E}\left(\Tr\Phi^2\right)^2\label{eq:eqcd_lag}.
\end{equation}
 The fields $A_i=A_i^a T^a$ now live in three dimensions and $D_i=\partial_i-ig_{3\text{d}}A_i$, where $g_{3\text{d}}^2=g^2 T$ at LO. The 
$\lambda_{E}$ term is a remnant of the four-$A^0$-interaction that arises in two-loop level in full-QCD perturbation theory.

This procedure can be iterated, integrating out the field, $\Phi$ or equivalently the soft scale, $gT$. We are then left with 
three-dimensional pure Yang-Mills theory, coined Magnetostatic QCD or MQCD, which only depends on the ultrasoft scale $g^2T$. 
At present day, perturbative matching onto these dimensionally reduced theories 
is a mature, well-understood technology, with the EFT couplings having been computed up to $\mathcal{O}(g^6)$ 
\cite{Laine:2005ai,Kajantie:1997tt,Ghisoiu:2015uza}. Moreover, it turns out \cite{Braaten:1995jr} that the
contribution to the pressure can be written as 
\begin{equation}
    P^G=P^G_{\text{hard}}+P^G_{\text{EQCD}}+P^G_{\text{MQCD}}.
\end{equation}
In other words, one can compute in the three different theories to obtain a result for the pressure up to $\mathcal{O}(g^6)$, which 
is free of IR divergences. Such an achievement marks a true \emph{tour de force} of the dimensional reduction framework. 
The hard contribution was computed in \cite{Kajantie:2002wa}\footnote{At present day, there 
are still some terms missing at $\mathcal{O}(g^6)$, which originate from the hard scale.} with the soft contribution 
obtained perturbatively in \cite{Vuorinen:2003fs,Kajantie:2003ax} and the MQCD contribution in \cite{Hietanen:2004ew,DiRenzo:2006nh} on the lattice. Indeed, 
a detail that we have glossed over up until this point is that these dimensionally reduced theories can be 
studied non-perturbatively with the use of three-dimensional lattice simulations; with the disappearance of the time direction, 
one does not need to worry about complications arising from analytically continuing back to real time. Nevertheless, as these EFTs
are naturally derived within the ITF, their applicability has been historically restricted to the computation of thermodynamic 
quantities. This is what makes their recent deployment to the study of real-time, dynamical quantities
all the more impressive \cite{CaronHuot:2008ni}, to be discussed in Sec.~\ref{sec:chtrick}. 

\subsection{The Real Time Formalism}\label{sec:rtf}

From the previous section alone, it 
is easy to see why one would like to have a formalism of thermal field theory in which time is kept real from the outset. 
One can start on this path by analytically continuing the Euclidean propagator Eq.~\eqref{eq:euc_prop} to real time \cite{Dolan:1973qd}.
This yields, for a scalar field
\begin{equation}
    D_{F}(K)=\frac{i}{K^2-m^2+i\varepsilon}+2\pi\delta(K^2-m^2)n_{\text{B}}(|k^0|).
\end{equation}
Despite the appealing feature of this expression, specifically, that there is a clear separation 
of the vacuum and thermal parts, doing perturbation theory with this propagator alone 
 results in so-called pinch singularities, where products of identical delta functions appear in loop integrals.

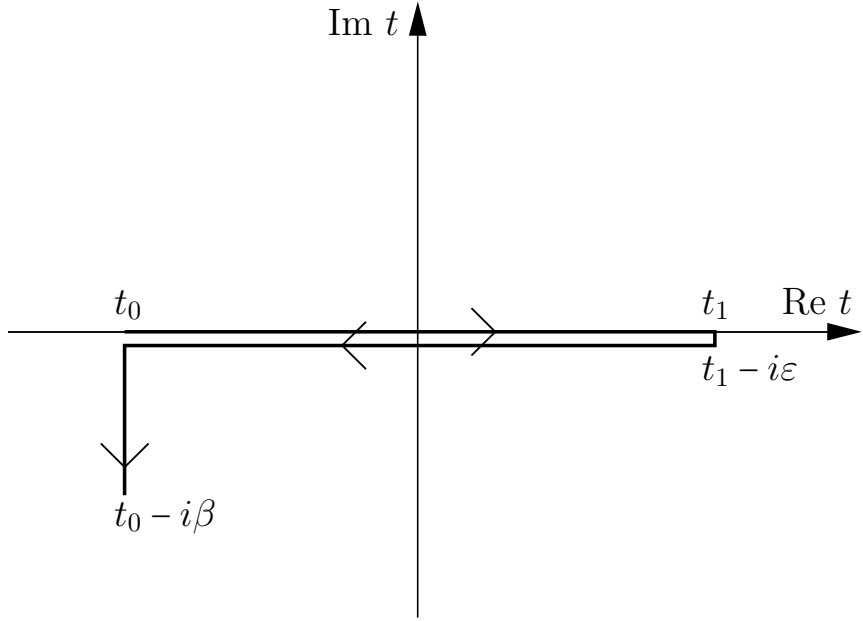
\begin{figure}[t]
    \begin{center}
        \includegraphics[width=0.8\textwidth]{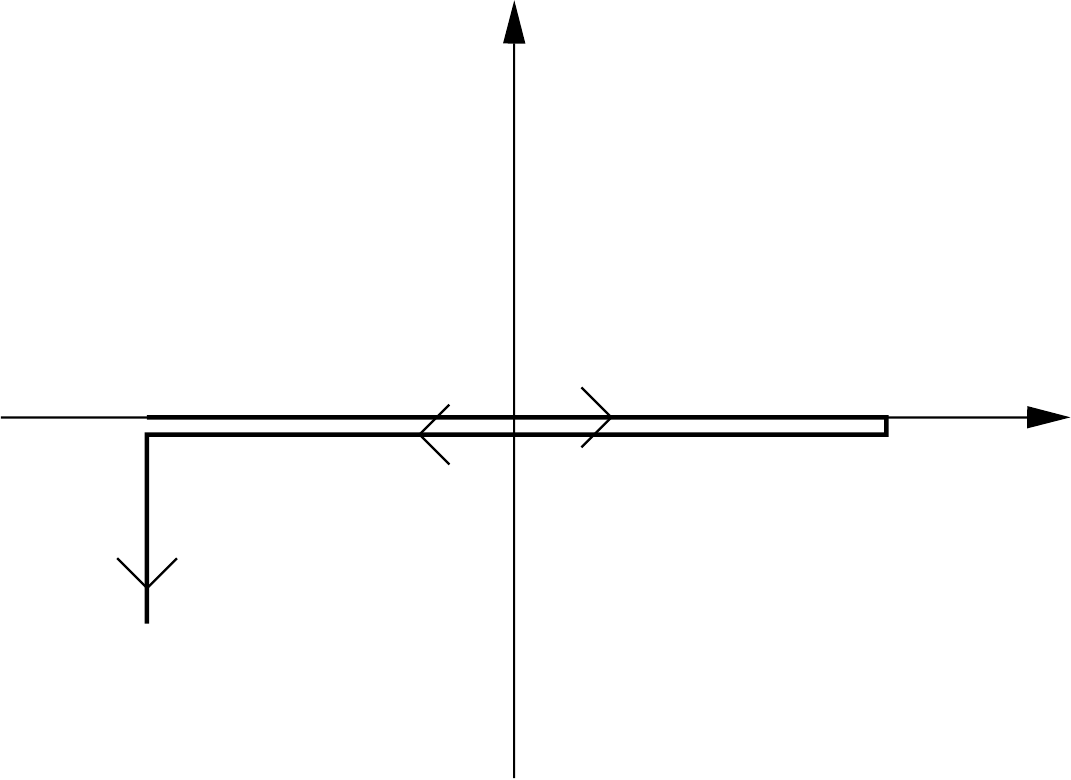}
        \put(-200,220){\Large{Im $t$}}
        \put(-30,115){\Large{Re $t$}}    
        \put(-60,115){\Large{$t_1$}}
        \put(-60,90){\Large{$t_1 -i\varepsilon$}}
        \put(-280,115){\Large{$t_0$}}
        \put(-280,35){\Large{$t_0-i\beta$}}
    \end{center}
    \caption{Illustration of the Schwinger-Keldysh contour in the complex t-plane. Image taken from \cite{Ghiglieri:2020dpq}.}
    \label{fig:sk}
\end{figure}
The starting point of the RTF instead consists of going back to Eq.~\eqref{eq:func_z} and deforming the contour in 
the following way: as is depicted in Fig.~\ref{fig:sk}, one begins from (real) $t_0$ and 
integrates along the real axis to $t_1$. The contour is then deformed below the real axis by 
an amount $-i\varepsilon$ before being brought back across the imaginary axis to $t_0 -i\varepsilon$. Finally, the 
path follows a vertical straight line from $t_0 -i\varepsilon$ to $t_0-i\beta$. The choice of path is known
as the \emph{Schwinger-Keldysh contour}, named after its authors \cite{Schwinger:1960qe,Keldysh:1964ud}. 
It turns out that in taking $t_0\to -\infty$ and $t_1\to +\infty$, contributions coming 
from integrating over the vertical portions of the contour factorise, appearing as overall multiplicative constants \cite{NIEMI1984105,Niemi:1983ea} 
and, per Eq.~\eqref{eq:expec} are not relevant for the computation of correlation functions. Note that this in contrast to the 
ITF, where the time $t_0$ is arbitrary. This can be seen by observing that the equilibrium density operator in Eq.~\eqref{eq:rho_eq} commutes 
with the Hamiltonian, implying that the equilibrium state is time translation invariant. 
 
The price to pay for insisting on a framework where time is kept real is the so-called doubling of degrees of freedom, 
 which becomes apparent when we write the generating functional
 \begin{equation}
    Z[j_1,j_2]=\int\mathcal{D}\varphi_1\mathcal{D}\varphi_2e^{iS[\varphi_1]-iS[\varphi_2]-\int d^4X\Big(j_1(X)\varphi_1(X)-j_2(X)\varphi_2(X)\Big)}\label{eq:z_sk},
 \end{equation} 
with 
\begin{equation}
    j_1(X)=j(t,\vec{\mathbf{x}})\quad\quad j_2(X)=j(t-i\varepsilon,\vec{\mathbf{x}})
\end{equation}
and functional differentiation defined by
\begin{equation}
    \frac{\delta j_a(X)}{\delta j_b(X')}=\delta_{ab}\delta^4(X-X').
\end{equation}
The ``1'' fields can be thought of living on the upper horizontal branch in Fig.~\ref{fig:sk} and are associated 
with physical particles. Conversely, the ``2'' fields live on the lower branch and act like ghosts in the sense that the 
are only to be included as internal lines when calculating time-ordered amplitudes. The vertices from type 2 fields
 come with an extra minus sign because of the different signs of the actions in Eq.~\eqref{eq:z_sk}. As a result of the doubling of degrees of freedom, 
the scalar theory's propagator appears as a matrix in 1,2 space \cite{Bellac:2011kqa}
\begin{align}
    \mathbf{D}_{ij}&=\frac{\delta}{\delta j_i}\frac{\delta}{\delta j_j}Z[j_1,j_2]\bigg\vert_{j_1=0,\,j_2=0}\nonumber
    \\\mathbf{D}&=\left(\begin{array}{cc} \langle\varphi_1\varphi_1\rangle& \langle\varphi_1\varphi_2\rangle\\\langle\varphi_2\varphi_1\rangle &\langle\varphi_2\varphi_2\rangle\end{array}\right)
    =\left(\begin{array}{ccc} D_{F}&&D_{>}\\D_{<}&& D_{\bar{F}}\end{array}\right).
\end{align}
The $11$ and $22$ entries are the time-ordered and anti-time-ordered propagators 
\begin{align}
    D_{F}(t_1,t_0)&=\Theta(t_1-t_0)\langle\varphi(t_1)\varphi(t_0)\rangle+\Theta(t_0-t_1)\langle\varphi(t_0)\varphi(t_1)\rangle \label{eq:to_pos}
    \\D_{\bar{F}}(t_1,t_0)&=\Theta(t_0-t_1)\langle\varphi(t_1)\varphi(t_0)\rangle+\Theta(t_1-t_0)\langle\varphi(t_0)\varphi(t_1)\rangle\label{eq:ato_pos}
\end{align}
respectively, where we choose to suppress the spatial argument for the moment. The off-diagonal entries are the cut or Wightman propagators, 
which describe physical correlations in the medium. They can be written 
in terms of the spectral function, 
\begin{equation}
    \rho_{\text{B}}(t_1,t_0)=\langle[\varphi(t),\varphi(0)]\rangle=D_{>}(t_1,t_0)-D_{<}(t_1,t_0).
\end{equation}
We also find it useful to define the retarded and advanced propagators in terms of the spectral function
\begin{align}
    D_{R}(t_1,t_0)&=\Theta(t_1-t_0)\rho_{\text{B}}(t_1,t_0)\label{eq:r_pos}
    \\D_{A}(t_1,t_0)&=-\Theta(t_0-t_1)\rho_{\text{B}}(t_1,t_0)\label{eq:a_pos}
\end{align}
so that we can write 
\begin{equation}
    \rho_{\text{B}}(t_1,t_0)=D_{R}(t_1,t_0)-D_{A}(t_1,t_0).
\end{equation}
Note that the propagators $D_{R}(t_1,t_0),\,D_{A}(t_1,t_0)$ are known as causal propagators.

\begin{figure}[t]
    \begin{center}
        \includegraphics[width=0.8\textwidth]{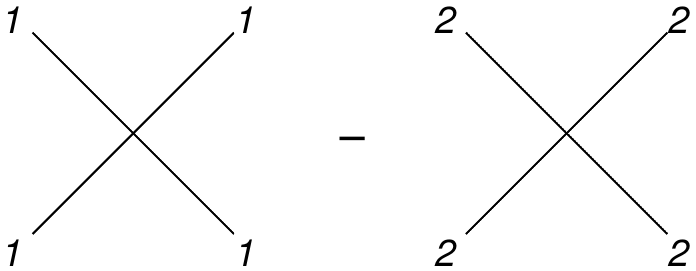}
    \end{center}
    \caption{Vertices appearing in the RTF for a scalar field with a quartic interaction in the 1,2 basis. 
    Vertices connecting propagators associated with the 2 fields come with a relative change of sign 
    because of the extra minus in Eq.~\eqref{eq:z_sk}. Image taken from \cite{Ghiglieri:2020dpq}.}
    \label{fig:phi4_12}
\end{figure}
Many of these objects are integral to TFT calculations, in contrast to the situation at zero 
temperature, where in order to do perturbation theory, all one really needs is the time-ordered propagator.
This is because, at zero temperature, one is able to utilise Lehmann-Symanzik-Zimmermann (LSZ) reduction in order to relate
 vacuum expectation values of time-ordered operators to S-matrix elements and hence, experimentally accessible quantities
 such as particle cross-sections. Such a manoeuvre is predicated on the assumption that the particles taking part in such a process 
 correspond to asymptotic states, created infinitely far away in the past or future. Yet 
 at finite temperature, random interactions induced by the medium do not preserve any state, implying 
 that S-matrix elements do not play such a central role. 

For a generic density operator, only three of the five correlators are independent. However, in equilibrium, the Kubo-Martin-Schwinger (KMS) relation \cite{Kubo:1957mj,PhysRev.115.1342}\footnote{
In addition, at equilibrium the correlators only depend on the time difference $t_1-t_0\equiv t$.} provides another constraint
\begin{equation}
    D_{>}(t)=D_{<}(t+i\beta),
\end{equation}
which in momentum space reads
\begin{equation}
    D_{>}(k^0)=\int dt e^{ik^0t}D_>(t)=e^{\beta k^0}D_{<}(k^0).
\end{equation}
Moreover, the cut propagators can be expressed in terms of the spectral function
\begin{align}
    D_{>}(K)&=\left(1+n_{\text{B}}\left(k^0\right)\right)\rho_{\text{B}}(K)
    \\D_{<}(K)&=n_{\text{B}}\left(k^0\right)\rho_{\text{B}}(K)
\end{align}
where we have reinstated the full momentum dependence, as the above relations will be used repeatedly throughout this thesis. 
Notice that for the case of fermions, there is a sign difference, i.e 
\begin{align}
    S_{>}(K)&=\left(1-n_{\text{F}}\left(k^0\right)\right)\rho_{\text{F}}(K)
    \\S_{<}(K)&=-n_{\text{F}}\left(k^0\right)\rho_{\text{F}}(K)
\end{align}
The spectral functions are related to the causal propagators by 
\begin{align}
    \rho_{\text{B}}(K)&=D_{R}(K)-D_{A}(K)
    \\\rho_{\text{F}}(K)&=S_{R}(K)-S_{A}(K)
\end{align}
and finally, we can write the momentum space analogues of Eqs.~\eqref{eq:to_pos}, ~\eqref{eq:ato_pos} as
\begin{align}
    D_{F}(K)=\frac{1}{2}\Big[D_{R}(K)+D_{A}(K)\Big]+\left(\frac{1}{2}+n_{\text{B}}\left(k^0\right)\right)\rho_{\text{B}}(K)\label{eq:feynman_prop_mom}
    \\D_{\bar{F}}(K)=-\frac{1}{2}\Big[D_{R}(K)+D_{A}(K)\Big]+\left(\frac{1}{2}+n_{\text{B}}\left(k^0\right)\right)\rho_{\text{B}}(K).
\end{align}
We also take this opportunity to define the symmetric \emph{rr}-propagator
\begin{equation}
    D_{rr}(K)\equiv\frac{1}{2}(D_{>}(K)+D_{<}(K))=\left(\frac{1}{2}+n_{\text{B}}\left(k^0\right)\right)\rho_{\text{B}}(K),\label{eq:grr}
\end{equation}
whose label will now be elaborated on.

While the ``1,2'' basis that we have used so far is easy to understand in terms of the Schwinger-Keldysh contour from Fig.~\ref{fig:sk}, it 
is not unique. We write down here the propagators associated with the so-called ``\emph{r/a}'' basis \cite{Keldysh:1964ud,Chou:1984es}, which 
is more convenient to work with for certain calculations. We start by defining
\begin{equation}
    \varphi_{r}\equiv\frac{1}{2}\left(\varphi_1+\varphi_2\right),\quad\quad \varphi_{a}\equiv\varphi_1-\varphi_2,
\end{equation}
which produces a propagator matrix
\begin{equation}
    \mathbf{D}=\left(\begin{array}{cc} \langle\varphi_r\varphi_r\rangle& \langle\varphi_r\varphi_a\rangle\\\langle\varphi_a\varphi_r\rangle &\langle\varphi_a\varphi_a\rangle\end{array}\right)
    =\left(\begin{array}{ccc} D_{rr}&&D_{R}\\D_{A}&& 0 \end{array}\right).
\end{equation}
The vanishing of the \emph{aa} propagator to all orders in perturbation theory can be seen by plugging the definitions Eqs.~\eqref{eq:to_pos}, ~\eqref{eq:ato_pos} into
$\langle\varphi_{a}\varphi_{a}\rangle$ and using that \\$\Theta(t_1-t_0)+\Theta(t_0-t_1)=1$. 

We take a moment to clarify our conventions in the \emph{r/a} formalism, as some of its features can appear quite confusing at first glance. 
As opposed to the 1,2 basis, where vertices 
are restricted to connect fields of the same type, there is mixing in the \emph{r/a} formalism. To elaborate on what is meant by this, consider the quartic interaction, 
translated to the \emph{r/a} basis, i.e 
\begin{equation}
    \frac{1}{4!}\left(\varphi_1^4-\varphi_2^4\right)=\frac{1}{2^2}\frac{1}{3!}\varphi_a^3\varphi_r+\frac{1}{3!}\varphi_r^3\varphi_a.\label{eq:vert_ra}
\end{equation}
\begin{figure}[t]
    \begin{center}
        \includegraphics[width=0.8\textwidth]{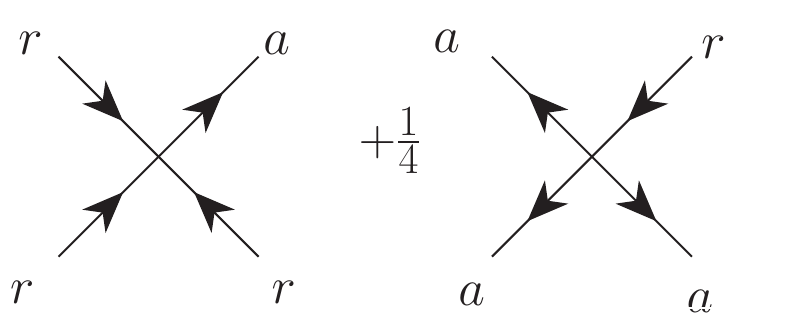}
    \end{center}
    \caption{Vertices appearing in the RTF for a scalar field with a quartic interaction in the \emph{r/a} basis.
    Image taken from \cite{Ghiglieri:2020dpq}.}
    \label{fig:phi4_ra}
\end{figure}
It is not difficult to check that this structure will persist even in QCD; there are only vertices with an odd number of \emph{a} fields, thus 
leaving us with only \emph{rra} and \emph{aaa} assignments. For 
vertices with more than one \emph{a}-field, there is an extra factor of $1/2$ for each extra \emph{a} line.
This extra prefactor is indicated in Fig.~\ref{fig:phi4_ra}. 
 
We emphasise that the labels in Fig.~\ref{fig:phi4_ra} refer to the nature of the field 
at the vertex itself. In other words, each of the propagators with an \emph{r} label in Fig.~\ref{fig:phi4_ra} could be either $\langle\varphi_r\varphi_r\rangle$ or $\langle\varphi_r\varphi_a\rangle$\footnote{
On the other hand, propagators with an \emph{a} label are fixed to be $\langle\varphi_a\varphi_r\rangle$, since $\langle\varphi_a\varphi_a\rangle$ is always zero.}. We further 
stress that the arrows in Fig.~\ref{fig:phi4_ra} refer to the direction of causality (as opposed to the direction 
of momentum). The causal arrow associated with retarded (advanced) propagators 
always goes in the same (opposite) direction as the momentum flow. For \emph{rr} propagators, causality instead flows in both directions, outward 
towards the vertices that the propagator is attached to.

\begin{figure}[t]
    \begin{center}
        \includegraphics[width=0.8\textwidth]{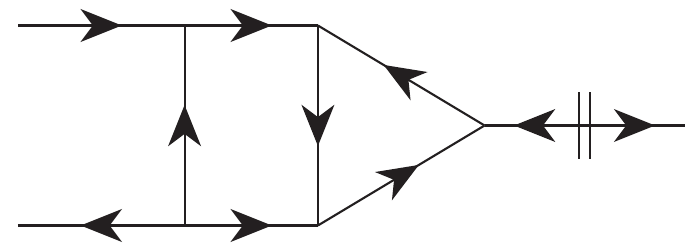}
        \put(-105,75){\Large{$t_0$}}
        \put(-175,115){\Large{$t_1$}}
    \end{center}
    \caption{Example of a diagram that vanishes due to causality. The line on the far-right 
    with a cut going through it is an \emph{rr} propagator.
    Image taken from \cite{Ghiglieri:2020dpq}.}
    \label{fig:closed_causation_loop}
\end{figure}
Despite the appearance of additional baggage, it is this transparency of causality flow that renders the \emph{r/a} basis useful. In particular, 
closed loops of causality, which cause diagrams to vanish are visibly manifest in this formalism. A closed loop of
causality is present when vertices at $t_0$, $t_1$, are connected by a product of (both) retarded and advanced propagators as is shown 
in Fig.~\ref{fig:closed_causation_loop}. Their vanishing can then be explicitly seen by noting that the step functions in Eqs.~\eqref{eq:r_pos}, ~\eqref{eq:a_pos} 
cannot be simultaneously satisfied.

Before concluding this section, for future reference, we write down both the 
free causal and \emph{rr} gluon propagators in Coulomb gauge \cite{Ghiglieri:2020dpq}. As was already mentioned, 
we forego commenting on the details related to gauge fixing at finite temperature
\begin{align}
    G^{00}_{R/A}(K)&=\frac{i}{k^2},\quad\quad G^{ij}_{R/A}(K)=\frac{i(\delta^{ij}-\hat{k}^i\hat{k}^j)}{K^2\pm i\varepsilon k^0}\label{eq:free_causal_c}
    \\G^{00}_{rr}(K)&=0,\quad\quad G^{ij}_{rr}(K)=(\delta^{ij}-\hat{k}^i\hat{k}^j)\left(\frac{1}{2}+n_{\text{B}}\left(|k^0|\right)\right)2\pi\delta(K^2)\label{eq:free_rr_c}
\end{align}
and Feynman gauge
\begin{align}
    G^{\mu\nu}_{R/A}(K)&=-\frac{i\eta^{\mu\nu}}{K^2\pm i\varepsilon k^0}=\eta^{\mu\nu}G_{R/A}(K)\label{eq:free_causal_f}
    \\G^{\mu\nu}_{rr}(K)&=-\eta^{\mu\nu}\left(\frac{1}{2}+n_{\text{B}}\left(k^0\right)\right)2\pi\varepsilon(k^0)\delta(K^2)=\eta^{\mu\nu}G_{rr}(K)\label{eq:free_rr_f},
\end{align}
where we note that the appearance of the delta function in the \emph{rr} propagators is due to the presence of the 
free spectral function 
\begin{equation}
    \rho_{\text{B}}(K)=2\pi\varepsilon(k^0)\delta(K^2).\label{eq:spec_free}
\end{equation}
For completeness, we also list the free causal and \emph{rr} propagators for fermions 
\begin{align}
    S_{R/A}(K)&=\frac{i\slashed{K}}{K^2\pm i\varepsilon k^0}\label{eq:free_causal_fermion}
    \\S_{rr}(K)&=\slashed{K}\left(\frac{1}{2}-n_{\text{F}}\left(k^0\right)\right)2\pi\varepsilon(k^0)\delta(K^2)\label{eq:free_rr_fermion},
\end{align}
where  
\begin{equation}
    \rho_{\text{F}}(K)=2\pi\varepsilon(k^0)\slashed{K}\delta(K^2).\label{eq:f_spec_free}
\end{equation}
is the free fermionic spectral function.
These expressions, along with the vertices specified in App.~\ref{sec:conventions} and the Hard Thermal Loop 
propagators to be given in the next section, supply the toolkit, which we draw from to perform real-time, perturbative 
calculations throughout the rest of the thesis.

\subsection{Hard Thermal Loops}\label{sec:htl}
In Sec.~\ref{sec:itf}, we highlighted the Linde problem \cite{Linde:1980ts} and mentioned that 
it can be in some sense bypassed through the use of dimensional reduction, which
provides a route towards the computation of IR finite thermodynamic quantities. Another method, which rather cures a naive
 modelling of the interaction between hard and soft modes is
supplied by Hard Thermal Loop (HTL) Effective Theory, first derived in the ITF
by Braaten and Pisarski \cite{Braaten:1989mz,Braaten:1991gm}, Frenkel and Taylor \cite{Frenkel:1989br,Frenkel:1991ts}
and Taylor and Wong \cite{Taylor:1990ia}. We choose to introduce the HTL in the RTF \cite{CaronHuot:2007nw}
so that the formulas presented here can be directly applied to the calculations that will follow in later sections.

HTLs are regions of phase space in loop diagrams
where the internal (loop) momentum is hard, i.e $\sim T$ and the momenta of the external legs is soft, i.e $\ll T$. 
It turns out that these loops can be added at no parametric cost and thus need to be 
resummed. To understand why, consider the retarded vacuum polarisation diagram with a hard thermal loop 
but soft external gluon legs. Using the Feynman rules from the previous section, taking $P\gg K$, where $P$ is the loop 
momentum and $K$ the external momentum \cite{Bellac:2011kqa}
\begin{equation}
    \Pi^{\mu\nu}_{R}(K)=-m_{D}^2\int\frac{d\Omega_v}{4\pi}\bigg(\delta^{\mu}_{0}\delta^{\nu}_{0}+v^{\mu}v^{\nu}\frac{k^0}{v\cdot K+i\varepsilon}\bigg),\label{eq:htl_self_c}
\end{equation}
where $m_D^2=g^2T^2(N_c+T_F N_F)/3$ is the leading order Debye mass, $v\equiv P/p^0=(1,\vec{\mathbf{p}}/p^0)$
and we are working in the Coulomb gauge. The key point 
is that, for $K\sim gT$, the denominator of the retarded free propagator from Eq.~\eqref{eq:free_causal_c} is parametrically 
of the same order as the two-point function above, thus clarifying the necessity of what is known 
as HTL resummation. This issue persists in diagrams with additional (soft) external gluons, implying the need 
for a reorganisation of perturbation theory.

At the level of the Lagrangian, this reorganisation is  
obtained by adding to the QCD Lagrangian the irrelevant operator 
\begin{equation}
    \delta\mathcal{L}_g=\frac{m_D^2}{2}\Tr\int\frac{d\Omega_v}{4\pi}F^{\mu\alpha}\frac{v_{\alpha}v_{\beta}}{(v\cdot D)^2}F^{\beta}_{\mu}.
\end{equation}
This operator generates all HTL correlators with $n\geq 2$ external (soft) gluons. We take this opportunity to define the condensate
\begin{equation}
    Z_g\equiv\frac{1}{d_R}\Big\langle F^{\mu\alpha}\frac{v_{\alpha}v_{\beta}}{(v\cdot D)^2}F^{\beta}_{\mu}\Big\rangle\label{eq:zg_def}
\end{equation}
The $D$ in the denominator is a covariant derivative, signifying the operator's nonlocality. This is a common feature 
of EFTs, which comes from integrating out the hard loop momentum, $P$. Note that this operator is manifestly gauge invariant.
It is of course necessary to perform an analogous resummation for diagrams with external soft quarks; all fermionic HTLs with two external, soft quark lines and an arbitrary 
number of soft external gluons can be generated with\footnote{It turns out that there are no fermionic HTL, i.e no 
amplitudes proportional to $g^2 T^2$ with more than two external fermion lines.} 
\begin{equation}    
    \delta\mathcal{L}_f=i\frac{m_{\infty\,q}^2}{2}\bar{\psi}\int\frac{d\Omega_v}{4\pi}\frac{\slashed{v}}{v\cdot D}\psi
\end{equation}
where $m_{\infty q}^2=g^2C_{F}T^2/4$ is the asymptotic mass of the quarks. Moreover, 
\begin{equation}
    Z_{f}\equiv\frac{1}{2d_{R}}\Big\langle\bar{\psi}\frac{v_{\mu}\gamma^{\mu}}{v\cdot D}\psi\Big\rangle\label{eq:zf_def}
\end{equation}
We are now going to shed some light on the physical scales that have popped up throughout the previous two sections by 
analysing the structure of the poles associated with the retarded HTL gluon propagator. After 
doing the angular integration in Eq.~\eqref{eq:htl_self_c}, we arrive at 
\begin{eqnarray}
    \label{htllong}
    G^{00}_{R}(K)&=&G^{L}_{R}(K)=\frac{i}{\displaystyle k^2+m_D^2\left(1-\frac{k^0}{2k}\ln\frac{k^0+k+ i\varepsilon}{k^0-k+ i\varepsilon}\right)},\\
    \nonumber G^{ij}_{R}(K)&=&(\delta^{ij}-\hat k^i\hat k^j)G^T_R(K)=
    \left.\frac{i(\delta^{ij}-\hat k^i\hat k^j)}
         {\displaystyle k_0^2-k^2-\frac{m_D^2}{2} \left(\frac{k_0^2}{k^2}
           -\left(\frac{k_0^2}{k^2}-1\right)\frac{k^0}{2k}
           \ln\frac{k^0{+}k}{k^0{-}k}\right)}\right\vert_{k^0=k^0+ i\varepsilon}.\\
    && 	\label{htltrans}
\end{eqnarray}
Let us first look at the static limit, i.e $k^0\to 0$
\begin{equation}
    G^{L}_{R}(K)(0,k)=\frac{i}{k^2+m_{D}^2},\quad\quad G^{T}_{R}(0,k)=-\frac{i}{k^2}.
\end{equation}
Here lies the physical interpretation associated with the Debye mass, $m_{D}$: static 
chromoelectric fields are screened because, at distances larger than the inverse Debye mass $\sim 1/gT$, they vanish exponentially. Static 
chromomagnetic fields are not screened in the HTL effective theory. However, as was mentioned in Sec.~\ref{sec:itf}, 
these non-perturbative or magnetic modes can be described with the use of the MQCD, which 
is just 3D (Euclidean) Yang Mills theory. From LQCD calculations \cite{Karsch:1998tx,Cucchieri:2000cy}, it 
has been confirmed that a mass is non-perturbatively generated at a distance $1/g^2T$, thus confining chromomagnetic modes at that scale.
It can also happen that magnetic gluons are \emph{dynamically screened}, that is, in a frequency-dependent way \cite{Weldon:1982aq}.
Examples of such a phenomenon occur, in the calculation of the energy loss of a heavy fermion in a QED plasma \cite{Braaten:1991jj}
and transport coefficients in a QCD plasma \cite{Baym:1990uj}.

For $K^2>0$ both the longitudinal and transverse propagators possess so-called
\emph{plasmon poles}, which represent collective plasma excitations at the scale $gT$. In general, one can only 
compute them numerically but at vanishing $k$, the distinction 
between longitudinal and transverse modes vanishes. The resulting dispersion relation then gives the plasma frequency, 
$\omega_{P}=\pm m_{D}/\sqrt{3}$, which, by definition describes the oscillation frequency 
for vanishing wave vectors, namely, spatially uniform oscillations.

For spacelike $K^2$, both propagators evidently take on complex values due to the presence of the logarithms. This 
is in contrast with the situation at zero-temperature where, according to the optical theorem, the only process
that is kinematically allowed to take place is pair production, in the forward timelike region. Indeed, the 
leading contribution in $T$ reflects the fact that the soft gluon can scatter off the on-shell 
thermal particles that already exist in the bath and in doing so, become damped. This phenomenon,
known as \emph{Landau damping}\footnote{Landau damping is responsible for the LO soft contribution to $\hat{q}$, 
to be discussed in Sec.~\ref{sec:soft_jq}.}, is another example of the dynamical screening phenomenon just mentioned. The computation of the gluon damping rate
 at vanishing momentum, along with the proof of its gauge invariance represents
one of the first successes of the HTL approach \cite{Pisarski:1990ds,Braaten:1990it}.

In the limit where $k\gg m_{D}$, the residue associated with the longitudinal pole 
vanishes exponentially \cite{Pisarski:1989wb}. However, the transverse modes survive with unitary residue 
and a pole at $k^0(k\gg m_{D})=\sqrt{k^2+m_{\infty g}^2}$, where $m_{\infty g}=m_{D}/\sqrt{2}$ is the asymptotic mass of the gluon 
\cite{Kalashnikov:1979cy,Weldon:1982aq}. This is nothing other than the dispersion relation for a gluon with momentum $\gtrsim T$, i.e 
those which populate the plasma (or jets, with $K \gg T$), which is not necessarily 
surprising, since we are probing the region where the HTL 
result agrees with the full one-loop result, where $k^0\sim k\gtrsim T$ and $k^0-k \ll T$. The takeaway message, which has been demonstrated in a somewhat roundabout way is that particles moving with large momentum (with 
respect to the plasma) have their dispersion relation shifted. As they approach the lightcone, this shift becomes 
a constant ``mass'', which we define as the \emph{asymptotic mass}. To leading order in $g$, we have
\begin{equation}
    m_{\infty g}^2=\Pi^{T}_{R}(k^0=k)=\frac{m_{D}^2}{2}.
\end{equation}
The expression above can be generalised and related to the condensates from Eqs.~\eqref{eq:zg_def}, ~\eqref{eq:zf_def} \cite{CaronHuot:2008uw}
\begin{align}\label{eq:asy_mass}
    m_{\infty\,q}^2&=g^2C_{F}(Z_g+Z_f),
    \\m_{\infty\,g}^2&=g^2(C_{A}Z_g+2T_{F}N_{F}Z_f).
\end{align}
In the context of jet propagation, these relations are understood to apply only in the limit where the energy of the jet, $E$ 
is large with respect to the temperature and can therefore, in the EFT language, be integrated out. Subsequent terms in the $T/E$ expansion (not shown 
above) can then be safely neglected\footnote{For the case where $E\gtrsim T$, higher orders in the 
$T/E$ expansion become relevant. In addition, higher order terms may contain operator that mix the gluonic and fermionic condensates.}. 
Indeed, we capitalise on the relation provided by Eq.~\eqref{eq:asy_mass} to compute higher order corrections 
to the asymptotic mass in Sec.~\ref{sec:nlo_asym} and Ch.~\ref{ch:asym_mass}\footnote{In fact, we will only compute 
corrections to $Z_g$, since its fermionic counterpart does not receive classical corrections. The reason for this is 
given in Sec.~\ref{sec:chtrick}.}

We refrain from performing an analysis of the fermionic HTL, which can be found, for example in \cite{Kapusta:2006pm,Bellac:2011kqa}.

To finish this section, we give the retarded HTL propagator in Feynman gauge, to be used in Sec.~\ref{sec:nlo_asym}
\begin{align}
    G^R_{\mu\nu}(K)&=\frac{i}{K^2+\frac{m_D^2 K^2}{k^2}\Big(1-\frac{k^0}{2k}\ln\frac{k^0+k}{k^0-k}\Big)}P^L_{\mu\nu}+
    \frac{i}{K^2-\frac{m_{D}^2}{2}\Big((\frac{k^0}{k})^2+(1-(\frac{k^0}{k})^2)\frac{k^0}{2k}\ln\frac{k^0+k}{k^0-k}\Big)}P^T_{\mu\nu}\nonumber
    \\&-\frac{i}{K^2}\frac{K_{\mu}K_{\nu}}{K^2}.\label{eq:feyn_htl}
\end{align}
The projectors are defined as
\begin{align}
    P_{L}^{\mu\nu}(K)&=-\eta^{\mu\nu}+\frac{K^{\mu}K^{\nu}}{K^2}-P_T^{\mu\nu}
    \label{eq:lprojector}
    \\P_{T}^{\mu\nu}(K)&=-\eta^{\mu\nu}+u^{\mu}u^{\nu}-\frac{K^{\mu}K^{\nu}
    -(K\cdot u)K^{\mu}u^{\nu}-(K\cdot u)K^{\nu}u^{\mu}+(K\cdot u)^2 u^{\mu}u^{\nu}}{k^2}
    \label{eq:tprojector}
    \\(u^{\mu})&=(1,0,0,0)
    \label{eq:bathframe}
\end{align}
where $u^{\mu}$ specifies the frame of the thermal bath.

\section{Thermal Correlators on the Lightcone}\label{sec:chtrick}

Something which is not so evident from the previous dicussion is that full-blown analytical computations with HTL effective theory, 
using propagators and vertices, (the latter of which have not been included in Sec.~\ref{sec:htl}) are extremely 
laborious and cumbersome. Take as an example the NLO computation of the heavy quark diffusion constant, $\kappa$ \cite{Caron-Huot:2007zhp,simonguy}, 
for which a numerical determination turns out to be neccesary.

Be that as it may, an insight from Caron-Huot \cite{CaronHuot:2008ni} has paved the way for the non-perturbative determination of contributions 
from the soft and ultrasoft scales to certain thermal correlators on the lightcone. Here, we present a sketch of 
Caron-Huot's result before applying it to calculations of $\mathcal{C}(k_{\perp})$ and $m_{\infty}$ in the following sections.

Following \cite{Ghiglieri:2013gia,Ghiglieri:2015zma}, let us start by considering the retarded correlator 
\begin{equation}    
    G_{R}^{\mu\nu}(K)=\int_{X}e^{i(k^+x^- +k^-x^+ -\mathbf{k}\cdot\mathbf{x})}G^{\mu\nu}_{R}(X).
\end{equation}
It was realised in \cite{CaronHuot:2008ni} that this Fourier transform provides analytic continuation (for fixed $k^-, \mathbf{k}$) for $k^+$ into the upper half-plane.
This is so because the retarded response function only has support in the forward lightcone, where $2x^+x^-\geq x_{\perp}^2$. In 
particular the exponential factor $\exp(ik^+x^-)$ will always be tamed in forward lightcone, where $x^->0$. This application 
of causality will be used repeatedly throughout this thesis.

In order to see how these kinds of arguments can be used to connect the evaluation of soft corrections with EQCD,
 consider the symmetric correlator, keeping $X$ general for the moment
\begin{align}
    G^{\mu\nu}_{rr}(X)&=\int_{K}e^{iK\cdot X}G^{\mu\nu}_{rr}(K)
    \\&=\int_{K}e^{iK\cdot X}\left(\frac{1}{2}+n_{\text{B}}(k^0)\right)\rho^{\mu\nu}_{\text{B}}(K),
\end{align}
where we have used Eq.~\eqref{eq:grr} in going to the second line. Upon shifting $\tilde{k}^z=k^z-k^0(x^0/x^z)$, we find
\begin{align}
    G^{\mu\nu}_{rr}(X)&=\int_{K}e^{-i(x^z\tilde{k}^z+\mathbf{x}\cdot\mathbf{k})}\left(\frac{1}{2}+n_{\text{B}}(k^0)\right)\rho^{\mu\nu}_{\text{B}}\Big(k^0,\tilde{k}^z+k^0(x^0/x^z),\mathbf{k}\Big)\nonumber
    \\&=\int_{K}e^{-i(x^z\tilde{k}^z+\mathbf{x}\cdot\mathbf{k})}\left(\frac{1}{2}+n_{\text{B}}(k^0)\right)\Big(G^{\mu\nu}_{R}(k^0,\tilde{k}^z+k^0(x^0/x^z),\mathbf{k})\nonumber
    \\& -G^{\mu\nu}_{A}(k^0,\tilde{k}^z+k^0(x^0/x^z),\mathbf{k})\Big).\label{eq:ch_shift}
\end{align}
At this point, we can try to do the $k^0$ integration, taking advantage of the analycity of the causal propagators;
we integrate the retarded (advanced) term by closing a contour in the upper (lower)
half-plane, only picking up contributions from the poles associated with the statistical function, situated at $k^0=i\omega_n$, the 
Matsubara frequencies.
However, in order to perform this maneuver, we need to be sure that the shift performed earlier does not affect the analytical 
properties of the causal propagators. At the level of the bare retarded propagator, for instance (in Feynman gauge)
\begin{equation}
    G^{\mu\nu}_R(k^0,k^z,\mathbf{k})\longrightarrow G^{\mu\nu}_R(k^0,\tilde{k}^z+k^0(x^0/x^z),\mathbf{k})=\frac{ig^{\mu\nu}}{(k^0 +i\varepsilon)^2-(\tilde{k}^z+\frac{x^0}{x^z}k^0)^2-k_{\perp}^2}, \label{eq:beforeinter}
\end{equation}
giving rise to the poles
\begin{equation}
    k^0_p=\frac{
    \tilde{k}^z\frac{x^0}{x^z}\pm\sqrt{{\tilde{{k}^z}}^2+k_{\perp}^2\left(1-\frac{{x^0}^2}{{x^z}^2}\right)}}{1-\frac{{x^0}^2}{{x^z}^2}}.\label{eq:ret_poles}
\end{equation}
Thus, we need to demand that $|x^0/x^z|\leq1$\footnote{There is no problem for the case where $|x^0/x^z|=1$ either; one finds from
Eq.~\eqref{eq:beforeinter} that $k^0$ parameterises analytic continuation to the upper $k^+$ half plane.}.
 
By doing the $k^0$ integration in Eq.~\eqref{eq:beforeinter}, we arrive at the expression with a sum over Matsubara frequencies, $\omega_n=2\pi nT$
\begin{equation}
G^{\mu\nu}_{rr}(X)=T\sum_{n}\int_{k}e^{-i(x^z\tilde{k}^z+\mathbf{x}\cdot\mathbf{k})}G^{\mu\nu}_{E}(\omega_n,\tilde{k}^z+i\omega_n(x^0/x^z),\mathbf{k})
\end{equation}
where we have used that the retarded Green's function is determined by 
an analytic continuation of the Euclidean function $iG^{\mu\nu}_{R}(k^0,\vec{\mathbf{k}})=G^{\mu\nu}_{E}(i\omega_{E}=k^0+i\varepsilon,\vec{\mathbf{k}})$ .

At this point, it perhaps seems as if we have not made much progress, since we have just traded an integral for an infinite sum.
 Nevertheless, the power of this technique becomes evident if one is interested in computing the soft contribution to this 
 symmetric correlator. In this case, the zero mode dominates so that one may write, renaming $\tilde{k}^z\to k^z$
\begin{equation}
G^{\mu\nu}_{rr}(X)=T\int_{k}e^{-i(x^zk^z+\mathbf{x}\cdot\mathbf{k})}G^{\mu\nu}_{E}(0,k^z,\mathbf{k}),
\end{equation}
only keeping the $\omega_n=0$ term. Remarkably, $G_{E}(0,k^z,\mathbf{k})$ is just the Euclidean correlation function of three-dimensional theory EQCD, 
which was discussed in Sec.~\ref{sec:dim_red}, 
signifying the time independence of the zero mode contribution. This realisation greatly simplified the analytical calculation of $\mathcal{O}(g)$ 
corrections to $\mathcal{C}(k_{\perp})$ (and hence $\hat{q}(\mu_{\perp}))$, which were then completed in \cite{CaronHuot:2008ni}. 
But the major implication associated with this breakthrough is that one may compute such a correlator with LQCD without having to worry about 
analytical continuation back from imaginary time to real time. Moreover, lattice EQCD computations can provide a non-perturbative determination of the
contributions from the soft scale, $gT$ in addition to those 
from the ultrasoft scale $g^2T$. We will discuss this further in Sec.~\ref{sec:np_classical}.

To give an intermediate summary, we have seen how the troubling IR structure of finite-temperature QCD can 
be treated with two methods: dimensional reduction and HTL effective theory. While it is perhaps not surprising 
that there exists some equivalence between these two methods, Caron-Huot's work provides a concrete, robust mapping 
between the two: for thermal correlators on the lightcone, which are usually computed with HTL because of their time-dependence, it is 
possible to compute their zero-mode contribution with EQCD.

\begin{figure}[h]
	\centering
	\includegraphics[width=1\textwidth]{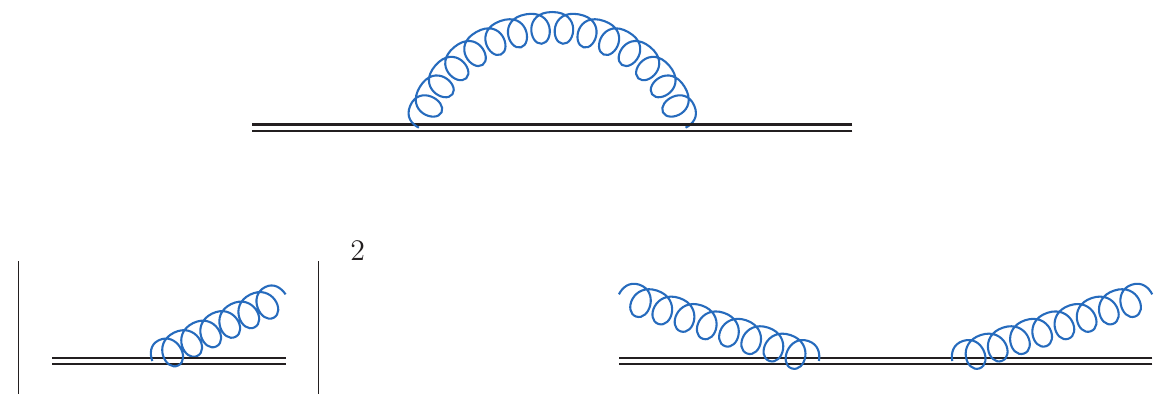}
	\caption{Schematic decomposition of a one-loop jet propagator (with an internal soft gluon) into its 
    real and imaginary parts. In the main text, 
    we use this as a device to give some intuition behind the 
    transverse scattering rate and the asymptotic mass. Note that one of the gluons needs to be soft in order to give 
    a non-zero contribution to the scattering rate; a cut through two hard gluons would yield a kinematically 
    impossible process.}
    \label{fig:re_im}
\end{figure}

\section{Classical corrections to Jet Quenching}\label{sec:soft_jq}
We proceed to demonstrate the power of the tools that have been introduced in the previous sections through a computation of the LO 
soft contribution to the transverse scattering rate, $\mathcal{C}(k_{\perp})$, 
which was introduced in Sec.~\ref{sec:rrate} as well as the LO soft contribution to the asymptotic mass, $m_{\infty}$. 

It will nevertheless be insightful to first consider the top diagram from Fig.\ref{fig:re_im}, representing 
the one-loop correction to a quark or gluon jet propagator, where the 
blue gluon represents an HTL propagator. By separating 
the imaginary and real parts of that diagram, we can schematically 
distinguish two kinds of jet-medium interactions:
\begin{itemize}
    \item By the optical theorem, the imaginary part of the top diagram from Fig.\ref{fig:re_im} 
        is given by cutting through the loop in that diagram, putting the cut propagators on-shell and 
        then squaring the process defined by the cut, as is shown in Fig.\ref{fig:re_im} - bottom left.
        This a \emph{true scattering} process, one where the jet actually exchanges transverse momentum 
        with a medium constituent and these kinds of effects are captured by the 
        transverse scattering rate. In the Wilson loop picture, which will be introduced shortly, 
        Fig.\ref{fig:re_im} - bottom left can be thought of as corresponding to Fig.~\ref{fig:leading_order} - left.
    \item Conversely, Fig.\ref{fig:re_im} - bottom right  corresponds to \emph{forward scattering}\footnote{The forward scattering process depicted on the 
    bottom right of Fig.~\ref{fig:re_im} also includes a contribution where the gluons are crossed.}. Scattering 
        off a medium constituent causes the jet to temporarily changes direction before it scatters again, forcing it back into its
        original momentum state, which induces a phase shift. It is this phase shift that induces a dispersion correction, otherwise 
        known as the asymptotic mass. For the case of a quark jet, there is an additional diagram 
        not shown in Fig.~\ref{fig:re_im} with the quark line cut.
\end{itemize}
Such a distinction should shed some light on the form of Eq.~\ref{eq:ham_ready}. There, the imaginary 
part of the Hamiltonian contains the interaction terms. Having provided some intuition using the points 
made above, we are now ready to start on the leading order calculations of $\mathcal{C}(k_{\perp}),\,m_{\infty}$.
We take this opportunity to explain the various calculational steps in some detail as many such 
manoeuvres will be used throughout Chs.~\ref{chap:qhat_chap} and ~\ref{ch:asym_mass}.

\subsection{Leading Order Soft Contribution to \texorpdfstring{$\mathcal{C}(k_{\perp})$}{TEXT}}\label{sec:scatt_lo_soft}
We wish to compute the transverse scattering rate associated with a parton passing through a medium of constant length $L_M$.
As was done throughout the previous chapter, the parton is assumed to propagate eikonally in the $x^+$ direction with momentum $P=(E,0,0)$ and $E\gg T$, the 
temperature of the plasma.

\begin{figure}[h]
	\centering
	\includegraphics[width=1\textwidth]{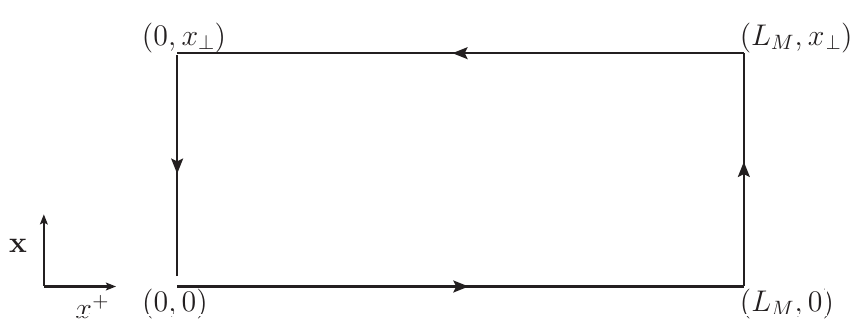}
	\caption{Wilson line from Eq.~\eqref{eq:Wloopdef}. The horizontal lines can be interpreted as the hard parton in the amplitude (bottom) 
    and the conjugate amplitude (top). The side rails are inserted for the sake of maintaining gauge invariance but will not contribute 
    in, for instance, the Coulomb gauge. The small gap in the bottom left corner of the Wilson loop in Fig.~\ref{fig:wloop_full}
    is intended to emphasise that the trace is over colour only; the fields themselves cannot be moved using the cyclicity of the trace. Reproduced from \cite{Ghiglieri:2015zma}.}
    \label{fig:wloop_full}
\end{figure}

The route of our computation begins by looking at the expectation value of the Wilson loop \cite{CasalderreySolana:2007qw,DEramo:2012uzl,Benzke:2012sz}, 
defined in the $x^{-}=0$ plane as
\begin{equation}
    \langle W(x_{\perp})\rangle=\frac{1}{N_{c}}\Tr\langle [0,x_{\perp}]_{-}\mathcal{W}^{\dagger}(x_{\perp})[x_{\perp},0]_{+}\mathcal{W}(0)\rangle, \label{eq:Wloopdef},
    \end{equation}
    where the trace is over colour. The parton's eikonal nature allows us to describe its propagation with the Wilson line $\mathcal{W}$ (the horizontal lines in Fig.~\ref{fig:wloop_full}), which incorporates an infinite 
    number of gluon exchanges between the parton and the medium
    \begin{equation}
        \mathcal{W}(x_{\perp})=\mathcal{P}\exp\Big(ig\int_{-\frac{L_M}{2}}^{\frac{L_M}{2}}dx^{+}A^{-}(x^{+},x_{\perp})\Big).\label{eq:wline_def}
    \end{equation}
    In order to preserve gauge invariance the ``side rails'' also need to be included \cite{Benzke:2012sz}
    \begin{equation}
        [x_{\perp},y_{\perp}]_{\pm}=\mathcal{P}\exp\Big(-ig\int_{1}^{0}ds(y_{\perp}-x_{\perp})\cdot A_{\perp}(\pm \frac{L_M}{2},x_{\perp}+(y_{\perp}-x_{\perp})s)\Big),\label{eq:side_rails}
    \end{equation}
    although in non-singular gauges, they will not contribute\footnote{See Appendix~\ref{app:siderails} for an
     explicit demonstration of this point.}. 
    The fields in the Wilson loop are to be understood as transforming in the representation, $R$ of the jet particle and as path-ordered. 
    Thus, in rough analogy with Fig.~\ref{fig:gluon_emission_lpm}, one may think of the eikonalised hard quark in the amplitude as 
    $\mathcal{W}(0)$ and $\mathcal{W}^{\dagger}(x_{\perp})$ as its conjugate-amplitude counterpart.
    The zero-subtracted Fourier transform of $\mathcal{C}_{R}(k_{\perp})$, defined in Eq.~\eqref{eq:scatt_ker_pos} 
    is then obtained by expanding
    \begin{equation}
        \lim_{L_M \to \infty}\langle W(x_{\perp})\rangle=\exp(-\mathcal{C}_{R}(x_{\perp})L_M) \label{eq:cdef}
    \end{equation}
in $g$. In order to declutter notation, we remove the $R$ subscript, i.e $\mathcal{C}(k_{\perp})\equiv\mathcal{C}_{R}(k_{\perp})$ and do 
the same for its Fourier transform. Similarly, $\hat{q}(\mu_{\perp})\equiv\hat{q}_{R}(\mu_{\perp})$ (unless otherwise stated).
Schematically, we expect such an expansion to have the following structure
\begin{equation}
    \lim_{L_M \to \infty}\langle W(x_{\perp})\rangle= 1-\left(\mathcal{C}_{\text{LO}}(x_{\perp})+\mathcal{C}_{\text{NLO}}(x_{\perp})+...\right)L_M 
    +\frac{\left(\mathcal{C}_{\text{LO}}(x_{\perp})+\mathcal{C}_{\text{NLO}}(x_{\perp})+...\right)^2}{2}L_M^2+...\label{eq:wloopexpansion}
\end{equation}
where the $\mathcal{C}_{\text{NLO}}$ term (in the vacuum) is suppressed by $\mathcal{O}(g^2)$ compared to $\mathcal{C}_{\text{LO}}$. By expanding 
$\lim_{L_M \to \infty}\langle W(x_{\perp})\rangle$ in $g$, we can then read off $\mathcal{C}(x_{\perp})$ to whatever order 
we are interested in, by looking at the coefficient multiplying $L_M$ in Eq.~\eqref{eq:wloopexpansion}. See App.~\ref{app:wloop} for a discussion on how to expand $\mathcal{W}$, while properly taking care of path-ordering.
\begin{figure}[h]
	\centering
	\includegraphics[width=1\textwidth]{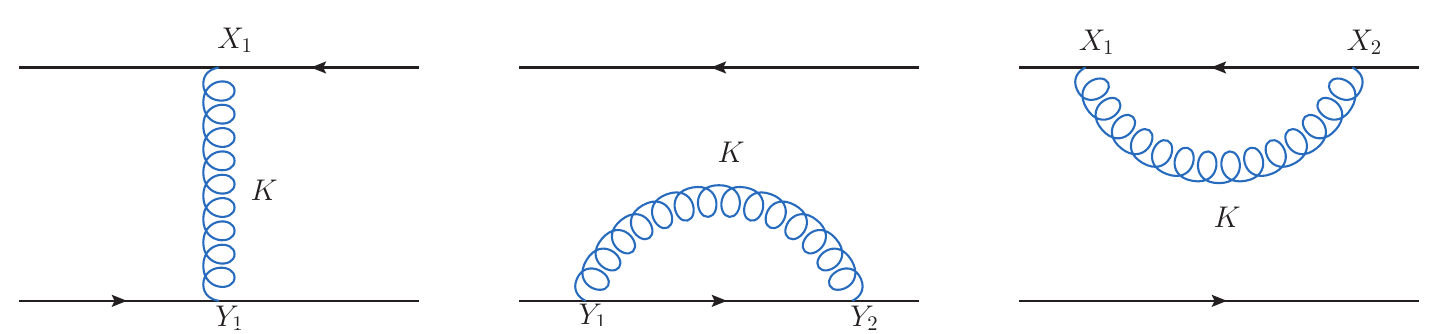}
	\caption{Three diagrams contributing to $\mathcal{C}(x_{\perp})$ at LO. The lower
    Wilson line is time-ordered, admitting only
    type ``1'' vertices whereas
    the lower one is anti-time-ordered and admits only type ``2'' vertices. In lightcone coordinates $X_1=(x_1^+,0,\mathbf{x}),\,X_2=(x_2^+,0,\mathbf{x})$
    and $Y_1=(y_1^+,0,0)$. We colour the exchanged gluon as blue in order to denote that it is soft, i.e its momenta is $K\sim gT$ and thus 
    must be described with an HTL propagator.}
    \label{fig:leading_order}
\end{figure}

Starting from Eq.~\eqref{eq:cdef}, we recover three terms at the first non-trivial order, corresponding to the three diagrams in Fig.~\ref{fig:leading_order}. We denote 
the leftmost diagram as ``real'' and the other two ``virtual''\footnote{At higher orders, 
there can also be ``partially virtual'' diagrams. In the simplest case, this class of diagrams
has one gluon loop attached to one of the $\mathcal{W}$ lines with 
another stretching between the two. An example is given in the left diagram of Fig.~\ref{fig:virtual}.}. From left to right
they read, respectively
\begin{align}
    g^{2}C_{R}\int_{-\frac{L_M}{2}}^{\frac{L_M}{2}}dx_{1}^{+}\int_{-\frac{L_M}{2}}^{\frac{L_M}{2}}dy_{1}^{+}&\langle A^{-}(x_{1}^{+},x_{\perp})A^{-}(y^{+},0)\rangle\nonumber
    \\&=g^{2}C_{R}\int_{-\frac{L_M}{2}}^{\frac{L_M}{2}}dx_{1}^{+}\int_{-\frac{L_M}{2}}^{\frac{L_M}{2}}dy_{1}^{+}\int_{K}e^{-ik^{-}(x_{1}^{+}-y_{1}^{+})}e^{i\mathbf{k}\cdot \mathbf{x}}G_{>}^{--}(k),\label{eq:realterm}
    \\-g^{2}C_{R}\int_{-\frac{L_M}{2}}^{\frac{L_M}{2}}dy_{2}^{+}\int_{-\frac{L_M}{2}}^{y_{2}^{+}}dy_{1}^{+}&\langle A^{-}(y_{2}^{+},0)A^{-}(y_{1}^{+},0)\rangle\nonumber
    \\&=-g^{2}C_{R}\int_{-\frac{L_M}{2}}^{\frac{L_M}{2}}dy_{2}^{+}\int_{-\frac{L_M}{2}}^{y_{2}^{+}}dy_{1}^{+}\int_{K}e^{-ik^{-}(y_{2}^{+}-y_{1}^{+})}G_{F}^{--}(k), \label{eq:vtermone}
    \\-g^{2}C_{R}\int_{-\frac{L_M}{2}}^{\frac{L_M}{2}}dx_{2}^{+}\int_{-\frac{L_M}{2}}^{x_{2}^{+}}dx_{1}^{+}&\langle A^{-}(x_{1}^{+},x_{\perp})A^{-}(x_{2}^{+},x_{\perp})\rangle\nonumber
    \\&=-g^{2}C_{R}\int_{-\frac{L_M}{2}}^{\frac{L_M}{2}}dx_{2}^{+}\int_{-\frac{L_M}{2}}^{x_{2}^{+}}dx_{1}^{+}\int_{K}e^{-ik^{-}(x_{2}^{+}-x_{1}^{+})}G_{\tilde{F}}^{--}(k).\label{eq:vtermtwo}
\end{align}
 Before combining the three terms, it is useful to rewrite the first one  
 \begin{align}
    \text{Eq.}\eqref{eq:realterm}&=g^{2}C_{R}\int_{-\frac{L_M}{2}}^{\frac{L_M}{2}}dx_{1}^{+}\int_{-\frac{L_M}{2}}^{\frac{L_M}{2}}dy_{1}^{+}\int_{K}e^{-ik^{-}(x_{1}^{+}-y_{1}^{+})}e^{i\mathbf{k}\cdot \mathbf{x}}G_{>}^{--}(k)\Big(\Theta(x_{1}^{+}-y_{1}^{+})+\Theta (y_{1}^{+}-x_{1}^{+})\Big)\nonumber
    \\&=g^{2}C_{R}\Big[\int_{-\frac{L_M}{2}}^{\frac{L_M}{2}}dx_{1}^{+}\int_{-\frac{L_M}{2}}^{x_{1}^{+}}dy_{1}^{+}\int_{K}e^{-ik^{-}(x_{1}^{+}-y_{1}^{+})}e^{i\mathbf{k}\cdot \mathbf{k}}G_{>}^{--}(k)\nonumber
    \\&+\int_{-\frac{L_M}{2}}^{\frac{L_M}{2}}dy_{1}^{+}\int_{-\frac{L_M}{2}}^{y^{+}_{1}}dx_{1}^{+}\int_{K}e^{-ik^{-}(x_{1}^{+}-y_{1}^{+})}e^{i\mathbf{k}\cdot \mathbf{x}}G_{>}^{--}(k)\Big]\nonumber
    \\&=g^{2}C_{R}\Big[\int_{-\frac{L_M}{2}}^{\frac{L_M}{2}}dx_{1}^{+}\int_{-\frac{L_M}{2}}^{x_{1}^{+}}dy_{1}^{+}\int_{K}e^{-ik^{-}(x_{1}^{+}-y_{1}^{+})}e^{i\mathbf{k}\cdot \mathbf{x}}G_{>}^{--}(k)\nonumber
    \\&+\int_{-\frac{L_M}{2}}^{\frac{L_M}{2}}dy_{1}^{+}\int_{-\frac{L_M}{2}}^{y^{+}_{1}}dx_{1}^{+}\int_{K}e^{-ik^{-}(x_{1}^{+}-y_{1}^{+})}e^{i\mathbf{k}\cdot \mathbf{x}}G_{>}^{--}(k)\Big]\label{eq:real}
\end{align}
The second term above then can be combined with the first by relabelling $x_1^+\to y_1^+, y_1^+\to x_1^+$ and using invariance in the transverse plane. Upon adding Eqs.~\eqref{eq:real}, ~\eqref{eq:vtermone} and ~\eqref{eq:vtermtwo},
we obtain\footnote{A slighter quicker way to arrive at Eq.~\eqref{eq:cinter} would be to use that the $A^-$ fields 
are spacelike separated and therefore commute, implying that $G_{rr}(K)=G_>(K)=G_{<}(K)$.}
\begin{equation}
    \mathcal{C}(x_{\perp})_{\text{LO}}L_{M}=2g^{2}C_{R}\int_{K}(1-e^{i\mathbf{k}\cdot\mathbf{x}})\int_{-\frac{L_M}{2}}^{\frac{L_M}{2}}dx_{1}^{+}\int_{-\frac{L}{2}}^{x_{1}^{+}}dy_{1}^{+}e^{-ik^{-}(x_{1}^{+}-y_{1}^{+})}G^{--}_{rr}(k),\label{eq:cinter}
\end{equation}
where we have also used Eq.~\eqref{eq:grr}. In order to make the position integration converge, we need to add
a small imaginary part to $k^-$ so that the exponential decays for $x_1^+\gg y_1^+$, i.e 
\begin{align}
    \int_{-\frac{L_M}{2}}^{\frac{L_M}{2}}dx_{1}^{+}\int_{-\frac{L_M}{2}}^{x_{1}^{+}}dy_{1}^{+}e^{-i(k^{-}-i\varepsilon)(x_{1}^{+}-y_{1}^{+})}
    &=\int_{-\frac{L_M}{2}}^{\frac{L_M}{2}}dx_{1}^{+}\frac{-i}{k^- -i\varepsilon}\nonumber
    \\ &=\frac{-i}{k^- -i\varepsilon}L_M,
\end{align}
where we have neglected the term with $e^{-\varepsilon L_M}$ factor. This yields, for the scattering kernel in position space 
\begin{equation}
    \mathcal{C}(x_{\perp})_{\text{LO}}=2g^{2}C_{R}\int_{\mathbf{k}}(1-e^{i\mathbf{k}\cdot\mathbf{x}})\int\frac{dk^+dk^-}{(2\pi)^2}\frac{-i}{k^- -i\varepsilon}G^{--}_{rr}(k).
\end{equation}
Using Eq.~\eqref{eq:scatt_ker_pos}, we can then easily read off $\mathcal{C}(k_{\perp})$. Indeed, since the contribution from real
diagrams always multiply the factor $e^{i\mathbf{k}\cdot\mathbf{x}}$ in $\mathcal{C}(x_{\perp})$, 
it is only necessary to compute the real diagrams in order to extract $\mathcal{C}(k_{\perp})$; the $x_{\perp}$-independent 
term contributes to probability conservation. We will make use of this simplification
throughout Ch.~\ref{chap:qhat_chap}.
 
Using $G_{rr}(K)$ is even in $K$ we then get
\begin{equation}
\mathcal{C}(k_{\perp})_{\text{LO}}=2\pi g^2 C_{R}\int\frac{dk^+ dk^-}{(2\pi)^2}\delta(k^-)\left(\frac{1}{2}+n_{\text{B}}(k^0)\right)\rho_{\text{B}}^{--}(K).\label{eq:afterinter}
\end{equation}
At this stage, we can deduce from the presence of the statistical function that $\mathcal{C}(k_{\perp})_{\text{LO}}$ receives \emph{classical 
contributions}. These classical contributions originate from the $gT$ and $g^2T$ scales where the exchange of gluons between the jet and the medium has a large occupation number
\footnote{The contribution from 
the thermal scale, $k_{\perp}\sim T$ has been computed in \cite{Arnold:2008vd} but turns out to be subleading depending
 on the exact value of $k_{\perp}$. See Eq.~\eqref{eq:scat_hard}. In any case, we concentrate 
on the LO contribution from the soft scale here because it gives us an opportunity to showcase the power of the Euclidean methods introduced in Sec.~\ref{sec:chtrick}.}. 
Such a condition is satisfied when the exchanged gluon has energy $\ll T$ so that the statistical function can be approximated 
as $n_{\text{B}}(k^0)\approx T/k^0\gg 1$.

We additionally assume that $k_{\perp}\gtrsim gT$. This is then 
the exact scaling associated with the gluon exchanges that dominate the multiple scattering regime (see Sec.~\ref{sec:rrate}).

As was discussed in Sec.~\ref{sec:htl} propagators with soft momenta running through them must be described by HTL propagators, which 
resum an infinite number of internal hard thermal loops. Going back to Eq.~\eqref{eq:afterinter}, we 
can make use of the ideas from Sec.~\ref{sec:chtrick}, shifting $k^z\rightarrow k^z+k^0$\footnote{Our choice 
of lightcone coordinates (see App.~\ref{sec:conventions}) ensures that the change 
of variables $\{k^+,k^-\}\rightarrow\{k^0,k^z\}$ comes with a unit Jacobian.}
\begin{equation}
    \mathcal{C}(k_{\perp})_{\text{LO}}=g^2 C_{R}\int\frac{dk^0}{(2\pi)}\left(\frac{1}{2}+n_{\text{B}}(k^0)\right)\rho_{\text{B}}^{--}(K)\Big\vert_{k^z=0}.
\end{equation}
The analytical properties of $G_{rr}^{--}(K)$ in the $k^0$ plane 
effectively allow us to let $k^0\rightarrow i\omega_n$ and trade $\int dk^0/2\pi\to iT\sum_{n}$, where we recall that the sum is over the Matsubara modes with $\omega_n=2\pi nT$
and the definitions of the HTL propagators (in Coulomb gauge) from Eqs.~\eqref{htllong}, \eqref{htltrans}
\begin{align}
    \mathcal{C}(k_{\perp})_{\text{LO}}&=iTg^2 C_{R}\sum_{n}\bigg(\frac{i}{-\omega_n^2+k_{\perp}^2+m_{D}^2\left(1-\frac{i\omega_n}{2\sqrt{-\omega_n^2+k_{\perp}^2}}\ln\frac{i\omega_n+\sqrt{-\omega_n^2+k_{\perp}^2}}{i\omega_n-\sqrt{-\omega_n^2+k_{\perp}^2}}\right)}\nonumber
    \\&+\frac{k_{\perp}^2}{-\omega_n^2+k_{\perp}^2}\frac{i}{-\omega_n^2-k_{\perp}^2-\frac{m_D^2}{2}\left(\frac{-\omega_n^2}{-\omega_n^2+k_{\perp}^2}-\left(\frac{-\omega_n^2}{-\omega_n^2+k_{\perp}^2}+1\right)\frac{i\omega_n}{2\sqrt{-\omega_n^2+k_{\perp}^2}}\ln\frac{i\omega_n+\sqrt{-\omega_n^2+k_{\perp}^2}}{i\omega_n-\sqrt{-\omega_n^2+k_{\perp}^2}}\right)}\bigg).\label{eq:abcd}
\end{align}
Given that we are only interested in the soft contribution, we may discard all but the zero mode
\begin{equation}
    \mathcal{C}(k_{\perp})_{\text{LO}}=Tg^2 C_{R}\bigg(-\frac{1}{k_{\perp}^2+m_{D}^2}
    +\frac{1}{k_{\perp}^2}\bigg)=\frac{g^2TC_{R}m_{D}^2}{k_{\perp}^2(k_{\perp}^2+m_{D}^2)}.
\end{equation}
Reassuringly, we have recovered Eq.~\eqref{eq:scattkerhtl}, which was first calculated by Aurenche, Gelis and Zakaret \cite{Aurenche:2002pd}
using a sum rule. Assuming the transverse momentum is cut off such that $T\gg\mu_{\perp}\gg m_{D}$, we then arrive at the LO soft contribution to the transverse momentum broadening coefficient, $\hat{q}(\mu_{\perp})$ with the use 
of Eq.~\eqref{eq:qhatdef}, which immediately yields Eq.~\eqref{eq:qhat_soft_lo}. In Ch.~\ref{chap:qhat_chap}, we will will build on this result by calculating double logarithmic corrections to $\hat{q}(\mu_{\perp})$.

\begin{figure}[h]
	\centering
	\includegraphics[width=1\textwidth]{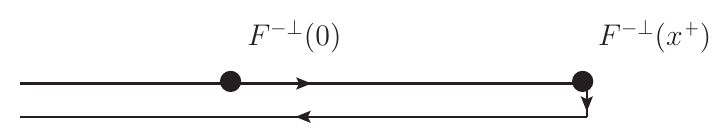}
	\caption{Pictoral depiction of the Wilson line in Eq.~\eqref{eq:zg_wline}. Reading Eq.~\eqref{eq:zg_wline}
     from right to left, the lightlike Wilson line starts at $x'^+=-\infty$, with $F^{-\perp}$ insertions at $x'^+=0$ and $x'^+=x^+$ before
     being brought back to $x'^+=-\infty$.}
    \label{fig:zg_bnl}
\end{figure}

\subsection{Leading Order Soft Contribution to \texorpdfstring{$Z_{g}$}{TEXT}}\label{sec:nlo_asym}
We have seen that the asymptotic mass, $m_{\infty}$ appears as a shift in the hard parton's dispersion relation due to forward 
scattering with the medium. Through Eq.~\eqref{eq:asy_mass}, we moreover understand how the masses 
can be computed in terms of $Z_g$ and $Z_f$. At leading order, they both receive a contribution
from the scale, $T$
\begin{align}\label{eq:condensates_lo}
    Z_{f}^{\text{LO}}&=\frac{T^2}{12},
    \\Z_{g}^{\text{LO}}&=\frac{T^2}{6}.
\end{align}
At $\mathcal{O}(g)$, $Z_g$ then gets a contribution from the soft scale, $gT$\footnote{This is however not the case 
for $Z_f$, which can be understood as a manifestation of Pauli Blocking.}. The purpose of this section will be to show how 
this contribution can be computed using the Wilson line lying along the lightcone
\begin{equation}
    Z_g=\frac{1}{d_R C_{R}}\int_{0}^{\infty}dx^+\,x^+\Tr\langle U_{R}(-\infty;x^+)v_{\mu}F^{\mu\nu}(x^+)U_{R}(x^+;0)v_{\rho}F^{\rho}_{\nu}(0)U_{R}(0;-\infty)\rangle.\label{eq:zg_wline}
\end{equation}
where
\begin{equation}
        \label{eq:4d:Wilson:line:expl}
          U_{R} (x^+;0) = \mathcal{P} \; \exp \left( i g \int_0^{x^+} d y^+ A^{-a}(y^+) T^a_R \right) 
\end{equation}
Per our discussion in Sec.~\ref{sec:chtrick}, these kinds of corrections are suitable for calculation in EQCD. 
We choose here to instead compute it using HTL theory, as it will nicely set up 
the calculation in full QCD that will follow in Ch.~\ref{ch:asym_mass}.
\begin{figure}[h]
	\centering
	\includegraphics[width=0.7\textwidth]{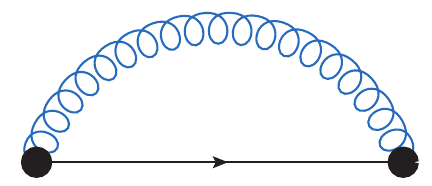}
	\caption{Diagram contributing to $Z_g$ at NLO. In keeping with our convention, the blue gluon is soft and therefore
    needs to be represented with an HTL propagator.}
    \label{fig:zg_lo_soft}
\end{figure}
Expanding Eq.~\eqref{eq:zg_wline} to first order in $g$, we find\footnote{There is no effect from the Wilson line Eq..~\eqref{eq:4d:Wilson:line:expl} at this 
order in the perturbative expansion.} (see App.~\ref{sec:conventions} for our conventions with respect to lightcone coordinates)
\begin{equation}
    Z_g^{\text{NLO}} = -2\int_0^\infty d x^+\,x^+ \left\langle(\partial^{-}A^{x}(x^+,0,0)-\partial^{x}A^{-}(x^{+},0,0))(\partial^{-}A^{x}(0,0,0)-\partial^{x}A^{-}(0,0,0))\right\rangle,
\end{equation}
where we have used invariance in the transverse plane to write $F^{-x}F^{-x}+F^{-y}F^{-y}=2F^{-x}$.
Being sandwiched between two 1 vertices, the relevant gluon propagator should be a time-ordered $G_{F}$.
However, from using that $x^+\geq 0$ and Eq.~\eqref{eq:feynman_prop_mom}, we understand that we can just 
as well work with $G_>$
\begin{align}
    Z_g ^{\text{NLO}}&=-2\int_0^\infty d x^+\int_{K}x^+e^{-ik^-x^+}(k^{-2}G_{>}^{xx}(K)-2k^{x}k^{-}G_{>}^{x-}(K)+{k^{x}}^{2}G_{>}^{--}(K))\nonumber
     \\&=2\int_{K}\frac{1}{(k^{-}-i\epsilon)^2}(k^{-2}G_{>}^{xx}(K)-2k^{x}k^{-}G_{>}^{x-}(K)+{k^{x}}^{2}G_{>}^{--}(K))\nonumber
     \\&=2\int_{K}\frac{1+n_{\text{B}}(k^0)}{(k^{-}-i\epsilon)^2}(k^{-2}(G_{R}^{xx}(K)-G_{A}^{xx}(K))-2k^{x}k^{-}(G_{R}^{x-}(K)-G_{A}^{x-}(K))\nonumber
     \\&+{k^{x}}^{2}(G_{R}^{--}(K)-G_{A}^{--}(K)))\nonumber
       \\&=2\int_{K}\frac{1+n_{\text{B}}(k^0)}{(k^{-}-i\epsilon)^2}(F_{R}(K)-F_{A}(K))\label{eq:zg_nlo_mess}
\end{align}
where
\begin{align}
    F_{R/A}(K)&=k^{-2}G^{xx}_{R/A}(K)-2k^{x}k^{-}\Big(G^{0x}_{R/A}(K)-G^{zx}_{R/A}(K)\Big)\nonumber
    \\&+{k^x}^{2}\Big(G^{00}_{R/A}(K)+G^{zz}_{R/A}(K)-2G^{z0}_{R/A}(K)\Big),
\end{align}
and the causal HTL propagators are given in the Feynman gauge in Eq.~\eqref{eq:feyn_htl}.
The pesky $1/(k^- -i\epsilon)^2$ factor makes it so that we cannot immediately trade 
the $k^0$ integral for a Matsubara sum, as was done in Eq.~\eqref{eq:abcd}. Instead, 
we must first do shift i.e $k^z\rightarrow k^z+k^0$\footnote{This is equivalent to the shift performed in Eq.~\eqref{eq:ch_shift}
but for lightlike separation.}. Then we can let $k^0\rightarrow i\omega_n$, trading the frequency integral for a sum over the Matsubara modes. Since we are once again 
 only interested in the soft contribution, we are permitted to only keep zero mode term, which leaves us with
\begin{equation}
    Z_g^{\text{NLO}}=T\int_{k}\frac{1}{(k^z+i\varepsilon)^2}\left(\frac{2{k^z}^2}{k^2}-\frac{k_{\perp}^2}{k^2+m_D^2}+\frac{k_{\perp}^2}{k^2}\right).
\end{equation}
In DR, because they are scaleless, the first and third terms do not contribute. Integrating the second term yields
\begin{equation}
    Z_g^{\text{NLO}}=-\frac{Tm_D}{2\pi},\label{eq:zg_nlo}
\end{equation}
 a negative correction with respect to Eq.~\eqref{eq:condensates_lo}. We interpret this as representing 
 a reduction in the thermal mass due to the screening of infrared gauge modes.
 
 \subsection{Non-Perturbative Determination of Classical Corrections}\label{sec:np_classical}
 To conclude this chapter, we provide a brief discussion on the program inspired by the 
 ideas \cite{CaronHuot:2008ni} laid out in Sec.~\ref{sec:chtrick}, that is, to
 determine using lattice EQCD corrections to $\mathcal{C}(k_{\perp})$ and $Z_g$.
 
A flurry of papers \cite{Panero:2013pla,Laine:2013lia,Moore:2019lgw,Moore:2021jwe} on the computation 
non-perturbative (NP) determination of $\mathcal{C}(k_{\perp})$ have emerged in the last decade or so. This
effort has culminated with the complete determination \cite{Schlichting:2021idr} of the NP
kernel and its impact on the in-medium splitting rate\footnote{This is essentially
the quantity that we have calculated, albeit to LL accuracy in Sec.~\ref{sec:hoa}.}. 
\begin{figure}[t]
    \begin{center}
        \includegraphics[width=0.4\textwidth]{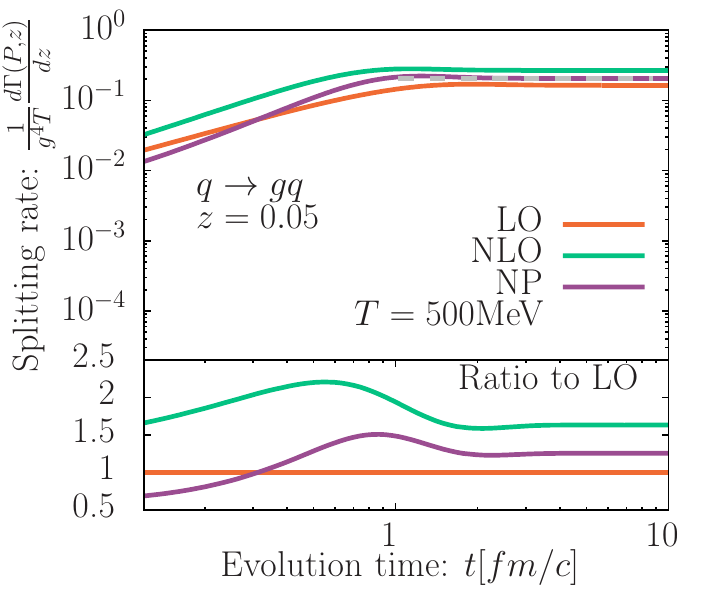}
        \includegraphics[width=0.4\textwidth]{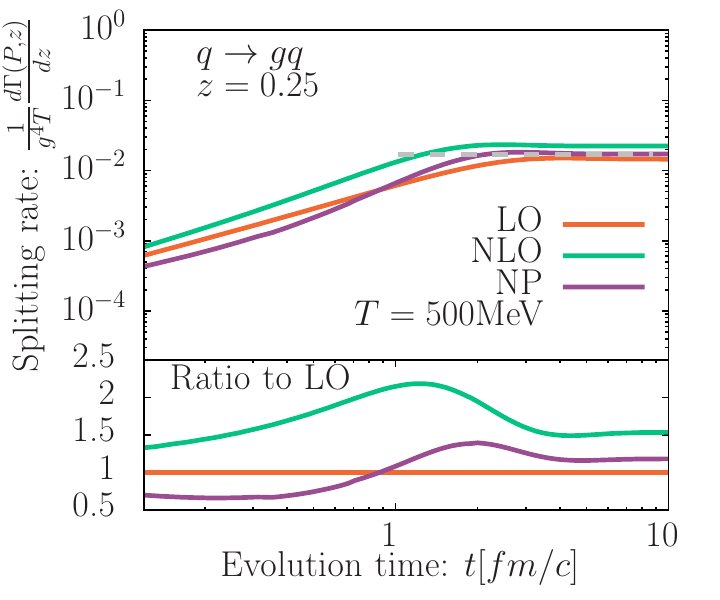}
        \includegraphics[width=0.4\textwidth]{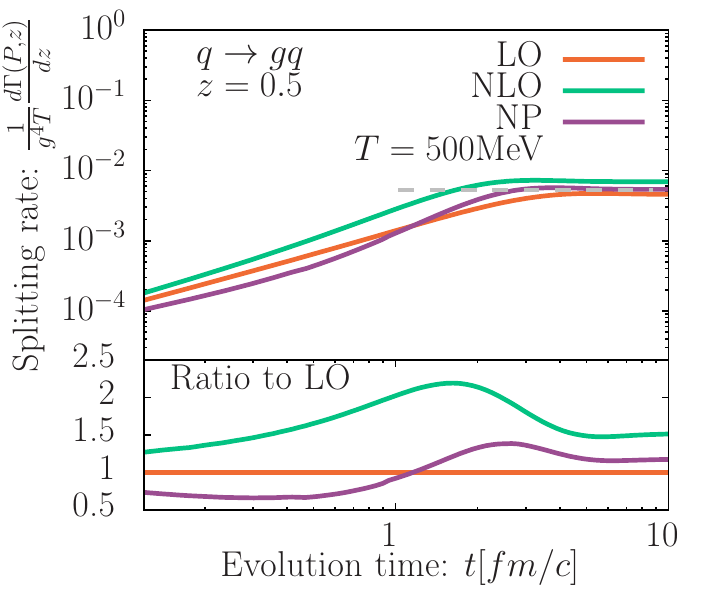}
    \end{center}
    \caption{Comparison of non-perturbative determination of in-medium splitting with both 
    LO and NLO kernel, evolved in time. Figures taken directly from \cite{Schlichting:2021idr}.}
    \label{fig:s_s}
\end{figure}
In that paper, the NP scattering rate is essentially used as input for Eq.~\eqref{eq:defimplfulla}, suitably 
generalised to deal with a medium of finite length (see \cite{CaronHuot:2010bp}), whose solution, obtained numerically
is then used to calculate the in-medium splitting rate. This is what is shown in Fig.~\ref{fig:s_s} as a function 
of time for $3$ different values of splitting fraction $z$ (note that we called the splitting fraction $x$ in Ch.~\ref{ch:eloss}).
In particular, the solution with the NP kernel is compared to the LO kernel and the NLO kernel, computed in \cite{CaronHuot:2008ni}.
We point out that for certain values of the parameter space, the NP kernel can be $50\%$ larger than the LO
value, implying that these higher-order corrections have the potential to be quantitatively impactful. 
An eventual goal, which we have started working towards in \cite{Ghiglieri:2022gyv}, to be detailed 
in the next chapter, is to understand how these NP corrections to $\mathcal{C}(k_{\perp})$ (or equivalently, $\hat{q}(\mu)$)
compete with the logarithmically enhanced corrections coming in at the next order in $g$.

An effort to determine the NP corrections to $Z_g$ has largely progressed in parallel \cite{Moore:2020wvy,Ghiglieri:2021bom} 
to the NP calculation of $\mathcal{C}(k_{\perp})$, as much of the technology developed for the lattice 
computation is directly applicable to both cases. However, extracting the NP correction is not 
as simple as just blindly computing Eq.\eqref{eq:zg_wline} on the lattice. For one, as detailed in \cite{Moore:2020wvy} it 
is not possible to integrate Eq.\eqref{eq:zg_wline} up to arbitrarily large $x^+$ since at some point, the signal-to-noise 
ratio from the numerical integration stifles a robust determination of the correlator. To overcome this issue, 
for large separations the correlator data must be fitted to a model, based on the expected exponential faloff from 
electrostatic and magnetostatic screening. Another issue is the high energy behaviour of Eq.~\eqref{eq:zg_wline}; at some point 
the lattice determination becomes limited by lattice spacing and discretisation effects. Fortunately however, it is in this (short distance) region where a perturbative 
EQCD calculation should in principle agree with the lattice determination. In this way, one must perform a 
\emph{matching procedure}\footnote{This should not be confused with the matching procedure mentioned in Sec.~\ref{sec:eft}, used
to determine the Wilson coefficients in the effective theory.} to fix the short-distance evaluation of the NP corrections. 

Furthermore, since EQCD possesses divergences for energies approaching the thermal scale, one 
needs to check that those UV divergences cancel against corresponding IR divergences coming from full QCD. This 
latter step has already been completed for $\mathcal{C}(k_{\perp})$ but remains to be done for $Z_g$ and 
will compose the majority of Ch.~\ref{ch:asym_mass}'s material. Indeed, the perturbative EQCD 
calculation in \cite{Ghiglieri:2021bom} does not include such a matching onto the 4D theory; 
they subtract off the divergent EQCD behaviour from their results. The following schematic equation, taken 
from \cite{Ghiglieri:2021bom} should help to set the stage for the calculation 
to be done in that chapter, where the different rows correspond to the order in $g$ at which the 
corrections are present.
\begin{align}
    &
    \hphantom{\bigg[\quad}
    \mathrm{scale\;}T
    &&
    \mathrm{scale\;}gT
    &&
    \mathrm{scale\;}g^2T
    &
    \nonumber
    \\
    Z_{g} =
      &\bigg[\quad
        \frac{T^2}{6}
      - \frac{T\mu_{\text{h}}}{\pi^2}
      &&
      &&
      &\bigg]
    \nonumber\\
    + &\bigg[\quad
      &&
      - \frac{T m_{\text{D}}}{2\pi}
      + \frac{T\mu_{\text{h}}}{\pi^2}
      &&
    &\bigg]
    \nonumber\\
    + &\bigg[\quad
        c^{\ln}_{\text{hard}} \ln\frac{T}{\mu_{\text{h}}}
      + c_{\text{T}}
      &&
      + c^{\ln}_{\text{hard}} \ln\frac{\mu_\text{h}}{m_{\text{D}}} 
      + c^{\ln}_{\text{soft}} \ln\frac{m_{\text{D}}}{\mu_{\text{s}}}
      + c_{gT}
      &&
      + c^{\ln}_{\text{soft}} \ln\frac{\mu_\text{s}}{g^2T}
      + c_{gT^2}    &\bigg]
    \nonumber\\[2mm]
    +&\,\mathcal{O}(g^3)\;.&
    &&\label{eq:zg_corrections}
\end{align} 
In the first row, there is only a contribution from the scale $T$, which we have previously written down in Eq.~\eqref{eq:condensates_lo}.
In the second row, there is only a classical contribution, which we calculated using HTL theory in the previous section (see Eq.~\eqref{eq:zg_nlo}).
In that calculation, we did not recover the part of the correction depending on the cutoff, $\mu_{\text{h}}$, which 
cancels against the same correction but with opposite sign at LO. This is 
indeed consistent with our previous assertation that such corrections are never found when using DR as a regulator. 

At $\mathcal{O}(g^2)$, there are contributions from the hard, soft and ultrasoft scales. In Ch.~\ref{ch:asym_mass}, 
we will essentially focus on calculating the $c^{\ln}_{\text{hard}}\ln T/\mu_{\text{h}}$ and $c_{\text{T}}$ terms\footnote{In 
particular, all of the other coefficients at that order are implicitly included in the lattice calculation.}, showing that 
the former cancels against the term $c^{\ln}_{\text{hard}}\ln \mu_{\text{h}}/m_{\text{D}}$, which was calculated in \cite{Ghiglieri:2021bom}.
Since we will use DR there, these logarithmic divergences will show up as $1/\epsilon$ poles.
The NP determination of $Z_g$ will then be able to stand on its own, void of any divergences. Our 
long-term goal is then to calculate the entire $\mathcal{O}(g^2)$ from the scale $T$, which, along with its 
$Z_f$ counterpart will allow for the complete evaluation of the asymptotic mass' $\mathcal{O}(g^2)$ corrections.
 
Something which has not been made so clear up this point is why we need not worry about further matching 
between the EQCD and full QCD at higher orders in $g$. The answer to this question 
is provided by the fact that EQCD is a \emph{super-renormalisable} theory; if one demands 
that quantities be cut off in the UV by $\mu_{\text{h}}$, at each order in $g$, the corrections 
will go as a different power of $\mu_{\text{h}}$. In other words, we can expect the UV behaviour 
at $\mathcal{O}(g^n)$ to be $\mu_{\text{h}}^{2-n}$ so that after the leading linear 
and subleading logarithmic divergences, we only produce negative powers in the UV cutoff.
\newpage
\chapter{Quantum Corrections to Jet Broadening}\label{chap:qhat_chap}

\begin{figure}[t]
    \begin{center}
        \includegraphics[width=0.4\textwidth]{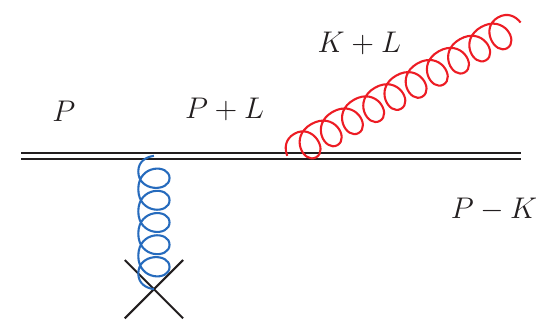}
        \includegraphics[width=0.4\textwidth]{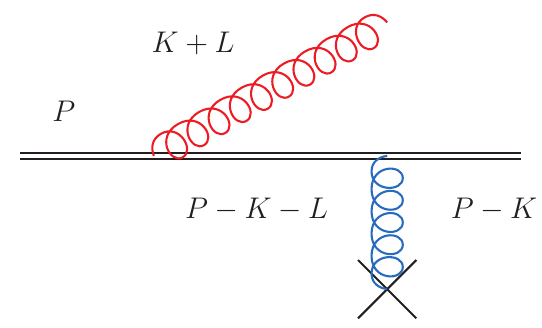}
        \includegraphics[width=0.4\textwidth]{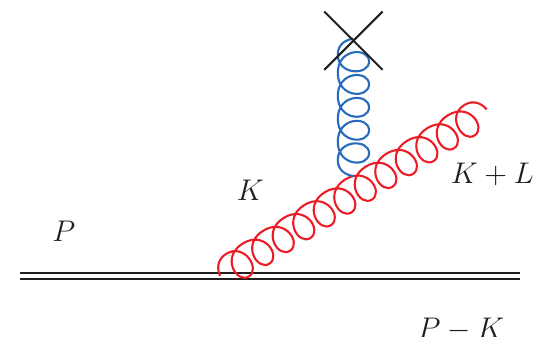}
    \end{center}
    \caption{Diagrams contributing to radiative momentum broadening 
    in the single-scattering regime. The double solid line represents 
    the hard jet parton (quark or gluon), curly lines are gluons and the 
    cross is a quark or gluon scattering centre in the medium.}
    \label{fig:dl_diags}
\end{figure}

In Caron-Huot's original work \cite{CaronHuot:2008ni} where he computed the $\mathcal{O}(g)$ correction to $\hat{q}(\mu_{\perp})$, 
he anticipated the potential presence of corrections at one further order in $g$ coming from gluons that are collinear to the hard 
parton. Indeed, in computing $\hat{q}(\mu_{\perp})$ to $\mathcal{O}(g^2)$, one starts to encounter diagrams corresponding to \emph{radiative corrections}
\cite{Wu:2011kc,Liou:2013qya,Blaizot:2013vha}; 
in contrast to the LO calculation in Sec.~\ref{sec:scatt_lo_soft} where 
just a soft gluon is exchanged with the medium, diagrams 
here also contain bremsstrahlung and are shown 
in Fig.~\ref{fig:dl_diags}. The authors \cite{Liou:2013qya,Blaizot:2013vha} showed that 
these corrections possess double logarithmic 
and single logarithmic\footnote{We do not discuss the single logarithmic corrections in this thesis.} enhancements,
where the argument of the double logarithm was found to be $L_M/\tau_{\text{min }}$ with $L_M$ the length 
of the medium and $\tau_{\text{min}}\sim 1/T$ an IR cutoff to be elucidated further upon discussion.

While the original calculations concretely demonstrated the presence of these (logarithmic) corrections, they did not 
strive to quantify their arguments precisely. Although difficult, the latter goal is of paramount importance, as it of 
course controls whether the logarithms are large compared to the factor of $g^2$ that precedes them 
and the (non-perturbatively calculated) classical corrections that start to come in at $\mathcal{O}(g)$. 
It has been argued in both papers \cite{Liou:2013qya,Blaizot:2013vha} that the double logarithmic corrections 
are sourced from the single-scattering regime, where this bremsstrahlung is triggered by a 
single, harder\footnote{That is, hard with respect to the soft scatterings 
that characterise the multiple scattering regime.} scattering between the jet and the medium. By tackling 
this problem using TFT, we will be able to rigorously judge and furthermore quantify how the thermal scale 
is relevant for these corrections.

The approach of \cite{Blaizot:2013vha} makes clear that these corrections arise from taking into account the recoil 
during the medium-induced emission of a gluon. There, evolution equations describing medium-induced gluon 
splitting are derived and studied. Additional terms in these equations coming from keeping track of the transverse momentum 
that the jet picks up from the medium are then interpreted as corrections to $\hat{q}(\mu{\perp})$.

While both of these calculations arrive at essentially the same conclusions, we choose 
to instead sketch the calculation done in \cite{Liou:2013qya}. We make contact between this 
calculation and our Wilson loop setup, introduced in Sec.~\ref{sec:scatt_lo_soft}, showing 
how we can reproduce their result. From there, we explore the neighbouring regions of phase space and 
in doing so, identify which of them have the potential to possess double logarithmic enhancements. 
The chapter closes with a demonstration of how the 
 double-logarithmic region of phase space is situated with respect to 
the region from which the classical corrections \cite{CaronHuot:2008ni} emerge. A strict use of Coulomb gauge is adopted throughout this chapter.

\begin{figure}[t]
    \begin{center}
        \includegraphics[width=14cm]{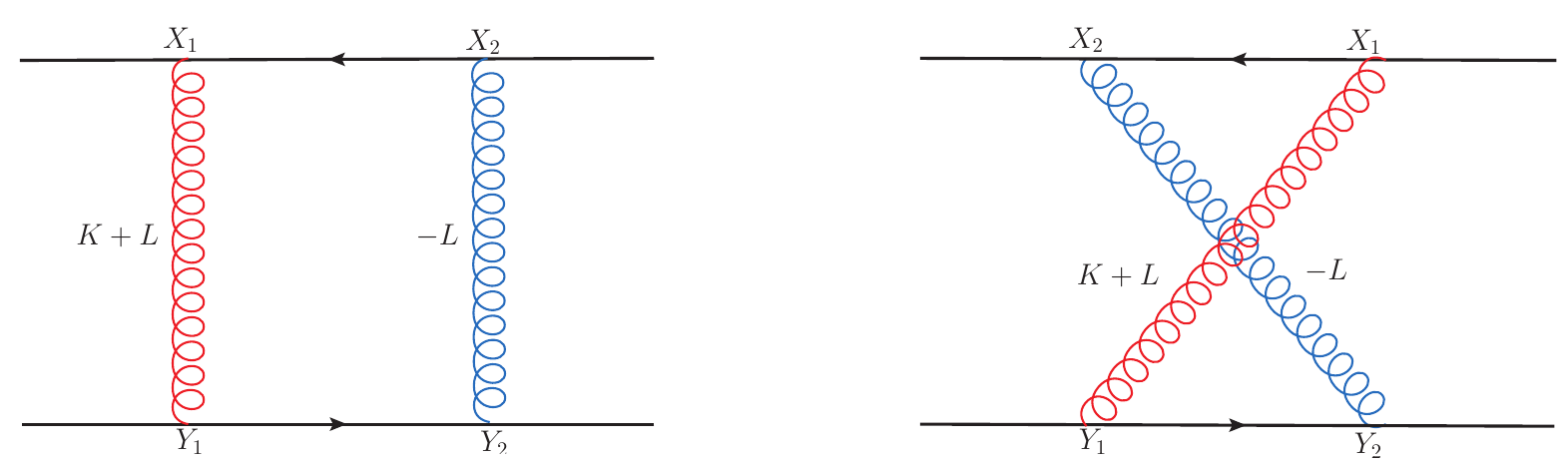}
        \includegraphics[width=5cm]{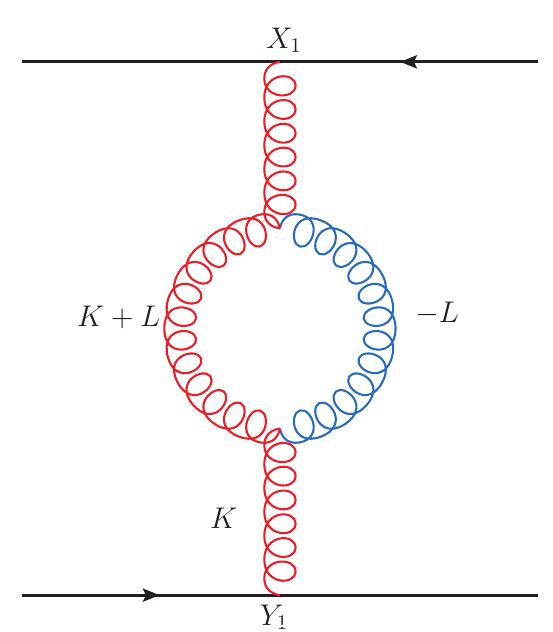}
        \includegraphics[width=7cm]{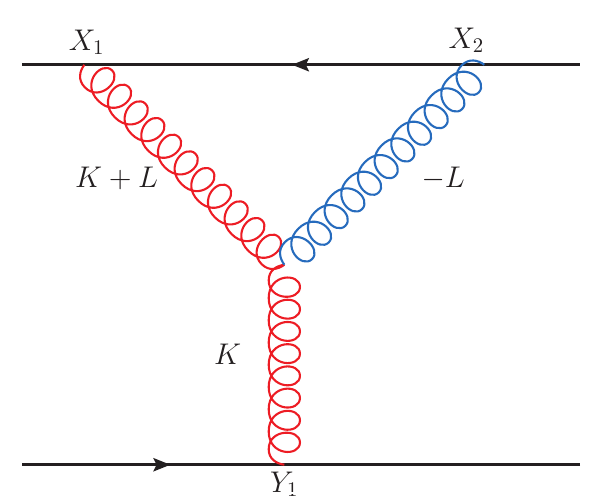}
    \end{center}
    \caption{Diagrams giving rise to the single-scattering
    regime. The two horizontal solid lines are the Wilson lines $\mathcal{W}$.
     Red gluons are coft, representing bremsstrahlung while the blue ones are HTL-resummed and soft. See the main text for 
     the precise scalings.
     Momenta flow from bottom to top.}
    \label{fig:real}
\end{figure}
This chapter is based on the work \cite{Ghiglieri:2022gyv}.

\section{Wilson Loop Setup at \texorpdfstring{$\mathcal{O}(g^2)$}{TEXT}}\label{sec:wloop_gsq}
In this section we will build upon the groundwork laid in Sec.~\ref{sec:scatt_lo_soft}, demonstrating how 
our Wilson loop setup can be applied to compute higher order corrections to $\mathcal{C}(k_{\perp})$ 
and hence $\hat{q}({\mu_{\perp}})$. It turns out that the diagrams from Fig.~\ref{fig:real}
are the ones that we will need to compute in order 
to capture the aforementioned double logarithmic corrections. Therefore, 
a good place to start is to try and understand how these diagrams correspond to
those shown in Fig.~\ref{fig:dl_diags} -- the single scattering processes.
 
As was also the case for the LO evaluation, 
the horizontal black lines in Fig.~\ref{fig:real} are respectively $\mathcal{W}$ and $\mathcal{W}^{\dagger}$ Wilson lines representing the hard jet parton in the amplitude and 
conjugate amplitude. In our setup, the $\mathcal{W}$ ($\mathcal{W}^{\dagger}$) line corresponds to a hard jet parton 
with initial (final)
momentum $P=(E,0,0)$ with $E\gg T$, the temperature of the plasma\footnote{Strictly speaking, the Wilson line 
is derived in the $E\to\infty$ limit.}. The gluons exchanged with the medium
are, as before, soft and thus are HTL-resummed. The red gluons denote collinear bremsstrahlung 
that is harder than the blue gluons but still soft in comparison to the leading jet parton\footnote{As is clearly visible from
Eq.~\eqref{eq:lmw_bdim_dlog}, one of the logarithmic divergences is collinear, with the other one being soft.}. 
Borrowing the terminology 
of \cite{Becher:2015hka}, we call these modes ``coft''. They are instead described by bare propagators. 
We concretely define the coft scaling here, for 
general four-momentum
\begin{equation}
Q\sim\tilde{\lambda}E(1,\lambda^2,\lambda)\quad\quad \lambda,\tilde{\lambda}\ll 1
\end{equation}
This scaling naturally incorporates the collinearity (with respect to the jet axis) of the radiation as well as the 
fact that its $+$ component is much smaller than the energy, $E$ of the jet. We hold off on choosing explicit 
scales from $\lambda,\,\tilde{\lambda}$ until a later point. Calculations are always set up so that the cut coft propagator has momentum $K+L$ and the HTL one momentum $L$. In this way,
$K$ is always the momentum exchanged between the two Wilson lines. 

But how can we interpret these diagrams in terms of those shown in Fig.~\ref{fig:dl_diags}. Well, cuts are 
understood to go horizontally through the middle of each diagram in Fig.~\ref{fig:real}. 
It then transpires that the first top two diagrams in Fig.~\ref{fig:real} correspond to the square
of the top two in Fig.~\ref{fig:dl_diags} and their interference. The bottom left diagram in Fig.~\ref{fig:real}
corresponds to the square of the bottom in Fig.~\ref{fig:dl_diags}. Finally the bottom right in Fig.~\ref{fig:real} corresponds 
to the interference of the top two with the bottom one in Fig.~\ref{fig:dl_diags}. 

Still, it is perhaps not clear why there are only $4$ diagrams in Fig.~\ref{fig:real} as opposed to $6$. This 
apparent discrepancy can be clarified by re-examining Eq.~\eqref{eq:wloopexpansion}, which we repeat here for convenience
\begin{equation}
    \lim_{L_M \to \infty}\langle W(x_{\perp})\rangle= 1-\left(\mathcal{C}_{\text{LO}}(x_{\perp})+\mathcal{C}_{\text{NLO}}(x_{\perp})+...\right)L_M 
    +\frac{\left(\mathcal{C}_{\text{LO}}(x_{\perp})+\mathcal{C}_{\text{NLO}}(x_{\perp})+...\right)^2}{2}L_M^2+...\,.
\end{equation}
It emerges that only two-particle irreducible (2PI) diagrams appear in the coefficient of $L_{M}$ above, implying that they are the 
only class of diagrams that we need consider \cite{Peter:1997me,Berwein:2012mw} in order to compute 
$\mathcal{C}_{\text{NNLO}}(x_{\perp})$\footnote{Note 
that $\mathcal{C}_{\text{NLO}}(x_{\perp})$ was computed in \cite{CaronHuot:2008ni}. That calculation involves the 
computation of same diagrams that we are considering here, but with different scalings -- all gluons are soft and 
are thus represented with HTL propagators there.} and hence $\mathcal{C}_{\text{NNLO}}(k_{\perp})$. Other diagrams
appear in higher order terms in the $L_{M}$ expansion.

This naturally begs the question: why are we considering the two-particle reducible top left diagram of Fig.~\ref{fig:real}?
To this end, note that the colour factor associated with this diagram, which we denote as $\text{II}$ to reflect its topology is given as
\begin{equation}
    C_{\text{II}}=\frac{1}{N_{c}}\Tr[T^{b}T^{d}T^{c}T^{a}]\delta^{ab}\delta^{cd}=\frac{1}{N_{c}}\Tr[T^{a}T^{a}T^{c}T^{c}]=C_{R}^{2},
\end{equation}
where the prefactor of $1/N_c$ comes from Eq.~\eqref{eq:Wloopdef} and the generators are in the representation $R$ of $\text{SU}(N_c)$ under which the 
hard parton transforms. The colour factor of the diagram to its right, which we denote as $\text{X}$
possesses a non-Abelian piece
\begin{equation}
    C_{\text{X}}=\frac{1}{N_{c}}\Tr[T^{d}T^{b}T^{c}T^{a}]\delta^{ab}\delta^{cd}=\frac{1}{N_{c}}\Tr[(T^{a}T^{c}+[T^{c},T^{a}])T^{c}T^{a}]=C_{R}^{2}-\frac{C_{R}C_{A}}{2}.
\end{equation}
It is precisely this non-abelian piece that survives once we add the two diagrams together, as we will now see. In order
to clean up the notation, we write $\delta\mathcal{C}=\mathcal{C}^{\text{log}}_{\text{NNLO}}$\footnote{
To be clear, we specify that $\delta{C}$ only refers to the part of the NNLO scattering kernel that (possibly) contains 
a double logarithmic enhancement.} so that by using Eqs.~\eqref{eq:Wloopdef}, ~\eqref{eq:wline_def}
and the expansion in terms of the scattering kernel, we arrive at
\begin{align}
    \delta\mathcal{C}&(x_{\perp})^{\text{\cal{R}}}_{\mathrm{II+X}}=-2g^{4}\int_{-\frac{L_M}{2}}^{\frac{L_M}{2}}dx_{2}^{+}\int_{-\frac{L_M}{2}}^{\frac{L_M}{2}}dx_{1}^{+}
    \int_{-\frac{L_M}{2}}^{\frac{L_M}{2}}dy_{2}^{+}\int_{-\frac{L_M}{2}}^{\frac{L_M}{2}} dy_{1}^{+}G_{>}^{--}(x_{1}^{+}-y_{1}^{+},x_{\perp})G_{>}^{--}(x_{2}^{+}-y_{2}^{+},x_{\perp})\times\nonumber
    \\&\times\Theta(y_{2}^{+}-y_{1}^{+})\Bigg(C_{R}^{2}\Theta(x_{2}^{+}-x_{1}^{+})+\Big(C_{R}^{2}-\frac{C_{R}C_{A}}{2}\Big)\Theta(x_{1}^{+}-x_{2}^{+})\Bigg)\nonumber
    \\&=-2g^{4}\int_{-\frac{L_M}{2}}^{\frac{L_M}{2}}dx_{2}^{+}\int_{-\frac{L_M}{2}}^{\frac{L_M}{2}}dx_{1}^{+}
    \int_{-\frac{L_M}{2}}^{\frac{L_M}{2}}dy_{2}^{+}\int_{-\frac{L_M}{2}}^{\frac{L_M}{2}} dy_{1}^{+}G_{>}^{--}(x_{1}^{+}-y_{1}^{+},x_{\perp})G_{>}^{--}(x_{2}^{+}-y_{2}^{+},x_{\perp})\times\nonumber
    \\&\times\Theta(y_{2}^{+}-y_{1}^{+})\Bigg(-\frac{C_{R}C_{A}}{2}\Theta(x_{1}^{+}-x_{2}^{+})+C_{R}^{2}\Bigg)
\end{align}
where we have added a factor of $2$ in front to account for the diagrams where the scalings are inverted. In addition, the \cal{R} superscript 
specifies that we have included the contribution from real diagrams. As was discussed in Sec.~\ref{sec:scatt_lo_soft}, real diagrams are the only
ones needed in order to determine $\delta\mathcal{C}(k_{\perp})$ and hence $\delta\hat{q}(\mu_{\perp})$. Both propagators 
are of type $>$ because they are stretching from the bottom Wilson line, which only admits  ``1'' vertices to the top one, which 
only admits ``2'' vertices.

Coming back 
to the expression above, it is interesting to observe that, upon symmetrising the
term proportional to $C_R^2$ in $x_1\leftrightarrow x_2 \,,\,y_1\leftrightarrow y_2$ and then adding them, we arrive at 
\begin{equation}
    \frac{1}{2}g^{4}C_{R}^{2}\int_{-\frac{L}{2}}^{\frac{L}{2}}dx_{1}^{+}\int_{-\frac{L}{2}}^{\frac{L}{2}}dy_{1}^{+}G^{--}_{>}(x_{1}^{+}-y_{1}^{+},x_{\perp})\int_{-\frac{L}{2}}^{\frac{L}{2}}dx_{2}^{+}\int_{-\frac{L}{2}}^{\frac{L}{2}}dy_{2}^{+}G^{--}_{>}(x_{2}^{+}-y_{2}^{+},x_{\perp}),
\end{equation}
which is indeed the square of the (real) diagram computed in Sec.~\ref{sec:scatt_lo_soft}; the position integrals will give an
$L_M^2$ factor, signifying that this diagram is part of the exponentiation of the tree-level contribution. 
In any case, the term proportional to $C_R^2$ 
will inevitably be forced to vanish due to the presence of a $\delta\left(\left(\mathbf{k}+\mathbf{l}\right)^2+m_{\infty}^2\right)$ in the coft propagator, 
signalling that this process is kinematically impossible.

To briefly summarise, we have argued that even though diagram $\text{II}$ does not contribute to $\delta\mathcal{C}(k_{\perp})$, it cancels 
the abelian piece of diagram $\text{X}$, leaving us with only a non-abelian contribution. For the other diagrams in Fig.~\ref{fig:real},
 we will not need to consider these kinds of extra diagrams, since their contributions do not possess an additional abelian piece.

Getting back to the matter at hand, one can do the position integrals and extract 
\begin{align}
    \delta\mathcal{C}(k_{\perp})_{\mathrm{II+X}}&=
  C_{R}C_{A}g^{4} \int\frac{dk^{+}dk^-}{(2\pi)^2}\int_{L}
  \frac{1}{(l^{-}-i\epsilon)^{2}}G_{>}^{--}(K+L)G_{>}^{--}(-L)2\pi\delta(k^-). \label{eq:II_X_before_expansion}
\end{align}
A factor of 2 has been added to account
for the inverted scaling. 
The $\delta(k^-)$ arises from the $x^+$ integrations, which essentially enforce the eikonalised 
hard parton to remain on-shell after the emission.
We postpone continuing with the evaluation of this diagram until the next section, where we 
assume specific scalings for $K,\,L$. 

We now proceed to the self-energy diagram on the bottom left in Fig.~\ref{fig:real}.
Its contribution reads
\begin{align}
    &\delta\mathcal{C}(k_{\perp})_{\mathrm{self}}=g^{4}C_{R}C_{A}
    \int\frac{dk^{+}}{2\pi}\int_{L}G_{R}^{-\rho}(K)\Big(g_{\gamma\sigma}(2L{+}K)_{\rho}
    -g_{\sigma\rho}(2K{+}L)_{\gamma}+g_{\rho\gamma}(K{-}L)_{\sigma}\Big)\nonumber
    \\&\times G_{>}^{\sigma\delta}(K+L)G_{>}^{\gamma\alpha}(-L)
    \Big(g_{\delta\beta}(2K{+}L)_{\alpha}-g_{\alpha\delta}(2L{+}K)_{\beta}-g_{\alpha\beta}(K{-}L)_{\delta}\Big) 
    G_{A}^{\beta-}(K)\Big\vert_{k^{-}=0}.
    \label{eq:aftercut}
\end{align}
Cuts going through the $K$ propagators are again vanishing, hence their retarded/advanced
assignments only, which is consistent with the cutting rules for Wightman functions 
\cite{Caron-Huot:2007zhp,Ghiglieri:2020dpq}.

Finally, the Y-shaped diagram on the bottom right  of Fig.~\ref{fig:real}, together with
its symmetry-related counterparts yield
\begin{align}
    \delta\mathcal{C}(x_{\perp})^{\text{\cal{R}}}_{\mathrm{Y}}=&2g^{4}C_{R}C_{A}\int 
    \frac{dk^+d^2k_{\perp}}{(2\pi)^{3}}\int_{L}
     \frac{i}{l^{-}+i\epsilon}e^{i\mathbf{k}\cdot \mathbf{x} }G_{>}^{-\delta}(K+L)G_{>}^{-\alpha}(-L)\nonumber
    \\&\hspace{1.2cm}\times\Big(g_{\alpha\delta}(2L+K)_{\beta}-g_{\delta\beta}(2K+L)_{\alpha}
    +g_{\beta\alpha}(K-L)_{\delta}\Big)G_{F}^{\beta-}(K)\Big\vert_{k^{-}=0}.
\end{align}
At this point, it is useful to write the Feynman propagator as 
\begin{equation}
    \label{eq:feynman}
    G_{F}(K)=\frac{1}{2}\Big(G_{R}(K)+G_{A}(K)\Big)+G_{rr}(K).
\end{equation}    
Since $G_{rr}$ is the average of the bare cut propagators and we have already set $k^{-}=0$, 
it vanishes. That leaves us with the average of the retarded and advanced bare Green's functions, 
which when evaluated at $k^{-}=0$ are the same. We can then
extract, as was done with the previous diagrams
\begin{align}
    \delta\mathcal{C}(k_{\perp})_{\mathrm{Y}}=&-2g^{4}C_{R}C_{A}\int \frac{dk^{+}}{2\pi}
    \int_{L}
    \frac{i}{l^{-}+i\epsilon}G_{>}^{-\delta}(K+L)G_{>}^{-\alpha}(-L)
    \nonumber
   \\&\hspace{1.2cm}\times\Big(g_{\alpha\delta}(2L+K)_{\beta}-g_{\delta\beta}(2K+L)_{\alpha}+
   g_{\beta\alpha}(K-L)_{\delta}\Big)G_{R}^{\beta-}(K)\Big\vert_{k^{-}=0}.\label{eq:y_before}
\end{align}
We find this to be a good stopping point for the evaluation of these diagrams.
 The computation will be resumed following a brief review of the calculation from \cite{Liou:2013qya}, which will 
 shed some light on how we can enforce single scattering kinematics, allowing 
 us to assign precise scalings to the momenta $K,\,L$.
 
 \section{Double Logarithms from \texorpdfstring{\cite{Liou:2013qya,Blaizot:2013vha}}{TEXT}}\label{sec:dlog_lit}

In \cite{Liou:2013qya}, the double logarithmic corrections to $\hat{q}(\mu_{\perp})$ emerge
through the use of the standard dipole formalism \cite{Kovchegov:2012mbw}
\begin{equation}
    \label{eq:N1term}
    \delta\mathcal{C}(k_{\perp})^{\mathcal{N}=1}=4\alpha_s C_R\int\frac{dk^+}{k^+}\int^{\rho_{\perp}}\frac{d^2l_{\perp}}{(2\pi)^2}\mathcal{C}_\text{LO}(l_{\perp})\frac{l_{\perp}^2}{k_{\perp}^2(\mathbf{k}+\mathbf{l})^2},
\end{equation}
where $l_{\perp}^2/(k_{\perp}^2(\mathbf{k}+\mathbf{l})^2)$ is the standard dipole factor and
$2C_R/k^+$ the soft limit of the $g\leftarrow  R$ DGLAP splitting function. Finally, 
$\mathcal{C}_{\text{LO}}(l_{\perp})$ is the leading order scattering rate from a gluon source.

Eq.~\eqref{eq:N1term} is in principle just the first, $N=1$ term in the opacity expansion (see Secs.~\ref{sec:single}, ~\ref{sec:opacity}). 
 Within Eq.~\eqref{eq:N1term}, we can identify 
the leading order value of the transverse momentum broadening coefficient i.e 
\begin{equation}
    \hat{q}_{\text{LO}}(\rho_{\perp})=\int^{\rho_{\perp}}\frac{d^2l_{\perp}}{(2\pi)^2}l_{\perp}^2\mathcal{C}_\text{LO}(l_{\perp}).
\end{equation}
Neglecting the dependence on the UV cutoff, $\rho_{\perp}$, thereby setting $\hat{q}_{\text{LO}}(\rho_{\perp})\rightarrow \hat{q}_0$
 effectively enforces the Harmonic Oscillator Approximation. In addition, as observed in \cite{Liou:2013qya}, 
 the requirement that the transverse momentum carried away by the bremsstrahlung is larger than that picked up 
 in a typical collision with the medium, $\vert\mathbf{k}+\mathbf{l}\vert\gg l_{\perp}$, implements
 single scattering kinematics, where the $\mathcal{N}=1$ term dominates the expansion. This leaves us with
 \begin{equation}
    \delta\mathcal{C}(k_{\perp})
   _{\text{\cite{Liou:2013qya,Blaizot:2013vha}}}
    =4\alpha_s C_{R}\hat{q}_0\int\frac{dk^{+}}{k^+}
    \frac{1}{k_{\perp}^{4}}.\label{eq:lit_hoa}
\end{equation}
Taking the second moment (with respect to $k_{\perp}$) will then give us the correction to $\hat{q}(\mu_{\perp})$ that we desire 
\begin{equation}
   \delta\hat{q}
   _{\text{\cite{Liou:2013qya,Blaizot:2013vha}}}
   (\mu_{\perp})=4\alpha_sC_{R}\hat{q}_0\int^{\mu}\frac{d^2k_{\perp}}{k_{\perp}^2}\int\frac{dk^+}{k^+}.\label{eq:lmw_bdim_dlog}
\end{equation}
It is clear from above how the double logarithm emerges --  one comes from the collinear
$d^2k_{\perp}/k_{\perp}^2$ divergence and the other comes from the soft $dk^+/k^+$ divergence. 
This explains why, at the level of our Wilson loop computation, bremsstrahlung possesses coft scaling. 

\begin{figure}[t]
	\centering
	\includegraphics[width=\textwidth]{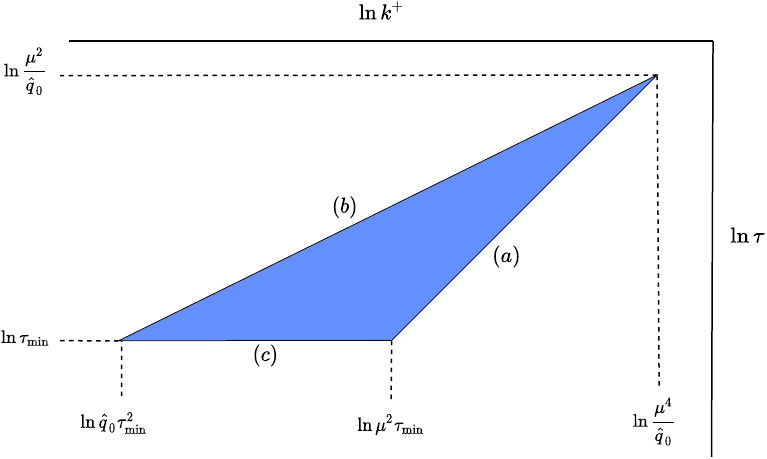}
	\caption{Schematic depiction of the bounds on the integral in Eq.. 
    The $(b)$ side of the triangle is given by $\tau=\sqrt{k^+/\hat{q}_0}$ whereas the
    $(a)$ one is given by $\tau=k^+/\mu_{\perp}^2$. The logarithmic axes allow
    one to easily read the result of the integral Eq.~\eqref{eq:BDIMzeroT} straight from the figure, up to 
    the prefactor $\alpha_s C_R\hat{q}_0/\pi$. Our labelling for the three boundaries follows that 
    of \cite{Liou:2013qya}.
    }
    \label{fig:bdimtriangle}
\end{figure}

For the sake of convenience, we choose to integrate over the formation time, $\tau=k^+/k_{\perp}^2$
instead of $k_{\perp}$. The integration limits, which bound the triangle shown in Fig.~\ref{fig:bdimtriangle}
are then \cite{Liou:2013qya}\footnote{The integration limits from \cite{Blaizot:2013vha} are slightly different to those 
in \cite{Liou:2013qya}. In particular, \cite{Blaizot:2013vha} cut off the energy of the radiated gluon
with the energy $E$ of the parent hard jet parton, whereas we instead take $k^+<\mu_{\perp}^4/\hat{q}_0<E$, which is equivalent
to what is done in \cite{Liou:2013qya}. In any case, this will not have an impact
on the arguments that follow.}:
\begin{itemize}
    \item $\tau<\sqrt{k^+/\hat{q}_0}$, represented by line $(b)$ in Fig.~\ref{fig:bdimtriangle}. 
    In the deep LPM regime, $k_{\perp}^2\sim\hat{q}_0\tau_{\text{LPM}}$. The former
    constraint then emerges upon demanding that $\tau<k_{\perp}^2/\hat{q}_0$ and solving for $\tau$.
    Effectively, this prevents the formation time from getting sufficiently large to lead us into 
    the multiple scattering regime, which will cut off the double-logarithmic phase space:
    the collinear $dk_\perp^2/k_\perp^2$ log can exist as long as the initial and final
    states can propagate along straight lines for sufficiently long times 
    before and after the single scattering \cite{Arnold:2008zu}.
   \item $\tau>k^+/\mu_{\perp}^2$, represented by line $(a)$. This condition on the formation time corresponds
    to enforcing the UV cutoff on transverse momentum, i.e. $k_{\perp}<\mu_{\perp}$.   
    In the original derivation of \cite{Liou:2013qya} this $\mu_{\perp}$ cutoff is identified with the saturation scale
    $Q_s^2\equiv \hat{q}_0 L_{M}$. 
    If instead $\mu_{\perp}> Q_s$ the boundaries of the 
    double-logarithmic region change, as shown in \cite{Blaizot:2019muz}. 
    \item $\tau>\tau_{\text{min}}$, represented by line $(c)$. 
    This  IR cutoff $\tau_{\text{min}}$ is intended to render 
    the end result finite. The motivation for this boundary comes from requiring that the 
    duration of the single scattering does not become
    comparable to the formation time of the radiated gluon, as the former is treated as instantaneous 
    within the collinear-radiation framework. In \cite{Liou:2013qya,Blaizot:2013vha} this duration was assumed to be
    proportional to the inverse temperature, leading to $\tau_{\text{min}}\sim 1/T$. We will return to this 
    assumption in the next section. 
\end{itemize}  

Adding these limits explicitly,
\begin{equation}
    \delta\hat{q}
    _{\text{\cite{Liou:2013qya,Blaizot:2013vha}}}
    (\mu_{\perp})=\frac{\alpha_s C_R}{\pi}\hat{q}_0\int_{\tau_{\text{min}}}^{\mu_{\perp}^2/\hat{q}_0}\frac{d\tau}{\tau}
    \int_{\hat{q}_0\tau^2}^{\mu_{\perp}^2\tau}\frac{dk^+}{k^+},
    \label{eq:BDIMzeroT}
\end{equation}
the double-logarithmic correction,
\begin{equation}
    \delta\hat{q}
    _{\text{\cite{Liou:2013qya,Blaizot:2013vha}}}
    (\mu_{\perp})=\frac{\alpha_s C_R}{\pi}\hat{q}_0\ln^2\frac{\mu_{\perp}^2}{\hat{q}_0\tau_{\text{min}}}
    \label{eq:BDIMresultzeroT}
\end{equation}
immediately follows. This is the form of \cite{Blaizot:2013vha}; that of \cite{Liou:2013qya} arises by 
choosing $\mu^2_{\perp}=\hat{q}_0 L_M$, so that the argument of the double logarithm becomes the familiar $L_M/\tau_{\text{min}}$.
The single-logarithmic corrections arising from a more sophisticated analysis of the
regions neighbouring the three boundaries of the triangle shown in Fig.~\ref{fig:bdimtriangle}
have been presented in \cite{Liou:2013qya}.

\section{Adapting \texorpdfstring{\cite{Liou:2013qya,Blaizot:2013vha}}{TEXT} Corrections to a Weakly Coupled QGP}\label{sec:adapt}
The double-logarithmic phase space we just sketched, as per \cite{Liou:2013qya,Blaizot:2013vha}, is derived  
for a medium described  by a random color field with a non-zero
two-point function for the $A^-$ component, i.e.
\begin{equation}
    \label{bdmimmedium}
    \left\langle A^{-a}(\mathbf{q},x^+) A^{-b}(\tilde{\mathbf{q}},\tilde{x^{+}}) \right\rangle=\delta^{ab}\delta(x^+-\tilde{x}^{+})
    n(x^+)(2\pi)^2\delta^{(2)}(\mathbf{q}-\tilde{\mathbf{q}})\frac{g^4}{q_{\perp}^4}
\end{equation}
This corresponds for instance to the time-honoured 
parameterisation of a medium of static scattering centres of density $n$ and is consistent with modelling the  
medium with the GW kernel, Eq.~\eqref{eq:scattkergw}. However, a weakly coupled
QGP contains more medium effects than those captured by these instantaneous, space-like interactions; in 
particular, as soon as the light-cone frequency ($k^+$) range overlaps with the temperature scale, one needs
to account for the Bose-Einstein stimulated emission of the radiated gluon and, at negative $k^+$, for the 
absorption of a gluon from the medium. 
We could naively account for these effects by 
amending Eq.~\eqref{eq:BDIMzeroT} into 
\begin{equation}
    \delta \hat{q}(\mu_{\perp})=\frac{\alpha_s C_R}{\pi}\hat{q}_0\int_{\tau_{\text{min}}}^{\mu_{\perp}^2/\hat{q}_0}\frac{d\tau}{\tau}
    \int_{\hat{q}_0\tau^2}^{\mu_{\perp}^2\tau}\frac{dk^+}{k^+}\left(1+2n_{\text{B}}(k^+)\right).
    \label{eq:BDIMnonzeroT}
\end{equation}
The factor of two in Eq.~\eqref{eq:BDIMnonzeroT} accounts for 
stimulated emission and for absorption, which has been reflected in the positive-frequency range. 

Let us now perform a careful, parametric analysis of 
Fig.~\ref{fig:bdimtriangle}, taking into account some relevant timescales listed in Table.~\ref{tab:timescales}\footnote{The reason 
the parametric dependence of $\tau_{\text{LPM}}$ in Table.~\ref{tab:timescales} differs to that from Eq.~\eqref{eq:lpm_ft}, is that 
in Ch.~\ref{ch:eloss}, we were always considering the energy of the radiation to be much larger than the temperature.}, in order 
to probe the robustness of the $\tau_{\text{min}}\sim 1/T$ assumption, made by the authors in \cite{Liou:2013qya,Blaizot:2013vha}. In 
doing so, we come across some potential issues:
\begin{itemize}
    \item This choice of $\tau_{\text{min}}$ is consistent for the rarer, harder exchanges between the jet and medium but fails 
    for softer scatterings, which take place over a longer duration. In particular, the duration-dependent 
    nature of the IR cutoff seems rather troubling.
    \item We also note that, for frequencies parametrically smaller
    than the temperature, which $\tau_{\text{min}}\sim1/T$ allows, the bottom-left corner of the triangle 
    becomes ill-defined, since one can have $\sqrt{k^+/\hat{q}_0}<1/g^2T$.
    \item Furthermore, the factorisation into soft and collinear logarithms undergirding 
    Eq.~\eqref{eq:BDIMzeroT} is predicated 
    on a collinear expansion $k^+\gg k_\perp$, i.e. $\tau k^+\gg 1$ \cite{Iancu:2014kga}.
    This $\ln \tau>-\ln k^+$ line in principle excludes parts of Fig.~\ref{fig:bdimtriangle}:
    the $(b)-(c)$ vertex is below it, since $\hat{q}_0\tau_{\text{min}}^3\sim g^4\ll 1$ for $\tau_{\text{min}}\sim 1/T$.
    The exclusion stretches to the $\ln \tau=-\ln k^+$ line,
    which crosses lines $(c)$ and $(b)$ if $\mu_{\perp}\tau_{\text{min}}>1$, i.e. $\mu_{\perp}>T$.
    If instead $\mu_{\perp}<T$, it crosses lines $(a)$ and $(b)$. 
\end{itemize}
Can the issues above be addressed by going back to Eq.~\eqref{eq:BDIMzeroT} and adapting the limits there or do we need to include the statistical 
function, as in Eq.~\eqref{eq:BDIMnonzeroT}? If $T$ is much smaller than the lower boundary on the energy integration, 
which from Fig.~\ref{fig:bdimtriangle}
is $\hat{q}_0\tau_{\text{min}}^2$, then the statistical function will be exponentially suppressed 
and we will of course recover Eq.~\ref{eq:BDIMresultzeroT}. We have thus sharpened the previous question: is 
$\hat{q}_0\tau_{\text{min}}^2\gg T$ or rather $\tau_{\text{min}}\gg\sqrt{T/\hat{q}_0}\sim1/(g^2 T)$
consistent with single scattering?

\begin{table}[ht]
    \centering
    \begin{tabular}{|c|c|c|c|}
    \hline
      \text{Quantity}& \text{Parametric Dependence} & \text{Symbol} \\
      \hline
        \text{Hard scattering duration} & $1/T$ & -- \\
        \text{Soft scattering duration} & $1/(gT)$ & -- \\
        \text{Soft scattering mean free time} & $1/(g^2T)$ & $t_{\text{el}}$\\
        \text{LO} $\hat{q}(\mu_{\perp})$ in HOA & $g^4T^3$ & $\hat{q}_0$\\
        \text{LPM} formation time & $\gtrsim 1/(g^2T)$ & $\tau_{\text{LPM}}$\\
        \hline
    \end{tabular}
    \caption{Some timescales associated with a weakly coupled QGP. By hard scattering, we mean $2\to 2$ exchange
    between the jet and medium, which exchange momentum $\sim T$. By soft scattering, we mean $2\to 2$ exchange
    between the jet and medium, which exchamge momentum $\sim gT$.}
    \label{tab:timescales}
\end{table}

Well, $t_{\text{el}}\sim1/g^2T$ is the mean free time between the frequent soft,  $k_{\perp}\sim gT$ scattering between the jet and medium 
constituents. The resummation of these scatterings on this $1/g^2T$ timescale makes it so that
 $\tau\gtrsim 1/g^2T$ in the \emph{multiple scattering regime}, thus answering negatively our previous question. Moreover, in what is to follow, 
we identify a \emph{strict single-scattering} regime, also giving rise to double-logarithmic corrections, where the formation time is \emph{a priori} $\tau\ll 1/g^2T$. 
We can therefore conclude that it is not reasonable to put such a demand on $\tau_{\text{min}}$ and that the $k^+$ 
integral in Eq.~\eqref{eq:BDIMnonzeroT} should run over the thermal scale.

Furthermore, our position is that the definition of single-scattering 
from \cite{Liou:2013qya,Blaizot:2013vha} is not exclusively that of a single scattering: the collinear divergence gets cut off
when $k_{\perp}$ becomes small enough ($\tau$ large enough) that multiple scatterings start to contribute
within a formation time, and that this happens before entering the deep LPM regime. However, at double-logarithmic accuracy one cannot distinguish between a
$ k_{\perp}^2\gg \hat{q}_0\tau$ and a $ k_{\perp}^2\sim \hat{q}_0\tau$ boundary, and 
in the HOA, one cannot disentangle multiple soft scatterings from a single harder one.

To shed some light on the issues raised above, our strategy is the following: 
we introduce an intermediate regulator $\tau_{\text{int}}$, 
with $1/gT \ll \tau_{\text{int}}\ll 1/g^2T$. The lower boundary will be discussed soon, whereas 
the upper boundary makes it so that for $\tau >\tau_{\text{int}}$ 
we thus include parts of the strict single scattering regime,
as well as the ``few scatterings'' regime just mentioned. We can then rewrite 
Eq.~\eqref{eq:BDIMnonzeroT} with the new integration boundaries, which 
we have argued in favour of 
\begin{equation}
    \delta \hat{q}(\mu_{\perp})^{\mathrm{few}}=\frac{\alpha_s C_R}{\pi}\hat{q}_0
    \int_{\tau_{\text{int}}}^{\mu_{\perp}^2/\hat{q}_0}\frac{d\tau}{\tau}
    \int_{\hat{q}_0\tau^2}^{\mu_{\perp}^2\tau}\frac{dk^+}{k^+}\left(1+2n_{\text{B}}(k^+)\right).
    \label{eq:BDIMnonzeroTred}
\end{equation}
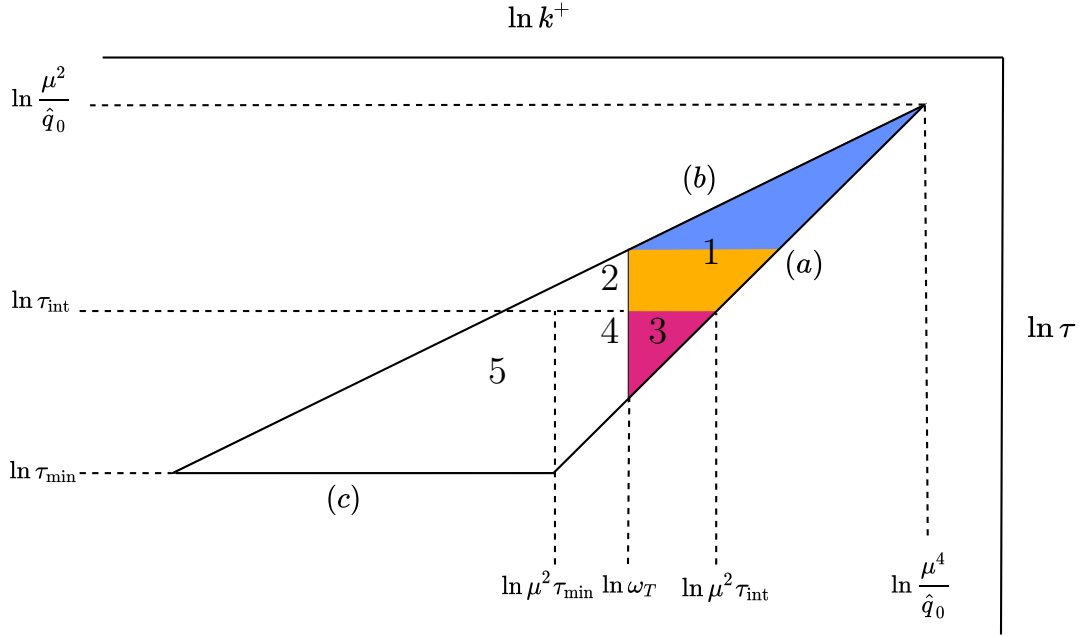
\begin{figure}[t]
	\centering
	\includegraphics[width=\textwidth]{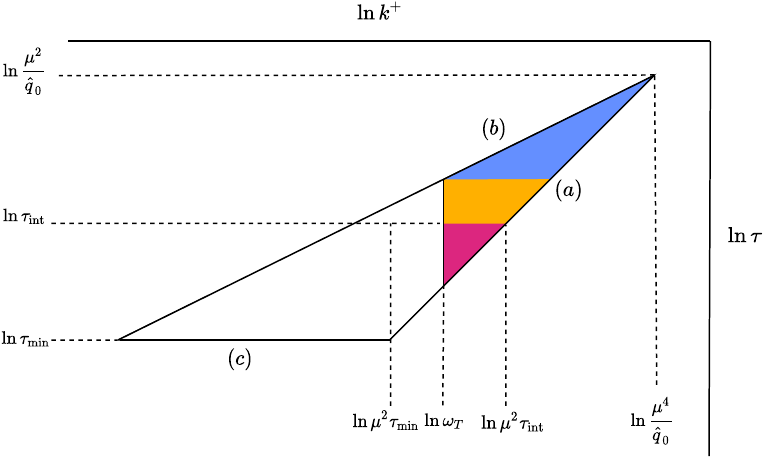}
    \put(-140,140){\Large{1}}
    \put(-178,130){\Large{2}}
    \put(-160,110){\Large{3}}
    \put(-178,110){\Large{4}}
    \put(-220,95){\Large{5}}
	\caption{A pictorial representation of how the phase space of the radiated gluon is partitioned 
    once thermal effects are taken into account. See the main text 
    for the explanation of the different subregions.} 
    \label{fig:triangle}
\end{figure}
To calculate Eq.~\eqref{eq:BDIMnonzeroTred}, the additional complication that we need 
to overcome (with respect to the calculation of Eq.~\eqref{eq:BDIMzeroT}) is dealing
with the statistical function in the $k^+$ integrand. In order to demonstrate 
how this can be done, let us for the moment consider the integral $\int_{\nu_{\text{IR}}}^{\nu_{\text{UV}}}dk^+n_{\text{B}}(k^+)/k^+$
with $\nu_{\text{UV}}\gg T\gg\nu_{\text{IR}}$. The idea is then to introduce an intermediate regulator, $\epsilon$
in the spirit of dimensional regularisation
\begin{align}
    \int^{\nu_{\text{UV}}}_{\nu_{\text{IR}}}\frac{dk^{+}}{k^{+}}n_{\text{B}} (k^{+})
    &=\lim_{\epsilon\to 0}\left[\int^{\nu_{\text{UV}}}_{0}\frac{dk^{+}}{k^{+}}k^{+\epsilon} n_{\text{B}} (k^{+})
    -\int^{\nu_{\text{IR}}}_{0}\frac{dk^{+}}{k^{+}}k^{+\epsilon}  n_{\text{B}} (k^{+})\right]\nonumber\\
    &
    = \lim_{\epsilon\to 0}\left[\int^{\infty}_{0}\frac{dk^{+}}{k^{+}}k^{+\epsilon}  n_{\text{B}} (k^{+})
    -\int^{\nu_{\text{IR}}}_{0}\frac{dk^{+}}{k^{+}}k^{+\epsilon}  \left(\frac{T}{k^+}-\frac{1}{2}\right)\right]
    +\mathcal{O}\left(\frac{\nu_{\text{IR}}}{T},e^{-\frac{\nu_{\text{UV}}}{T}}\right)\nonumber \\
    &= \lim_{\epsilon\to 0}\left[T^\epsilon \zeta(\epsilon)\Gamma(\epsilon) - 
    \left(\frac{T\nu_{\text{IR}}^{\epsilon}}{\nu_{\text{IR}}(\epsilon-1)}-\frac{\nu_{\text{IR}}^{\epsilon}}{2\epsilon}\right)\right]
    +\mathcal{O}\left(\frac{\nu_{\text{IR}}}{T},e^{-\frac{\nu_{\text{UV}}}{T}}    \right)\nonumber\\
    &=\frac{T}{\nu_{\text{IR}}}+\frac{1}{2}\ln\frac{\nu_{\text{IR}} e^{\gamma_E}}{2 \pi  T} 
    +\mathcal{O}\left(\frac{\nu_{\text{IR}}}{T},e^{-\frac{\nu_{\text{UV}}}{T}}    \right)\,. \label{eq:kplusintthermal}
\end{align}
The main advantage is that it allows us to use the known analytically-continued integrations
of the Bose--Einstein distribution in terms of the Riemann $\zeta$ and Euler $\Gamma$ functions.
Moreover, we take this opportunity to define the thermal scale $\omega_{T}\equiv 2\pi T \exp(-\gamma_{E})$ 
with $\gamma_{E}$ the Euler--Mascheroni constant.

In terms of Eq.~\eqref{eq:BDIMnonzeroTred}, let us look at the thermal part first, i.e
\begin{equation}
    \int_{\tau_{\text{int}}}^{\frac{\mu_{\perp}^2}{\hat{q}_0}}\frac{d\tau}{\tau}
    \int_{\hat{q}_0\tau^2}^{\mu_{\perp}^2\tau}\frac{dk^+}{k^+}n_{\text{B}}(k^+) = 
    \int_{\hat{q}_0\tau_{\text{int}}^2}^{\mu_{\perp}^2\tau_{\text{int}}}\frac{dk^+}{k^+}n_{\text{B}}(k^+)
    \int_{\tau_{\text{int}}}^{\sqrt{\frac{k^+}{\hat{q}_0}}}\frac{d\tau}{\tau}
    +\int^{\frac{\mu_{\perp}^4}{\hat{q}_0}}_{\mu_{\perp}^2\tau_{\text{int}}}\frac{dk^+}{k^+}n_{\text{B}}(k^+)
    \int_{\frac{k^+}{\mu_{\perp}^2}}^{\sqrt{\frac{k^+}{\hat{q}_0}}}\frac{d\tau}{\tau}.
    \label{eq:BDIMnonzeroTredstart}
\end{equation}
In our $\hat{q}_0\tau_{\text{int}}^2\ll \omega_{T} \ll \mu_{\perp}^2 \tau_{\text{int}}$ hierarchy, the second integral is exponentially suppressed.
For the first one, we can proceed as follows
\begin{align}
    \int_{\hat{q}_0\tau_{\text{int}}^2}^{\mu_{\perp}^2\tau_{\text{int}}}\frac{dk^+}{k^+}n_{\text{B}}(k^+)
    \int_{\tau_{\text{int}}}^{\sqrt{\frac{k^+}{\hat{q}_0}}}\frac{d\tau}{\tau}=
    \frac{1}{2}\int_{\hat{q}_0\tau_{\text{int}}^2}^{\mu_{\perp}^2\tau_{\text{int}}}\frac{dk^+}{k^+}n_{\text{B}}(k^+)
    \ln\frac{k^+}{\hat{q}_0\tau_{\text{int}}^2}.
\end{align}
We can then exploit that 
\begin{equation}
    \label{polygamma}
    \int_{0}^{\infty}\frac{dk^+}{k^+}k^{+\epsilon}n_{\text{B}}(k^+)\ln\frac{k^+}{T}=
    T^{\epsilon} \Gamma (\epsilon) \left(\zeta '(\epsilon)+\zeta (\epsilon) \psi(\epsilon)\right),
\end{equation}
where $\psi(x)$ is the digamma function. This, together with Eq.~\eqref{eq:kplusintthermal}, leads to 
\begin{align}
    \int_{\hat{q}_0\tau_{\text{int}}^2}^{\mu_{\perp}^2\tau_{\text{int}}}\frac{dk^+}{k^+}n_{\text{B}}(k^+)
    \int_{\tau_{\text{int}}}^{\sqrt{\frac{k^+}{\hat{q}_0}}}\frac{d\tau}{\tau}
   =&\frac{T}{2\hat{q}_0\tau_{\text{int}}^2}+\frac{1}{8}\left(- \ln^2\frac{\omega_{T}}{\hat{q}_0\tau_{\text{int}}^2}
  + \gamma_E^2 - \frac{\pi^2}{4}  +2\gamma_1\right)\nonumber\\
  &+\mathcal{O}\left(\frac{\hat{q}_0\tau_{\text{int}}^2}{T},
  e^{-\frac{\mu_{\perp}^2\tau_{\text{int}}}{T}}\right),
\end{align}
which in turn can be added to the straightforward vacuum part to give
\begin{align}
    \delta \hat{q}(\mu_{\perp})^{\mathrm{few}}=
    \frac{\alpha_s C_R}{2\pi}\hat{q}_0\bigg\{\frac{2T}{\hat{q}_0\tau_{\text{int}}^2}+
    \ln^2\frac{\mu_{\perp}^2}{\hat{q}_0 \tau_{\text{int}}} 
    -\frac{1}{2}\ln^2\frac{\omega_{T}}{\hat{q}_0\tau_{\text{int}}^2} + \frac{\gamma_E^2}{2} - \frac{\pi^2}{8}  +\gamma_1
    +\ldots
  \bigg\}.\label{eq:few_before}
\end{align}
At double-logarithmic accuracy, we then have
\begin{equation}
   \delta \hat{q}(\mu_{\perp})_\mathrm{dlog}^{\mathrm{few}}=\frac{\alpha_s C_R}{2\pi}\hat{q}_0\bigg\{\ln^2\frac{\mu_{\perp} ^2}{\hat{q}_0 \tau_{\text{int}}} 
    -\frac{1}{2}\ln^2\frac{\omega_{T}}{\hat{q}_0\tau_{\text{int}}^2}
    \bigg\}\quad \text{with}\;  
    \frac{\omega_{T}}{\mu_{\perp}^2}\ll\tau_{\text{int}}\ll\sqrt{\frac{\omega_{T}}{\hat{q}_0}}.
    \label{eq:BDIMresultnonzeroT}
\end{equation}
Graphically, if we  take $k^+=\omega_{T}$ as a vertical line in Fig.~\ref{fig:triangle}, the first term
in Eq.~\eqref{eq:BDIMresultnonzeroT}
comes from integrating over the entire ``1+2'' triangle, whereas the second term comes
from subtracting the smaller 2 triangle, which therefore does not contribute to 
double-logarithmic accuracy. Furthermore, 
this $k^+=\omega_{T}$ line intersects line $(b)$ at $\sqrt{\omega_{T}/\hat{q}_0}\sim 1/g^2T$, thus 
excluding the range where the formation time estimate becomes unreliable.

We also remark that, for $\tau_{\text{int}}>\tau_{\text{min}}$, the horizontal $\tau=\tau_{\text{int}}$ line
intersects the diagonal sides of the triangle at $k^+=\hat{q}_0\tau_{\text{int}}^2$ (line $(b)$) and 
$k^+=\mu_{\perp}^2 \tau_{\text{int}}$ (line $(a)$). The
form~\eqref{eq:BDIMresultnonzeroT} arises when the temperature scale $\omega_{T}$ falls in between these
two values, so that $\hat{q}_0\tau_{\text{int}}^2\ll \omega_{T} \ll \mu_{\perp}^2 \tau_{\text{int}}$, resulting in the range 
of validity expressed there. 

This is where the size of $\mu_{\perp}$ with respect to the medium scales enters. 
At leading order, i.e. without considering these radiative corrections, choosing $gT\ll \mu_{\perp} \ll T$
ensures that only the contributions of the soft modes \cite{Aurenche:2002pd} are included, whereas 
$T\ll \mu_{\perp}$ also includes the thermal-mode contribution \cite{Arnold:2008vd}. If we allow 
for radiative corrections and further require a strict single-scattering contribution,
we shall show that $\mu_{\perp}\gg \sqrt{g}T$ is necessary, so as to include 
the so-called \emph{semi-collinear processes} of \cite{Ghiglieri:2013gia,Ghiglieri:2015ala}.\footnote{%
\label{foot_small}
A good way of understanding why  $k_{\perp}\sim gT$ is not large enough,
thus moving us to the first available scaling $k_{\perp}\sim \sqrt{g}T$, $\mu_{\perp}>\sqrt{g}T$, 
is that for $gT<\mu_{\perp}<\sqrt{g}T$ the lines $(a)$ 
and $(b)$ intersect  at frequencies that are not parametrically 
larger than the temperature. The entire double-log triangle would 
then find itself at thermal or subthermal frequencies, effectively removing double-logarithmic enhancements.}
Finally, $\frac{\omega_{T}}{\mu_{\perp}^2}\ll\tau_{\text{int}}$ should be complemented by the duration bound $\tau_{\text{int}}\gg\tau_{\text{min}}$: 
for $ \mu_{\perp}\ll T$ this is automatically satisfied, whereas for $\mu_{\perp}\gtrsim T$
it will then be a further constraint to be imposed. In other words,
for $\mu_{\perp}<\sqrt{\omega_{T}/\tau_{\text{min}}}\sim T$ the vertical $k^+=\omega_{T}$ line intersects boundaries $(b)$ and $(a)$, 
whereas for $\mu_{\perp}>\sqrt{\omega_{T}/\tau_{\text{min}}}\sim T$ it would intersect $(b)$ and $(c)$. 

We assume $\sqrt{g}T\ll\mu_{\perp}\ll T$ for all remaining sections in this chapter, except for Sec.~\ref{sub:largemu}.

\section{Double Logarithmic Corrections from the Wilson Loop Setup}\label{sec:reproduce}
To briefly recapitulate, in Sec.~\ref{sec:dlog_lit} we described how the double logarithmic corrections
to $\hat{q}(\mu_{\perp})$ were calculated in \cite{Liou:2013qya}, taking care to highlight the 
boundaries of the region of phase space that those authors argued should give rise to these corrections. 
Whereas that calculation was done for a medium characterised by a random colour field, we 
showed in Sec.~\ref{sec:adapt} how, for the case of a dynamical medium, one can and in fact, one 
must account for contributions from the thermal scale. 

We resume here the computation started in Sec.~\ref{sec:wloop_gsq}. Our plan is then to show how 
the latter result can be obtained through our setup, while also arguing in favour 
of the necessity to include corrections from an additional region of phase space, which we coin 
the \emph{strict single-scattering regime}. To start, we draw on what we have 
learned regarding the nature of the double logarithmic enhancements in the previous 
sections to specify scalings for the coft and soft propagators introduced in Sec.~\ref{sec:wloop_gsq}.

 In order to 
ensure that the energy integration runs over the thermal scale, we let $\tilde{\lambda}=T/E$. Taking
$\lambda=\sqrt{g}$ then fixes the transverse component to $\sim\sqrt{g}T$ (see footnote~\ref{foot_small}) so that 
\begin{equation}
    (K+L)\sim (T, gT,\sqrt{g}T).\label{eq:coft_scaling}
\end{equation}
Before specifying the $L$ scaling, let us go back to the computation of $\delta\mathcal{C}(k_{\perp})=
\delta\mathcal{C}(k_{\perp})_{\mathrm{II+X}}+\delta\mathcal{C}(k_{\perp})_{\text{self}}+\delta\mathcal{C}(k_{\perp})_{\mathrm{Y}}$ in Sec.~\ref{sec:wloop_gsq}.
Once the Lorentz algebra is done, the coft propagator appears in all of the diagrams as 
\begin{equation}
    G^{--}_>(K+L)=G^{L}_{>}(K+L)+\frac{(\mathbf{k}+\mathbf{l})^2}{(\vec{\mathbf{k}}+\vec{\mathbf{l}})^2}G^{T}_{>}(K+L).
\end{equation}
In Coulomb gauge, the longitudinal component of the (bare) coft propagator vanishes.
The transverse component then reads, expanded in $g$, assuming Eq.~\eqref{eq:coft_scaling} and that 
$k^+\gg l^+$\footnote{If instead $k^+\sim l^+$, the resultant corrections would no longer 
be logarithmically enhanced.}
\begin{align}
    G_{>}^{T}(K+L)&=2\pi\varepsilon(k^{+}+l^{+})\Big(1+n_{\text{B}}(k^{+}+l^{+})\Big)\delta((K+L)^{2}-m_{\infty}^{2})\bigg\vert_{k^-=0}\nonumber
    \\&\approx
    2\pi\varepsilon(k^{+})\Big(1+n_{\text{B}}(k^{+})\Big)\frac{\delta(l^{-}-\frac{(\mathbf{k}+\mathbf{l})^2+m_{\infty}^2}{2k^+})}{2\abs{k^{+}}},\label{eq:transverse}
\end{align}
where $\varepsilon(x)$ is $1$ for $x>0$ and $-1$ for $x<0$. Upon implementing the 
single scattering condition, $k_{\perp}\gg l_{\perp}$ highlighted in Sec.~\ref{sec:dlog_lit}
\footnote{As $K^2\sim gT^2$, $m_{\infty}\sim gT$ is also neglected in this expansion.}
\begin{equation}
    G_{>}^{T}(K+L)\sim2\pi\varepsilon(k^{+})\Big(1+n_{\text{B}}(k^{+})\Big)\frac{\delta(l^{-}-\frac{k_{\perp}^2}{2k^+})}{2\abs{k^{+}}}.
\end{equation}
Presented like this, the delta function suggests the interpretation of $l^-\sim gT$ as an inverse formation time.
Moreover, the smallest momentum that can be acquired from the medium through a single scattering is, at LO $l_{\perp}\sim gT$\footnote{We
 remind the reader (see Sec.~\ref{sec:scatt_kernel}) that even though there are more frequent scatterings with $l_{\perp}\lesssim gT$, 
 they are suppressed because of the extra factor of $l_{\perp}^2$ in the numerator of $\mathcal{C}(l_{\perp})$.}, 
which, in turn, fixes $l^+\sim gT$.
The scaling of the soft modes thus read
\begin{equation}
    L\sim (gT, gT,gT).
\end{equation}
It has been shown that the interaction of the coft and soft modes
give rise to so-called \cite{Ghiglieri:2013gia,Ghiglieri:2015ala} semi-collinear processes.
These semi-collinear processes contribute to regions $3$ and $4$ in Fig.~\ref{fig:triangle}, which make up 
the strict single scattering regime. Here, the characteristic formation time, $1/l^-\sim gT$ overlaps 
with the duration of the soft, $l_{\perp}\sim gT$ scattering.

We can now add contributions from the three diagrams, i.e Eqs.~\eqref{eq:II_X_before_expansion}, ~\eqref{eq:aftercut} and ~\eqref{eq:y_before} 
before expanding the combination in the scaling that we have just motivated to get
\begin{align}
    \delta\mathcal{C}(k_{\perp})_{\text{semi}}&=2g^{4}C_{R}C_{A}\int\frac{dk^{+}}{2\pi}\int\frac{d^{4}L}{(2\pi)^{3}}\Big(1+n_{\text{B}} (k^+)\Big)
    \frac{\delta(l^{-}-\frac{k_{\perp}^{2}}{2k^{+}})}{k_{\perp}^{4}k^{+}}\frac{T}{l^{+}+\frac{l^{-}}{2}}
    \Bigg(l_{\perp}^{2}\rho^{--}(l)\nonumber
    \\&+2\rho^{T}(l)\Big({l^{-}}^{2}-\frac{l^{+}l^{-}l_{\perp}^{2}}{l^2}\Big)\Bigg).\label{eq:together}
\end{align}
In this scaling where $l^+\sim gT$ is soft, we only need to keep the zero mode contribution\footnote{This is 
made manifest by the fact that we have already made the replacement $\left(1+n_{B}\left(l^0\right)\right)\rightarrow T/l^0$.}. 
We are therefore permitted to make use of the ideas discussed in Sec.~\ref{sec:chtrick}; the $l^+$ integrand 
has after all already been encountered in the calculation of the LO scattering rate in Sec.~\ref{sec:scatt_lo_soft}
\footnote{In terms of doing the $l^+$ integration in Eq.~\eqref{eq:together}, one might worry about what happens when $l^0=l^+ +l^-/2=0$.
 However, this is only a spurious pole since the spectral functions are linear in $l^0$ as $l^0\to 0$, implying 
 that they can be dealt with using a Principal Value prescription.}. Doing the $l^+$ integration yields
 \begin{align}
    \delta\mathcal{C}(k_{\perp})_{\text{semi}}&=2g^{4}C_{R}C_{A}T\int\frac{dk^{+}}{2\pi}\int\frac{d^{2}l_{\perp}}{(2\pi)^{2}}\int dl^{-}\Big(1+n_{\text{B}} (k^+)\Big)
    \frac{\delta(l^{-}-\frac{k_{\perp}^{2}}{2k^{+}})}{k_{\perp}^{4}k^{+}}\nonumber
    \\&\times\Bigg(-\frac{l_{\perp}^{2}}{l_{\perp}^{2}+{l^{-}}^{2}+m_{D}^{2}}+\frac{l_{\perp}^{4}}{(l_{\perp}^{2}+{l^{-}}^{2})^{2}}\nonumber
    +\frac{l_{\perp}^{2}{l^{-}}^{2}}{(l_{\perp}^{2}+{l^{-}}^{2})^{2}}+\frac{2{l^{-}}^{2}}{l_{\perp}^{2}+{l^{-}}^{2}}\Bigg)\nonumber
    \\&=2g^{4}C_{R}C_{A}T\int\frac{dk^{+}}{2\pi}\int\frac{d^{2}l_{\perp}}{(2\pi)^{2}}
    \int dl^{-}\Big(1+n_{\text{B}} (k^+)\Big)\frac{\delta(l^{-}-\frac{k_{\perp}^{2}}{2k^{+}})}
    {k_{\perp}^{4}k^{+}}\nonumber
    \\&\times\Bigg(\frac{m_{D}^{2}l_{\perp}^{2}}{(l_{\perp}^{2}+{l^{-}}^{2})(l_{\perp}^{2}
    +{l^{-}}^{2}+m_{D}^{2})}+\frac{2{l^{-}}^{2}}{l_{\perp}^{2}+{l^{-}}^{2}}\Bigg)\label{eq:lplusintegral}.
\end{align}
Despite having derived the above result assuming that $l^-\sim gT$, it is instructive 
to see what happens when we take the limit where $l^-\ll l^+,\,l_{\perp}$. In position space, 
this scaling would then correspond to a process (between the jet and medium), which takes place 
instantaneously in the $x^+$ direction. This instantaneous approximation 
is consistent with the assumptions made in \cite{Liou:2013qya,Blaizot:2013vha}. It is 
therefore not surprising that by taking this limit, we recover
 \begin{equation}
    \delta\mathcal{C}(k_{\perp})_{\text{semi},\,l^{-}\to 0}=2g^4C_{R}C_{A}T\int\frac{dk^{+}}{2\pi}\int\frac{d^{2}l_{\perp}}{(2\pi)^{2}}
    (1+n_{\text{B}}(k^+))\frac{1}{k_{\perp}^4k^+}\frac{m_{D}^2l_{\perp}^2}{l_{\perp}^2(l_{\perp}^2+m_{D}^2)},
 \end{equation}
 which after taking the second moment with respect to $k_{\perp}$ reduces to Eq.~\eqref{eq:BDIMnonzeroTred}. 
 
\begin{figure}[t]
    \begin{center}
        \includegraphics[width=6cm]{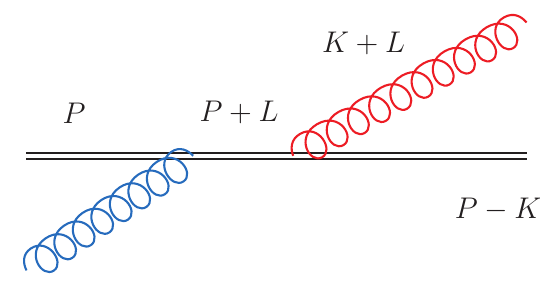}
    \end{center}
    \caption{One of the extra diagrams that appear once the duration
    of the jet-medium interaction is comparable to the gluon formation time. 
    The blob represents a resummed Hard Thermal Loop: the $L$ gluon is thus a time-like 
    plasmon. Diagrams corresponding to the other processes of Fig.~\ref{fig:dl_diags},
    as well as the plasmon emission/gluon absorption crossings are not shown.}
    \label{fig:semidiag}
\end{figure}

This slight detour highlights the fact that investigating these semi-collinear processes with $l^-\sim l^+,\,l_{\perp}$
allows us to go beyond the instantaneous approximation; $L$ is no longer restricted to spacelike values, thus opening 
the phase space for the absorption and emission of soft, timelike plasmons (see Sec.~\ref{sec:htl}), as shown in
Fig.~\ref{fig:semidiag}. Furthermore, as $l^-$ is larger we understand explicitly that the processes' associated 
formation time is smaller (comparable to the duration of the elastic scatterings), signifying that 
we are accessing the strict single scattering regime, with $\tau\ll\tau_{\text{int}}$ (see Fig.~\ref{fig:triangle}).

Let us write 
\begin{equation}
    \label{resultfromsemi2}
    \delta\mathcal{C}(k_{\perp})_\mathrm{semi}= \frac{g^2 C_R}{\pi k_{\perp}^4}\int\frac{dk^+}{k^+}\int dl^- \delta\left(l^- -\frac{k_{\perp}^2}{2k^+}\right)
    (1+n_{\text{B}}(k^+))
    \hat{q}\left(\rho_{\perp};l^-\right)
\end{equation}
with $\hat{q}(\rho_{\perp};l^-)$ is a modified (adjoint) $\hat{q}$ \cite{Ghiglieri:2013gia,Ghiglieri:2015ala}
that also accounts for the $l^-$-dependence, i.e.
\begin{equation}
    \label{defqhatde}
   \hat{q}(\rho_{\perp};l^-) =g^2 C_A T\int^{\rho_{\perp}}\frac{d^2l_{\perp}}{(2\pi)^2}
     \left[
         \frac{m_D^2l_{\perp}^2}{
         (l_{\perp}^2+l^{-2})(l_{\perp}^2+l^{-2} +m_D^2)}+2\frac{ l^{-2}}{l_{\perp}^2+l^{-2}} 
     \right].
\end{equation}
As is shown in detail in App.~\ref{app:AX}, the second term appears 
in the LO computation of the hard contribution to $\hat{q}(\mu_{\perp})$. It's 
presence here is perhaps unsurprising; the computation \cite{Arnold:2008vd}
also considers  elastic $2\leftrightarrow2$ scatterings exchanging $k_{\perp}\gg gT$.
As that calculation treats this gluon with bare propagators, it does not properly account 
for its soft dynamics, as encoded by HTL resummation. Hence, the semi-collinear 
limit of \cite{Arnold:2008vd} should be subtracted, to avoid double counting it. We are then left with
\begin{equation}
    \label{defqhatdesubtr}
    \hat{q}(\rho_{\perp};l^-)_{\text{subtr}} \equiv\hat{q}(\rho_{\perp};l^-)-
    \hat{q}(\rho_{\perp};l^-)
    _{\text{\cite{Arnold:2008vd}}} 
    =g^2 C_A T
    \int^{\rho_{\perp}}\frac{d^2l_{\perp}}{(2\pi)^2}
         \frac{m_D^2l_{\perp}^2}{
         (l_{\perp}^2+l^{-2})(l_{\perp}^2+l^{-2} +m_D^2)}.
\end{equation}
Doing the $l_{\perp}$ integration, assuming that $\rho_{\perp}\ll k_{\perp}\ll\mu_{\perp}$, while 
at the same time taking $\rho_{\perp}\gg l^-,\,m_{D}$ yields
\begin{equation}
    \label{qhatdecalc}
   \hat{q}(\rho_{\perp};l^-)_{\text{subtr}}
   = \alpha_s C_A T
   \bigg\{\underbrace{m_D^2 \ln\left(\frac{\rho_{\perp}^2}{m_D^2}\right)}_{\mathrm{HO}}\,
    -\underbrace{l^{-2}\ln\left(1+\frac{m_D^2}{l^{-2}}\right)-m_D^2\ln\left(1+\frac{l^{-2}}{m_D^2}\right)}_{l^-\,\text{-- dependent}}
    \bigg\}
   .
\end{equation}

\subsection{HOA Contribution to Strict Single Scattering}\label{sec:hoa_ss}
Let us for the moment only concentrate on the first term.
The UV logarithmic divergence, characterised by $\rho_{\perp}$ signals the fact that 
there lies another region of phase space bordering this one that could also 
give rise to double logarithmic corrections to $\delta\hat{q}(\mu_{\perp})$. We will come back to 
this issue. Applying the HOA to the first term above, 
we get
\begin{equation}
    \hat{q}_{0}(\rho_{\perp})=\alpha_sC_{A}Tm_D^2\ln\left(\frac{\rho_{\perp}^2}{m_{D}^2}\right)\to \hat{q}_0,
\end{equation}
where we have inferred the value of $\hat{q}_0$ from the single-scattering condition, $\rho_{\perp}\ll\mu_{\perp}$.
This approximation is what has been done to arrive at Eq.~\eqref{eq:lit_hoa}. To 
get the double logarithmic correction, we then only need to do the $k^+$ and $k_{\perp}$ integrals. 
As was done in Sec.~\ref{sec:dlog_lit}, we instead do the integration in terms of the formation time, $\tau$
\begin{equation}
    \delta\hat{q}_{\text{semi}}^{\text{HO}}(\mu_{\perp})=\frac{\alpha_s C_R}{\pi}\hat{q}_0
    \int_{k^+_{\text{IR}}}^{\mu_{\perp}^2\tau_{\text{int}}}\frac{dk^+}{k^+}\left(1+2n_{\text{B}}\left(k^+\right)\right)\int_{k^+/\mu_{\perp}^2}^{\tau_{\text{int}}}\frac{d\tau}{\tau}
    \label{eq:sss_before}
\end{equation}
The integration limits match those that come from integrating over regions $3$ and $4$
of Fig.~\ref{fig:triangle}\footnote{Technically, the integration here assumes a generic 
IR cutoff for the energy, satisfying $gT\ll k_{\text{IR}}\ll T$. Equating $k_{\text{IR}}=\mu_{\perp}^2\tau_{\text{min}}$ 
would precisely match the boundaries of regions $3$ and $4$ in Fig.~\ref{fig:triangle}. Our reason for 
keeping $k_{\text{IR}}$ generic is that it will allow us to make cleaner contact with the region, from 
which the classical corrections emerge in Sec.~\ref{sec:class}.}. The integration 
here can be dealt with in exactly the same way as for Eq.~\eqref{eq:BDIMnonzeroTred}. The result 
reads
\begin{align}
    \label{semievaltaufinal}
    \delta\hat{q}_\mathrm{semi}^{\text{HO}}(\mu_{\perp})&= \frac{\alpha_s C_R }{2\pi} \hat{q}_0
    \bigg[ \frac{4T \ln \left(\frac{\mu_{\perp}^2
   \tau_{\text{int}}}{k^{+}_{\text{IR}} e}\right)}{k^{+}_{\text{IR}}}+
   \left( \ln
   ^2\left(\frac{\mu_{\perp}^2 \tau_{\text{int}}}{\omega_{T}}\right)-2 \gamma _1+\frac{\pi ^2}{4}- \gamma_E
   ^2\right)\bigg]\nonumber
   \\&+\mathcal{O}\left(k^{+}_{\text{IR}},e^{-\mu_{\perp}\tau_{\text{int}}^2/T}\right),
\end{align}
where $\gamma_{n}$ is the nth Stieltjes constant.
The terms above depending on the IR cutoff, $k^{+}_{\text{IR}}$ indeed suggest sensitivity to the 
region giving rise to classical corrections \cite{CaronHuot:2008ni}. We will elaborate on this 
further in Sec.~\ref{sec:class}. For completeness, we specify the double logarithmic contribution
\begin{equation}
    \delta\hat{q}_\mathrm{semi\,dlog}^{\text{HO}}(\mu_{\perp})=\frac{\alpha_s C_{R}}{2\pi}\hat{q}_0\ln^2\frac{\mu_{\perp}^2\tau_{\text{int}}}{\omega_T}
    \label{eq:ss_ho_final}
\end{equation}
We now add this result to that from in the previous section, Eq.~\eqref{eq:BDIMresultnonzeroT}, the double logarithmic 
contribution coming from integrating over the region for which $\tau\geq\tau_{\text{int}}$
\begin{equation}
    \delta\hat{q}(\mu_{\perp})_{\text{dlog}}=\delta\hat{q}^{\text{few}}_{\text{dlog}}(\mu_{\perp})+\delta\hat{q}_\mathrm{semi\,dlog}^{\text{HO}}(\mu_{\perp})
    =\frac{\alpha_s C_{R}}{4\pi}\hat{q}_0\ln^2\frac{\mu_{\perp}^4}{\hat{q}_0\omega_T}.\label{eq:dlog_final}
\end{equation}
Upon adding the two contributions together, it is reassuring to observe the disappearance of the intermediate cutoff, $\tau_{\text{int}}$.
Moreover, with respect to Eq.~\eqref{eq:BDIMresultzeroT}, the IR regulator $\tau_{\text{min}}$ has been replaced with the 
thermal scale, $\omega_T$. This is one of the main results of this thesis; by carefully 
including the contribution from the thermal scale, we have bypassed many of the issues associated with $\tau_{\text{min}}$, which 
were highlighted in Sec.~\ref{sec:adapt}. As we will demonstrate in the next section, 
Eq.~\eqref{eq:dlog_final} turns out to be the dominant contribution to overall double logarithmic correction to $\hat{q}(\mu_{\perp})$.

It is interesting to note that the correction above corresponds (up to the prefactor of $\alpha_sC_{R}\hat{q}_0/\pi$) 
to the area of the sum of the ``1'' and ``3'' regions from Fig.~\ref{fig:triangle}. This observation 
that neither region ``2'' nor ``4 ''contribute makes 
transparent the fact that, up to double-logarithmic accuracy, the effect of a populated medium can be understood 
as cutting off the phase space with a vertical, $k^+>\omega_T\sim T$ line instead of a horizontal $\tau>\tau_{\text{min}}$ 
line.

Why does the thermal scale have such an effect? To start, we note that 
for $k^+\ll T$
\begin{equation}
    2n_{\text{B}}(k^+)\approx\frac{2T}{k^+}-1.
\end{equation}
Looking to Eq.~\eqref{eq:sss_before}, we note that in this limit, the part of the $k^+$ integrand giving rise to 
the logarithmic enhancement will be washed away; the vacuum part -- $1$ -- cancels against the first 
quantum correction coming from the expanded $n_{\text{B}}(k^+)$\footnote{Analogous cancellations
 of this kind are known in the literature: they are discussed briefly and in general terms in \cite{Heinz:1986kz}. Similar 
 cancellations also appear in the power corrections to Hard Thermal Loops, which receive both 
 vacuum and thermal contributions\cite{Manuel:2016wqs,Carignano:2017ovz}. These are separately IR-divergent but their sum is IR-finite.}. This is exactly what is exhibited in 
Eq.~\eqref{semievaltaufinal}, where $k_{\text{IR}}^+$ only appears in the power law term\footnote{
Despite the presence of $k_{\text{IR}}^+$ in a single logarithm in Eq.~\eqref{semievaltaufinal}, that term
    can still be considered benign as it multiplies a power of $k_{\text{IR}}^+$. As was discussed in Sec.~\ref{sec:eft},
    corrections that depend on powers of the cutoff are unphysical and will inevitably cancel against other 
    power law corrections from other regions. We explicitly show how this works with this 
    power law correction in Sec.~\ref{sec:class}.}.

\subsection{\texorpdfstring{$l^-$}{TEXT} -- Dependent Contrbution to Strict Single Scattering}
We now turn to the calculation of the $l^-$ -- dependent terms in Eq.~\eqref{qhatdecalc}. As we 
will see, this $l^-$ dependence renders the integration a little more challenging. Before beginning its evaluation, 
we use the delta function in Eq.~\eqref{resultfromsemi2} to rewrite everything 
in terms of the formation time, i.e $\tau=k^+/k_{\perp}^2=1/(2l^-)$. The starting point is 
\begin{align}
    \mathcal{I}(\mu_{\perp})\equiv
    \int_{k^{+}_{\text{IR}}}^{\mu_{\perp}^2\tau_{\text{int}}}\frac{dk^+}{k^+}
    \left(\frac{1}{2}+n_{\text{B}}(k^+)\right) \int_{k^+/\mu_{\perp}^2}^{\tau_{\text{int}}}\frac{d\tau}{\tau}
    \left[-\frac{1}{4\tau^{2}}\ln\left(1+4m_D^2\tau^2\right)-
    m_D^2\ln\left(1+\frac{1}{4m_D^2\tau^2}\right)\right].
\end{align}
The $\tau$ integration yields
\begin{align}
    \mathcal{I}(\mu_{\perp})=
    \int_{k^{+}_{\text{IR}}}^{\mu_{\perp}^2\tau_{\text{int}}}\frac{dk^+}{k^+}
    \left(\frac{1}{2}+n_{\text{B}}(k^+)\right) \bigg[&\frac{m_D^2}{2} \mathrm{Li}_2
    \left(-\frac{\mu_{\perp}^4}{4k^{+2}m_D^2}\right)
    -\frac{m_D^2}{2}\ln\left(1+\frac{\mu_{\perp}^4}{4k^{+2}m_D^2}\right) \nonumber \\
    &-\frac{\mu_{\perp}^4}{8k^{+2}}\ln\left(1+\frac{4k^{+2}m_D^2}{\mu_{\perp}^4}\right)
    +\mathcal{O}\left(\frac{1}{\tau_{\text{int}}^2}\right) \bigg],
\end{align}
where $\mathrm{Li}_2$ is the dilogarithm and we have expanded for $m_D\tau_{\text{int}}\gg 1$, recalling that $1/gT\ll \tau_{\text{int}}\ll 1/g^2T$.
The vacuum ($1/2$) part needs to be integrated as is, without further expansions: while close to 
the IR boundary we could exploit that $\mu_{\perp}^2/(2k^+m_D)\gg 1$, that would not be true 
close to the UV boundary, where instead $\mu_{\perp}^2/(2k^+m_D)\ll 1$. The integral is however
not problematic and yields
\begin{align}
    \mathcal{I}_\mathrm{vac}(\mu_{\perp})=&
    -\frac{1}{6} m_D^2 \bigg[\ln ^3\left(\frac{\mu_{\perp} ^2}{2 k^{+}_{\text{IR}}
    m_D}\right)+\frac{3}{2} \ln ^2\left(\frac{\mu_{\perp} ^2}{2 k^{+}_{\text{IR}}
    m_D}\right)+\frac{1}{4} \left(6+\pi ^2\right) \ln
    \left(\frac{\mu_{\perp} ^2}{2 k^{+}_{\text{IR}} m_D}\right)+\frac{1}{8} \left(6+\pi
    ^2\right)\bigg]\nonumber\\
    &+\mathcal{O}\left(\frac{1}{\tau_{\text{int}}^2},(k^{+}_{\text{IR}})^2\right),\label{Ivacfinal}
\end{align}
where we have again expanded in the cutoffs. For the thermal part, we can on 
the other hand expand for $\mu_{\perp}^2/(2k^+m_D)\gg 1$, as the effective UV cutoff
introduced by Boltzmann suppression makes the region where $\mu_{\perp}^2/(2k^+m_D)\lesssim 1$ exponentially
suppressed. Hence
\begin{align}
    \mathcal{I}_{T}(\mu_{\perp})=-m_D^2
    \int_{k^{+}_{\text{IR}}}^{\infty}\frac{dk^+}{k^+}
   n_{\text{B}}(k^+)\bigg[&\ln^2\frac{\mu_{\perp}^2}{2k^{+}m_D}
    +\ln\frac{\mu_{\perp}^2}{2k^{+}m_D}
    +\frac{1}{2}\left(1+\frac{\pi^2}{6}\right)\bigg]+\ldots,
\end{align}
where the dots stand for higher-order terms in the expansions in the cutoffs and for the exponentially
suppressed term arising from approximating the UV cutoff to infinity. This integral can be done with 
Eqs.~\eqref{eq:kplusintthermal}, \eqref{polygamma} and
\begin{equation}
    \label{polygamma2}
    \int_{0}^{\infty}\frac{dk^+}{k^+}k^{+\epsilon}n_{\text{B}}(k^+)\ln^2\frac{k^+}{T}=
    T^{\epsilon} \Gamma (\epsilon) \left[\zeta ''(\epsilon)+2 \psi(\epsilon)
   \zeta '(\epsilon)+\zeta (\epsilon) \left(\psi (\epsilon)^2+\psi
   ^{(1)}(\epsilon)\right)\right],
\end{equation}
where $\psi^{(1)}$ is the polygamma function of order 1. Using these three integrals we find
\begin{align}
    \mathcal{I}_T(\mu_{\perp})=&-\frac{m_D^2 T \left( \ln^2 \left(\frac{\mu_{\perp} ^2}{2 k^{+}_{\text{IR}}
    m_D}\right)- \ln \left(\frac{\mu_{\perp} ^2}{2 k^{+}_{\text{IR}}
    m_D}\right)+\frac{\pi ^2}{12}+\frac32\right)}{ k^{+}_{\text{IR}}}\nonumber\\
    &+\frac{m_D^2}{6}\bigg\{
           \ln ^3\left(\frac{\mu_{\perp} ^2}{2 k^{+}_{\text{IR}}
        m_D}\right)- \ln ^3\left(\frac{\mu_{\perp} ^2}{2 m_D
        \omega_T}\right)
        +
        \frac32\ln^2\left(\frac{\mu_{\perp} ^2}{2 k^{+}_{\text{IR}} m_D}\right)-\frac32 \ln ^2\left(\frac{\mu_{\perp}
        ^2}{2 m_D \omega_T}\right)\nonumber \\
        &+
        \frac14\left(6+\pi ^2\right) \ln \left(\frac{\mu_{\perp} ^2}{2 k^{+}_{\text{IR}}
        m_D}\right)+\ln \left(\frac{\mu_{\perp} ^2}{2 m_D
        \omega_T}\right)\left[6 \gamma _1 -\pi ^2 +3 \gamma_E ^2 -\frac32 \right]\nonumber \\
        &
        -6 \gamma_E  \gamma _1-3 \left(-
        \gamma _1+ \gamma _2+\frac{\pi ^2}{8}\right)+\frac{1}{2} (3-4 \gamma_E ) \gamma_E ^2    
    \bigg\}+\ldots\,.\label{Itherm}
\end{align}
Upon summing Eqs.~\eqref{Ivacfinal} and \eqref{Itherm} we see that 
all logarithmic dependence on $k^{+}_{\text{IR}}$ vanishes, yielding
\begin{align}
    \mathcal{I}(\mu_{\perp})=&-\frac{m_D^2 T \left( \ln^2 \left(\frac{\mu_{\perp} ^2}{2 k^{+}_{\text{IR}}
    m_D}\right)- \ln \left(\frac{\mu_{\perp} ^2}{2 k^{+}_{\text{IR}}
    m_D}\right)+\frac{\pi ^2}{12}+\frac32\right)}{ k^{+}_{\text{IR}}}\nonumber\\
    &+\frac{m_D^2}{6}\bigg\{
        - \ln ^3\left(\frac{\mu_{\perp} ^2}{2 m_D
        \omega_T}\right)
       -\frac32 \ln ^2\left(\frac{\mu_{\perp}^2}{2 m_D \omega_T}\right)
        +\ln \left(\frac{\mu_{\perp} ^2}{2 m_D
        \omega_T}\right)\left[6 \gamma _1 -\pi ^2 +3 \gamma_E ^2 -\frac32 \right]\nonumber \\
        &
        -6 \gamma_E  \gamma _1-3 \left(-
        \gamma _1+ \gamma _2\right)+\frac{1}{2} (3-4 \gamma_E ) \gamma_E ^2    
        -\frac{1}{4} \left(3+2\pi
        ^2\right)\bigg\}+\ldots\,.\label{Ifinal}
\end{align}
Upon reinstating the proper prefactor we have
\begin{equation}
    \delta\hat{q}_\mathrm{semi}^{l^--\mathrm{dep}}(\mu_{\perp})=\frac{2\alpha_s^2 C_R C_A T}{\pi} \mathcal{I}(\mu_{\perp}),
    \label{semievaltaufinalnontot}
\end{equation}
so that
\begin{equation}
    \delta\hat{q}_\text{semi dlog}^{l^--\mathrm{dep}}(\mu_{\perp}) =-\frac{\alpha_{s}^2C_{R}C_{A}Tm_{D}^2}{3\pi}
    \ln^3\frac{e^{1/2}\mu_{\perp}^2}{2\omega_T m_{D}}.
    \label{eq:lm_final}
\end{equation}
Let us compare this result to Eq.~\eqref{eq:dlog_final}. Reinstating the Coulomb logarithm 
in the latter, it becomes, up to its overall prefactor, 
\begin{equation}
\delta\hat{q}_\mathrm{semi\,dlog}^{\text{HO}}(\mu_{\perp})\propto\ln\frac{\rho_{\perp}^2}{m_{D}^2}\ln^2\frac{\mu_{\perp}^2}{\sqrt{\hat{q}_0\omega_T}}.
\end{equation}
Eq.~\eqref{eq:lm_final} in parametrically smaller compared to $\delta\hat{q}_\mathrm{semi\,dlog}^{\text{HO}}(\mu_{\perp})$. This
 can be seen by noting that $\mu_{\perp}^2/\sqrt{\hat{q}_0\omega_T}$ is larger by $1/g$ than $\mu_{\perp}^2/(m_{D}\omega_T)$, 
 which is instead comparable to $\rho_{\perp}^2/m_{D}^2$, since $\rho_{\perp}\ll\mu_{\perp}$. We thus emphasise that 
 effects analysed in the semi-collinear evaluation, such as the relaxation of the instantaneous 
approximation are not relevant at double-logarithmic accuracy in the HOA.

 \section{Extension to \texorpdfstring{$\mu_{\perp}>T$}{TEXT}}
\label{sub:largemu}

Up until this point, we have taken $T\gg\mu_{\perp}\gg\sqrt{g}T$, the first range where double-logarithm corrections appear. We now move on to consider what happens
if $\mu_{\perp}\gg T$ in terms of the structure of the relevant portion of phase space. 

If we start from Fig.~\ref{fig:triangle} and proceed to increase $\mu_{\perp}$, line $(a)$
gets shifted to the right, making the original $(a)-(b)-(c)$ triangle larger. As we 
argued previously, once $\mu_{\perp}>\sqrt{\omega_{T}/\tau_{\text{min}}}\sim T$ the vertical $k^+=\omega_{T}$ line intersects
lines $(b)$ and $(c)$ rather than $(b)$ and $(a)$. Hence, nothing would change 
for the part of the evaluation for $\tau>\tau_{\text{int}}$, as given by Eq.~\eqref{eq:BDIMresultnonzeroT}.
On the other hand, the evaluation of the 
strict single scattering regime, for $\tau<\tau_{\text{int}}$, needs to be amended to account for this. It 
must also incorporate that for $\mu_{\perp}\gtrsim T$ we are including strict single
scatterings where $l_{\perp}$ is approaching or possibly exceeding the $T$ scale. In 
principle, one would thus need to account for the fact that the coefficient 
of the leading-log (harmonic oscillator) $\hat{q}_0$ is not constant. As shown 
in \cite{Arnold:2008vd} and summarized in Sec.~\ref{sec:scatt_kernel}, 
$\hat{q}_0(\rho\ll T)$ and $\hat{q}_0(\rho\gg T)$ differ at leading-log by
15\% for $N_f=3$ and $N_c=3$, so in a first approximation we may neglect this effect. We leave 
a more precise evaluation of semi-collinear processes in this region to future work.

\begin{figure}[t]
	\centering
	\includegraphics[width=\textwidth]{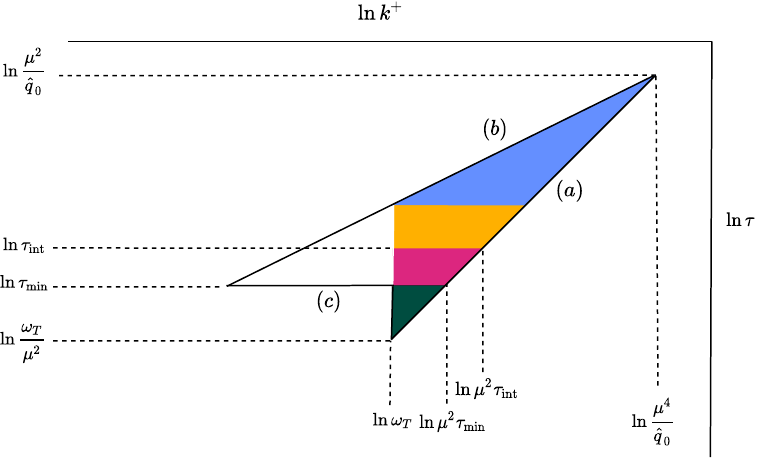}
    \put(-150,130){\Large{1}}
    \put(-205,115){\Large{2}}
    \put(-185,95){\Large{3}}
    \put(-205,95){\Large{5}}
    \put(-185,80){\Large{6}}
	\caption{Structure of phase space once the cutoff on $\mu_{\perp}$ is allowed to increase to 
    values much larger than the temperature. Notably, region ``6'' in dark green
    emerges. It cannot be well described by our setup due to the associated small formation time.}
    \label{fig:largemu}
\end{figure}
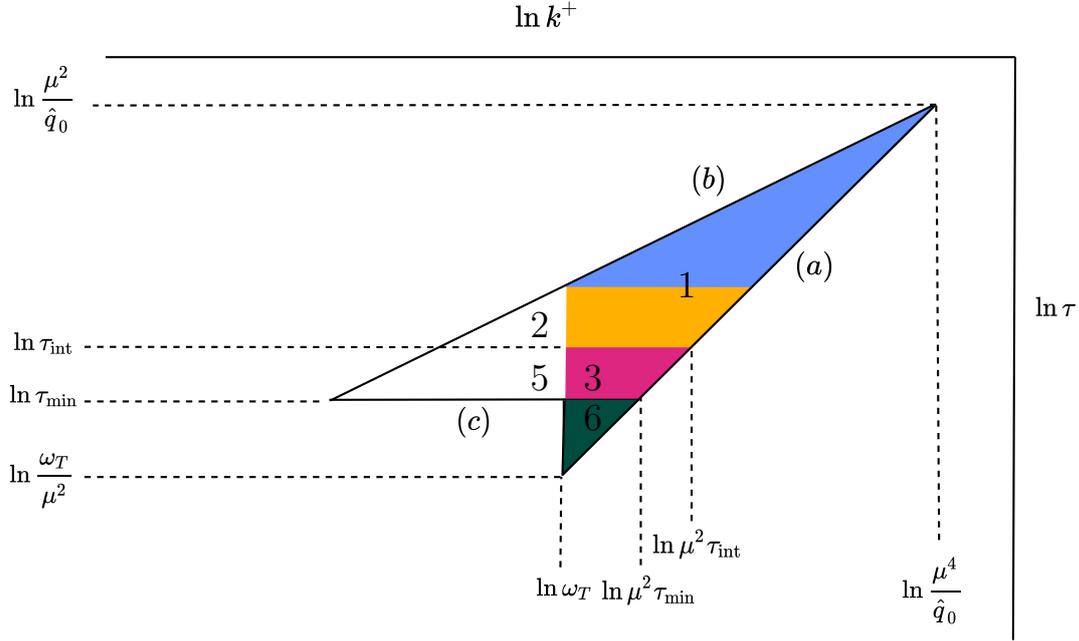
Under this approximation, we can simply extend Eq.~\eqref{eq:dlog_final} to the present case. 
This would however include the dark green triangle ``6'' in Fig.~\ref{fig:largemu},
which lies under the $\tau=\tau_{\text{min}}$ line. It is not clear whether this sharp line 
needs to be included, and only an evaluation like the one we show in detail in the previous section,
but extended to larger $\mu_{\perp}$ can address the effect of the relaxation of the instantaneous 
approximation in this setting. The expectation from the previous 
$\mu_{\perp}<T$ results is that this effect should be subleading. We however choose
to proceed conservatively here and subtract 
that slice from our result. Moreover, parts of this triangle lie below the $\tau k^+=1$ line,
further motivating its subtraction.
In our double-logarithmic approximation, this corresponds again,
for the reasons explained at the end of Sec.~\ref{sec:hoa_ss}, to subtracting off the area of the dark green triangle.
This leads to 
\begin{align}
    \delta \hat{q}(\mu_{\perp}\gg T)_\mathrm{dlog}
    &=\frac{\alpha_s C_R}{2\pi}\hat{q}_0(\rho_{\perp}\ll\mu_{\perp})\left[\frac{1}{2}\ln^2\left(\frac{\mu ^4}{\hat{q}_0 \omega_{T}}\right)-
    \ln^2\left(\frac{\mu_{\perp}^2\tau_{\text{min}}}{\omega_{T}}\right) \right]\nonumber \\
    &=\frac{\alpha_s C_R}{2\pi}\hat{q}_0(\rho_{\perp}\ll\mu_{\perp})\left[\ln^2\left(\frac{\mu ^2}{\hat{q}_0 \tau_{\text{min}}}\right)-
    \frac{1}{2} \ln^2\left(\frac{\omega_{T}}{\hat{q}_0 \tau_{\text{min}}^2}\right) \right] .
     \label{eq:dlogfinalhard}
 \end{align}
Unsurprisingly, the subtraction of the ``6'' triangle from the ``1+3+6''   triangle 
on the first line is equal to that of the unshaded ``2+5'' triangle from the original $(a)-(b)-(c)$ triangle
on the second line. This implies that, for $\mu_{\perp}>T$, our result is closer to the original
one of \cite{Liou:2013qya}.

This can be better appreciated by noting that the vertical line 
at $k^+=\mu_{\perp}^2\tau_{\text{min}}$ cuts the $(a)-(b)-(c)$ triangle into two triangles of equal area. Hence,
for $\omega_{T}>\mu_{\perp}^2\tau_{\text{min}}$, i.e. $\mu_{\perp}\lesssim T$ our result~\eqref{eq:dlog_final} is less than half of the original 
double logarithm~\eqref{eq:BDIMresultzeroT}, 
whereas for $\mu_{\perp}>\sqrt{\omega_{T}/\tau_{\text{min}}}$ our result~\eqref{eq:dlogfinalhard} is more than half of it.

\section{Analysis of Neighbouring Regions}

In the previous section, we showed that the dominant 
contribution to double logarithmic corrections to $\hat{q}(\mu_{\perp})$ in a weakly coupled QGP
are given by the result Eq.~\eqref{eq:dlog_final}. On the way to arriving at this result, we took
the HOA; the dependence on $\rho_{\perp}$, the UV cutoff for $l_{\perp}$ in the first term of  Eq.~\eqref{qhatdecalc}
signals sensitivity to a neighbouring region with larger $l_{\perp}$, which could also potentially possess a double logarithmic enhancement.
In this section, we explore a couple of adjacent regions of phase space and comment on whether or not they are relevant 
with respect to our ultimate pursuit.
\subsection{\texorpdfstring{$l_{\perp}\sim\sqrt{g}T,\,l^{+}\sim l^{-}\sim gT$}{TEXT}}

Since we now have $L^{2}\gg m_{D}^{2}$, we can start by expanding $G_{rr}^{--}(L)$, which will inevitably appear in our evaluation
\footnote{This can be understood as a consequence of the fact that in Fig.~\ref{fig:real}, the only 
non-zero contributions are those with horizontal cuts going through the HTL propagators, which produce either a 
$G_{>}^{--}(L)$ or $G_{<}^{--}(L)$. The cut propagators can then always be traded for an $rr$ propagator, since $l^0\ll T$ (see Eq.~\eqref{eq:grr}).}
\begin{align}
    G_{rr}^{--}(L)&=\Big(\frac{1}{2}+n_{\text{B}} (l^{0})\Big)\Big(\rho_{\text{B}}^{L}(l)+\frac{l_{\perp}^{2}}{l^2}\rho_{\text{B}}^{T}(l)\Big)\nonumber
    \\&\approx\frac{T}{l^{0}}\Big(\rho_{\text{B}}^{L}(l)+\frac{l_{\perp}^{2}}{l^2}\rho_{\text{B}}^{T}(l)\Big)\nonumber.
\end{align}
The expansion ($l^0\ll T$) is also valid for the scaling presented in the previous section.
However, looking in more detail at the spectral function, we see that
\begin{align}
    \rho^{L}_{\text{B}}(l)&=G_{R}^{L}(l)-G_{A}^{L}(l)\nonumber
    \\&=\frac{i}{l^2+\Pi_{R}^{L}(l)}-\frac{i}{l^2+\Pi_{A}^{L}(l)}\nonumber
    \\&\approx\frac{i}{l^2}-\frac{i\Pi_{R}^{L}(l)}{l^4}-\frac{i}{l^2}+\frac{i\Pi_{A}^{L}(l)}{l^4}\nonumber
\end{align}
where the causal HTL propagators have been defined in Eqs.~\eqref{htllong}, ~\eqref{htltrans}. 
Let us now further unpack the causal self energies
\begin{align}
    \Pi_{R/A}^{L}(l)&=m_{D}^{2}\Big(1-\frac{l^{0}\pm i\varepsilon}{2l}\log\frac{l^{0}+l\pm i\varepsilon}{l^{0}-l\pm i\varepsilon}\Big)\nonumber
    \\&=m_{D}^{2}\Big(1-\frac{l^{0}\pm i\varepsilon}{2l}\Big(\log\Big\vert\frac{l^{0}+l}{l^{0}-l}\Big\vert+i\arctan\frac{\mp 2l\varepsilon}{L^2}\Big)\Big)\nonumber
    \\&=m_{D}^{2}\Big(1-\frac{l^{0}}{2l}\Big(\log\Big\vert\frac{l^{0}+l}{l^{0}-l}\Big\vert\mp i\pi\Theta(-L^2)\Big)\Big)
\end{align}
where in going to the last line we have taken $\varepsilon\to 0$. When taking the difference of the retarded and advanced self energies, only
the imaginary term survives and so
\begin{align}
    \rho^{L}_{\text{B}}(l)&=\frac{m_{D}^{2}\pi l^{0}}{l^{5}}\Theta(-L^2)\nonumber
    \\&\approx\frac{m_{D}^{2}\pi l^{0}}{l_{\perp}^{5}}
\end{align}
where we have used that $L$ is spacelike in going to the second line. Similarly, the transverse spectral function can be computed to give\footnote{In this case, 
it is not only the imaginary term that survives when $\Pi^{L}_{R}-\Pi^{L}_{A}$ computed. However, the surviving real part will be proportional 
to $\delta(L^{2})\approx\delta(l_{\perp}^{2}) $, which will always be zero in this scaling.}
\begin{align}
    \rho_{\text{B}}^{T}(l)&=\frac{m_{D}^{2}\pi l^{0}}{2lL^{4}}\Big(1-\frac{{l^0}^{2}}{l^{2}}\Big)\Theta(-L^2)
    \\&\approx\frac{m_{D}^{2}\pi l^{0}}{2l_{\perp}^{5}}
\end{align}
Next, we can go back to Sec.~\ref{sec:wloop_gsq}, add up the contributions 
from the three diagrams and expand in $g$ in the usual coft 
scaling, $K\sim (T, gT, \sqrt{g} T)$, while 
at the same time taking into account our new scaling for $L$
\begin{align}
    \delta\mathcal{C}(k_{\perp})_{l_{\perp}\sim\sqrt{g} T}&=g^{4}C_{R}C_{A}\int\frac{dk^{+}}{2\pi}\int\frac{dl^{+}d^{2}l_{\perp}}{(2\pi)^3}\frac{1+n_{\text{B}}(k^{+})}{2k^+}\frac{4l_{\perp}^{2}}{k_{\perp}^{2}(\mathbf{k}+\mathbf{l})^{2}}G^{--}_{rr}(-L)\nonumber
    \\&=g^{4}C_{R}C_{A}\int\frac{dk^{+}}{2\pi}\int\frac{dl^{+}d^{2}l_{\perp}}{(2\pi)^3}\frac{1+n_{\text{B}} (k^{+})}{2k^+}\frac{4l_{\perp}^{2}}{k_{\perp}^{2}(\mathbf{k}+\mathbf{l})^{2}}\frac{3Tm_{D}^{2}}{2l_{\perp}^{5}}.
\end{align}
The integrand is independent of $l^{+}$ and so after the integration has been done, the result will be proportional to a power of the cutoff on $l^{+}$. As is always the case 
when one is computing integrals using the method of regions, this cutoff term will be cancelled by the integration done in some other adjacent region. However, we know that in the 
$L\sim(gT,gT,gT)$ scaling, adopted in the previous section, there are no terms that depend on the $l^{+}$ cutoff. We can
 therefore conclude that this region is not adjacent to that region, implying that we need not concern ourselves with it.
 We have checked that something similar happens with the scaling $(l^+,l^-,l_{\perp})\sim(\sqrt{g}T,gT,gT)$ and 
 so we discard this region as well.
\subsection{\texorpdfstring{$l_{\perp}\sim l^{+}\sim\sqrt{g}T,l^{-}\sim gT$}{TEXT}}\label{sec:mult_sing}

We proceed similarly to the last section by first calculating the expanded spectral functions with this new scaling. The spectral functions read
\begin{align}
    \rho^{L}_{\text{B}}(L)&=\frac{m_{D}^{2}\pi l^{0}}{l^{5}}\Theta(-L^2)\nonumber
    \\&\approx\frac{m_{D}^{2}\pi l^{+}}{({l^{+}}^{2}+l_{\perp}^{2})^{\frac{5}{2}}}
\end{align}
and
\begin{align}
    \rho_{\text{B}}^{T}(L)&=\frac{m_{D}^{2}\pi l^{0}}{2lL^{4}}\Big(1-\frac{{l^0}^{2}}{l^{2}}\Big)\Theta(l^{2}-{l^{0}}^{2})\nonumber
    \\&\approx\frac{m_{D}^{2}\pi l^{+}}{2l_{\perp}^{2}({l^{+}}^{2}+l_{\perp}^{2})^{\frac{3}{2}}}.
\end{align}
When using these to compute the scattering kernel, we see that this time, the integrand does depend on $l^+$
\begin{align}
    \delta\mathcal{C}(k_{\perp})_{l^+\sim l_{\perp}\sim\sqrt{g}T}&=g^{4}C_{R}C_{A}\int\frac{dk^{+}}{2\pi}\int\frac{dl^{+}d^{2}l_{\perp}}{(2\pi)^3}\frac{1+n_{\text{B}} (k^{+})}{2k^+}
    \frac{4l_{\perp}^{2}}{k_{\perp}^{2}(\mathbf{k}+\mathbf{l})^{2}}G^{--}_{rr}(-L)\nonumber
    \\&=g^{4}C_{R}C_{A}\int\frac{dk^{+}}{2\pi}\int\frac{dl^{+}d^{2}l_{\perp}}{(2\pi)^3}\frac{1+n_{\text{B}} (k^{+})}{2k^+}\frac{4l_{\perp}^{2}}
    {k_{\perp}^{2}(\mathbf{k}+\mathbf{l})^{2}}\frac{3Tm_{D}^{2}\pi}{2({l^{+}}^{2}+l_{\perp}^{2})^{\frac{5}{2}}}\nonumber
    \\&=2g^{4}C_{R}C_{A}\int\frac{dk^{+}}{2\pi}\int\frac{d^{2}l_{\perp}}{(2\pi)^2}\frac{1+n_{\text{B}} (k^{+})}{k^+}\frac{Tm_{D}^{2}}
    {k_{\perp}^{2}(\mathbf{k}+\mathbf{l})^{2}l_{\perp}^{2}}.\label{uvlperpnoncoll}
\end{align}
It is not difficult to check that if we expand the expression above for $k_{\perp}\gg l_{\perp}$, we recover Eq.~\eqref{eq:lplusintegral}, with the $l_{\perp}\gg l^-\sim m_{D}$ limit
taken there. The fact that these two expressions agree in these limits implies that share a common boundary and are thus adjacent.

The $d^2l_{\perp}$ integration would, in this scaling run over the $\mathbf{k}+\mathbf{l}\approx 0$ range, 
corresponding to a longer-lived bremsstrahlung. Given this longer lifetime, we can expect
multiple soft scatterings at a rate of $g^2T$ to occur, in addition to a single, harder 
scattering. Addressing this region properly would thus consist of dealing with LPM resummation 
beyond the HOA. It is for this reason that we have been forced 
to adopt the HOA and in doing so, are not able to compute
the double logarithmic contribution coming from this region.
We discuss a pathway to dealing with this issue in the next section.

\section{Momentum Broadening Beyond the HOA}\label{sec:beyond}

To briefly recap, up until this point, we have managed to 
compute double-logarithmic corrections to $\hat{q}(\mu_{\perp})$, with the dominant corrections 
coming from Eq.~\eqref{eq:dlog_final}. In arriving at that expression, we observed sensitivity to a neighbouring 
region (studied in Sec.~\ref{sec:mult_sing}) that allows for both single and multiple scattering, which requires 
a more involved treatment. We postponed dealing with this region by taking the HOA, inferring the coefficient of $\hat{q}_0$
from the single-scattering consideration $\mu_{\perp}\gg\rho_{\perp}$. In addition, we have had 
to impose by hand the boundary $(b)$, which is intended to partition the single and multiple scattering regimes.  
 This is in contrast to the boundary separating the double-logarithmic phase space 
from its classical counterpart, which, as we will demonstrate, is clearly 
cut off by the scale $\omega_{\text{T}}$. In this section, we present a way to proceed beyond these two hurdles, 
to self-consistently determine $\delta\hat{q}(\mu_{\perp})$.

To this end, we will derive, starting from the formalism of~\cite{Liou:2013qya,Iancu:2014kga},
an LPM resummation equation that is not restricted to the HOA.
By numerically solving that equation and performing the integrations over the two
logarithmic variables $k^+$ and $\tau$ (or equivalently $k^+$ and $k_{\perp}$)
one would then see the emergence of multiple scatterings cutting off the 
double-log and thus be able to determine how good of an approximation line $(b)$
is.
Let us then start from Eqs.~(6), (7), (11) and (12) in \cite{Liou:2013qya} (see also (55)),
which construct a framework for resumming multiple interactions in the HOA
and in the large-$N_c$ limit. Combining (11) and (12) yields, in the notation of \cite{Liou:2013qya}
\begin{align}
    S(x_{\perp})=-\alpha_{s} C_R\mathrm{Re}&\int \frac{d k^+}{{k^+}^3}\int_0^{L_{M}}d z_2 \int_0^{z_2}dz_1 
    \nabla_{\mathbf{b}_2}\cdot\nabla_{\mathbf{b}_1}\nonumber\\
   &\hspace{-1cm} \times \bigg[e^{-\hat{q}_{0\,R} x_{\perp}^2(L_{M}-z_2+z_1)/4}
    G_{\text{HOA}}(\mathbf{b}_2,z_2;\mathbf{b}_1,z_1)
    -\mathrm{vac}
        \bigg]\bigg\vert^{\mathbf{b}_2=\mathbf{x}}_{\mathbf{b}_2=0}\bigg\vert^{\mathbf{b}_1=\mathbf{x}}_{\mathbf{b}_1=0},\label{LMWstart}
\end{align}
where we have already undone the large-$N_c$ approximation 
by replacing $N_c/2$, the original large-$N_c$ limit of $C_F$, with 
$C_R$. $\hat{q}_{0\,R}$, with $R=F,G$ denotes the specific partonic broadening coefficients
for a quark or gluon source.
The $d k^+$ frequency integration is understood over the positive frequencies
of the radiated gluon.
The propagator $G_{\text{HOA}}$ is the 
Green's function of the following Schr\"odinger equation (see (55) there)
\begin{equation}
    \label{LMWschroHO}
    \bigg\{ i\partial_z +\frac{\nabla^2_{\mathbf{b}}}{2 k^+}
    +\frac{i}{4}\bigg[\hat{q}_{0 R} x_{\perp}^2+\frac{\hat{q}_{0A}}{2}\big(b_{\perp}^2+(\mathbf{b}-\mathbf{x})^2-x_{\perp}^2\big)\bigg]
    \bigg\}G_{\text{HOA}}(\mathbf{b},z;\mathbf{b}_1,z_1)=0,
\end{equation}
with 
\begin{equation}
    G_{\text{HOA}}(\mathbf{b},z_1;\mathbf{b}_1,z_1)=\delta^{(2)}(\mathbf{b}-\mathbf{b}_1).
\end{equation}
``vac'' denotes the subtraction of the vacuum term $G_0$, i.e. the solution of Eq.~\eqref{LMWschro}
with a vanishing $\hat{q}_0$. The double vertical bars at the end 
of Eq.~\eqref{LMWstart} signify that the expression in brackets should be understood as
\begin{align}
    e^{-\frac{\hat{q}_{0\,R} x_{\perp}^2}{4}(L_{M}-z_2+z_1)}\bigg[ &G_{\text{HOA}}(\mathbf{x},z_2;\mathbf{x},z_1)+G_{\text{HOA}}(0,z_2;0,z_1)
   \nonumber\\
    &  -G_{\text{HOA}}(\mathbf{x},z_2;0,z_1)-G_{\text{HOA}}(0,z_2;\mathbf{x},z_1)\bigg]-\mathrm{vac}.  \label{LMWbars}
\end{align}
Physically, these four terms can be understood as the two positive, virtual terms, where the gluon
is emitted and reabsorbed within the amplitude (both $\mathbf{b}=0$) or conjugate amplitude (both $\mathbf{b}=\mathbf{x}$)
minus the real terms, where the gluon is emitted on one side and absorbed on the other side of the 
cut.

As a first step, we can 
identify their $S$-matrix element with
 our $\langle W(x_{\perp})\rangle$.\footnote{To this end, it suffices
 to note that the broadening probability is given in our 
 framework by the Fourier transform of $\langle W(x_{\perp})\rangle$ and  in 
 theirs by that of $S$ --- see their Eq.~(1). For a clear explanation of how the 
 broadening probability is related to $\mathcal{C}(k_{\perp})$, see App. A1 of 
 \cite{Arnold:2009mr}.}
Hence we can equate 
\begin{equation}
S(x_{\perp})=\exp[-L_{M}(\mathcal{C}(x_{\perp})+\delta\mathcal{C}(x_{\perp}))]\approx
e^{-L_{M}\mathcal{C}(x_{\perp})}(1-L_{M}\delta\mathcal{C}(x_{\perp})).
\label{eq:LMWdict}
\end{equation}
Secondly, Eq.~\eqref{LMWstart} is in the harmonic-oscillator approximation, i.e.
\begin{equation}
    \label{eq:HOexplicit}
    \mathcal{C}_{R}(x_{\perp})\stackrel{{\text{HOA}}}{=}\frac{\hat{q}_{0\,R}}{4}x_{\perp}^2,
\end{equation}
We can then undo this approximation so that Eq.~\eqref{LMWschroHO} becomes
\begin{equation}
    \label{LMWschro}
    \bigg\{i\partial_z  +\frac{\nabla^2_{\mathbf{b}}-m_{\infty\,g}^2}{2k^+}
    +i\bigg[\mathcal{C}_R(x_{\perp})+\frac{1}{2}\big(\mathcal{C}_A(b_{\perp})+\mathcal{C}_A(\vert\mathbf{b}-\mathbf{x}\vert)-\mathcal{C}_A(x_{\perp})\big)\bigg]
    \bigg\}G(\mathbf{b},z;\mathbf{b}_1,z_1)=0,
\end{equation}
where we have also introduced the gluon's asymptotic mass, with $m_{\infty\,g}^2=m_D^2/2$ \footnote{\label{foot:mass}%
This mass term is necessary when $\nabla^2_{b_{\perp}}$, the transverse momentum of the gluon,
becomes of order $g^2T^2$; it can be neglected in the deep LPM regime, where 
typical transverse momenta are larger, $k_{\perp}^2\sim\sqrt{\hat{q}_0k^+}$, with $k^+ \gg T$.}. To extract a leading-order
determination of radiative correction from this formalism, it should suffice
to use, for $\mathcal{C}_F(x_{\perp}),\,\mathcal{C}_A(x_{\perp})$ the Fourier transform of the smooth kernel provided by Eq.~\eqref{hardqcdresum}.

Finally, we can also account for the effect of a populated medium by considering the effects of 
stimulated emission and absorption, i.e $\int d k^+\Theta(k^+)\to \int dk^+ (1/2+n_{\text{B}}(k^+))$. 
In a longitudinally uniform medium, $G$ is only a function of $t\equiv z_2-z_1$ so that we can
use our identification~\eqref{eq:LMWdict} to obtain, in the large-$L_{M}$ limit
\begin{align}
    \delta \mathcal{C}(x_{\perp})=\alpha_{s} C_R\mathrm{Re}\int \frac{dk^+}{k^{+3}}\left(\frac{1}{2}+n_{\text{B}}(k^+)\right)
    \int_0^{L_{M}}d t \,
    \nabla_{\mathbf{b}_2}\cdot\nabla_{\mathbf{b}_1}\bigg[&e^{\mathcal{C}_R(x_{\perp})t}
    G(\mathbf{b}_2,\mathbf{b}_1;t)\nonumber\\
    &
   -\mathrm{vac}
        \bigg]\bigg\vert^{\mathbf{b}_2=\mathbf{x}}_{\mathbf{b}_2=0}\bigg\vert^{\mathbf{b}_1=\mathbf{x}}_{\mathbf{b}_1=0}.
        \label{LMWC}
\end{align}
We observe that the source-specific part of the scattering kernel in the Hamiltonian~\eqref{LMWschro} 
and the amplification factor $e^{\mathcal{C}_R(x_{\perp})t}$ above can be eliminated by noting that if
$\tilde G(\mathbf{b},\mathbf{b}_1;t)\equiv e^{\mathcal{C}_R(x_{\perp})t} G(\mathbf{b},\mathbf{b}_1;t)$ is a Green's function of the operator
\begin{equation}
    \label{LMWschroshift}
    \bigg\{ i\partial_t  +\frac{\nabla^2_{\mathbf{b}}-m_{\infty\,g}^2}{2k^+}
    +\frac{i}{2}\big(\mathcal{C}_A(b_{\perp})+\mathcal{C}_A(\vert\mathbf{b}-\mathbf{x}\vert)-\mathcal{C}_A(x_{\perp})\big)\bigg\}\tilde G(\mathbf{b},\mathbf{b}_1;t)=0,
\end{equation}
then $G(\mathbf{B},\mathbf{B}_1;t)$ is a Green's function of 
Eq.~\eqref{LMWschro}. Hence
\begin{align}
    \delta \mathcal{C}(x_{\perp})=\alpha_{s} C_R\mathrm{Re}\int \frac{dk^+}{k^{+3}}\left(\frac{1}{2}+n_{\text{B}}(k^+)\right)\int_0^{L_{M}}d t \,
    \nabla_{\mathbf{b}_2}\cdot\nabla_{\mathbf{b}_1}\bigg[&
    \tilde G(\mathbf{b}_2,\mathbf{b}_1;t)\nonumber\\
    &
   -\mathrm{vac}
        \bigg]\bigg\vert^{\mathbf{b}_2=\mathbf{x}}_{\mathbf{b}_2=0}\bigg\vert^{\mathbf{b}_1=\mathbf{x}}_{\mathbf{b}_1=0}.
        \label{LMWCshift}
\end{align}
This reformulation makes transparent the fact that the Hamiltonian only contains 
the purely non-abelian $\mathcal{C}_A$. That is because before or after both
emission vertices, there are only the two source lines for the hard jet parton
 and one-gluon 
exchanges between them are resummed into $\exp\big(-\mathcal{C}_R(x_{\perp})(L_{M}-t)\big)$.
In the time region between the two emission vertices, the hard jet parton and 
conjugate hard jet parton lines are no longer a color singlet, with corresponding
color factor $C_R$, but rather an octet, with color factor 
$C_R-C_A/2$. This corresponds to the 
$\mathcal{C}_R(x_{\perp})-\mathcal{C}_A(x_{\perp})/2$ combination in Eq.~\eqref{LMWschro}. But we should 
remove the overall $\exp\big(-\mathcal{C}_R(x_{\perp})L_{M}\big)$ damping, which, as per our 
dictionary~\eqref{eq:LMWdict}, leads
to the outright disappearance of the $\mathcal{C}_R(x_{\perp})$ part, corresponding to the 
fact that those exchanges would happen also in the absence of the radiated 
gluon. Up to the statistical factors and thermal masses, Eqs.~\eqref{LMWschroshift} 
and ~\eqref{LMWCshift} agree with~\cite{Iancu:2014kga}.

For a medium that is isotropic in the azimuthal direction, we can use 
the reflection symmetry $\mathbf{b}\to \mathbf{x}-\mathbf{b}$ of Eq.~\eqref{LMWschroshift}
to simplify Eq.~\eqref{LMWCshift} into
\begin{align}
    \delta \mathcal{C}(x_{\perp})=-2\alpha_{s} C_R\mathrm{Re}\int \frac{dk^+}{k^{+3}}\left(\frac{1}{2}+n_{\text{B}}(k^+)\right)
    \int_0^{L_{M}}d t \,
    \nabla_{\mathbf{b}_2}\cdot\nabla_{\mathbf{b}_1}\bigg[&
    \tilde G(\mathbf{b}_2,\mathbf{b}_1;t)\nonumber\\
    &
   -\mathrm{vac}
        \bigg]\bigg\vert^{\mathbf{b}_2=\mathbf{x},\mathbf{b}_1=0}_{\mathbf{b}_2=0,\mathbf{b}_1=0}.
        \label{LMWCshiftfinal}
\end{align}

Eq.~\eqref{LMWschroshift} is, together with Eq.~\eqref{LMWCshiftfinal}, the main 
result of this section. We anticipate that its solution would  allow for
a much better understanding of how the double logarithm is cut off by the transition from single
to multiple scatterings as a function of the energy $k^+$ 
of the radiated gluon. Furthermore, Eq.~\eqref{LMWschroshift} goes beyond the HOA 
and would thus allow us to properly deal with the region (discussed in Sec.~\ref{sec:mult_sing}) where $\mathbf{l}+\mathbf{k}$ become small and 
in turn develop sensitivity to multiple scatterings.
However, Eq.~\eqref{LMWschroshift} is not easy to solve, as it would require 
generalizing the methods of \cite{CaronHuot:2010bp,Andres:2020vxs,Andres:2020kfg,Schlichting:2021idr}
to the extra dependence on $x_\perp$ of $\tilde{G}$. As a first step,
one could consider using the improved opacity expansion, already mentioned in Sec.~\ref{sec:recent} to capture 
the qualitative aspects of the transition from the HO approximation to 
including rarer harder scatterings.

We shall leave the full or approximate solution of Eq.~\eqref{LMWschroshift} to future work. 
We continue this section by providing a non-trivial consistency check, namely that 
the single-scattering term in Eqs.~\eqref{LMWschroshift} and \eqref{LMWCshiftfinal} agrees with Eq.~\eqref{eq:N1term}. The 
calculation here proceeds in exactly the same way as for the single-scattering rate, derived in Sec.~\ref{sec:single}.
Following the same steps, we arrive at
\begin{align}
    \delta \mathcal{C}(x_{\perp})^{N=1}=2\alpha_{s} C_R\int \frac{dk^+}{k^+}&\left(1+2n_{\text{B}}(k^+)\right)\int_{\mathbf{p}}\int_{\mathbf{l}} \,
    \mathcal{C}_A(l_{\perp})\frac{(1-e^{i\mathbf{x}\cdot\mathbf{p}})(1+e^{i\mathbf{x}\cdot\mathbf{l}})}
    {p_{\perp}^2+m_{\infty\,g}^2}\nonumber
    \\&\times\bigg[\frac{p_{\perp}^2}{p_{\perp}^2+m_{\infty\,g}^2}
    -\frac{\mathbf{p}\cdot(\mathbf{p}+\mathbf{l})}{
    (\mathbf{p}+\mathbf{l})^2+m_{\infty\,g}^2}
        \bigg].
    \label{eq:multiscat}
\end{align}

We take a moment here to compare the equation above with Eq.~\eqref{eq:op_ls}, with the $L_{M}\to\infty$ limit taken 
there. Even though the latter 
equation is a radiation splitting rate, we can immediately understand that it resembles the former, which computes 
corrections to the scattering kernel due to radiative processes. However, the product $(1-e^{i\mathbf{x}\cdot\mathbf{p}})(1+e^{i\mathbf{x}\cdot\mathbf{l}})$ 
above makes manifest that the above equation includes real, probability-conserving as well as virtual processes in comparison
to Eq.~\eqref{eq:op_ls}, which only contains real processes. To make contact with Eq.~\eqref{eq:N1term}, we need to extract the real-process contribution, 
$\delta\mathcal{C}(k_{\perp})^\mathrm{real}_\mathrm{single}$.
This just corresponds to Fourier-transforming $\mathbf{x}$ into $\mathbf{k}$ and 
only keeping terms proportional to $\exp( i\mathbf{x}\cdot\mathbf{p})$ and 
$\exp( i\mathbf{x}\cdot(\mathbf{p}+\mathbf{l}))$, yielding
\begin{align}
    \delta\mathcal{C}(k_{\perp})^\mathrm{real}_\mathrm{single}
=&4\alpha_{s} C_{R}\int dk^+\frac{\frac{1}{2}+n_{\text{B}}(k^+)}{k^+}\int_{\mathbf{l}}
\mathcal{C}_g(l_{\perp}) \left[\frac{\mathbf{k}}{k_{\perp}^2+m_{\infty\,g}^2}-
\frac{\mathbf{k}+\mathbf{l}}{(\mathbf{k}{+}\mathbf{l})^2+m_{\infty\,g}^2}\right]^2,
\label{LWMsingle}
\end{align}
which, in the $m_{\infty\,g}\to0$ and ($1/2+n_{\text{B}}(k^+)\to \Theta(k^+)$) limits
agree with Eq.~\eqref{eq:N1term}. In App.~\ref{sub:virtual} we compute the aforementioned virtual contributions diagrammatically and 
in doing so, check that they are consistent with Eq.~\eqref{eq:multiscat}.

\section{Connection to the Classical Contribution}\label{sec:class}
\begin{figure}[t]
    \begin{center}
        \includegraphics[width=12cm]{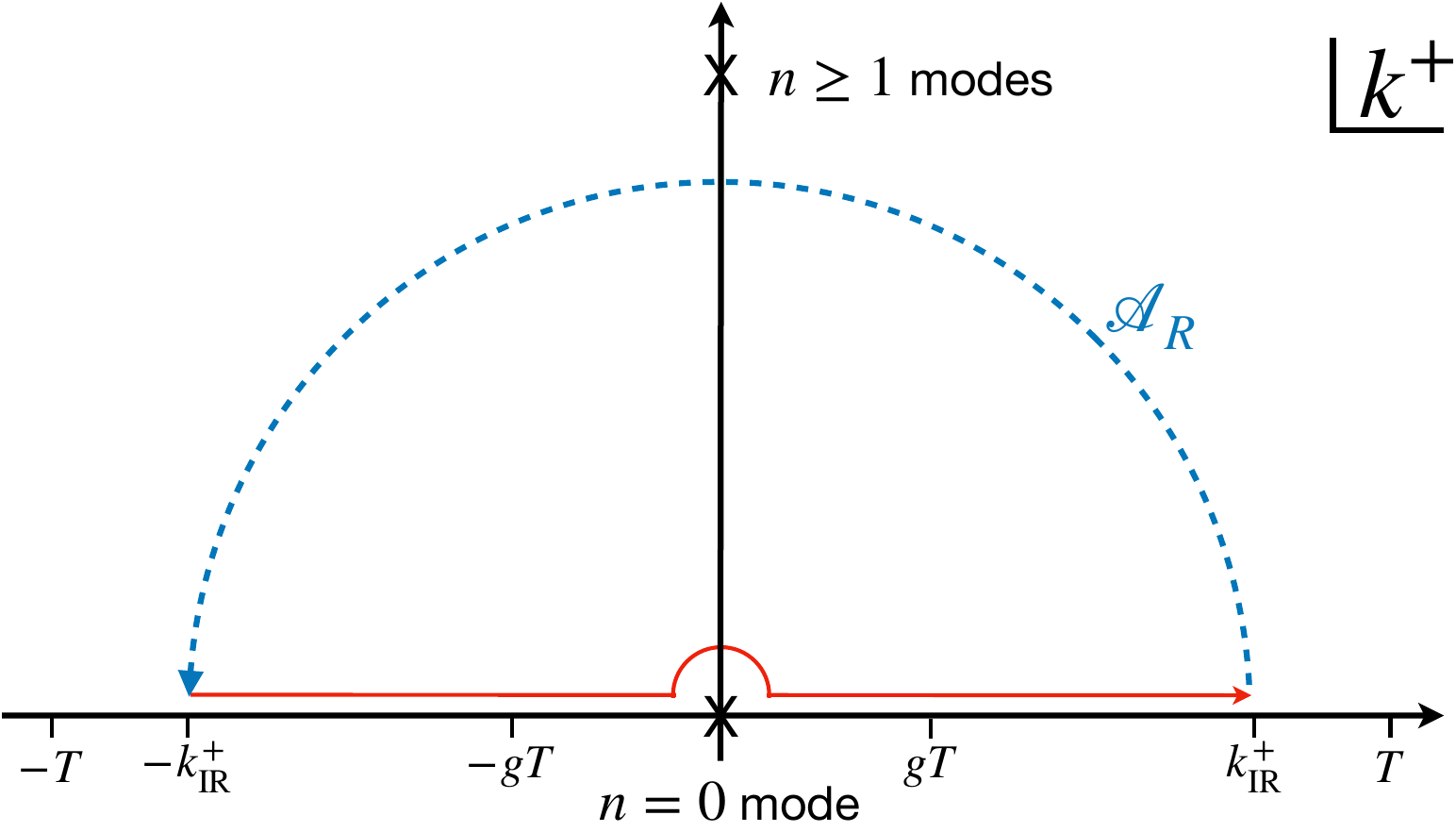}
    \end{center}
    \caption{The contour discussed in the main text 
    for the evaluation of the retarded part of the 
    soft contribution to $\delta\mathcal{C}(k_{\perp})$. The corresponding 
    advanced part is not shown. The cross at the origin is 
    Matsubara zero mode, hence the deformation of the (red) integration 
    contour there, which gives rise to the 
    Euclidean contribution.}
    \label{fig:matsubara}
\end{figure}
When computing double logarithmic corrections to $\hat{q}(\mu_{\perp})$ in Sec.~\ref{sec:reproduce}, 
we observed corrections depending on a power law of the IR energy cutoff, $k^+_{\text{IR}}$. 
Power law corrections of this kind will
 always cancel against those coming from a neighbouring region of phase space and in some sense, can 
 be disregarded. In this instance however, we take advantage of this fact to further clarify how
 the region of phase space from which the quantum corrections emerge is connected to the region giving 
 rise to the classical corrections. In particular, we set out to show how the first term in Eq.~\eqref{semievaltaufinal} 
  is also produced (but with an opposite sign) in the calculation of Caron-Huot \cite{CaronHuot:2008ni}. We 
  do not show how power law corrections from either Eq.~\eqref{eq:few_before} or Eq.~\eqref{Ifinal} cancel 
  against their classical counterparts.
  
But how should this work in practice? Well, diagrams such as 
those in Fig.~\ref{fig:real} are precisely those evaluated 
in \cite{CaronHuot:2008ni} with $K\sim L\sim gT$. On that side,
the calculation should produce a $\sim 1/k^+_{\text{IR}}$ term. There, $k^+_{\text{IR}}$ is a 
UV cutoff, which can thus be safely sent to infinity, resulting 
in the absence of such corrections from \cite{CaronHuot:2008ni}.

As we have discussed in Sec.~\ref{sec:chtrick}, based on ideas from the same paper,
the classical contribution is dominated by the Matsubara zero mode, which can be mapped to EQCD. 
In terms of doing the $k^+$ integration in the 4D theory, this consists of making 
the replacement $ 1/2+n_{\text{B}}(k^+)\to T/k^+$ before closing 
the integration contour, $\mathcal{A}$ at a radius, $k^+_{\text{IR}}$ between the zero and first Matsubara modes.
This is moreover consistent with 
our calculation, in the sense that $gT\ll k^+_{\text{IR}}\ll T$. Any function that falls to zero faster 
than $1/k^+$ on this arc will only give rise to inverse powers of $k^+_{\text{IR}}$, which could be safely neglected in the derivation 
\cite{CaronHuot:2008ni}. The contour of interest is depicted in Fig.~\ref{fig:matsubara}. 

Rather than starting from Caron-Huot's calculation, we instead begin from around Eq.~(F.49) \cite{Ghiglieri:2015ala}, 
which provides the NLO calculation of the longitudinal momentum diffusion constant,
$\hat{q}_L(\mu_{\perp})\equiv \langle q_z^2\rangle/L_{M}$, also conveniently carried out in Coulomb gauge. To make 
the connection with its transverse counterpart, one can multiply by $(k_{\perp}/k^+)^2$ under the integral sign\footnote{
See \cite{Ghiglieri:2022gyv} for more details.}.
This yields
\begin{eqnarray}
    \label{qhatarc}
      \delta\hat{q}\Big\vert_{\mathrm{soft}}^\mathrm{arcs}(\mu_{\perp}) \!\! &=&
    2g^4C_R C_A T
    \! \int_{\mathcal{A}_R} \! \frac{dk^+}{2\pi}
    \! \int^{\mu_{\perp}} \! \frac{d^2k_{\perp}}{(2\pi)^2}
    \! \int_{L}
    \\ && \times  \nonumber
    \bigg\{
    \frac{k_{\perp}^2}{k^{+2}}G^{--}_{rr}(L)\pi\delta(l^-)
    \left(\frac{\mathbf{k}}{k_{\perp}^2+m_\infty^2} -\frac{\mathbf{k}+\mathbf{l}}{(\mathbf{k}+\mathbf{l})^2+m_\infty^2}\right)^2
   +\mathcal{O}\left(\frac{1}{k^{+3}}\right)\bigg\} +\mathcal{A}_A.
    \end{eqnarray}
We can now perform some of the $L$ integrations using the reduction to EQCD 
discussed in Sec.~\ref{sec:chtrick}, finding
\begin{eqnarray}
    \label{qhatarcint}
      \delta\hat{q}\Big\vert_{\mathrm{soft}}^\mathrm{arcs}(\mu_{\perp}) \!\! &=&
    g^4C_R C_A T^2
    \! \int_{\mathcal{A}_R} \! \frac{dk^+}{2\pi}
    \! \int^{\mu_{\perp}} \! \frac{d^2k_{\perp}}{(2\pi)^2}
    \! \int_{\mathbf{l}}
    \\ && \times  \nonumber
    \bigg\{
    \frac{k_{\perp}^2}{k^{+2}}\frac{m_D^2}{l_{\perp}^2(l_{\perp}^2+m_D^2)}
    \left(\frac{\mathbf{k}}{k_{\perp}^2+m_\infty^2} -\frac{\mathbf{k}+\mathbf{l}}{(\mathbf{k}+\mathbf{l})^2+m_\infty^2}\right)^2
    +\mathcal{O}\left(\frac{1}{k^{+3}}\right)\bigg\} +\mathcal{A}_A.
    \end{eqnarray}
If we take the $l_{\perp},m_D\ll k_{\perp}$ approximation, consistently
with our treatment across the $k^+_{\text{IR}}$ boundary, we find
\begin{eqnarray}
    \label{qhatarcintsemi}
      \delta\hat{q}\Big\vert_{\mathrm{soft}}^\mathrm{arcs}(\mu_{\perp})\!\! &=&
    g^4C_R C_A T^2
    \! \int_{\mathcal{A}_R} \! \frac{dk^+}{2\pi}
    \! \int^{\mu_{\perp}} \! \frac{d^2k_{\perp}}{(2\pi)^2}
    \! \int_{\mathbf{l}}
    \frac{1}{k^{+2}k_{\perp}^2}\frac{m_D^2}{l_{\perp}^2+m_D^2}
 +\mathcal{A}_A.
    \end{eqnarray}
We regulate the $l_{\perp}$ integration using the HOA
consistently with what we did in Sec.~\ref{sec:hoa_ss}, we find
\begin{eqnarray}
    \label{qhatarcintsemiHO}
      \delta\hat{q}\Big\vert_{\mathrm{soft}}^\mathrm{arcs\,HO}(\mu_{\perp}) \!\! &=&
    g^2C_R  T\hat{q}_0
    \! \int_{\mathcal{A}_R} \! \frac{dk^+}{2\pi}
    \! \int^{\mu_{\perp}} \! \frac{d^2k_{\perp}}{(2\pi)^2}
    \frac{1}{k^{+2}k_{\perp}^2}
 +\mathcal{A}_A.
    \end{eqnarray}
We can then rewrite the $k_{\perp}$ integration as a $\tau$ one, with the same 
boundaries of Eq.~\eqref{eq:sss_before}, i.e.
\begin{eqnarray}
    \label{qhatarcintsemiHOlm}
      \delta\hat{q}\Big\vert_{\mathrm{soft}}^\mathrm{arcs\,HO}(\mu_{\perp}) \!\! &=&
    g^2C_R T\hat{q}_0
    \! \int_{\mathcal{A}_R} \! \frac{dk^+}{2\pi}
    \! \int_{k^+/\mu_{\perp}^2}^{\tau_{\text{int}}} \!
    \frac{d\tau}{4\pi}
    \frac{1}{k^{+2}\tau}
 +\mathcal{A}_A.
    \end{eqnarray}
The contribution on the arc can be carried out as follows, understanding the 
arc to be at constant $\vert k^+\vert = k^+_{\text{IR}}$. Changing variables to $k^+=k^+_{\text{IR}} e^{it}$ then gives
\begin{equation}
    \int_{\mathcal{A}_R} \! \frac{dk^+}{2\pi}\frac{1}{k^{+2}}
    \ln\left(\frac{\mu_{\perp}^2\tau_{\text{int}}}{k^+}\right)=
    \frac{i}{2\pi k^+_{\text{IR}}} \int_\pi^{0} dt e^{-i t}
    \bigg[\ln\left(\frac{\mu_{\perp}^2\tau_{\text{int}}}{k^+_{\text{IR}}}\right)
    -it\bigg]= \frac{-1}{\pi k^+_{\text{IR}}}\bigg[\ln\left(\frac{\mu_{\perp}^2\tau_{\text{int}}}{k^+_{\text{IR}}}\right)
    -1-i\frac\pi2\bigg].
\end{equation}
The imaginary part cancels against an opposite one from $\mathcal{C}_A$. We then have
\begin{align}
      \delta\hat{q}\Big\vert_{\mathrm{soft}}^\mathrm{arcs\,HO}(\mu_{\perp}) \!\! =&
    2\frac{g^2C_R  T^2\hat{q}_0}{4\pi}
    \frac{-1}{\pi k^+_{\text{IR}}}\bigg[\ln\left(\frac{\mu_{\perp}^2\tau_{\text{int}}}{k^+_{\text{IR}}}\right)
    -1\bigg]
   \nonumber \\
   =&-2\frac{\alpha_s C_R  T\hat{q}_0}{\pi k^+_{\text{IR}}}
  \bigg[\ln\left(\frac{\mu_{\perp}^2\tau_{\text{int}}}{k^+_{\text{IR}}}\right)
  -1\bigg]
  .\label{qhatarcintsemiHOalmostfinal}
    \end{align}
This is precisely opposite to the
$T/k^+_{\text{IR}}$-proportional part of Eq.~\eqref{semievaltaufinal}, as 
we set out to show. 

We provide a final comment 
with respect to how the double-logarithmic phase space is cut off. The $k^+$ 
integration is clearly cut off around the scale $\omega_T$ because below that scale, there is the 
cancellation between the vacuum term and the first quantum correction, 
as was explained explicitly around (...), which renders the $k^+$ integrand $\sim1/{k^+}^2$.
But can one understand how the $\tau$ integration transitions to power law behaviour? Well, 
we can understand from the calculation above that there is no 
analogous scale to $\omega_T$ since a $1/\tau$ integrand persists for $k^+\ll T$. Instead, $\omega_T$ 
 cuts off the $\tau$ logarithm in a somewhat subtle way; in approaching the $k^+_{\text{IR}}$ vertical line from the 
 right, the $\tau$ integration gives rise to single logs that are always multiplied by powers of $k^+_{\text{IR}}$. As we have just shown, 
 these single logs cancel against identical terms, which are produced from approaching the $k^+_{\text{IR}}$ vertical line from the 
 left.

\chapter{Quantum Corrections to Forward Scattering}\label{ch:asym_mass}
\begin{figure}[ht]
	\centering
    \includegraphics[width=\textwidth]{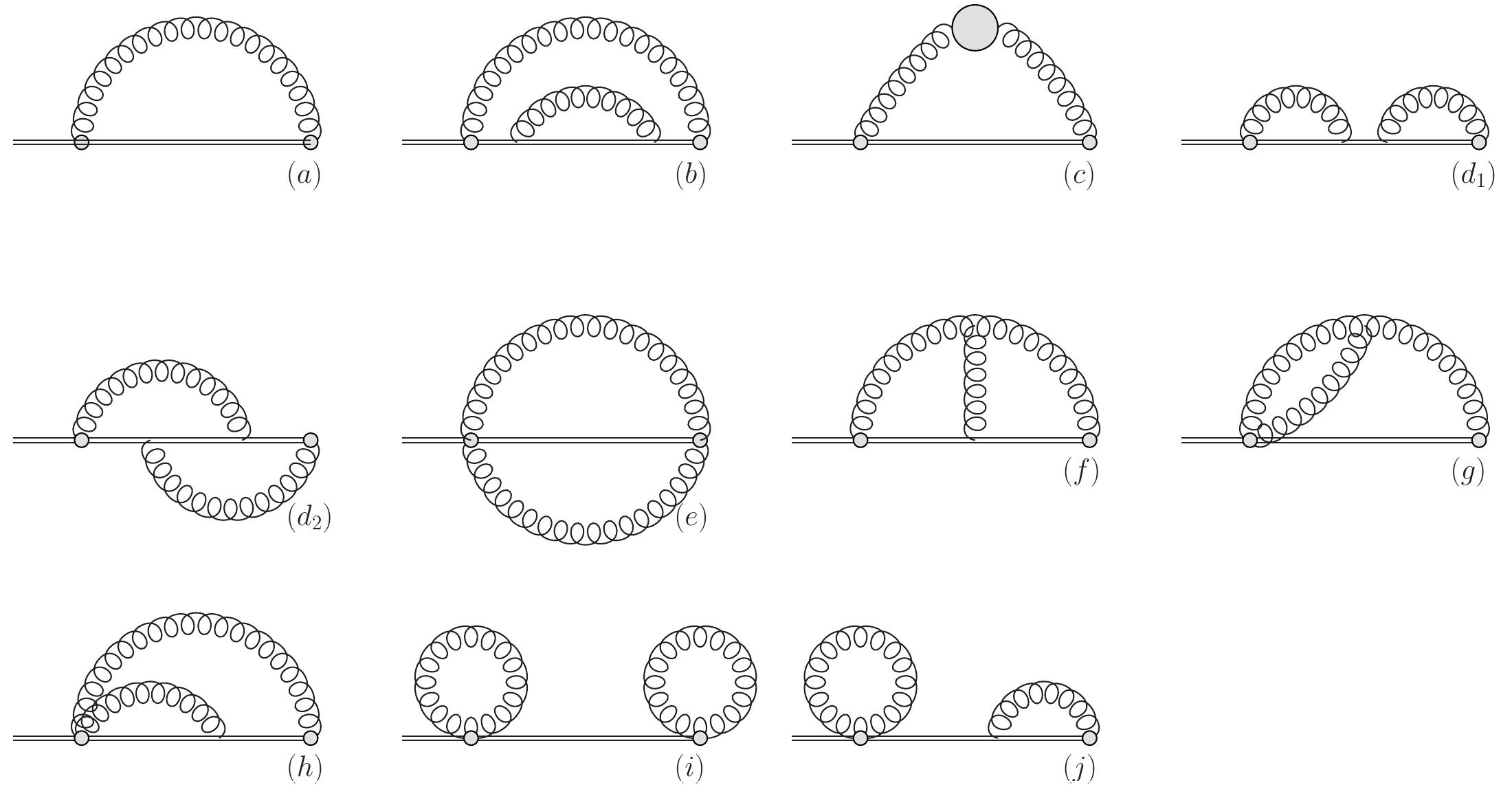}
	\caption{Diagrams relevant to the leading order and next-to-leading 
    order evaluation of $Z_g$. Eq.~\eqref{eq:zg_wline}. External grey blobs
    denote $F^{-\perp}$ insertions whereas internal grey blobs denote self energies. The solid 
    lines represent Wilson lines and the curly lines a gauge boson.}
    \label{fig:amass_all_diagrams}
\end{figure}

This chapter maintains the theme of computing higher order corrections 
to quantities that serve as input to jet energy loss calculations. In particular, we work towards
the calculation of the $\mathcal{O}(g^2)$ corrections to the asymptotic mass, $m_{\infty}$ with 
two goals in mind:
\begin{itemize}
    \item Perform the matching between perturbative finite-temperature QCD and perturbative EQCD in order 
    to cancel EFT divergences coming from the EQCD side. As was discussed at the end of Sec.~\ref{sec:np_classical}, this matching procedure
    will rid the non-perturbative determination of the classical corrections to $m_{\infty}$ of any UV divergences. Recalling that\footnote{
        The $Z_g,\,Z_f$ operators in Eqs.~\eqref{eq:mq_recall}, ~\eqref{eq:mg_recall} are strictly 
        speaking not equal to each other -- they are equal up to an overall colour factor. 
        See, for instance, Eq.~\eqref{eq:zg_wline}. In what follows, 
        we compute for a jet particle in colour representation, $R$.
    }
    \begin{align}
        m_{\infty\,q}^2&=g^2C_{F}(Z_g+Z_f),\label{eq:mq_recall}
        \\m_{\infty\,g}^2&=g^2(C_{A}Z_g+2T_{F}N_{F}Z_f),\label{eq:mg_recall}
    \end{align}
    we remind the reader that the operator $Z_f$ is not as sensitive as $Z_g$ to classical corrections, meaning that we need 
    only complete this matching for the $Z_g$ operator. This calculation makes up the content of Secs.~\ref{sec:full_qcd} and 
    ~\ref{sec:eqcd_details} and is furthermore presented in \cite{Ghiglieri:2023cyw}. In the latter, we take some time to 
    introduce the terminology from \cite{Moore:2020wvy,Ghiglieri:2021bom} in order to shed some more 
    light on the context of the matching calculation.
    \item Complete the $\mathcal{O}(g^2)$ evaluation of $Z_g$. We begin this endeavour in Sec.~\ref{sec:full}. 
\end{itemize}
Before embarking on these goals, in Sec.~\ref{sec:setup}, we introduce the various two-loop diagrams that arise at NLO
and argue why the majority of them are zero in Feynman gauge, which 
we adopt throughout this chapter. In Sec.~\ref{sec:proof}, we proceed to demonstrate a further cancellation, 
which leaves us only needing to concern ourselves with diagram $(c)$ from Fig.~\ref{fig:amass_all_diagrams}.

\section{Calculation Setup}\label{sec:setup}
We dive straight into the computation of the various diagrams, listed in Fig.~\ref{fig:amass_all_diagrams}.
While in theory, the vacuum result should be subtracted from each diagram, we neglect this term 
throughout the chapter, as it vanishes in DR anyway.

Diagram $(a)$ just gives the LO contribution to $Z_g$, which we rewrite here for convenience \cite{Kalashnikov:1979cy,Weldon:1982aq}
\begin{equation}
    Z_g^{(a)}=\frac{T^2}{6}.\label{eq:a}
\end{equation}
We now proceed to the arduous task of computing all of the NLO diagrams.

\subsection{Diagram \texorpdfstring{$(c)$}{TEXT}}\label{sec:c}

In Sec.~\ref{sec:nlo_asym}, we computed the EQCD counterpart of diagram $(a)$. Diagram $(c)$ has
the same structure, albeit with an extra self-energy insertion. We can thus pick
up from Eq.~\eqref{eq:zg_nlo_mess}, writing  
\begin{align}
        Z_g ^{(c)}&=-2\int_0^\infty d x^+\int_{K}x^+e^{-ik^-x^+}(k^{-2}G_{>}^{xx}(K)-2k^{x}k^{-}G_{>}^{x-}(K)+{k^{x}}^{2}G_{>}^{--}(K))\nonumber
         \\&=2\int_{K}\frac{1}{(k^{-}-i\varepsilon)^2}(k^{-2}G_{>}^{xx}(K)-2k^{x}k^{-}G_{>}^{x-}(K)+{k^{x}}^{2}G_{>}^{--}(K))\label{cdiagprops}
         \\&=2\int_{K}\frac{1+n_{\text{B}}(k^0)}{(k^{-}-i\varepsilon)^2}(k^{-2}(G_{R}^{xx}(K)-G_{A}^{xx}(K))-2k^{x}k^{-}(G_{R}^{x-}(K)-G_{A}^{x-}(K))\nonumber
         \\&+{k^{x}}^{2}(G_{R}^{--}(K)-G_{A}^{--}(K)))\nonumber
           \\&=2\int_{K}\frac{1+n_{\text{B}}(k^0)}{(k^{-}-i\varepsilon)^2}(F_{R}(K)-F_{A}(K))
\end{align}
where, in contrast to Eq.~\eqref{eq:zg_nlo_mess},
\begin{align}
    F_{R/A}(K)&=\frac{i}{(K^2+i\varepsilon k^0)^2}\big(k^{-2}\Pi^{xx}_{R/A}(K)-2k^{x}k^{-}(\Pi^{0x}_{R/A}(K)-\Pi^{zx}_{R/A}(K))\nonumber
    \\&+{k^x}^{2}(\Pi^{00}_{R/A}(K)+\Pi^{zz}_{R/A}(K)-2\Pi^{z0}_{R/A}(K))\big).
\end{align}
It is then useful to decompose the self-energies in terms of their longitudinal 
and transverse components
\begin{equation}
    \Pi^{R/A}_{\mu\nu}(K)=P^L_{\mu\nu}(K)\Pi^{R/A}_{L}(K)+P^T_{\mu\nu}(K)\Pi^{R/A}_{T}(K)\label{eq:selfenergydecomp}
 \end{equation}
where $P^L_{\mu\nu}$ and $P^T_{\mu\nu}$ are defined in Eqs.~\eqref{eq:lprojector} and ~\eqref{eq:tprojector} respectively.
Making use of those same equations then leads to
\begin{align}
    F_{R/A}(K)=&=\frac{i}{(K^2+i\varepsilon k^0)^2}\bigg[k^{-2}\bigg(\Pi_{L}^{R}(K)\frac{k_{\perp}^2}{k^2}\bigg)\nonumber
        \\&-2k^xk^-\bigg(\Pi_{L}^{R/A}(K)\Big(\frac{k^xk^z}{q^2}-\frac{k^xk^-}{K^2}\Big)-\Pi_{T}^{R/A}(K)\frac{k^xk^z}{q^2}\bigg)\nonumber
        \\&+{k^x}^{2}\bigg(\Pi_{L}^{R/A}(K)\Big(1-\frac{k_{\perp}^2}{k^2}+\frac{k^2-2k^0k^z+{k^z}^{2}}{K^2}\Big)+\Pi_{T}^{R/A}(K)\frac{k_{\perp}^2}{k^2}\bigg)\bigg]\nonumber
        \\&=\frac{i}{(K^2+i\varepsilon k^0)^2}\bigg[k^{-2}\Pi_{T}^{R/A}(K)+\frac{{k^x}^{2}K^2}{k^2}(\Pi_{L}^{R/A}(K)-\Pi_{T}^{R/A}(K))\bigg],
\end{align}
which in turn results in\footnote{Technically, as the $K$ integration is done in DR 
we should make the replacement $2k^{-2}\rightarrow(D-2)k^{-2}$ where $D=4-2\epsilon$. Once we get 
to the stage where we are actually evaluating divergent integrals, we will be careful to rewrite everything 
in the appropriate $d$-dimensional form.}
\begin{align}
    Z_g^{(c)}&=i\int_{K}\frac{1+n_\mathrm{B}(k^0)}{(k^--i\varepsilon)^2}\bigg[
    \frac{2k^{-2}\Pi_T^R(K)}{(K^2+i\varepsilon k^0)^2}+ \frac{k_\perp^2(\Pi_L^R(K)-\Pi_T^R(K))}{k^2(K^2+i\varepsilon k^0)}
    -\frac{2k^{-2}\Pi_T^A(K)}{(K^2-i\varepsilon k^0)^2}\nonumber
    \\&- \frac{k_\perp^2(\Pi_L^A(K)-\Pi_T^A(K))}{k^2(K^2-i\varepsilon k^0)}
    \bigg].
    \label{zcexpr}
\end{align}
It is convenient to divide up the terms as follows
\begin{equation}
    Z_g^{(c)}\equiv Z_{g\,k^{-2}}^{(c)}+Z_{g\,\text{L}-\text{T}}^{(c)}.\label{eq:split}
\end{equation}
We find this to be a good stopping point, before matching the IR divergences of the expression above to those 
coming from the EQCD calculation in Sec.~\ref{sec:full_qcd}.

\begin{figure}[ht]
	\centering
    \includegraphics[width=0.7\textwidth]{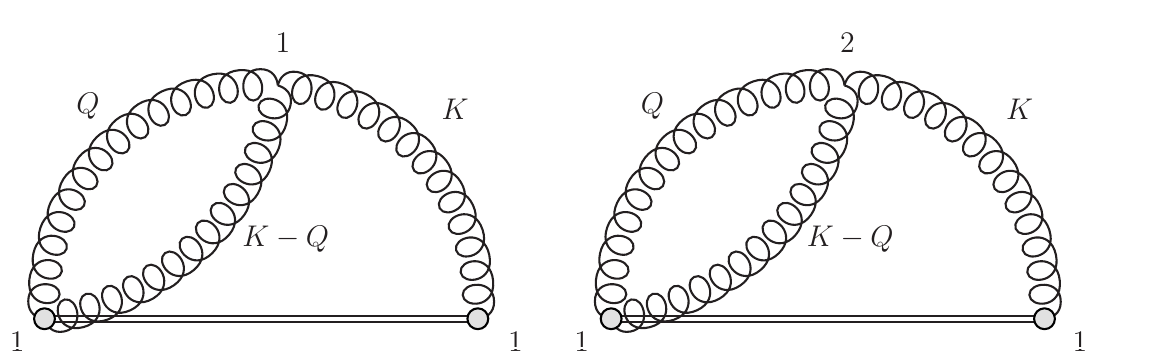}
	\caption{Different topologies associated with diagram $(g)$ with momentum flowing from left to right. The ``1, 2'' labels indicate the
     branch of the Schwinger-Keldysh contour from which the vertices are being pulled.}
    \label{fig:g_ass}
\end{figure}

\subsection{Diagram \texorpdfstring{$(g)$}{TEXT}}\label{sec:g}
For the computation of diagram $(g)$, working in the 1/2 basis there are two possible thermal
assignments, shown in Fig.~\ref{fig:g_ass}. We start with the left diagram
\begin{align}
    Z_g^{(g)\,1}& = - \frac{2}{d_A} \int_0^\infty   dx^+ x^+ 
    \bigg\langle \Big[\partial^- A^{x,a}_1(x^+) - \partial^x A^{-,a}_1(x^+) \Big] 
    g f^{abc} A^{-,b}_1(0) A^{x,c}_1(0) \bigg\rangle^1 \nonumber
    \\&=- \frac{2g f^{abc}}{d_A} \int_0^\infty   dx^+ x^+ \int_{K}e^{-k^- x^+}
    \bigg\langle \Big[-ik^- A^{x,a}_1 + k^x A^{-,a}_1 \Big] 
     A^{-,b}_1 A^{x,c}_1(0) \bigg\rangle^1,
\end{align}
where the 1 subscripts on the gauge fields indicate that they are located 
on the time-ordered branch of the Schwinger-Keldysh contour (see Fig.~\ref{fig:sk}). The 
1 superscript on the angled bracket indicates that the thermal expectation value 
is to be evaluated with the three-gluon vertex of type 1. Upon 
evaluating the expectation value, we get 
\begin{align}
    Z_g^{(g)\,1} &= -\frac{2gf^{abc}}{d_{A}} \int_0^\infty dx^+ x^+  \int_{K,Q} e^{-i k^- x^+} \; \left(g f^{abc} \; V_{\alpha\beta\gamma}(Q,K-Q,-K)\right) \notag \\
    & \times \Big[ \left(-\eta^{-\alpha} G_{F}(Q)\right) \left(-\eta^{x\beta} G_{F}(K-Q)\right)
    \left(-i k^- \left(-\eta^{x\gamma} G_{F}(K)\right) + i k^x \left(-\eta^{-\gamma} G_{F}(K)\right) \right) \Big]\nonumber
    \\&= - 2 i g^2 C_{A} \int_0^\infty dx^+ x^+ \int_{K,Q} e^{-i k^- x^+} \;  \notag 
    \Big[G_{F}(Q) G_{F}(K-Q) 
    k^- G_{F}(K) V_{+xx}(Q,K-Q,-K)  \notag \\
    &\hspace{5pt}
    - G_{F}(Q) G_{F}(K-Q) k^x G_{F}(K) V_{+x+}(Q,K-Q,-K) \Big] 
\end{align}
From Eq.~\eqref{eq:three_gluon}, we see that the second term vanishes. Moreover,
\begin{equation}
    \int_0^\infty dx^+ x^+ e^{-i k^- x^+} = -\frac{1}{(k^- - i \varepsilon)^2},
\end{equation}
so that 
\begin{equation}
    Z_g^{(g)\,1}= 2i g^2C_{A} \int_{K,Q}
    G_{F}(Q)G_{F}(K-Q)G_{F}(K) \left( \frac{k^- q^-}{(k^- - i \varepsilon)^2} - 2 \right) .
    \label{eq:gto}
\end{equation}
Now, let us repeat the same procedure for the vertex assigned to the 2-branch of the Schwinger-Keldysh contour (Fig.\ref{fig:g_ass}-right)
\begin{align}
    Z_g^{(g)\,2} &= - \frac{2}{d_{A}}\int_0^\infty d x^+ x^+ \bigg\langle \Big[ \partial^- A^{x,c}_1(x^+) - \partial^x A^{-,c}_1(x^+) \Big] g f^{abc} A^{-,b}_1(0) A^{x,c}_1(0) \bigg\rangle^{2} \notag \\
    &= - \frac{2}{d_{A}} \int_0^\infty d x^+ x^+ f^{abc} g^2 \int_{K,Q} e^{-i k^- x^+}\left(-f^{abc} V_{\alpha\beta\gamma}(Q,K-Q,-K)\right) \notag \\
    &\times\Big[ \left(-\eta^{-\alpha} G_{<}(Q)\right) \left(-\eta^{x\beta}G_{<}(K-Q)\right)
    \left( -i k^-\left(-\eta^{x\gamma} G_{<}(K)\right) + i k^x \left(-\eta^{-\gamma}G_{<}(K)\right) \right) \Big] \notag \\
    &= -2i g^2 C_{A} \int_{K,Q} G_{<}(Q) G_{<}(K-Q) G_{>}(K) \left( \frac{k^- q^-}{(k^- - i \varepsilon)^2} - 2 \right) .
    \label{eq:gcut}
\end{align}
To get the full contribution to $Z_g^{(g)}$, we need to add Eqs.~\eqref{eq:gto} and Eq.~\eqref{eq:gcut} along 
with the mirror image of each diagram from Fig.\ref{fig:g_ass}. While the mirror of Fig.\ref{fig:g_ass}-left gives 
the same contribution as Eq.~\eqref{eq:gto}, to get the mirror of Fig.\ref{fig:g_ass}-right, we should take 
Eq.~\eqref{eq:gcut} and make the replacement $G_{<}(Q) G_{<}(K-Q) G_{>}(K)\rightarrow G_{>}(Q) G_{>}(K-Q) G_{<}(K)$. This leaves
us with 
\begin{align}
    Z^{(g)}_g&=2ig^2C_{A}\int_{K,Q}\left(\frac{q^-}{k^- -i\varepsilon}-2\right)\Big[2G_{F}(Q)G_{F}(K-Q)G_{F}(K)-G_{>}(Q)G_{>}(K-Q)G_{<}(K)\nonumber
    \\&-G_{<}(Q)G_{<}(K-Q)G_{>}(K)\Big]\label{eq:zgg_before}
\end{align}
In order to simplify things, we employ the amputated self-energy-like objects defined in App.~\ref{app:amputated}, which 
allow us to rewrite Eq.~\eqref{eq:zgg_before} as
\begin{align}
    Z^{(g)}_g&=2ig^2C_{A}\int_{K,Q}\left(\frac{q^-}{k^- -i\varepsilon}-2\right)\Big[\Big(2\widetilde{\Pi_{aa}}(K)+\widetilde{\Pi_{R}}(K)+\widetilde{\Pi_{A}}(K)\Big)G_{F}(K)\nonumber
    \\& -\widetilde{\Pi_{>}}(K)G_{<}(K)-\widetilde{\Pi_{<}}(K)G_{>}(K)\Big].\label{eq:zg_middle}
\end{align}
In arriving at the above expression, one must invoke that the closed loops 
of causality vanish (see Sec.~\ref{sec:rtf}). Specifically, this argument applies to the terms containing $G_{R}(Q)G_{A}(K-Q),\,G_{A}(Q)G_{R}(K-Q)$. 
It then suffices to use the KMS-like relations in App.~\ref{app:amputated} to arrive at 
\begin{equation}
    Z^{(g)}_g=4ig^2C_{A}\int_{K,Q}\left(\frac{q^-}{k^- -i\varepsilon}-2\right)\left(1+n_{\text{B}}(k^0)\right)\left(\widetilde{\Pi_{R}}(K)G_{R}(K)-\widetilde{\Pi_{A}}(K)G_{A}(K)\right).
    \label{eq:zgg}
\end{equation}
We halt with the evaluation at this stage. Indeed, we will see 
that this contribution cancels with the one coming from diagram $(f)$ in Sec.~\ref{sec:proof_diagram}.

\begin{figure}[ht]
	\centering
    \includegraphics[width=0.7\textwidth]{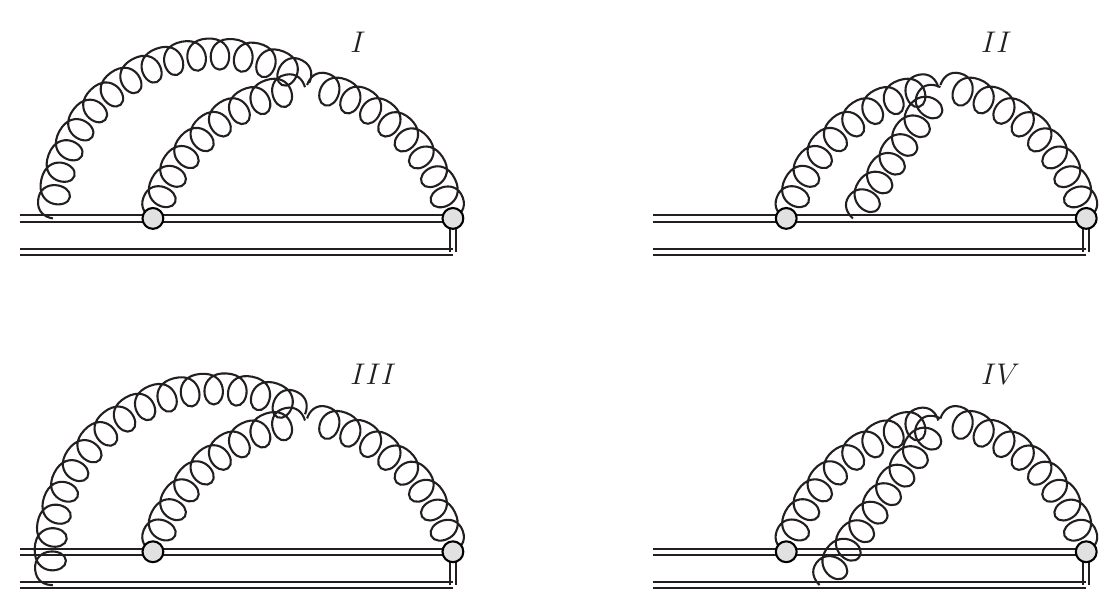}
    \caption{Different topologies associated with diagram $(f)$ with momentum flowing from left to right. 
    In all cases, we take the propagator attached to the Wilson line (as opposed to the $F$ operator insertions) to have momentum $K-Q$.}
    \label{fig:f_ass}
\end{figure}

\subsection{Diagram \texorpdfstring{$(f)$}{TEXT}}\label{sec:f}

This is the only non-vanishing diagram where the gluons are being pulled from the Wilson line itself.
The corresponding contributions to this diagram are listed in Fig.~\ref{fig:f_ass}. 
We then have $8$ possible assignments overall, since each of these $4$ subdiagrams 
can have either a $1$ or $2$ labelling for the three-gluon vertex.

Since it is an $A^{-}$ field that is being pulled from the Wilson line 
(see Eq.~\eqref{eq:4d:Wilson:line:expl}), we can conclude that in each subdiagram, 
the only contribution to the three-gluon vertex will be proportional to $\eta_{xx}$. 
We proceed to first compute the contribution from $Z_g^{f\,I}$, pointing to Fig.~\ref{fig:f_ass} for clarification about the 
labelling
\begin{align}
    Z_g^{(f)\,I}&=\frac{-2}{d_{R}C_{R}}(ig)\int_{0}^{\infty} dx^+ x^+\int_{-\infty}^{0} dy^+
    \bigg[\frac{i f^{cba}}{4}\bigg\langle\partial^{-}A^{a,x}(x^+)\partial^{-}A^{b,x}(0)A^{c,-}(y^+)\bigg\rangle^1\nonumber
    \\&+\frac{i f^{acb}}{4}\bigg\langle\partial^{-}A^{a,x}(x^+)\partial^{-}A^{b,x}(0)A^{c,-}(y^+)\bigg\rangle^2 \bigg]
\end{align}
where the two factors in parentheses out in front come from the colour trace and the Wilson line respectively. Then, Fourier transforming in a way
that is consistent with the labelling convention mentioned in Fig.~\ref{fig:f_ass}, we find
\begin{align}
    Z_g^{(f)\,I}&=-\frac{2ig}{d_{R}C_{R}}\int_{0}^{\infty} dx^+ x^+\int_{-\infty}^{0} dy^+ \int _{K,Q}k^{-}q^{-} e^{-ik^{-}x^{+}}e^{-i(q^{-}-k^{-})y^{+}}\nonumber
    \\&\times\bigg[\frac{i f^{cba}}{4}\bigg\langle A^{a,x}(K) A^{b,x}(Q) A^{c,-}(K-Q)\bigg\rangle^1
    +\frac{i f^{acb}}{4}\bigg\langle A^{a,x}(K) A^{b,x}(Q) A^{c,-}(K-Q)\bigg\rangle^2\bigg]\nonumber
    \\&=-\frac{2ig}{d_R C_R} \int_{K,Q}\frac{q^-}{(k^{-}-i\varepsilon)(k^{-}-q^{-}-i\varepsilon)}\nonumber
    \\&\times\bigg[\frac{i f^{cba}}{4}\bigg\langle A^{a,x}(K) A^{b,x}(Q) A^{c,-}(K-Q)\bigg\rangle^1
    +\frac{i f^{acb}}{4}\bigg\langle A^{a,x}(K) A^{b,x}(Q) A^{c,-}(K-Q)\bigg\rangle^2\bigg]\nonumber
    \\&=ig^2C_{A}\int_{K,Q}\frac{q^-(k^- + q^-)}{(k^{-}-i\varepsilon)(k^{-}-q^{-}-i\varepsilon)}\Big[-G_{F}(Q)G_{F}(K)G_{F}(K-Q)\nonumber
    \\&+G_{>}(Q)G_{<}(K)G_{>}(K-Q)\Big].
\end{align}
Note that the thermal assignments are separated by a minus: this is because, in the second, we are pulling a 2 vertex from the action. 
We can immediately then write down the expression for the contribution
of the subdiagram with the gluon instead attached to the Wilson line directly below
\begin{align}
    Z_g^{(f)\,III}&=-ig^2C_{A}\int_{K,Q}\frac{q^-(k^- + q^-)}{(k^{-}-i\varepsilon)(k^{-}-q^{-}-i\varepsilon)}\Big[-G_{F}(Q)G_{F}(K)G_{<}(K-Q)\nonumber
    \\&+G_{>}(Q)G_{<}(K)G_{\bar{F}}(K-Q)\Big].
\end{align}
The minus sign out in front can be attributed to the fact that in this case, the limits of the $y^+$ integral are switched. 
The thermal assignments are of course different in this subdiagram, since the lower Wilson line is anti-time-ordered.
Combining the two subdiagrams then yields, after some algebra

\begin{align}
     Z_g^{(f)\,I}+Z_g^{(f)\,III}&=-ig^2C_{A}\int_{K,Q}\frac{q^-(k^- + q^-)}{(k^- -i\varepsilon)(k^- - q^- -i\varepsilon)}
     G_{R}(K-Q)\Big[G_{rr}(Q)G_{R}(K)\nonumber
     \\&+G_{rr}(K)G_{A}(Q)+\frac{1}{2}\left(G_{R}(K)G_{R}(Q)+G_{A}(K)G_{A}(Q)\right)\Big]\label{eq:one_p_three}
\end{align}
the last two terms are scaleless and therefore vanish in dimensional regularisation.
 
Moving onto the subdiagram where the gluon is attached to the line between the two $F$ insertions, we have
\begin{align}
    Z_g^{(f)\,II}&=\frac{-2}{d_{R}C_{R}}(ig)\int_{0}^{\infty} dx^+ x^+\int_{0}^{x^+} dy^+ \int _{K,Q}k^{-}q^{-} e^{-ik^{-}x^{+}}e^{-i(q^{-}-k^{-})y^{+}}
   \\&\times\bigg[ \frac{i f^{cba}}{4}\bigg\langle A^{a,x}(K) A^{b,-}(K-Q) A^{c,x}(Q)\bigg\rangle^1
   +\frac{i f^{bca}}{4}\bigg\langle A^{a,x}(K) A^{b,-}(K-Q) A^{c,x}(Q)\bigg\rangle^2\bigg]\nonumber
    \\&=\frac{2g}{d_{R}C_{R}}\int_{K,Q}\frac{1}{q^{-}-k^{-}}\left(\frac{q^-}{k^- -i\varepsilon} -\frac{k^-}{q^- -i\varepsilon}\right)\nonumber
    \\&\times\bigg[ \frac{i f^{cba}}{4}\bigg\langle A^{a,x}(K) A^{b,-}(K-Q) A^{c,x}(Q)\bigg\rangle^1
    +\frac{i f^{bca}}{4}\bigg\langle A^{a,x}(K) A^{b,-}(K-Q) A^{c,x}(Q)\bigg\rangle^2\bigg]\nonumber
    \\&=-ig^2C_{A}\int_{K,Q}(k^- + q^-)\left(\frac{1}{k^- -i\varepsilon} +\frac{1}{q^- -i\varepsilon}\right)\Big[-G_{F}(Q)G_{F}(K)G_{F}(K-Q)\nonumber
    \\&+G_{>}(Q)G_{<}(K)G_{>}(K-Q)\Big].
\end{align}
In this case, the corresponding diagram with the gluon attached below has the same sign, due to 
relative minuses coming from both the colour trace and the inversion of the limits on
the $y^+$ integral. Thus, 
\begin{align}
    Z_g^{(f)\,IV}&=-ig^2C_{A}\int_{K,Q}(k^- + q^-)\left(\frac{1}{k^- -i\varepsilon} +\frac{1}{q^- -i\varepsilon}\right)
    \Big[-G_{F}(Q)G_{F}(K)G_{<}(K-Q)\nonumber
    \\&+G_{>}(Q)G_{<}(K)G_{\bar{F}}(K-Q)\Big].
\end{align}
Together these two add to give, 
\begin{align}
     Z_g^{(f)\,II}&+Z_g^{(f)\,IV}=ig^2C_{A}\int_{K,Q}(k^- + q^-)\left(\frac{1}{k^- -i\varepsilon} +\frac{1}{q^- -i\varepsilon}\right)
     \Big\{G_{A}(K-Q)\Big[\frac{1}{2}G_{A}(Q)G_{R}(K)\nonumber
     \\&+2G_{rr}(K)G_{rr}(Q)+G_{rr}(K)G_{R}(Q)\Big]+2G_{rr}(K-Q)\Big[\frac{1}{2}G_{R}(K)G_{R}(Q)+G_{rr}(Q)G_{R}(K)\nonumber
     \\&+G_{rr}(K)G_{A}(Q)\Big]\Big\}\nonumber
     \\&=ig^2C_{A}\int_{K,Q}\frac{k^- + q^-}{k^- - i\varepsilon}\nonumber
     \\&\times\Big\{4\left(\frac{1}{2}+n_{\text{B}}(k^0)\right)\Big[G_{R}(K)\widetilde{\Pi_{R}}(K)-G_{A}(K)\widetilde{\Pi_{A}}(K)\Big]
     +2G_{R}(K)\widetilde{\Pi_{R}}(K)\nonumber
     \\&-G_{R}(K)\Big[G_{R}(K-Q)G_{rr}(Q)+\frac{1}{2}G_{R}(K-Q)G_R(Q)\Big]\Big\}
     \label{eq:zgfiipv}
\end{align}
where we have made use of the ``amputated self-energy" relations from App.~\ref{app:amputated}. We come 
back to Eqs.~\eqref{eq:one_p_three} and ~\eqref{eq:zgfiipv} in Sec.~\ref{sec:proof_correlator}.

\subsection{Diagrams \texorpdfstring{$(b),\,(d_1),\,(d_2),\,(e),\,(h),\,(i),\,(j)$}{TEXT}}\label{sec:b}
It suffices to write down the expression for diagram $(b)$
\begin{align}
    Z_g^{(b)}&=-\frac{\mathcal{2C}(ig)^2}{d_{R}C_{F}}\int_0^{\infty}dx^+x^+\int_0^{x^+}dy^+\int_0^{y^+}dz^+
    \Bigl\langle\Big(\partial^-A^x_{a}(x^+)-\partial^x A^-_{a}(x^+)\Big)\Big(\partial^-A^x_{a}(0)-\partial^x A^-_{a}(0)\Big)\Bigr\rangle\nonumber
    \\&\times\Bigl\langle A^-_b(y^+)A^-_b(z^+)\Bigr\rangle,
\end{align}
where $\mathcal{C}$ is some non-zero colour factor. In Feynman gauge, the $\Bigl\langle A^-_b(y^+)A^-_b(z^+)\Bigr\rangle$ factor 
above will inevitably produce an $\eta^{--}$, from which we can conclude that this diagram vanishes. The 
other diagrams $(b),\,(d_1),\,(d_2),\,(e),\,(h),\,(i),\,(j)$ then vanish for the same reason as they only 
contain terms proportional to either $\Bigl\langle A^- A^-\Bigr\rangle$ or $\Bigl\langle A^-A^x\Bigr\rangle$, the 
latter of which is also zero in Feynman gauge.

\section{Proof of \texorpdfstring{$Z_g^{(f)}+Z_{g}^{(g)}=0$}{TEXT}}\label{sec:proof}

We provide two arguments, supporting the claim that in Feynman gauge,
we need only consider the contribution from diagram $(c)$ in Fig.~\ref{fig:amass_all_diagrams}.

\subsection{Proof from the Correlator}\label{sec:proof_correlator}
Let us start by looking at Eq.~\eqref{eq:zg_def} and writing
\begin{equation}
     Z_g =  -\frac{1}{d_R C_R} \int_0^\infty  dx^+ x^+
     \,\Tr\,\Bigl\langle
       U_{R}(-\infty;x^+) F^{-\perp} (x^+)\,U_{R}(x^+;0)\,
       F^{-\perp}(0)U_{R}(0;-\infty)     \Bigr\rangle,
       \label{eq:deffundlorentz}
\end{equation}
where $F^{-\perp}F^{-\perp}$ is taken to mean $F^{-x}F^{-x}+F^{-y}F^{-y}$.
Continuing onwards, let us recast Eq.~\eqref{eq:zg_def} as a double integral,
using translation invariance
\begin{align}
     Z_g &=  
      -\frac{1}{d_{R}C_{R}} \int_0^\infty d x^+ \int_0^{x^+} d y^+ 
     \Tr\,\Bigl\langle
       U_{R}(-\infty;x^++y^+) F^{-\perp} (x^++y^+)\,U_{R}(x^++y^+;y^+)\,
       F^{-\perp}(y^+)\nonumber \\
       &\times U_{R}(y^+;-\infty)     \Bigr\rangle,\nonumber\\
       &=  
      -\frac{1}{d_{R}C_{R}} \int_0^\infty d x^+ \int_0^{x^+/2} \! d y^+ 
     \,\Tr\,\Bigl\langle
       U_{R}(-\infty;x^+) F^{-\perp} (x^+)\,U_{R}(x^+;y^+)\,
       F^{-\perp}(y^+)
      U_{R}(y^+;-\infty)     \Bigr\rangle,
       \label{eq:deffunddouble}
\end{align}
where we have shifted $x^+$ in going to the second line. We now rewrite $Z_g$ as
\begin{equation}
    Z_g\equiv Z_g^{\perp\perp}+Z_g^{\perp-}+Z_g^{-\perp}+Z_g^{--},
    \label{zgsplit}
\end{equation}
with the apex labelling to be elaborated on momentarily. By calling on a method introduced in~\cite{simonguy}: we rewrite
\begin{equation} 
F^{-\perp}=-\partial^\perp A^-+[D^-,A^\perp].
\end{equation} 
The equation of motion of the Wilson line (see Eq.~\eqref{eq:wloop_eom}), $D^-_{x^+} U(x^+;y^+)=0$ can then be used to rewrite 
the commutator as a total derivative, i.e.
\begin{equation}
\label{totald}
U_{R}(a;x^+)[D^-,A^\perp(x^+)]U_{R}(x^+;b)=\frac{d}{dx^+}[U_{R}(a;x^+)\,A^\perp(x^+)\,U_{R}(x^+;b)].
\end{equation}
This should shed light on the labelling in Eq.~\eqref{zgsplit}: the 
apexes there denote whether the $-\partial^\perp A^-$ (${}^\perp$) or the $[D^-,A^\perp]$
component (${}^-$) of the field strength tensor has been taken.

Starting with the first term from Eq.~\eqref{zgsplit}, 
\begin{equation}
    Z_g^{\perp\perp}=-\frac{1}{d_RC_R} \int_0^\infty  d x^+ x^+
     \,\Tr\,\Bigl\langle
       U_{R}(-\infty;x^+) \partial^{\perp}A^- (x^+)\,U_{R}(x^+;0)\,
       \partial^{\perp}A^-(0)U_{R}(0;-\infty)     \Bigr\rangle.
       \label{perpperp}
\end{equation}
Whereas for the second term, we have, employing Eq.~\eqref{totald}
\begin{align}
    Z_g^{\perp-}&=\frac{1}{d_{R}C_{R}} \int_0^\infty d x^+ 
     \,\Tr\,\Bigl\langle
       U_{R}(-\infty;x^+)\partial^{\perp}A^- (x^+)\,U_{R}(x^+;x^+/2)\,
       \Big[A^\perp(x^+/2)U_{R}(x^+/2;0)\nonumber
       \\&-U_{R}(x^+/2;0)A^\perp(0)\Big]
      U_{R}(0;-\infty)     \Bigr\rangle.
\end{align}
Translation invariance dictates that for the first term
\begin{align}
    &\Bigl\langle
       U_{R}(-\infty;x^+)\partial^{\perp}A^- (x^+)\,U_{R}(x^+;x^+/2)\,
       A^\perp(x^+/2)U_{R}(x^+/2;-\infty)\Bigr\rangle=
       \Bigl\langle
       U_{R}(-\infty;x^+/2)\nonumber\\
       &\times\partial^{\perp}A^- (x^+/2)\,U_{R}(x^+/2;0)\,
       A^\perp(0)U_{R}(0;-\infty)\Bigr\rangle.
\end{align}
By aptly redefining the integration variable for this term one finds that
\begin{align}
    Z_g^{\perp-}=\frac{1}{d_{R}C_{R}} \int_0^\infty d x^+ 
     \,\Tr\,\Bigl\langle
       U_{R}(-\infty;x^+)\partial^{\perp}A^- (x^+)\,U_{R}(x^+;0)\,
       A^\perp(0)
      U_{R}(0;-\infty)     \Bigr\rangle.
      \label{perpminus}
\end{align}
For its counterpart we have
\begin{align}
   Z_{g}^{-\perp} =  -
      \frac{1}{d_{R}C_{R}} \int_0^\infty  d y^+ 
     \,\Tr\,\Bigl\langle
      U_{R}(-\infty;2y^+) A^{\perp} (2y^+)\,U_{R}(2y^+;y^+)\,
      \partial^{\perp}A^-(y^+)
      U_{R}(y^+;-\infty)     \Bigr\rangle.
\end{align}
where we used the fact that, in a non-singular gauge, $A^\perp(\infty)=0$.
By applying translation invariance and relabeling $y^+$ into $x^+$ 
this becomes
\begin{align}
   Z_g^{-\perp} =  -
      \frac{1}{d_{R}C_{R}} \int_0^\infty dx^+ 
     \,\Tr\,\Bigl\langle
      U_{R}(-\infty;x^+) A^{\perp} (x^+)\,U_{R}(x^+;0)\,
      \partial^{\perp}A^-(0)      U_{R}(0;-\infty)     \Bigr\rangle.
      \label{minusperp}
\end{align}
Finally, for the term with both commutators, we find
\begin{align}
    Z_g^{--}=\frac{1}{d_{R}C_{R}} 
     \,\Tr\,\Bigl\langle
       U_{R}(-\infty;0)A^\perp (0)\,
       A^\perp(0)
      U_{R}(0;-\infty)     \Bigr\rangle,
      \label{minusminus}
\end{align}
where we have again set the boundary term at infinity to zero.

At this point, we recall that in Feynman gauge the ${}^{--}$ gluon propagator vanishes and 
furthermore, that a three-gluon vertex with more than one ${}^-$ gluon does as well.
Let us then consider Eq.~\eqref{perpperp}: at LO it must vanish in Feynman gauge.
At NLO it can only contribute to part of diagram $(c)$. Note however that the 
Wilson lines cannot contribute on their own for the same reasons outlined in Sec.~\eqref{sec:b}.
Moreover, there cannot be a contribution of the form of diagram $(g)$ as 
the $gA^{-}A^{\perp}$ part of the covariant derivative 
has been removed from the correlator through the previous manipulations.

Similar arguments can be safely applied to Eqs.~\eqref{perpminus} and \eqref{minusperp}:
they also have a vanishing $(f)$ contribution and they correspond to the mixed $G^{x-}_>(K)$
terms in Eq.~\eqref{cdiagprops}.

Eq.~\eqref{minusminus} does not vanish at LO: it corresponds precisely to diagram $(a)$.
At NLO it too contains a contribution to $(c)$, that proportional to $G^{xx}_>(K)$
terms in Eq.~\eqref{cdiagprops}. It also contains the only contribution directly 
involving the Wilson lines, as either of them can source an $A^-$ gluon 
that can connect with a three-gluon vertex to the two $A^\perp$ ones. However,
that would give rise to the following colour structure 
\begin{equation}    
    \mathcal{C}=f^{abc}A^{\perp\,a}A^{\perp\,b}A^{-\,c},
\end{equation}
which vanishes, due to the symmetric (anti-symmetric) nature of $A^{\perp\,a}A^{\perp\,b}$ ($f^{abc}$).
Given that, through this reorganisation, we only encounter non-zero contributions from 
diagram $(c)$, we conclude that the respective contributions from diagrams $(f)$ and $(g)$ must 
cancel each other. We further back up this argument with a direct diagrammatic proof in the next
section.

\subsection{Diagrammatic Proof}\label{sec:proof_diagram}

Coming back to Eq.~\eqref{eq:one_p_three}, we note that by
 suitably relabelling momenta, we can drastically simplify this expression: let us start by considering the second term and relabeling
$K-Q\rightarrow K,\\ Q\rightarrow Q-K\,,K\rightarrow Q$. This yields
\begin{align}
     Z_g^{(f)\,I}+Z_g^{(f)\,III}&=-ig^2C_{A}\int_{K,Q}\Big[\frac{q^-(k^- + q^-)}{(k^- -i\varepsilon)(k^- - q^- -i\varepsilon)}
     +\frac{(q^- - k^-)(2q^- - k^-)}{(q^- -i\varepsilon)(k^- -i\varepsilon)}\Big]\nonumber
     \\&\times G_{R}(K-Q)G_{rr}(Q)G_{R}(K)\nonumber
     \\&=-ig^2C_{A}\int_{K,Q}\Big[\frac{q^-(k^- + q^-)}{(k^- -i\varepsilon)(k^- - q^- -i\varepsilon)}
     +\frac{2q^{-}}{(k^- -i\varepsilon)}\Big]\nonumber
     \\&\times G_{R}(K-Q)G_{rr}(Q)G_{R}(K)
\end{align}
where in going to the second line, we have used that we can throw away any terms in the second term with a factor of $k^-$ in the numerator, 
since the $k^-$ integral can then be closed in the upper half plane. We then perform another relabelling of the momenta in the second term: 
$K-Q\rightarrow K\,,\,Q\rightarrow -Q\,,\,K\rightarrow K-Q$, 
under which the multiplied propagators are invariant
\begin{align}
     Z_g^{(f)\,I}+Z_g^{(f)\,III}&=-ig^2C_{A}\int_{K,Q}\Big[\frac{q^-(k^- + q^-)}{(k^- -i\varepsilon)(k^- - q^- -i\varepsilon)}
     -\frac{2q^-}{(k^- - q^- -i\varepsilon)}\Big]\nonumber
     \\&\times G_{R}(K-Q)G_{rr}(Q)G_{R}(K)\nonumber
     \\&=ig^2C_{A}\int_{K,Q}\frac{q^-}{k^- - i\varepsilon} G_{R}(K-Q)G_{rr}(Q)G_{R}(K).
     \label{eq:zfiplvurelab}
\end{align}
At this point, let us go back and look
at the expression coming from the evaluation of diagram $(g)$, Eq.~\eqref{eq:zgg} and relabel so that $Q\rightarrow K-Q\,,\,K-Q\rightarrow Q\,,\,K\rightarrow K$, yielding
\begin{equation}
    Z^{(g)}_g=-4ig^2C_{A}\int_{K,Q}\frac{k^- + q^-}{k^- -i\varepsilon}\left(1+n_{\text{B}}(k^0)\right)\left(\widetilde{\Pi_{R}}(K)G_{R}(K)-\widetilde{\Pi_{A}}(K)G_{A}(K)\right).
\end{equation}
which of course resembles what we have in Eq.~\eqref{eq:zgfiipv}. 
Let us then add both contributions to see if they cancel
\begin{align}
    Z_g^{(f)\,II}&+Z_g^{(f)\,V}+Z_{g}^{(g)}=ig^2C_{A}\int_{K,Q}\frac{k^- +q^-}{k^- - i\varepsilon}\Big\{2\widetilde{\Pi_{A}}(K)G_{A}(K)\nonumber
    \\&-G_{R}(K)G_{R}(K-Q)G_{rr}(Q)-\frac{1}{2}G_{R}(K)G_{R}(K-Q)G_{R}(Q)\Big\}.
    \label{eq:zgfpg}
\end{align}
The first term above is advanced in $k^-$, so we can just close the integration in the lower half plane to see that it vanishes. Likewise, for the $k^-/(k^- - i\varepsilon)$ proportional part of the second term, we can close the $k^-$ integration in the upper half plane to see that it does not contribute. 
The third term will vanish after a suitable relabelling of the momenta whereas 
the $q^-/(k^- - i\varepsilon)$ part of the second term cancels against what we found in Eq.~\eqref{eq:zfiplvurelab}.
\\\newline 
Alternatively, we could invoke the ideas from Sec.~\ref{sec:chtrick} since in the case of the second and third term, the propagators are lightlike separated and are thus retarded in both $k^-$ and $k^+$. The $1/(k^- -i\varepsilon)$ out in front means that we cannot close the $k^-$ in the upper half plane without picking up a contribution there. 
However, there are no such extra poles in $k^+$ and so we can therefore close the $k^+$ integration in the upper half-plane. 
\\\newline 
In any case, we undoubtedly conclude 
\begin{equation}
    Z_g^{(f)}+Z_{g}^{(g)}=0.
\end{equation}

\section{Matching from Full QCD}\label{sec:full_qcd}

In this section, we compute zero-mode contributions from the gluonic part of diagram $(c)$ and 
thus $Z_g$, before checking that they cancel with the corresponding EQCD 
contributions in the next section. Throughout, for consistency's sake, 
we set up the integration so that the $K$ integration is done in DR whereas the $Q$ 
integration is done in $4$ dimensions\footnote{We choose this scheme to avoid having to consider the $d$-dimensional equivalents 
of the one-loop self-energies, listed in App.~\ref{app:one_loop}.}.
It is useful in this context 
 to think of the contributions to $Z_{g}$, in diagram $(c)$ as a 
 double sum over Matsubara modes, one sum coming from 
 $k^0$ and the other coming from $q^0$.
 Since EQCD is a theory of soft modes, it only accounts for the 
 \emph{zero modes} of the full theory. It is then natural to 
 expect corresponding divergences on the full QCD side to show 
 up in the term of the double sum with both zero modes. Indeed, 
 this is \emph{physically} what happens. However, this double zero mode
  term will give no contribution in DR,
   as it is the result of a scaleless integral. Thus, 
   \emph{mathematically}, what happens is that the divergence gets shifted 
from the $n=0\,,\,n=0$ term to the $n=0\,,\,n\neq 0$ terms. 
We make a point to organise things so the divergence comes from the zero mode associated with the momentum 
$K$. 

We emphasise that we only need to do this matching for the gluonic contribution to $Z_g^{(c)}$ as EQCD 
only describes the dynamics of the gluonic zero modes.

\subsection{\texorpdfstring{$k^{-2}$}{TEXT} Piece}\label{sec:zero_kminus}

We begin by computing the $k^{-2}$  contribution, defined by Eqs.~\eqref{zcexpr} and ~\eqref{eq:split}. We moreover 
need the gluonic form of $\Pi_{T}^{R}(K)$, Eq.~\eqref{rettransselfglue} 
\begin{align}
    Z^{(c)}_{g\,k^{-2}\,G}&=i\frac{g^2C_{A}}{8\pi^2}\int_K\int_{0}^{\infty}dq\frac{(1+n_{B}(k^0))}{k^3}n_{B}(q)\Bigg[\frac{1}{(K^2+i\varepsilon k^0)^2}\Bigg(8qk(k^2+k_0^2)\nonumber
    \\&-K^2(4q^2+3k^2+k_0^2)\left(\ln\frac{k_0+k-2q+i\varepsilon}{k_0-k-2q+i\varepsilon}+\ln\frac{k_0-k+2q+i\varepsilon}{k_0+k+2q+i\varepsilon}\right)\nonumber
    \\&-4qk_0K^2\left(\ln\frac{k_0+k+i\varepsilon}{k_0-k+i\varepsilon}+\ln\frac{k_0-k+2q+i\varepsilon}{k_0+k+2q+i\varepsilon}-\ln\frac{k_0+k-2q+i\varepsilon}{k_0-k-2q+i\varepsilon}\right)\Bigg)-\text{adv.}\Bigg]
\end{align}
Let us then the write $k^0$ integral as a sum over the Matsubara modes. This leaves us with 
\begin{align}
    Z^{(c)}_{g\,k^{-2}\,G}&=-\frac{g^2C_{A}T}{8\pi^2}\sum_{n}\int_k\int_{0}^{\infty}dq\frac{1}{k^3}n_{B}(q)\frac{1}{(-\omega_n^2-k^2)^2}\Bigg(8qk(k^2-\omega_n^2)\nonumber
    \\&-(-\omega_n^2-k^2)(4q^2+3k^2-\omega_n^2)\left(\ln\frac{i\omega_n+k-2q}{i\omega_n-k-2q}+\ln\frac{i\omega_n-k+2q}{i\omega_n+k+2q}\right)\nonumber
    \\&-4iq\omega_n(-\omega_n^2-k^2)\left(\ln\frac{i\omega_n+k}{i\omega_n-k}+\ln\frac{i\omega_n-k+2q}{i\omega_n+k+2q}-\ln\frac{i\omega_n+k-2q}{i\omega_n-k-2q}\right)\Bigg).
    \label{gcqminusgmatsum}
\end{align}
Next, we extract the zero mode contribution
\begin{equation}
    Z^{(c)}_{g\,k^{-2}\,G\,n=0}=-\frac{g^2C_{A}T}{8\pi^2}\int_k\int_{0}^{\infty}dq\frac{1}{k^3}n_{B}(k)\frac{1}{k^4}\Bigg(8qk^3
    +2k^2(4q^2+3k^2)\ln\left\vert\frac{2q-k}{k+2q}\right\vert\Bigg).
\end{equation}
Before carrying out the integration in DR, it is first useful to rescale $k\rightarrow kq$ so that the integrations factorise, i.e
\begin{align}
    Z^{(c)}_{g\,k^{-2}\,G\,n=0}&=-\frac{g^2C_{A}T}{8\pi^2}\Big( \frac{\bar{\mu}^{2}e^{\gamma_{E}}}{4\pi} \Big)^\epsilon\Big( \frac{2\pi^{\frac{d}{2}}}{\Gamma\left(\frac{d}{2}\right)(2\pi)^d} \Big)
    \int_0^{\infty}dk\int_{0}^{\infty}dq\frac{k^{d-1}}{k^5}\frac{q^{d+2}}{q^5}n_{B}(q)\Bigg(8k\nonumber
    \\&+2(4+3k^2)\ln\left\vert\frac{2-k}{k+2}\right\vert\Bigg)\nonumber
    \\&=\frac{g^2C_{A}T^2}{32\pi^2}\bigg[-\frac{3}{\epsilon}-\left(7+6\ln\frac{\bar{\mu}}{4T}\right)\bigg],\label{IRpoleqmsquare}
\end{align}
where, as is specified in App.~\ref{sec:conventions}, DR is carried out in $d=3-2\epsilon$ dimensions.

\subsection{\texorpdfstring{$\Pi_{L}-\Pi_{T}$}{TEXT} Piece}\label{sec:zero_lmt}

Let us start by writing the full contribution
(i.e, that coming from all modes) using Eq.s~\eqref{zcexpr} and ~\eqref{retdiffgluonsimple}. 
 This yields 
\begin{align}
Z^{(c)}_{g\,L-T\,G}&=i\frac{g^2C_{A}}{16\pi^2}
    \int_K\int_{0}^{\infty}dq\frac{(1+n_{B}(k^0))}{k^3(k^--i\varepsilon)^2}n_{B}(q)\Bigg[\frac{k_{\perp}^2}{k^2(K^2+i\varepsilon k^0)}\Bigg(-8qk(3k_0^2-k^2)\nonumber
    \\&+K^2(12q^2-k^2+3k_0^2)\left(\ln\frac{k_0+k-2q+i\varepsilon}{k_0-k-2q+i\varepsilon}+\ln\frac{k_0-k+2q+i\varepsilon}{k_0+k+2q+i\varepsilon}\right)\nonumber
    \\&+12qk_0K^2\left(\ln\frac{k_0+k+i\varepsilon}{k_0-k+i\varepsilon}+\ln\frac{k_0-k+2q+i\varepsilon}{k_0+k+2q+i\varepsilon}-\ln\frac{k_0+k-2q+i\varepsilon}{k_0-k-2q+i\varepsilon}\right)\Bigg)-\text{adv.}\Bigg].
\end{align}
After doing a Caron-Huot shift, 
 i.e $k^z\rightarrow k^z+k^0$, we isolate the zero mode contribution  
by replacing $\int\frac{dk^0}{2\pi}$ with $iT\sum_{n}$ and then setting $\omega_{n}\rightarrow 0$. We are left with
\begin{align}
Z^{(c)}_{g\,L-T\,G\,n=0}&=-\frac{g^2C_{A}T}{16\pi^2}
    \int_k\int_{0}^{\infty}dq\frac{1}{k^3(k^z +i\varepsilon)^2}n_{B}(q)\frac{k_{\perp}^2}{k^2(-k^2)}\Bigg(-8qk(-k^2)\nonumber
    \\&+(-k^2)(12q^2-k^2)\left(\ln\frac{k-2q}{-k-2q}+\ln\frac{-k+2q}{k+2q}\right)\Bigg)\nonumber
    \\&=\frac{g^2C_{A}T}{16\pi^2}
    \int_k\int_{0}^{\infty}dq\frac{k_{\perp}^2}{k^7(k^z +i\varepsilon)^2}n_{B}(q)\Bigg(8qk^3-k^2(12q^2-k^2)\left(2\ln\bigg\vert\frac{2q-k}{k+2q}\bigg\vert\right)\Bigg).
\end{align}
In contrast to the previous piece, the integrand depends on the polar angle, $\theta$, which we deal with 
by making the substitution $x=\cos\theta$. The integral then yields
\begin{align}
Z^{(c)}_{g\,L-T\,G\,n=0}&=\frac{g^2C_{A}T}{16\pi^2}\left(\frac{\bar{\mu}^2 e^{\gamma_E}}{4\pi}\right)^\epsilon \left(\frac{2\pi^{\frac{d-1}{2}}}{\Gamma\left(\frac{d-1}{2}\right)(2\pi)^d}\right)\int_{0}^{\infty}dk\int_{-1}^{1}dx\int_0^{\infty}dq\,n_{B}(q)k^{d-8}
    \frac{(1-x^2)^{\frac{d-1}{2}}}{x^2}\nonumber
    \\&\Bigg(8qk^3-k^2(12q^2-k^2)\left(2\ln\bigg\vert\frac{2q-k}{k+2q}\bigg\vert\right)\Bigg)\nonumber
    \\&=\frac{g^2C_{A}T}{16\pi^2}\left(\frac{\bar{\mu}^2 e^{\gamma_E}}{4\pi}\right)^\epsilon \left(\frac{2\pi^{\frac{d-1}{2}}}{\Gamma\left(\frac{d-1}{2}\right)(2\pi)^d}\right)\int_{0}^{\infty}dk\int_{-1}^{1}dx\int_0^{\infty}dq T^{d-2}\frac{q^{d-7}}{e^q-1}k^{d-8}\frac{(1-x^2)^{\frac{d-1}{2}}}{x^2}\nonumber
    \\&\Bigg(8q^4k^3-2k^2q^2(12q^2-k^2q^2)\ln\bigg\vert\frac{2-k}{k+2}\bigg\vert\Bigg)\nonumber
    \\&=\frac{g^2C_{A}T^{d-1}}{16\pi^2}\left(\frac{\bar{\mu}^2 e^{\gamma_E}}{4\pi}\right)^\epsilon \left(\frac{2\pi^{\frac{d-1}{2}}}{\Gamma\left(\frac{d-1}{2}\right)(2\pi)^d}\right)\int_{0}^{\infty}dk\int_{-1}^{1}dx\int_0^{\infty}dq\frac{q^{d-3}}{e^q-1}k^{d-6}\frac{(1-x^2)^{\frac{d-1}{2}}}{x^2}\nonumber
    \\&\Bigg(8k-2(12-k^2)\ln\bigg\vert\frac{2-k}{k+2}\bigg\vert\Bigg).
\end{align}
In the first line, we have dropped the $i\varepsilon$ piece since the $x$ dependent part of the integrand is a $\beta$ function, the integral of which will converge for some $d$. Going from the second to the third line, we have rescaled $q\rightarrow qk$ and $k\rightarrow k/T$. 
Upon doing the integrations, we find
\begin{equation}
Z^{(c)}_{g\,L-T\,G\,n=0}=\frac{g^2C_{A}T^2}{32\pi^2}\left(-\frac{1}{\epsilon}+4-\ln\frac{\bar{\mu}^2 e^{\gamma_{E}}}{4T^2}-\psi\left(\frac{3}{2}\right)\right),\label{LmTzeromodediv}
\end{equation}
where $\psi(x)$ is the digamma function. Having extracted these IR divergences, we can now 
move on to identify their (UV) EQCD counterparts in the next section.

\section{Matching from EQCD}\label{sec:eqcd_details}

Before completing the matching of divergences between full QCD and EQCD, we take a moment to lay out some details of the 
perturbative EQCD (pEQCD) calculation 
of $Z_g$ from~\cite{Moore:2020wvy,Ghiglieri:2021bom}. From the EQCD Lagrangian, Eq.~\eqref{eq:eqcd_lag},
one derives the EQCD analog of the Wilson line, Eq.~\eqref{eq:4d:Wilson:line:expl}
\begin{equation}
    \tilde{U}_{R}(L;0)=\mathcal{P}\; \exp \left( i g_{\text{3d}} \int_0^{L} dz\left(A_z^a(z)+i\Phi^a(z)\right)  T^a_R \right).\label{eq:wline_eqcd} 
\end{equation}
In EQCD, since the field operators are classical, they commute, which allows us to write
\begin{equation}
    \frac{1}{d_{R}C_{R}}\Tr\langle\tilde{U}_{R}(-\infty;L)\mathcal{O}^{a}(L)T_{R}^{a}\tilde{U}_{R}(L;0)\mathcal{O}^{b}(0)T_{R}^{b}\tilde{U}_{R}(0;-\infty)\rangle
    =\frac{T_{R}}{d_{R}C_{R}}\langle\mathcal{O}^a(L)\tilde{U}^{ab}_{A}(L;0)\mathcal{O}^{b}(0)\rangle.
\end{equation}
Thus, the EQCD equivalent of $Z_g$ reads, given rotational invariance in the transverse plane 
\begin{equation}
    Z_{g}^{3\text{d}}=-\frac{2T_RT}{d_{R}C_{R}}\int_{0}^{L_{\text{min}}}dL\,L \Bigl\langle\Big(F^{a}_{xz}(L)+i(D_x\Phi(L))^a\Big)
    \tilde{U}^{ab}_{A}(L;0)\Big(F^{b}_{xz}(0)+i(D_x\Phi(0))^b\Big)\Bigr\rangle
\end{equation}
In terms of the lattice computation, it is 
more convenient to work with the quantities
\begin{align}
    \label{def:cond:EE}
      \langle EE\rangle &\equiv \frac{1}{2}
        \bigl\langle (D_x \Phi(L))^a\,\tilde{U}^{ab}_{A}(L;0)\,(D_x\Phi(0))^b \bigr\rangle
        \;,\\
    \label{def:cond:BB}
      \langle BB\rangle &\equiv \frac{1}{2} \bigl\langle F_{xz}^a(L)\,\tilde{U}^{ab}_{A}(L;0) \, F_{xz}^b(0) \bigr\rangle
        \;,\\
    \label{def:cond:EB}
      \langle EB\rangle &\equiv
          \frac{i}{2} \bigl\langle (D_x \Phi(L))^a\,\tilde{U}_{A}^{ab}(L;0) \, F_{xz}^b(0) \bigr\rangle
       \\&+ \frac{i}{2} \bigl\langle F_{xz}^a(L) \, \tilde{U}_{A}^{ab}(L;0) \, (D_x \Phi(0))^b \bigr\rangle
      \;,
\end{align}
which leaves us with 
\begin{align}
    Z_{g}^{3\text{d}}&=-\frac{4T_RT}{d_{R}C_{R}}\int_{0}^{L_{\text{min}}}dL\,L \Big(-\langle EE\rangle +\langle BB\rangle+i\langle EB\rangle  \Big)\nonumber
    \\&=\frac{4T_RT}{d_{R}C_{R}}\int_{0}^{L_{\text{min}}}dL\,L\langle FF\rangle \label{eq:zg3d},
\end{align}
where we have implicitly defined $\langle FF\rangle= \langle EE\rangle -\langle BB\rangle-i\langle EB\rangle$.
\begin{figure}[t]
    \centering
      \includegraphics[width=.5\textwidth]{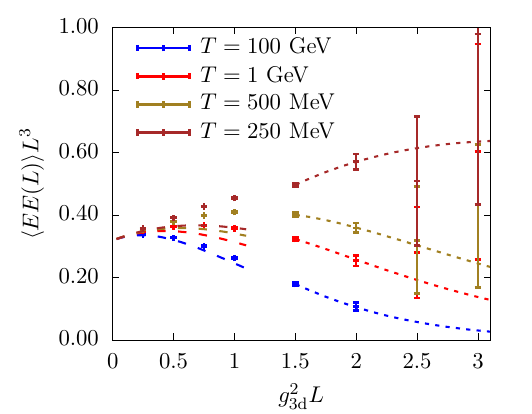}%
      \includegraphics[width=.5\textwidth]{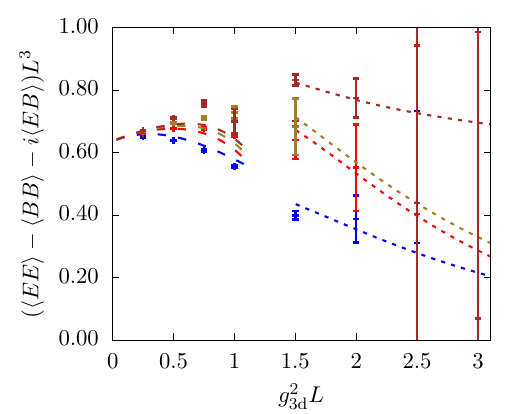}
    \caption{
          Continuum-extrapolated
        EQCD lattice data on $\langle EE\rangle$ (left) and
        $\langle FF\rangle$ (right). Short 
        dashes represent the long $L$-tail whole short dashes denote 
        the NLO pEQCD evaluation.  
        $L^3$ compensates the $c_0/L^3$ divergence. Figure 
        from~\cite{Ghiglieri:2021bom}.
        }
    \label{fig:comp_LO_NLO_latt}
\end{figure}

In Fig.~\ref{fig:comp_LO_NLO_latt}, the lattice EQCD evaluations of $\langle EE\rangle$ (left) and $\langle FF\rangle$ (right)
are shown. One cannot integrate up to arbitrarily large values of $L$ on the lattice, which is made apparent by the rather large 
error bars in that region. This IR region is addressed by a fitting ansatz, intended to reproduce the expected exponential falloff from electrostatic 
and magnetostatic screening. In the UV, where $m_{D}L\ll 1$, the lattice determination agrees well with the pEQCD determination, as expected. However, 
as one moves further into the deep UV, the lattice approach 
becomes impractical due to associated discretisation effects and one must rely on a perturbative 
evaluation. Such a perturbative evaluation does not, however, come from pEQCD; as an IR EFT of full QCD, 
EQCD exhibits unphysical behaviour for momenta $k\gg m_{D}$.

Because 
of the super-renormalisability of EQCD, in the UV, it is natural to expect that
\begin{equation}
    \langle FF\rangle_{m_{D}L\ll 1}=\frac{c_0}{L^3}+\frac{c_2g^2_{\text{3d}}}{L^2}+\frac{c_4g^4_{\text{3d}}}{L}+...\,.
\end{equation}
The first and second terms above are divergent when inserted into Eq.~\eqref{eq:zg3d}. The former would have 
come out of the calculation in Sec.~\ref{sec:nlo_asym} if we had employed a regularisation other than DR. It moreover 
cancels against a power law divergence coming from the LO calculation\footnote{Similarly, this divergence
would be present in, for instance, Eq.\eqref{eq:a} if we had used a regularisation other than DR. See Eq.~\eqref{eq:zg_corrections}.}. The latter was computed 
in \cite{Ghiglieri:2021bom} and, being a logarithmic divergence, does show up in DR. 

\begin{figure}[ht]
	\centering
    \includegraphics[width=0.7\textwidth]{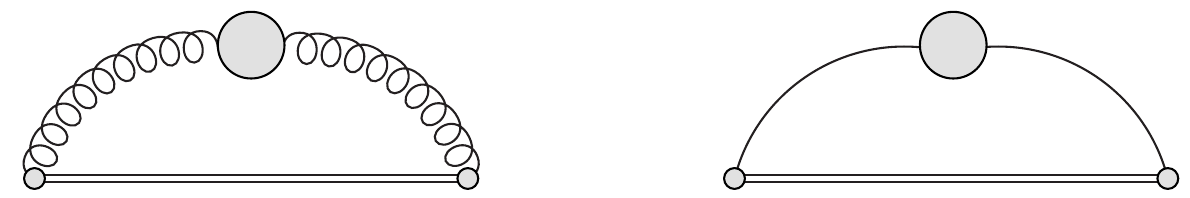}
    \caption{EQCD equivalent of diagram $(c)$. The labelling is identical 
    to that from Fig.~\ref{fig:amass_all_diagrams} except that here, curly lines 
    represent spatial gauge bosons, solid double lines are (adjoint) Wilson lines and solid lines are adjoint scalars.}
    \label{fig:c_eqcd}
\end{figure}
To arrive at a solution for $Z_{g}^{\text{3d}}$, accurate across 
all energy scales, one can imagine stitching the perturbative, intermediate and IR regions together
\begin{align}
    Z_{g}^{\text{3d}}\big\vert^\text{merge} =&\frac{4T}{d_{A}}\bigg\{ \int_0^{L_\text{min}} dL\,L\bigg[
    \langle FF\rangle_\text{NLO}-\frac{c_0}{L^3}-\frac{c_2 g_{\text{3d}}^2}{L^2}\bigg]+\int_{L_\text{min}}^{L_\text{max}} dL\,L\bigg[
    \langle FF\rangle_\text{lat}-\frac{c_0}{L^3}\bigg]\nonumber \\
    &
    \hspace{5mm}+\int_{L_\text{max}}^\infty dL\,L\bigg[
    \langle FF\rangle_\text{tail}-\frac{c_0}{L^3}\bigg]\bigg\}.
    \label{eq_subtr}
\end{align}
$L_{\text{min}}$ is taken to be $0.25/g_{3\text{d}}^2$, as can be seen in Fig.~\ref{fig:comp_LO_NLO_latt}.

Our goal in this section is essentially to 
check the validity of the NLO subtraction in the first term above. In particular, such a logarithmic UV divergence 
should cancel against the logarithmic IR divergence from the full QCD 
calculation that we computed in the previous section. Moreover, from the arguments of Sec.~\ref{sec:proof}, 
we know that this UV divergence must come from the EQCD analogue of diagram $(c)$, shown in Fig.~\ref{fig:c_eqcd}.
We first take the $\langle BB\rangle$ contribution from Appendix A.3 of \cite{Ghiglieri:2021bom}\footnote{
At the time of writing, an extra factor of $1/2$ is present in Eq.~\eqref{eq:eqcddivT} compared to its counterpart 
in \cite{Ghiglieri:2021bom}, due to a typo identified in the original work. For the same reason, an extra 
factor of $1/2$ is inserted in Eq.~\eqref{eq:eqcddivL}.} (Eq. A.14)
\begin{equation}
    (c)^{BB}_{\text{div}}=\frac{1}{4}\int_k\frac{e^{iLk_z}}{k^4}(k_{\perp}^2+2k_{z}^2)\delta^{ab}\Big(-\Pi^{T}_{A^a_iA^b_j\,\text{div}}(k)\Big).
    \label{eq:eqcddivT}
\end{equation}
where 
\begin{align}
    \Pi^{T}_{A^a_iA^b_j}(k)&=\delta^{ab}\frac{g^2_{\text{3d}}C_{A}}{16\pi}\left(-2m_{D}+\frac{14}{4}k+\frac{k^2+4m_{D}^2}{k}\arctan\frac{k}{2m_D}\right)\nonumber
    \\&\stackrel{k\gg m_{D}}{\approx}\delta^{ab}\frac{g^2_{3d}C_{A}}{16\pi}\left(-3\pi k\right)\equiv\Pi^{T}_{A^a_iA^b_j\,\text{div}}(k).\label{eq:bb_self}
\end{align}
We have expanded for $k\gg m_{D}$ in order to probe its UV, i.e, the region where EQCD is no longer applicable. The 
loop integration, performed to arrive at Eq.~\eqref{eq:bb_self} was 
done in $3$ dimensions, whereas the $k$ integration in Eq.~\eqref{eq:eqcddivT} will be done in $d$ 
dimensions -- this is consistent with the integration procedure followed on the full QCD side. The second term in Eq.~\eqref{eq:bb_self}, 
proportional to $2k_z^2$ should correspond to the $2k^{-2}$ term computed in Sec.~\ref{sec:zero_kminus}. Indeed, looking 
to Eq.~\eqref{eq:zg3d}, we find 
\begin{align}
    Z^{3d\,(c)}_{g\,k^{-2}\,G\,\text{div}}&=-\frac{4T}{d_{A}}\int_{0}^{L_{\text{min}}}dL\,L\,(c)^{BB}_{k^{-2}\,\text{div}}
    \nonumber \\
    &=\frac{3g^2_{3d}C_{A}T}{32\pi^2}\Big(\frac{1}{\epsilon}-2+3\gamma_E+2\ln{L_{\text{min}}\bar{\mu}}+\psi\left(\frac{1}{2}\right)\Big).
    \label{EQCDcqmsquare}
\end{align}
Given that $g^2_{3d}\sim g^2T$, it is clear that the $1/\epsilon$ term cancels the one from Eq.~\eqref{IRpoleqmsquare}.

As for the $L-T$ piece, we need the contribution from the $\langle EE \rangle$ correlator. Again, we 
extract this from Eq. A.8 of~\cite{Ghiglieri:2021bom} and expand towards the UV to arrive at 
\begin{align}
    \Pi_{\Phi^a\Phi^b}(k)&=\delta^{ab}\frac{g^2_{3d}C_{A}m_{D}}{16\pi}\left(-4-\frac{6C_{F}-C_{A}}{6C_{A}}\frac{\lambda_{E}}{g^2_{3d}}-8\frac{k^2-m_{D}^2}{k m_{D}}\arctan\frac{k}{m_D}\right)\nonumber
    \\&\stackrel{k\gg m_{D}}{\approx}\delta^{ab}\frac{g^2_{3d}C_{A}}{16\pi}\left(-4\pi k\right)\equiv\Pi_{\Phi^a\Phi^b\,\text{div}}(k).
\end{align}
From Eq. $A.13$ of \cite{Ghiglieri:2021bom}, we then write
\begin{equation}
    (c)^{EE}_{\text{div}}=\frac{1}{4}\int_k\frac{e^{iL k_z}}{k^4}k_{\perp}^2\delta^{ab}\Big(-\Pi^{T}_{\Phi^a\Phi^b\,\text{div}}(k)\Big).
    \label{eq:eqcddivL}
\end{equation}
In order to get the full $\Pi_{L}-\Pi_{T}$ EQCD contribution, we need to add to this the first term in Eq.~\eqref{eq:eqcddivT}, 
proportional to $k_{\perp}^2$. Inserting these into Eq.~\eqref{eq:zg3d}, 
\begin{align}
Z^{3d\,(c)}_{g\,L-T\,G\,\text{div}}&=-\frac{4T}{d_{A}}\int_{0}^{L_{\text{min}}}dL\,L\,\left((c)^{BB}_{k_{\perp}^{2}\,\text{div}}-(c)^{EE}_{\text{div}}\right)
\nonumber\\
&=\frac{g^2_{3d}C_{A}T}{32\pi^2}\bigg(\frac{1}{\epsilon}-3+3\gamma_E+2\ln{L_{\text{min}}\bar{\mu}}+\psi\left(-\frac{1}{2}\right)\bigg),
\label{EQCDcLTdiv}
\end{align}
which, reassuringly contains a $1/\epsilon$ pole that cancels exactly against the one from Eq.~\eqref{LmTzeromodediv}. Having 
confirmed the validity of the subtraction in Eq.~\eqref{eq_subtr}, we proceed to write 
\begin{equation}
    Z_g^\mathrm{match} =\frac{ g^2 C_{A} T^2}{4 \pi ^2}\bigg[ \ln (2L_\text{min} T)+ \gamma_E -\frac{3}{2}\bigg]\,,\qquad
    Z_g^\text{non-pert class}=Z_{g}^{\text{3d}}\big\vert^\text{merge}+Z_g^\text{match},
    \label{totmatch}
\end{equation}
where, from Eqs.~\eqref{IRpoleqmsquare}, ~\eqref{LmTzeromodediv}, ~\eqref{EQCDcqmsquare} and ~\eqref{EQCDcLTdiv}
\begin{align}
    Z_g^\text{match} &=Z^{(c)}_{g\,k^{-2}\,G\,n=0}+Z^{\text{3d}\,(c)}_{g\,k^{-2}\,G\,\text{div}}+ Z^{(c)}_{g\,L-T\,G\,n=0}+Z^{\text{3d}\,(c)}_{g\,L-T\,G\,\text{div}}\nonumber\\
    &=\frac{ g^2 C_A T^2}{8 \pi ^2}\bigg[2 \ln (2L_{\text{min}} T)+2 \gamma_E -3\bigg].
    \label{sumup}
\end{align}
Note that in computing $Z_g^\text{non-pert class}$, the $\ln L_{\text{min}}$ above cancels against the $\int dL \frac{c_2 g_{\text{3d}}^2}{L}$ term  in Eq.~\eqref{eq_subtr}.
For reasonable values of $g_{\text{3d}}$, extracted from \cite{Moore:2021jwe}, we find that
$Z_g^\text{non-pert class}$ provides a negative contribution to $Z_g$. However, we should 
not yet make any conclusive remarks about the convergence of $Z_g$ until all of its $\mathcal{O}(g^2)$
corrections have been obtained. We present the first steps 
towards a full determination of the $\mathcal{O}(g^2)$ corrections in the next section.

\section{Full Evaluation of Diagram \texorpdfstring{$(c)$}{TEXT}}\label{sec:full}

We conclude this chapter by completing the evaluation of diagram $(c)$ -- the self-energy diagram. The computation
is split as follows: the finite-temperature fermionic and gluonic self-energies are given in Secs.\ref{sec:ferm},
~\ref{sec:gluon} respectively and the vacuum fermionic and gluonic self-energies are presented together 
in Sec.~\ref{vac:cont}. We 
begin by computing the finite-temperature fermionic contribution, which, naively should be free of any divergences. We discover 
that such an expectation is not met and comment on sources of these 
divergences where appropriate. As was done in Sec.~\ref{sec:full_qcd}, we further 
divide up the finite-temperature self-energies into the $k^{-2}$ and $\Pi_{L}-\Pi_{T}$ pieces with the $Q$
integration performed in $4$ dimensions and the $K$ integration in $d$ dimensions. Many 
of the technical details associated with the integration are left to App.~\ref{sec:eval}.

\section{Finite-Temperature Fermionic Contribution}\label{sec:ferm}
We pick up directly from Eq.~\eqref{zcexpr}, with the finite-temperature gluon and fermion 
self-energies given in App.~\ref{app:one_loop}. The contribution from the $k^{-2}$-proportional part is 
computed in App.~\ref{sec:fermion_km} and reads
\begin{align}
  Z_{g\, k^{-2}\,F}^{(c)}=& \frac{g^2T_F N_FT^2}{96\pi ^2 }\bigg\{
    -3 \left(4 \ln (2) \ln \left(\frac{2}{A^8}\right)+2\ln (2)\right)+\frac{\pi ^2}{6}-8 \gamma_E  \ln 2\nonumber \\
   &+4.4713 -\frac{96\ln 2}
   {\pi^2}+3.7933 + 1.70118
  \bigg\}.\label{totqmsquarefermion}
\end{align}
From the absence of any $\epsilon$-dependent terms, it is evidently finite. 
We have cross-checked this result by numerically performing the sum over the 
Matusbara modes associated with $K$.

However, we find that the $\text{L}-\text{T}$ contribution, computed in App.~\ref{sec:fermion_lmt} is 
\begin{align}
    &Z_{g\,F\,\text{ L-T }} ^{(c)}=\frac{g^2 T_F N_F T^2}{48\pi^2}
    \bigg\{\frac{2}{\epsilon^2}
    +\frac{2\ln  \frac{\bar\mu \exp(\gamma_E+1/2) }{4\pi T}+2(\ln(2)-1) +2\ln\frac{\bar\mu A^{12}e^{-1/2}}{16 \pi T} }{\epsilon}\nonumber \\
   &+ 2 \ln^2\frac{\bar\mu \exp(\gamma_E+1/2) }{4\pi T}-4 \gamma _1+\frac{\pi ^2}{4}
    -2 \gamma_E^2+\frac32
    \nonumber \\
    &+2(\ln(2)-1)\bigg[
   2\ln\frac{\bar\mu\exp(\gamma_E+1/2)}{4\pi T}\bigg]\nonumber \\
   &
    +2 \ln^2\frac{\bar\mu A^{12}e^{-1/2}}{16 \pi T}\nonumber\\
    &+48\ln (A) (\gamma_{E} +\ln (2 \pi )-6\ln(A))
    +\frac{12 \zeta
   ''(2)}{\pi ^2}+\frac{1}{2}
   \text{Li}_2\left(\frac{2}{3}\right)+\frac14\text{Li}_2\left(\frac{3}{4}\right)
   \nonumber\\
    &-\frac{1}{2}-\frac{11}{2} \ln
   ^2(2)-2 \ln (\pi ) \ln (4 \pi )-2 \gamma_E  \left(\gamma_E +\ln \left(4 \pi
   ^2\right)\right)-\frac{1}{2}\ln (3) \coth ^{-1}(7)
    \nonumber \\
    &
    \underbrace{-0.438129}_{k<2q}+\underbrace{1.97382}_{k>2q}
    \bigg\}.
    \label{finalLTfermion}
\end{align}
In contrast to the $k^{-2}$ piece, this contribution is divergent -- the 
$1/\epsilon^2$ pole signals the presence of a double logarithmic divergence. We can understand that
one of these divergences coming from the region where $k>2q$, implying it is of a UV nature. From only the 
expression given above, the origin of
the other logarithmic divergence is not at all apparent. However, by looking to, for instance, 
Eq.~\eqref{zcexprlogssym}, we note the potential for a logarithmic divergence in $k$ when $k^0-k\ll k^0+k$. This 
is precisely the region of phase space that is sensitive to \emph{collinear physics}, where $K^{\mu}\sim T$ with $K^2$ small. We 
thus identify the second logarithmic as a collinear one. We further take this opportunity to hammer home the point that these divergences 
\emph{are not} cancelled by opposing ones coming from the soft 
sector, as EQCD is a theory of gluons only.

\subsection{Finite-Temperature Gluonic Contribution}\label{sec:gluon}

We start with the finite-temperature gluonic $k^{-2}$-proportional part, with 
the details left to App.~\ref{sec:gluon_km}. In this case, 
we must be mindful not to double count the zero mode contribution already 
obtained in Eq.~\eqref{IRpoleqmsquare}. In doing so, we identify 
a finite contribution that comes from the non-zero Matsubara modes 
\begin{equation}
    Z^{(c)}_{g\,k^{-2}\,G\,n\neq 0\,\text{finite}}=-\frac{g^2C_{A}T^2}{32\pi^2}\times0.00380658,
    \label{eq:non_zero}
\end{equation}
which we have obtained numerically. In addition, we isolate a divergent contribution from the 
non-zero Matsubara modes, given by 
\begin{align}
    Z^{(c)}_{g\,k^{-2}\,G\,n\neq 0\,\text{div}}&=
    \frac{g^2 C_A T^2}{48 \pi ^2 } \left(\frac{1}{\epsilon}
    -2 \ln\frac{4\pi  T}{A^{12}\bar \mu} 
    -2.44147 -2.44091\right).
    \label{finaldivpartgqmsquare}
\end{align}
Unlike the finite-temperature fermionic $k^{-2}$ piece, the gluonic part is divergent -- one from the zero-mode 
and another from the non-zero modes. While we have already discussed the former at length, the latter
is of UV nature and we comment further on how it is dealt with at the end of Sec.~\ref{vac:cont}.

A similar computation, contained in App.~\ref{sec:gluon_lmt} shows that 
the finite-temperature gluonic $\text{L}-\text{T}$ reads 
\begin{align}
    &Z_{g\,\text{G L-T }} ^{(c)}=\frac{g^2 C_A T^2}{48\pi^2}
    \bigg\{\frac{2}{\epsilon^2}
    +\frac{2\ln  \frac{\bar\mu \exp(\gamma_E+1/2) }{4\pi T}+2(\ln(2)-1) +2\ln\frac{\bar\mu A^{12}e^{-1/2}}{8 \pi T} }{\epsilon}\nonumber \\
   &+ 2 \ln^2\frac{\bar\mu \exp(\gamma_E+1/2) }{4\pi T}-4 \gamma _1+\frac{\pi ^2}{4}
    -2 \gamma_E^2+\frac32
    \nonumber \\
    &+2(\ln(2)-1)\bigg[
   2\ln\frac{\bar\mu\exp(\gamma_E+1/2)}{4\pi T}\bigg]\nonumber \\
   &
    +2 \ln^2\frac{\bar\mu A^{12}e^{-1/2}}{8 \pi T}\nonumber\\
    &+48\ln (A) (\gamma_{E} +\ln (2 \pi )-6\ln(A))
    +\frac{12 \zeta
   ''(2)}{\pi ^2}+\frac12
   \text{Li}_2\left(\frac{2}{3}\right)+\frac14\text{Li}_2\left(\frac{3}{4}\right)
   \nonumber\\
    &-\frac12-\frac{3}{2} \ln
   ^2(2)-2 \ln (\pi ) \ln (4 \pi )-2 \gamma_E  \left(\gamma_E +\ln \left(4 \pi
   ^2\right)\right)-\frac12\ln (3) \coth ^{-1}(7)
    \nonumber \\
    &
     +\underbrace{0.145277}_{k<2q}\underbrace{-5.13832}_{k>2q}
    \bigg\}.
    \label{finalLTgluon}
\end{align}
As was the case in Eq.~\eqref{finalLTfermion}, there is a $1/\epsilon^2$ pole; by adding 
Eq.~\eqref{finalLTfermion} and Eq.~\eqref{finalLTgluon} we find the divergent terms 
are proportional to $m_{D}^2$, signifying that the divergences present in both 
contributions have the same origin.

\subsection{Vacuum Contribution}\label{vac:cont}

Another piece, which was not needed in the matching calculation 
is given by the vacuum self-energy inserted between the two thermal propagators. The 
vacuum self-energy (composed of both the fermionic and gluonic pieces) is 
given in Eq.~\eqref{MSbarvac}. Upon plugging it into Eq.~\eqref{zcexpr}, we find
\begin{equation}
    Z_{g\text{ vac}}^{(c)}=i\frac{\alpha_s}{2\pi}\int_{K}[1+n_\mathrm{B}(k^0)]\bigg[
    \frac{-\ln\frac{k^2-(k^0+i\varepsilon)^2}{\bar\mu^2}\bigg(\frac53 C_A-\frac43 T_FN_F\bigg)
    +\frac{31}{9}C_A-\frac{20}{9}T_FN_F}{(K^2+i\varepsilon k^0)}
    -\text{adv}
    \bigg].
    \label{zcexprvac}
\end{equation}
To proceed, we rewrite the frequency integration as a Matsubara sum, i.e
\begin{equation}
    Z_{g\text{ vac}}^{(c)}=T\sum_n\frac{\alpha_s}{2\pi}\int_{k}
    \frac{-\ln\frac{k^2+\omega_n^2}{\bar\mu^2}\bigg(\frac53 C_A-\frac43 T_FN_F\bigg)
    +\frac{31}{9}C_A-\frac{20}{9}T_FN_F}{k^2+\omega_n^2}
    .
    \label{zcexprvacsum}
\end{equation}
This yields
\begin{align}
      Z_{g\text{ vac}}^{(c)}=&\frac{ g^2C_A T^2}{864 \pi ^2}  \bigg[30 \ln\frac{A^{12}\bar\mu}{4 \pi  T}+31
      \bigg]
  -\frac{g^2 T_F N_F T^2}{216 \pi ^2} \bigg[
   6 \ln\frac{A^{12}\bar\mu}{4\pi T}
   +5 
   \bigg]\nonumber\\
   =&\frac{g^2T^2}{48\pi^2}\bigg[\bigg(\frac53 C_A-\frac43 T_FN_F\bigg)\ln\frac{A^{12}\bar\mu}{4\pi T}+\frac{31}{18}C_A-\frac{10}{9}T_FN_F\bigg].
\end{align}
It is finite as the $\overline{MS}$ subtraction has already been performed in Eq.~\eqref{zcexprvac}. If we 
instead keep the $1/\epsilon$ poles in Eq.~\eqref{vacpol}, we find 
\begin{equation}
    Z_{g\text{ vac pole}}^{(c)}=T\sum_n\frac{\alpha_s}{2\pi}\int_{k}
    \frac{\frac{1}{\epsilon}\bigg(\frac53 C_A-\frac43 T_FN_F\bigg)}{k^2+\omega_n^2}
    =\frac{g^2T^2}{96\pi^2\epsilon}\bigg(\frac53 C_A-\frac43 T_FN_F\bigg).
    \label{zcexprvacsumpole}
\end{equation}
Upon adding this to Eq.~\eqref{finaldivpartgqmsquare}, we see that they 
match the $\beta$ function $C_{A},\,T_{F}N_{F}$ coefficients in \eqref{eq:beta_one}. We therefore 
understand that these UV divergences are taken care of by the usual UV zero-temperature renormalisation procedure.

\section{Recap and Discussion of Divergences}\label{sec:discussion}
With that, we conclude our evaluation of diagram $(c)$ and therefore the 
$\mathcal{O}(g^2)$ corrections to $Z_g$. Let us briefly lay out 
our findings:
\begin{itemize}
    \item In Sec.~\ref{sec:full_qcd}, we extracted $Z_g$'s (IR) zero-mode divergences. These divergences 
        are expected and are in fact needed as they cancel against UV divergences coming from EQCD at NLO, as we showed 
        in Sec.~\ref{sec:eqcd_details}. This is one of the main results of the chapter.
    \item We also identified what we call UV, vacuum divergences coming from Eqs.~\eqref{finaldivpartgqmsquare}, \eqref{zcexprvacsumpole}. 
    We have confirmed that these are subtracted off in the zero-temperature renormalisation.
    \item Lastly, we have found double-logarithmic UV, collinear divergences, contributing to both the gluonic and fermionic
        $\text{L}-\text{T}$ pieces in Eqs.~\eqref{finalLTfermion} and \eqref{finalLTgluon}. In contrast 
        to what has been discussed above, these divergences signify that our treatment here is
        incomplete. As they come from a region of phase space where collinear physics dominates, it is likely that some form of
        LPM resummation will be needed to cure these divergences. Further investigation on this matter is underway.
\end{itemize}

As a closing remark, we emphasise that, without a full determination of the 
$\mathcal{O}(g^2)$ corrections to $Z_g$, the arbitrariness of how 
the classical corrections are separated from the quantum, $\mathcal{O}(g^2)$ corrections 
prevents us from making concrete inferences regarding the numerical 
size of the non-perturbative evaluation and furthermore,
the convergence of the perturbative expansion for $Z_g$.
\newpage
\chapter{Conclusion and Outlook}\label{ch:conc}

A common theme that has permeated throughout this thesis is the computation of higher-order 
corrections to $\hat{q}$ and $m_{\infty}$, both of which serve as central inputs to 
jet energy loss calculations. In particular, the advancements contained in this work are tied to the 
exploration of quantum corrections at $\mathcal{O}(g^2)$ to these 
quantities and how they are related to classical corrections, present at $\mathcal{O}(g)$.

In Ch.~\ref{chap:qhat_chap}, we calculated double-logarithmic corrections to $\hat{q}$ in the 
setting of a weakly coupled QGP. We investigated the consequences of taking the thermal scale seriously in this context 
and provided clarification, previously lacking in the literature, 
on how the logarithmically enhanced region of phase space 
is connected to the classical one. Additionally, we were able to show how the inclusion of semi-collinear 
processes allows one to go beyond the instantaneous approximation made in the original works \cite{Liou:2013qya,Blaizot:2013vha}.

Our result for $\delta \hat{q}(\mu_{\perp})$ at leading double logarithmic accuracy is given by Eq.~\eqref{eq:ss_ho_final} and was 
obtained in the harmonic oscillator approximation. The necessity of implementing this approximation comes 
from the need to neglect a neighbouring region of phase space, where both single and multiple scattering can occur; a concrete 
understanding of this region is lacking in the sense that we do not know the form of the transverse scattering kernel there. 
In the future, we hope that this issue can be addressed by solving Eq.~\eqref{LMWCshiftfinal}. While solving this equation 
exactly would likely prove to be a formidable task, a first step could be to solve it 
approximately by using the improved opacity expansion \cite{Mehtar-Tani:2019tvy,Mehtar-Tani:2019ygg,Barata:2020rdn,Barata:2021wuf}.
 This could also serve to provide a more satisfying cutoff for the single-scattering 
 regime in the direction of multiple scattering -- boundary $(b)$.

We are eager to perform a similar computation and analysis of the single logarithmic corrections 
to $\hat{q}$. Indeed, next to leading logarithm corrections are often 
needed to fairly assess the quantitative impact of their leading logarithm counterparts -- see \cite{Arnold:2008zu}
for the collinear splitting rate and \cite{Arnold:2003zc} for QGP transport coefficients. Nevertheless, it does 
not make sense to move forward in this direction until all of the double logarithmic corrections have been computed.

We could also imagine applying what we have learned 
 to some of the recent developments in jet energy loss theory, mentioned in Sec.~\ref{sec:recent}. In more detail,
it would be interesting to investigate the impact of the thermal scale in the context of double gluon emission \cite{Arnold:2015qya,Arnold:2016kek,Arnold:2016mth,Arnold:2018yjd,Arnold:2018fjr,Arnold:2019qqc,
Arnold:2020uzm,Arnold:2021pin,Arnold:2022epx,Arnold:2022fku,Arnold:2022mby,Arnold:2023qwi}, given that the 
same \emph{universal} double logarithms also arise in that context.

Another fruitful direction 
could be to study the resummation of these logarithmic corrections, similarly to what was done in \cite{Iancu:2014kga,Iancu:2014sha,Caucal:2021lgf,Caucal:2022fhc}.
These authors have performed a resummation according to original triangle shown in Fig.~\ref{fig:bdimtriangle}, 
evolving $\hat{q}$ from some initial timescale $\tau_0$ to longer timescales 
by resumming longer-lived quantum fluctuations. Our evaluation of the $k^+\sim T$ and smaller $\tau$ regions could 
then be incorporated by evolving $\hat{q}$ through these regions, with the result then being 
passed on as an initial condition to the resummation equations given in \cite{Caucal:2021lgf,Caucal:2022fhc}.

In Ch.~\ref{ch:asym_mass}, we set out to complete the evaluation of $\mathcal{O}(g^2)$ corrections 
to $Z_g$ by computing the contribution from the thermal scale. We were able to show that the 
unphysical UV EQCD divergences, identified in \cite{Ghiglieri:2021bom}, cancel against IR divergences on the full QCD side,
thereby providing a finite result to the non-perturbative evaluation of the classical corrections, also 
calculated in \cite{Ghiglieri:2021bom}. When computing the rest of the corrections at $\mathcal{O}(g^2)$, 
we encountered two other kinds of divergences. The first kind comes from both the 
fermion and gluon vacuum contribution as well as the $k^{-2}$ finite-temperature gluonic piece. These 
are UV, vacuum divergences and are appropriately dealt with through the renormalisation of the 
zero-temperature theory.

The set of divergences, present in both the finite-temperature
gluonic and fermionic ``$\text{L}-\text{T}$'' pieces are more troublesome. They have a collinear origin, i.e, 
they come from a region of phase space where $K^{\mu}\sim T$ but with $K^2$ much smaller. The next course 
of action in this project is to correctly treat the region from which they emerge, 
which will likely involve having to take LPM resummation into account.
 Following this, we plan to compute the $\mathcal{O}(g^2)$ 
corrections to $Z_f$, which will complete the $\mathcal{O}(g^2)$ determination of the asymptotic masses. Since 
$Z_f$ is not as sensitive to classical corrections, this should be, in principle, a more straightforward task.

With all of the $\mathcal{O}(g^2)$ to $\hat{q}$ and $m_{\infty}$ in hand, a more long-term 
goal would be to assess the quantitative impact of these corrections on the in-medium splitting rate, as was 
done for the classical corrections to $\hat{q}$ in \cite{Moore:2021jwe,Schlichting:2021idr}. This would 
then pave the way for more phenomenologically-based studies \cite{Yazdi:2022bru,Shi:2022rja}, which 
we originally motivated the need for in the Introduction.

\newpage

\appendix
\chapter{Conventions}\label{sec:conventions}
We take here a moment to specify our notation, which is intended to be particularly suited for studying jet energy loss. We adopt the use 
of natural units ($\hbar=c=k_{B}=1$) and  work in the mostly 
minus metric, $(+,-,-,-)$.

We write four-momentums with capital letters, $Q=(p^0,\vec{\mathbf{q}})$, the modulus of three-momentums with lower case letters, $q$ and the modulus 
of transverse vectors as $q_{\perp}$. Position vectors are denoted similarly.
It will often be convenient to use light-cone coordinates such that $Q=(q^+,q^-,\mathbf{q})$ with the following convention
\begin{align}
   q^{+}&\equiv\frac{q^{0}+q^{z}}{2}=\bar{v}_{\mu}q^{\mu},\nonumber
   \\q^{-}&\equiv q^{0}-q^{z}=v_{\mu}q^{\mu},\nonumber
   \\q\cdot k&=q^{+}k^{-}+q^{-}k^{+}-\mathbf{q}\cdot\mathbf{k},\nonumber
\end{align}
where we have defined the two light-like reference vectors as 
\begin{align}
   \bar{v}^{\mu}&\equiv\frac{1}{2}(1,0,0,-1),\nonumber
   \\v^{\mu}&\equiv(1,0,0,1).\nonumber
\end{align}
We shall often deal with scalings where $q^-\ll q^+$, which then implies
$q^0\approx q^z\approx q^+$. For the sake of brevity, we introduce the following shorthand for integration over position and momenta in various dimensions
\begin{align}
    \int_{Q}&\equiv\int\frac{d^4Q}{(2\pi)^4}\nonumber
    \\\int_{q}&\equiv\int\frac{d^3q}{(2\pi)^3}\nonumber
   \\\int_{\mathbf{q}}&\equiv\int\frac{d^2q_{\perp}}{(2\pi)^2}\nonumber
   \\\int_{\mathbf{x}}&\equiv\int d^2x_{\perp}\nonumber
\end{align}
Our convention for the covariant derivative is consistent with that of Schwartz \cite{Schwartz:2014sze}
\begin{equation}
   D_{\mu}=\partial_{\mu}-igA_{\mu}\label{eq:cov_d},
\end{equation}
which in turn fixes the three-gluon vertex to be 
\begin{equation}\label{eq:three_gluon}
   \vcenter{\hbox{\includegraphics[width=3.5cm,height=3cm]{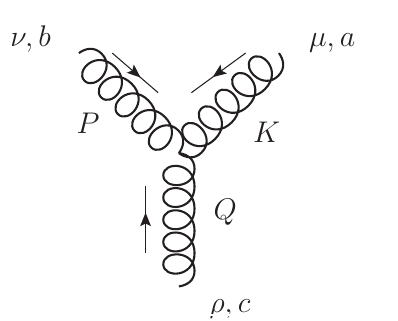}}}
   \qquad\qquad
   \begin{aligned}
      =gf^{abc}V^{\mu\nu\rho},
   \end{aligned}
\end{equation}
where 
\begin{equation}
   V^{\mu\nu\rho}=\eta^{\mu\nu}(K-P)^{\rho}+\eta^{\nu\rho}(P-Q)^{\mu}+\eta^{\rho\mu}(Q-K)^{\nu}.
\end{equation}
Finallly, we mention our convention for dimensional regularisation
\begin{equation}
   \int_q=\Big(\frac{\bar{\mu}^2e^{\gamma_{E}}}{4\pi}\Big)^{\epsilon}\int\frac{d^dq}{(2\pi)^d},\label{eq:dr_def}
\end{equation}
where we will often regularise integrals in $d=3-2\epsilon$ dimensions.
\chapter{Wilson Line Details}\label{app:wloop}
A Wilson line is a bi-local field that in general 
depends on two spacetime points
\begin{equation}
    \mathcal{V}(Y,Z)=\mathcal{P}\exp\Big(ig\int_{s_Y}^{s_Z}ds\frac{dX^{\mu}}{ds}A_{\mu}(X(s))\Big),
\end{equation}
where $A_{\mu}=A_{\mu}^aT^a$ is a (non-abelian) Lie-algebra-valued gauge (gluon) field and $\mathcal{P}$ is the path-ordering operator. $\mathcal{P}$
ensures that fields evaluated earlier on the path from $Y$ to $Z$ are placed to the right with respect to fields that are evaluated 
later on the path\footnote{Beware that in the more standard definition found in the literature, 
$\mathcal{P}$ orders colour matrices but not the fields, which are instead time-ordered.}. As well as transforming covariantly under gauge transformation, they naturally 
incorporate the resummation of gluons. We can expand the expression above by first writing 
\begin{equation}
    \mathbf{\mathcal{F}}(s)=\frac{dX^{\mu}}{ds}A_{\mu},
\end{equation}
where the bold font is meant to emphasis that these objects are matrices. Then, incorporating the usual definition 
of the matrix exponential expansion\footnote{The expression Eq.~\eqref{eq:v_exp} can be adapted to the case of a conjugate 
Wilson line by sending $i\to -i$ and enforcing instead an anti-path-ordering prescription.}
\begin{equation}
    \mathcal{V}(Y,Z)=\sum_{n=0}^{\infty}\left(\frac{ig}{n!}\right)^n\int_{s_Y}^{s_Z}ds_1\int_{s_Y}^{s_Z}ds_2...\int_{s_Y}^{s_Z}ds_n
    \mathcal{P}\Big[\mathbf{\mathcal{F}}(s_1)\mathbf{\mathcal{F}}(s_2)...\mathbf{\mathcal{F}}(s_n)\Big]\label{eq:v_exp}
\end{equation}
The above integration is done over an $n$-dimensional hypercube but one can divide this region into $n!$ sub-regions. To 
give an illustrative example, consider $n=2,\,s_Y=0,\,s_Z=1$
\begin{equation}
    \int_0^1ds_1\int_0^1ds_2\mathcal{P}\Big[\mathbf{\mathcal{F}}(s_1)\mathbf{\mathcal{F}}(s_2)\Big]
    =\int_0^1 ds_1 \int_0^{s_1} ds_2\mathbf{\mathcal{F}}(s_1)\mathbf{\mathcal{F}}(s_2)+\int_0^1 ds_2\int_0^{s_2}ds_1\mathbf{\mathcal{F}}(s_2)\mathbf{\mathcal{F}}(s_1).
\end{equation}
At this point, $s_1,\,s_2$ are just dummy variables, so we can write 
\begin{equation}
    \int_0^1ds_1\int_0^1ds_2\mathcal{P}\Big[\mathbf{\mathcal{F}}(s_1)\mathbf{\mathcal{F}}(s_2)\Big]
    =2\int_0^1 ds_1\int_0^{s_1}ds_2\mathbf{\mathcal{F}}(s_1)\mathbf{\mathcal{F}}(s_2),
\end{equation}
where such a procedure can be suitably generalised to the the $n$-dimensional case. In this way, one 
can eliminate the path-ordering rewrite Eq.~\eqref{eq:v_exp} as 
\begin{equation}
    \mathcal{V}(Y,Z)=\sum_{n=0}^{\infty}\left(ig\right)^n\int_{s_Y}^{s_Z}ds_1\int_{s_Y}^{s_1}ds_2...\int_{s_Y}^{s_{n-1}}ds_n
    \mathbf{\mathcal{F}}(s_1)\mathbf{\mathcal{F}}(s_2)...\mathbf{\mathcal{F}}(s_n).
\end{equation}
In practice, it is this formula that we make use of to calculate Wilson lines in this thesis.

Furthermore, let $s_Y=s_0,\,s_Z=s$ so that $Y=X(s_0)\,Z=X(s)$. Then, by taking a derivative with respect to $s$
\begin{equation}
    \frac{dx^\mu}{ds}\mathcal{V}(x(s),x(s_0))=ig\frac{dx^\mu}{ds}A_{\mu}(x(s))\mathcal{V}(x(s),x(s_0)),
\end{equation}
we can derive an equation of motion for the Wilson line, which reads
\begin{equation}
    \frac{dx^\mu}{ds}\Big(\frac{\partial}{\partial x^{\mu}}-ig A_{\mu}(x(s))\Big)\mathcal{V}(s,s_0)=0=\frac{dx^\mu}{ds}D_{\mu}\mathcal{V}(s,s_0),\label{eq:wloop_eom}
\end{equation}
where we have used the definition of the covariant derivative, Eq.~\eqref{eq:cov_d}.
\chapter{\texorpdfstring{$\hat{q}(\mu_{\perp})$}{TEXT} Computation}\label{ch:qhat_app}

\section{Side rails in Coulomb Gauge}\label{app:siderails}
\begin{figure}[h]
	\centering
	\includegraphics[width=\textwidth]{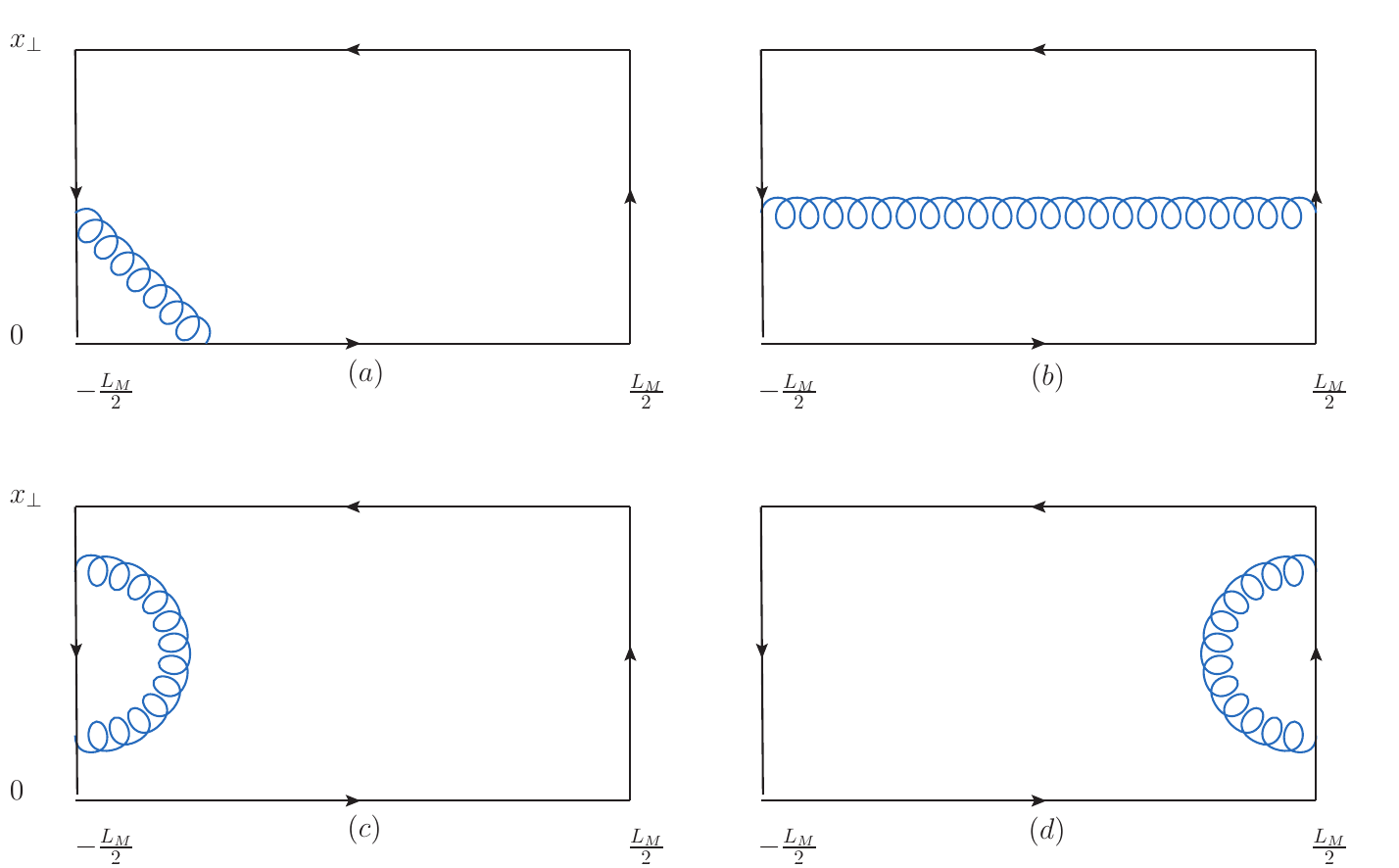}
	\caption{Diagrams obtained from expanding the Wilson loop to the first non-trivial order, where at 
    least one of the gluon legs are attached to a side rail. In the main text, we argue that none of these 
    diagrams are relevant for the computations performed in this thesis.}
    \label{fig:side_rails}
\end{figure}
Back in Sec.~\ref{sec:scatt_lo_soft}, we showed how to compute the leading order soft contribution to $\hat{q}(\mu_{\perp})$ using 
the Wilson loop defined in \cite{Benzke:2012sz}. While carrying out that computation, we only needed to 
consider the diagrams in Fig.~\ref{fig:leading_order} where gluons are attached to the horizontal Wilson lines.
In this Appendix, we argue why other diagrams, some of which are shown in Fig.~\ref{fig:side_rails} are not relevant 
for such that calculation, as well as the subsequent ones carried out in Ch.~\ref{chap:qhat_chap}.

We start with the top left diagram from Fig.~\ref{fig:side_rails}. Using Eqs.~\eqref{eq:wline_def}, \eqref{eq:side_rails}, we have
\begin{align}
(a)&=-g^2C_{R}\int_0^1 ds\int_{-\frac{L_M}{2}}^{\frac{L_M}{2}} dx^+x_{\perp}\langle G_{>}^{\perp -}(-\frac{L_{M}}{2}-x^+,0,y_{\perp}s)\rangle\nonumber
\\&=-g^2C_{R}\int_0^1 ds\int_{-\frac{L_M}{2}}^{\frac{L_M}{2}} dx^+\int_{K}e^{ik^-(\frac{L_M}{2}+x^+)}e^{i\mathbf{k}\cdot\mathbf{x}s}x_{\perp}\cdot\langle G_{>}^{\perp -}(K)\rangle,
\end{align}
where a sum over the $\perp=x,\,y$ directions is implied.
We recognise that $x^+$ integration can be rewritten in terms of a delta function, i.e
\begin{equation}
(a)=-g^2C_{R}\int_0^1 ds\int_{K}\delta(k^-)e^{ik^-\frac{L_M}{2}}e^{i\mathbf{k}\cdot \mathbf{x}s}x_{\perp}\cdot\langle G_{>}^{\perp -}(K)\rangle.
\end{equation}
Then, upon doing the $k^-$ integration, $L_{M}$ will vanish from the entire integrand, implying that this type of 
diagram will not produce a linear factor of $L_M$. Therefore, according to Eq.~\eqref{eq:wloopexpansion}, it will furthermore not 
contribute to $\mathcal{C}(x_{\perp})$.

Moving on to the rest of the diagrams from Fig.~\ref{fig:side_rails}, we note that their sum is analogous 
to the sum of the three diagrams from Fig.~\ref{fig:leading_order}, but in this case with all gluons attached to the side rails
\begin{equation}
    \mathcal{C}(x_{\perp})_{\text{side rails}}=-\frac{g^{2}C_{R}}{L_M}\int_{1}^{0}ds\int_{1}^{0}dt\int \frac{d^{4}k}{(2\pi)^{4}}e^{-i\mathbf{k}\cdot\mathbf{x}(s+t-1)}(1-e^{-ik^{-}L_M})x_{\perp}G^{\perp\perp}_{>}(k) x_{\perp}.
\end{equation}
The relevant part that we must pay attention to here is 
\begin{equation}
\lim_{L_M\to\infty}\frac{1}{L_M}\int dk^{-}(1-e^{-ik^{-}L_M})G^{\perp\perp}_{>}(k).\label{eq:doubledelta}
\end{equation}
where we have added the limit explicitly because of its particular importance here. For reasons that will soon become clear, it is convenient
to divide the integrand by ${k^{-}}^{2}$. Then, by making a change of variable $z=k^{-}L_M$ and noting that $G^{\perp\perp}_{>}(z/L_M)$ is regular as $L_M\to\infty$ (in the Coulomb
gauge) we can write
\begin{align}
    \lim_{L_M\to\infty}\frac{1}{L^{2}}\int dz\frac{1-e^{-iz}}{\frac{z^{2}}{L_M^{2}}}G^{\perp\perp}_{rr}(\frac{z}{L_{M}})=&\int dz\frac{1-e^{-iz}}{z^{2}}G^{\perp\perp}_{>}(0)\nonumber
    \\&=G^{\perp\perp}_{>}(0)\int dz\frac{1-\cos(z)}{z^{2}}\nonumber
    \\&=\pi G^{\perp\perp}_{>}(0)
\end{align}
where the final integral over $z$ can be calculated using residue methods\footnote{One can use a principal
value prescription to show that the odd part of the integrand vanishes since $\frac{\sin(z)}{z^{2}}$ is divergent as $z\to 0$.}.
We have just shown that (now it is clear why we divided the integrand by ${k^{-}}^{2}$) Eq. \ref{eq:doubledelta} is indeed proportional
to the integral of ${k^{-}}^{2}\delta(k^{-})$, which indeed justifies why we need not worry about these kinds of diagrams,
where gluons are attached to the side rails. This clarifies why, throughout the main body of the thesis, 
the only diagrams that are included are ones where gluons are attached to the horizontal, $\mathcal{W}$ Wilson lines.

\section{Hard Semi-Collinear Subtraction}\label{app:AX}
In this Appendix, we will show how the second term in Eq.~\eqref{defqhatde} 
is automatically taken into account if we add the hard contribution 
to $\hat{q}(\mu_{\perp})$ from \cite{Arnold:2008vd}. Indeed, that paper 
computes the contribution to $\hat{q}(\mu_{\perp})$ for $\sqrt{ET}\gg k_{\perp}\gg gT$ at leading order,
i.e. through elastic Coulomb scatterings with the light quarks and gluons of the medium.
The  incoming and outgoing momenta for such scatterers are named
$P_2$ and $P_2-Q$ there. We can then identify $P_2$ with our $L$ and $Q$ with $-K$.
In order to obtain the contribution to $\mathcal{C}(k_{\perp})$ and thence $\hat{q}(\mu_{\perp})$, \cite{Arnold:2008vd}
integrates over all values  and orientations of $p_2$. In so doing, it 
ends up including the slice where $p_2\sim gT$, $k_{\perp}\sim\sqrt{g}T$, which 
is precisely the semi-collinear scaling we investigated in Sec.~\ref{sec:reproduce}.
As explained there, we need to subtract this limit of \cite{Arnold:2008vd}, so 
as to avoid double countings.

Our starting point is Eq.~(3.8) of \cite{Arnold:2008vd}, where we have specialised to the case 
of scattering off a soft gluon ($n_{\text{B}}(l^0)\approx T/l^0$) 
and undone a couple of auxiliary integrals (see (3.7) there), 
as well as applied the dictionary just described. We then have
\begin{align}
    \delta\mathcal{C}(k_{\perp})^{\text{hard}}&=
    \frac{2g^{4}C_RC_{A}}{k_{\perp}^{4}}\int\frac{dk^{0}}{2\pi}\int\frac{dk^{z}}{2\pi}\int\frac{d^{4}L}
    {(2\pi)^{4}}(l^{0}-l^{z})^{2}\frac{T}{l^{0}}(1+n_{\text{B}} (l^0+k^0))\nonumber
    \\&\times 2\pi\delta(k^{0}-k^{z})2\pi\delta((L+K)^{2})2\theta(l^{0}+k^{0})2
    \theta(l^{0})2\pi\delta(L^{2}).
\end{align}
For a soft gluon $L\sim gT$  we can simplify the 
expression above as
\begin{align}
    \delta\mathcal{C}(k_{\perp})^{\text{hard}}&=
    \frac{2g^{4}C_RC_{A}}{k_{\perp}^{4}}\int\frac{dk^{+}}{2\pi}\int\frac{d^{4}L}
    {(2\pi)^{4}}\frac{Tl^{-2}}{l^++l^-/2}(1+n_{\text{B}} (k^+))\nonumber
    \\&\times 2\pi\delta(2k^+l^--k_{\perp}^2)2\theta(k^{+})2
    \theta(l^++l^-/2)2\pi\delta(L^{2}).
\end{align}
We can set up the $\delta$ function to fix $l^+$, i. e.
\begin{align}
    \delta\mathcal{C}(k_{\perp})^{\text{hard}}&=
    \frac{2g^{4}C_RC_{A}}{k_{\perp}^{4}}\int\frac{dk^{+}}{2\pi}\int\frac{d^{4}L}
    {(2\pi)^{4}}\frac{Tl^{-2}}{l^++l^-/2}(1+n_{\text{B}} (k^+))\frac{\theta(k^{+})}{k^+}\nonumber
    \\&\times 2\pi\delta\left(l^--\frac{k_{\perp}^2}{2k^+}\right)
    \frac{\theta(l^++l^-/2)}{\abs{l^-}}2\pi\delta\left(l^+-\frac{l_{\perp}^2}{2l^-}\right).
\end{align}
This finally yields
\begin{equation}
    \label{semihardfinal}
    \delta\mathcal{C}(k_{\perp})^{\text{hard}} = \frac{g^{2}C_R}{\pi k_{\perp}^{4}}
    \int\frac{dk^{+}}{k^+}(1+n_{\text{B}} (k^+))\theta(k^{+})\int\frac{d^2l_{\perp}}
    {(2\pi)^2}\frac{g^2 C_A T\;2l^{-2}}{l_{\perp}^2+l^{-2}}\bigg\vert_{l^-=\frac{k_{\perp}^2}{2k^+}}.
\end{equation}

It is encouraging to see the resemblance of this result to that of the second term of Eq.~\eqref{defqhatde} 
when plugged in Eq.~\eqref{resultfromsemi2}. 
However, to get the two results to match exactly, we need to 
compute the same result with same integral with instead $L\to L-K$ and $L+K\to L$ keeping 
$L$ soft, that is, a soft outgoing gluon scatterer. 
This second integral will give the negative $k^+$ contribution, and the integrals' sum will indeed yield 
\begin{equation}
    \label{resultfromsemi2app}
    \delta\mathcal{C}(k_{\perp})_\mathrm{semi}^{\text{hard}}= \frac{g^2 C_R}{\pi k_{\perp}^4}\int\frac{dk^+}{k^+}(1+n_{\text{B}}(k^+))
    \hat{q}
    _{\text{\cite{Arnold:2008vd}}}
    \left(\rho_{\perp};\frac{k_{\perp}^2}{2k^+}\right).
\end{equation}
\\\newline
\label{app:diagrams}
\begin{figure}[ht]
    \begin{center}
        \includegraphics[width=14cm]{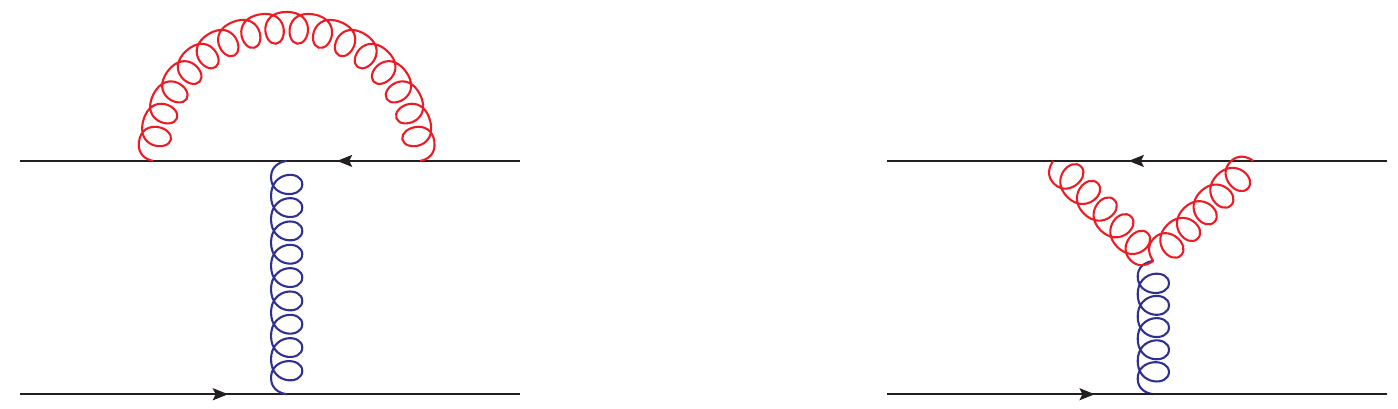}
    \end{center}
    \caption{ Diagrams whose cuts correspond to virtual processes.}
    \label{fig:virtual}
\end{figure}

\section{Partially Virtual Diagrams}\label{sub:virtual}
We start from the first diagram in Fig.~\ref{fig:virtual}. As was the case for the real II and 
X diagrams, the $C_R^2$-proportional, abelian part of this diagram vanishes 
with the counterparts, not shown in the figure, where the coft gluon does not straddle
the soft one. Accounting for this and for the symmetric diagrams with the coft gluon
attached to the other Wilson line we have
\begin{equation}
    \delta\mathcal{C}(l_{\perp})^\mathrm{virt}_{\widehat{\mathrm{T}}}=-g^{4}C_{R}C_{A}\int\frac{d^{4}K}{(2\pi)^{4}}
    \int\frac{dl^{+}}{2\pi}\frac{1}{(k^{-}+i\epsilon)^{2}}
G_{>}^{--}(L)G_{rr}^{--}(K)\Big\vert_{l^{-}=0},
\end{equation}
where we have labeled this diagram $\widehat{\mathrm{T}}$ following its topology. We note that 
the Wilson line integration forces $l^-=0$, differently from the real processes.
In this case too
we defer the evaluation of this expression until after the second diagram has been evaluated. 
Its contribution, together with that of its symmetric counterpart with two vertices on the bottom Wilson line, 
is\footnote{At first glance, it is not clear why we cannot have any other real-time 
assignments for this diagram. Indeed, one could imagine the
assignment $G_{F}(L)G_{>}(K+L)G_{>}(-L)$. 
We have checked that this leaves us with an expression that, at first order in the collinear expansion,
is odd in $k^+$ and thus integrates to zero.}
\begin{align}
    \delta\mathcal{C}(x_{\perp})^\mathrm{virt}_{{\mathrm{Y}}}=&\frac{g^{4}C_{R}C_{A}}{2}
    \int\frac{d^{4}K}{(2\pi)^{4}}\int\frac{d^4L}{(2\pi)^{4}}\frac{-i}{k^{-}-i\epsilon}
    e^{i\mathbf{l}\cdot \mathbf{x}} G_{>}^{\beta-}(L)2\pi\delta(l^-)\nonumber
    \\&\hspace{-2cm}\times\big(G_{F}^{-\delta}(K)G_{F}^{-\alpha}(K{+}L)-
    G_{\bar{F}}^{-\delta}(K)G_{\bar{F}}^{-\alpha}(K{+}L)\big)
    (g_{\delta\beta}(K{-}L)+g_{\beta\alpha}(2L{+}K)_{\delta}-g_{\alpha\delta}(2K{+}L)_{\beta}).
\end{align}

We can use the analogue of Eq.~\eqref{eq:feynman}
 for the anti-time-ordered propagator, i.e.
\begin{equation}
    G_{\bar{F}}(K)=-\frac{1}{2}\Big(G_{R}(K)+G_{A}(K)\Big)+G_{rr}(K)
    \label{eq:torderedprops}
\end{equation}
Using these definitions, we get 
\begin{align}
    \delta\mathcal{C}(l_{\perp})^\mathrm{virt}_{{\mathrm{Y}}}=&g^{4}C_{R}C_{A}\int\frac{d^{4}K}{(2\pi)^{4}}
    \int\frac{dl^{+}dl^-}{2\pi}\frac{i}{k^{-}-i\epsilon}G_{>}^{\beta-}(L)\delta(l^-)\nonumber
    \\&\hspace{-2.cm}\times\big(G_{rr}^{-\delta}(K)G_{R}^{-\alpha}(K+L)
    +G_{R}^{-\delta}(K)G_{rr}^{-\alpha}(K+L)\big)
    (g_{\delta\beta}(K{-}L)+g_{\beta\alpha}(2L{+}K)_{\delta}-g_{\alpha\delta}(2K{+}L)_{\beta}).
\end{align}
The virtual contribution is given by the sum of the two, that is
\begin{equation}
    \delta\mathcal{C}(l_{\perp})^\mathrm{virt} =  \delta\mathcal{C}(l_{\perp})^\mathrm{virt}_{\widehat{\mathrm{T}}}
    + \delta\mathcal{C}(l_{\perp})^\mathrm{virt}_{{\mathrm{Y}}}.
\end{equation}
In this case we do not enforce a semi-collinear scaling for the coft gluon, since, as we 
shall show, there is no double-logarithmic contribution. Furthermore,
the Wilson line integrations set $l^-=0$, thus making the interaction 
with the medium instantaneous. We rather assume $l^+ \sim l_{\perp}\sim k_{\perp}\sim g T\ll k^+$, 
leading to 
\begin{align}
    \delta\mathcal{C}(l_{\perp})^\mathrm{virt}=&2g^{4}C_{R}C_{A}\int\frac{dl^+}{2\pi}
    \int\frac{dk^+d^2k_{\perp}}{(2\pi)^{3}}
 \frac{\frac{1}{2}+n_{\text{B}} (k^{+})}{k^{+}}
    \frac{  G^{--}_{>}(l^+,l_{\perp})}{k_{\perp}^{2}+m_{\infty\,g}^{2}}
    \nonumber \\
    &\times\bigg[\frac{k_{\perp}^2+\mathbf{k}\cdot\mathbf{l}}{(\mathbf{k}+\mathbf{l})^{2}+m_{\infty\,g}^{2}}
    -\frac{k_{\perp}^2}{k_{\perp}^2+m_{\infty g}^2}\bigg].
    \label{partvirt_almostfinal}
\end{align}
The asymptotic masses at the denominator
have been included since we assumed $k_{\perp}\sim gT$. They arise from the $k^+\sim T$, 
$K^2\sim g^2T^2$ limit of the HTL propagators in Sec.~\ref{sec:htl}.
The $l^+$ integration can be carried out through the mapping to EQCD, as was demonstrated in Sec.~\ref{sec:scatt_lo_soft}. It leads to 
\begin{align}
    \delta\mathcal{C}(l_{\perp})^\mathrm{virt}=&2g^{4}C_{R}C_{A}
    \int\frac{dk^+d^2k_{\perp}}{(2\pi)^{3}}
 \frac{\frac{1}{2}+n_{\text{B}} (k^{+})}{k^{+}}
    \frac{Tm_{D}^{2}}{l_{\perp}^{2}(l_{\perp}^{2}+m_{D}^{2})}
    \frac{ 1}{k_{\perp}^{2}+m_{\infty\,g}^{2}}
    \nonumber \\
    &\times\bigg[\frac{k_{\perp}^2+\mathbf{k}\cdot\mathbf{l}}{(\mathbf{k}+\mathbf{l})^{2}+m_{\infty\,g}^{2}}
    -\frac{k_{\perp}^2}{k_{\perp}^2+m_{\infty\,g}^2}\bigg].
    \label{partvirt_final}
\end{align}
We recognise that this expression is proportional to the leading-order, adjoint soft 
scattering kernel $\mathcal{C}(l_{\perp})^\mathrm{LO}_\mathrm{soft}$, Eq.~\eqref{eq:scattkerhtl}. 
From this expression and its associated 
$\hat{q}$ contribution one can see that the $\vert\mathbf{k}+\mathbf{l}\vert\gg l_{\perp}$ single-scattering 
region is free of double-logarithmic enhancements, as found in \cite{Liou:2013qya}.
Furthermore, for $k_{\perp}\sim l_{\perp}\sim gT$, which we have used for our derivation, the 
lifetime of the virtual coft fluctuation is long, so that multiple soft scatterings 
can occur within it. Hence, Eq.~\eqref{partvirt_final} should just be 
considered as the $\mathcal{N}=1$ term in the opacity expansion of the virtual 
contribution to Eq.~\eqref{LMWCshiftfinal}. Indeed, we have checked that 
Eq.~\eqref{partvirt_final} can also be obtained from the virtual terms 
in Eq.~\eqref{eq:multiscat}, i.e. those proportional to $\exp(i\mathbf{x}\cdot\mathbf{l})$, further 
confirming the  soundness of Eqs.~\eqref{LMWschroshift} and \eqref{LMWCshiftfinal}.

\chapter{\texorpdfstring{$Z_g$}{TEXT} Computation}\label{app:amass_app}

This appendix contains some extra details relevant for the calculation 
in Ch.~\ref{ch:asym_mass}.

\section{Amputated Self-Energies}\label{app:amputated}
It is pointed out in \cite{Ghiglieri:2020dpq} that using KMS relations \cite{Kubo:1957mj,PhysRev.115.1342}, 
one can show that amputated self-energies of thermal propagators obey 
\begin{align}
   \Pi^{aa}(K)&=\left(\frac{1}{2}+n_{\text{B}}(k^0)\right)\Big[\Pi^{R}(K)-\Pi^{A}(K)\Big]
   \\\Pi^{>}(K)&=\left(1+n_{\text{B}}(k^0)\right)\Big[\Pi^{R}(K)-\Pi^{A}(K)\Big]
   \\\Pi^{<}(K)&=n_{\text{B}}(k^0)\Big[\Pi^{R}(K)-\Pi^{A}(K)\Big],
\end{align}
where 
\begin{align}
   \Pi^{R}(K)&=\Pi^{ar}
   \\\Pi^{A}(K)&=\Pi^{ra}.
\end{align}
We then define the amputated self-energy-like objects\footnote{Note that the $1/4$ in Eq.~\eqref{eq:ampselfen}
comes from the extra factor associated with the $aaa$ vertex from the \emph{r/a} basis. In particular, see Fig.~\ref{fig:phi4_ra}.}
\begin{align}
   \widetilde{\Pi^{T}}(Q)&\equiv G^{F}(K)G^{F}(Q-K),
   \\\widetilde{\Pi^{<}}(Q)&\equiv  G^{<}(K)G^{<}(Q-K),
   \\\widetilde{\Pi^{>}}(Q)&\equiv G^{>}(K)G^{>}(Q-K),
   \\\widetilde{\Pi^{R}}(Q)&\equiv G^{rr}(K)G^R(Q-K)+G^{rr}(Q-K)G^{R}(K),
   \\\widetilde{\Pi^{A}}(Q)&\equiv G^{rr}(K)G^A(Q-K)+G^{rr}(Q-K)G^{A}(K),
   \\\widetilde{\Pi^{aa}}(Q)&\equiv G^{rr}(K)G^{rr}(Q-K)+\frac{1}{4}G^{R}(Q-K)G^{R}(K) 
   +\frac{1}{4}D^{A}(Q-K)D^{A}(K).\label{eq:ampselfen}
\end{align}
Strictly speaking, there should be a $K$ integral on the right-hand side of the equations above. However, this 
would then greatly muddle the notation in and around Eq.~\eqref{eq:zgg}, for instance, where these identities are employed. 
In the end, it does not matter as we only use these relations to help combine the different assignments; 
in doing the actual loop integration, we always rewrite everything in terms of the propagators.

\section{One-Loop Self Energies}\label{app:one_loop}
In this appendix, we list some finite-temperature one-loop self energies that are needed 
for the computation in Ch.~\ref{ch:asym_mass}. We essentially 
copy them from Chapter $8$ of \cite{Kapusta:2006pm} but rewrite them 
here in such a way that is more suited to our computation.

To start, the gluonic part of the retarded, transverse self-energy reads
\begin{align}
    \Pi_T^R(K)_G&=\frac{g^2C_A}{16 \pi ^2 k^3}\int_0^\infty dq\,n_\mathrm{B}(q) \bigg[8  q k\left(k^2+k_0^2\right)\nonumber
    \\&-K^2 \left(4 q^2+4
   k^2+K^2\right)\left( \ln \left(\frac{k^0+k-2q+i\varepsilon}{k^0-k-2q+i\varepsilon}\right)
   +\ln \left(\frac{k^0-k+2q+i\varepsilon}{k^0+k+2q+i\varepsilon}\right)\right)\nonumber \\
   &-4 q k^0
   K^2 \left(2\ln\left(\frac{k^0+k+i\varepsilon}{k^0-k+i\varepsilon}\right)+
   \ln \left(\frac{k^0-k+2q+i\varepsilon}{k^0+k+2q+i\varepsilon}\right)-\ln 
   \left(\frac{k^0+k-2q+i\varepsilon}{k^0-k-2q+i\varepsilon}\right)\right)\bigg].
   \label{rettransselfglue}
\end{align}
The finite-temperature, gluonic part of the longitudinal retarded self-energy is instead
\begin{align}
   \Pi_L^R(K)_G =& - \frac{g^2 C_{A} K^2}{8 \pi ^2 k^3}
   \int_0^\infty dq \, n_\mathrm{B}(q)  \bigg[8 q k \nonumber \\
  & -  \left (4 q^2-2k^2+k_0^2\right) \bigg[\ln\frac{k^0+k-2q+i\varepsilon}{ k^0-k-2q+i\varepsilon}+\ln\frac{k^0-k+2q+i\varepsilon}{ k^0+k+2q+i\varepsilon}\bigg]\nonumber \\
   &-4 q k^0 \bigg[ 2\ln \frac{k^0+k+i\varepsilon }{k^0-k+i\varepsilon} +\ln\frac{k^0-k+2q+i\varepsilon}{ k^0+k+2q+i\varepsilon}-\ln\frac{k^0+k-2q+i\varepsilon}{ k^0-k-2q+i\varepsilon}\bigg]
   \bigg].
   \label{retlongglue}
\end{align}
It is also worthwhile to have the difference of these two quantities on hand

\begin{align}
   \Pi_L^R(K)_G-\Pi_T^R(K)_G =&  \frac{g^2 C_{A}}{16 \pi ^2 k^3}
   \int_0^\infty dq \, n_\mathrm{B}(q)  \bigg[-8 q k(3 k_0^2-k^2) \nonumber \\
  &  + K^2\left (12 q^2-k^2+3k_0^2\right) \bigg[\ln\frac{k^0+k-2q+i\varepsilon}{ k^0-k-2q+i\varepsilon}+\ln\frac{k^0-k+2q+i\varepsilon}{ k^0+k+2q+i\varepsilon}\bigg]\nonumber \\
   &+12 q k^0K^2 \bigg[ 2\ln \frac{k^0+k+i\varepsilon }{k^0-k+i\varepsilon} +\ln\frac{k^0-k+2q+i\varepsilon}{k^0+k+2q+i\varepsilon}-\ln\frac{k^0+k-2q+i\varepsilon}{ k^0-k-2q+i\varepsilon}\bigg].
   \bigg].
   \label{retdiffgluonsimple}
\end{align}

The fermionic part of the transverse self-energy is then

\begin{align}
    \Pi^R_T(K)_F =& \frac{g^2T_F N_F}{8 \pi ^2 k^3}\int_0^\infty dq\,n_\mathrm{F}(q) 
    \bigg[ 8  q  k\left(k^2+k_0^2\right)\nonumber \\
  &-K^2 \left(4 q^2+k^2+k_0^2\right)\bigg[\ln\frac{k^0+k-2q+i\varepsilon}{ k^0-k-2q+i\varepsilon}+\ln\frac{k^0-k+2q+i\varepsilon}{ k^0+k+2q+i\varepsilon}\bigg]\nonumber \\
   &-4 q k^0 K^2 \bigg[ 2\ln \frac{k^0+k+i\varepsilon }{k^0-k+i\varepsilon} +\ln\frac{k^0-k+2q+i\varepsilon}{ k^0+k+2q+i\varepsilon}-\ln\frac{k^0+k-2q+i\varepsilon}{ k^0-k-2q+i\varepsilon}\bigg]
   \bigg].
   \label{rettransselfquarksimple2}
\end{align}
The longitudinal counterpart is instead

\begin{align}
   \Pi_L^R(K)_F =& - \frac{g^2 T_F N_F K^2}{4 \pi ^2 k^3}
   \int_0^\infty dq \, n_\mathrm{F}(q)  \bigg[8  q  k \nonumber \\
  & -  \left (4 q^2-k^2+k_0^2\right) \bigg[\ln\frac{k^0+k-2q+i\varepsilon}{ k^0-k-2q+i\varepsilon}+\ln\frac{k^0-k+2q+i\varepsilon}{ k^0+k+2q+i\varepsilon}\bigg]\nonumber \\
   &-4 q k^0 \bigg[ 2\ln \frac{k^0+k+i\varepsilon }{k^0-k+i\varepsilon} +\ln\frac{k^0-k+2q+i\varepsilon}{ k^0+k+2q+i\varepsilon}-\ln\frac{k^0+k-2q+i\varepsilon}{ k^0-k-2q+i\varepsilon}\bigg]
   \bigg],
   \label{retlongquarksimple}
\end{align}
For further use we then have
\begin{align}
   \Pi_L^R(K)_F-\Pi_T^R(K)_F =&  \frac{g^2 T_F N_F }{8 \pi ^2 k^3}
   \int_0^\infty dq \, n_\mathrm{F}(q)  \bigg[-8  q k(3 k_0^2-k^2) \nonumber \\
  &  + K^2\left (12 q^2-k^2+3k_0^2\right) \bigg[\ln\frac{k^0+k-2q+i\varepsilon}{ k^0-k-2q+i\varepsilon}+\ln\frac{k^0-k+2q+i\varepsilon}{ k^0+k+2q+i\varepsilon}\bigg]\nonumber \\
   &+12 q k^0K^2 \bigg[ 2\ln \frac{k^0+k+i\varepsilon }{k^0-k+i\varepsilon} +\ln\frac{k^0-k+2q+i\varepsilon}{ k^0+k+2q+i\varepsilon}-\ln\frac{k^0+k-2q+i\varepsilon}{ k^0-k-2q+i\varepsilon}\bigg]
   \bigg],
   \label{retdiffquarksimple}
\end{align}

We additionally need the vacuum part of the self-energy:
The bare vacuum part is \cite{Peskin:1995ev}
\begin{equation}
    \Pi_L^R(K)_\text{vac}=\Pi_T^R(K)_\text{vac}=\frac{\alpha_s}{4\pi}K^2\bigg[
    \bigg(\frac1\epsilon -\gamma_E+\ln(4\pi) -\ln(-K^2)\bigg)\bigg(\frac53 C_A-\frac43 T_FN_F\bigg)
    +\frac{31}{9}C_A-\frac{20}{9}T_FN_F\bigg].
    \label{vacpol}
\end{equation}
Whereas in the $\overline{MS}$ scheme we have
\begin{equation}
    \Pi_L^R(K)_{\text{vac}\,\overline{MS}}=\Pi_T^R(K)_{\text{vac}\,\overline{MS}}=\frac{\alpha_s}{4\pi}K^2\bigg[
     -\ln\frac{k^2-(k^0+i\varepsilon)^2}{\bar\mu^2}\bigg(\frac53 C_A-\frac43 T_FN_F\bigg)
    +\frac{31}{9}C_A-\frac{20}{9}T_FN_F\bigg].
    \label{MSbarvac}
\end{equation}

\section{Details on Diagram \texorpdfstring{$(c)$}{TEXT} Evaluation}\label{sec:eval}
We pack many of the technical details related to the computation in Sec.~\ref{sec:full} into this appendix.

\subsection{Fermionic Contribution: \texorpdfstring{$k^{-2}$}{TEXT} Piece}\label{sec:fermion_km}

Looking to Eq.~\eqref{rettransselfquarksimple2}, we start with the simplest terms, i.e. those that 
do not multiply any log, i.e.
\begin{align}
    \Pi^R_T(K)_{F\,\text{no log}} =& \frac{g^2}{16 \pi ^2 k^3}\int_0^\infty dq\bigg[2T_F N_F n_\text{F}(q) 8  q k\left(k^2+k_0^2\right) \bigg]=
    \frac{g^2T_{F}N_{F}T^2}{12 k^2}\left(k^2+k_0^2\right),
   \label{rettransnolog}
\end{align}
Hence
\begin{equation}
    Z_{g\,k^{-2}\,F\,\text{ no log}}^{(c)}=i\frac{g^2T_{F}N_{F}T^2}{6}
    \int_{K}\Big(1+n_\mathrm{B}(k^0)\Big)\bigg[
    \frac{(1+k_0^2/k^2)}{(K^2+i\varepsilon k^0)^2}
    -\text{adv.}
    \bigg].
    \label{zcexprnolog}
\end{equation}
We can do the $k^0$ integration using contour methods: since the terms in the square brackets 
have the desired $\text{ret.}-\text{adv}$ structure, we can close the contour 
in the upper (lower) half-plane to integrate over the retarded (advanced) piece. 
We find
\begin{align}
   Z_{g\,k^{-2}\,F\,\text{ no log}}^{(c)}= &\frac{g^2T_{F}N_{F}T^2}{6}
    \int_{K}\Big(\frac{1}{2}+ n_\mathrm{B}(k^0)\Big)\bigg[\frac{i(1+k_0^2/k^2)}{(K^2+i\varepsilon k^0)^2}-\frac{i(1+k_0^2/k^2)}{(K^2-i\varepsilon k^0)^2}\bigg]\nonumber \\
    &=
    -\frac{g^2T_{F}N_{F}T^2}{6} T\sum_n\int_k \frac{1-\omega_{n}^2/k^2}{(-k^2-\omega_n^2)^2}=\frac{g^2T_{F}N_{F} T^2}{24\pi^2}
    \label{qmsquarenologpiece}
\end{align}
Note that in going from Eq.~\eqref{zcexprnolog}, we have rewritten $\big(1+n_\mathrm{B}(k^0)\big)\rightarrow\big(\frac{1}{2}+n_\mathrm{B}(k^0)\big)$, 
which is allowed, because the vacuum part is scaleless and vanishes in DR anyway.
The $1/2$ accompanying the statistical function is inserted, precisely because it is the proper
term for analytical continuation, and because $1/2+n_\mathrm{B}(k^0)$ is odd,
thus complementing the odd expression in square brackets. 

Now we move on to the terms with logarithms, starting with the ``HTL log'' contribution, which 
is obtained by expanding for $Q\gg K$ in Eq.~\eqref{rettransselfquarksimple2}
\begin{align}
    \Pi^R_T(K)_F \bigg\vert_\text{HTL log}=& -\frac{g^2T_F N_F}{ \pi ^2 k^3}\int_0^\infty dq\,n_\mathrm{F}(q) 
    q k^0 K^2  \ln \frac{k^0+k+i\varepsilon }{k^0-k+i\varepsilon} \nonumber \\
    =&-g^2T_F N_F T^2\frac{k^0 K^2}{ 12 k^3}
       \ln \frac{k^0+k+i\varepsilon }{k^0-k+i\varepsilon}.
   \label{rettransselfquarksimple2htl}
\end{align}
The contribution from this log to $Z_g$ reads
\begin{equation}
    Z_{g\,k^{-2}\,F\,\text{ HTL log}}^{(c)}=-i\frac{g^2T_F N_F T^2}{6}\int_{K}\Big(\frac{1}{2}+n_\mathrm{B}(k^0)\Big)\bigg[
    \frac{k^0}{  k^3(K^2+i\varepsilon k^0)}\ln \frac{k^0+k+i\varepsilon }{k^0-k+i\varepsilon}
    -\text{adv.}
    \bigg].
    \label{zcexprhtllog}
\end{equation}
The reason this term is dealt with separately is that the strategy of separating out 
a $\delta(K^2)$ contribution and a $\Theta$-function contribution from the imaginary 
part of the log in the retarded-advanced difference would in this case give rise
to two contributions that are separately divergent on the light-cone at $K^2=0$.

We can proceed by re-introducing the auxiliary angular integral and then using
analyticity to get the Matsubara modes, i.e.
\begin{align}
     Z_{g\,k^{-2}\,F\,\text{ HTL log}}^{(c)}=&-i\frac{g^2T_F N_F T^2}{6}\int_{K}\int_{-1}^1 dx\,\Big(\frac{1}{2}+n_\mathrm{B}(k^0)\Big)\bigg[
    \frac{k^0}{  k^2(K^2+i\varepsilon k^0)}\frac{1 }{k^0-kx+i\varepsilon}
    -\text{adv.}
    \bigg]\nonumber \\
    =&\frac{g^2T_F N_F T^2}{6}T\sum_n\int_k
    \int_{-1}^1 dx
    \frac{i\omega_n}{  k^2(-\omega_n^2-k^2)}\frac{1 }{i\omega_n-kx}\nonumber\\
    =&-\frac{g^2T_F N_F T^2}{6}T\sum_n\int_k
    \int_{-1}^1 dx
    \frac{i\omega_n}{  k^2(\omega_n^2+k^2)}\frac{-i \omega_n +k x }{\omega_n^2+k^2x^2}
    \label{zcexprhtllogmatsu}
\end{align}
If we keep the even part only of the integer sum we find
\begin{align}
     Z_{g\,F\,\text{HTL log}}^{(c)}=&-\frac{g^2T_F N_F T^2}{6}T\sum_n\int_k
    \int_{-1}^1 dx
    \frac{\omega_n^2}{  k^2(\omega_n^2+k^2)(\omega_n^2+k^2x^2)}\nonumber \\
    =&   -\frac{g^2T_F N_F T^2}{24\pi^2}\bigg[
    \frac{\ln (2)}{\epsilon}+\left(-\ln (4) \ln\left(2 \frac{4\pi T e^{-\gamma_E}}{\bar\mu}\right)+\frac{\pi
   ^2}{12}+2 \ln ^2(2)+\ln (4)\right)\bigg].
    \label{zcexprhtllogmatsufinal}
\end{align}
Let us now move on to the other $k^{-2}\Pi^T$-proportional terms in the fermion 
contribution, i.e.
\begin{align}
    Z_{g\, k^{-2}\,F\,\text{ non-HTL }}^{(c)}=&\frac{ig^2T_F N_F}{4 \pi ^2 }\int_{K}\bigg\{\frac{1+n_\mathrm{B}(k^0)}{k^3(K^2+i\varepsilon k^0)}
        \int_0^\infty dq\,n_\mathrm{F}(q) 
    \bigg[ \nonumber \\
  &- \left(4 q^2+k^2+k_0^2\right)\bigg(\ln\frac{k^0+k-2q+i\varepsilon}{ k^0-k-2q+i\varepsilon}+\ln\frac{k^0-k+2q+i\varepsilon}{ k^0+k+2q+i\varepsilon}\bigg)\nonumber \\
   &-4 q k^0  \bigg( \ln\frac{k^0-k+2q+i\varepsilon}{ k^0+k+2q+i\varepsilon}-\ln\frac{k^0+k-2q+i\varepsilon}{ k^0-k-2q+i\varepsilon}\bigg)
   \bigg]
    -\text{adv.}\bigg\}.
    \label{zcfermionqmsqa}
\end{align}
We can split this into a pole part ($\propto\delta(K^2)$) and a cut part
($\propto\mathbb{P}1/K^2$, with $\mathbb{P}$ the principal value).
The former reads
\begin{align}
    Z_{g\, k^{-2}\,F\,\text{pole}}^{(c)}=&\frac{g^2T_F N_F}{4 \pi ^2 }\int_{K}\frac{\frac{1}{2}+n_\mathrm{B}(\vert k^0\vert)}{k^3} 2\pi\delta(K^2)
        \int_0^\infty dq\,n_\mathrm{F}(q) 
    \bigg[ \nonumber \\
  &- \left(4 q^2+k^2+k_0^2\right)\bigg(\ln\left\vert\frac{k^0+k-2q}{ k^0-k-2q}\right\vert+\ln\left\vert\frac{k^0-k+2q}{ k^0+k+2q}\right\vert\bigg)\nonumber \\
   &-4 q k^0  \bigg( \ln\left\vert\frac{k^0-k+2q}{ k^0+k+2q}\right\vert-\ln\left\vert\frac{k^0+k-2q}{ k^0-k-2q}\right\vert\bigg)
   \bigg].    \label{zcfermionqmsqpole}
\end{align}
This becomes
\begin{align}
    Z_{g k^{-2}\,F\,\text{pole}}^{(c)}=&\frac{g^2T_F N_F}{2 \pi ^2 }\int_{k}\frac{\frac{1}{2}+n_\mathrm{B}(k)}{k^4}
        \int_0^\infty dq\,n_\mathrm{F}(q) 
    \bigg[ 
  \left(2 q^2+k^2\right)\ln\frac{ k+q}{\vert k-q\vert}
   +2  q k  \ln\frac{\vert k^2-q^2\vert}{q^2}
   \bigg].    \label{zcfermionqmsqpole2}
\end{align}
Let us split this into regions
\begin{align}
    Z_{g k^{-2}\,F\, \text{pole}}^{(c)}=&\frac{g^2T_F N_F}{2 \pi ^2 }\int_0^\infty dq\,n_\mathrm{F}(q)
    \int_{k}\frac{\frac{1}{2}+n_\mathrm{B}(k)}{k^4}  \nonumber \\
    &\times \bigg\{   \Theta(q-k)\bigg[ 
  \left(2 q^2+k^2\right)\ln\frac{ q+k}{q-k}
   +2  q k  \ln\left(1-\frac{k^2}{q^2}\right)\mp \frac{4qkT}{k\left(\frac{1}{2}
   +n_{\mathrm{B}}(k)\right)}
   \bigg]\nonumber\\
   &+
     \Theta(k-q)\bigg[ 
  \left(2 q^2+k^2\right)\ln\frac{ k+q}{ k-q}
   +2  q k  \ln\frac{k^2-q^2}{q^2}
   \bigg]\bigg\}.    \label{zcfermionqmsqpole3}
\end{align}
The strategy is as follows: on the second line we have introduced 
a subtraction term, $-4qk$, which renders the integration IR finite, 
so that we can do it numerically. We then add back $+4qkT/k\left(\frac{1}{2}
   +n_{\mathrm{B}}(k)\right)$ and do it in DR.
For the third line, we instead do the thermal part numerically and the 
vacuum part analytically in DR. We then find
\begin{align}
    &Z_{g k^{-2}\,F\,\text{pole}}^{(c)}=\frac{g^2T_F N_FT^2}{96\pi ^2 }\bigg[
   \underbrace{4.4713}_{q>k\text{ num}} \quad\underbrace{-\frac{96\ln 2}
   {\pi^2}}_{q>k\text{ subtr}}\quad\underbrace{+3.7933}_{k>q\text{ therm.}} \nonumber \\
   &+\frac{1}{\epsilon^2}+\frac{2\ln\frac{\bar\mu}{8\pi T}+24 \ln A+1}{\epsilon}
   +(48 \ln (A)+2) \ln \frac{\bar\mu}{8\pi T}+\frac{12 \zeta ''(2)}{\pi ^2}+2 \ln
   ^2\frac{\bar\mu}{8\pi T}-\frac{\pi ^2}{12}\nonumber\\
   &-2 \Big[2 \gamma_E  (-12 \ln (A)+\ln (2\pi ))-12 (1+2 \ln (2\pi )) \ln (A)-1+\gamma_E ^2+\ln (2) (\ln
   (8)-1)\nonumber
   \\ &+\ln (\pi ) \ln (4 \pi )\Big]
\bigg].    
\label{zcfermionqmsqfinal}
\end{align}
Here $\ln(A)$ is the logarithm of Glaisher's constant and the terms on the second, third and 
and fourth lines come from the DR-evaluated vacuum contribution for $k>q$.

Going back to Eq.~\eqref{zcfermionqmsqa} the cut piece on the other hand reads
\begin{align}
    Z_{g k^{-2}\,F\,\text{cut}}^{(c)}=&\frac{g^2T_F N_F}{4 \pi ^2 }\int_{K}(2\pi)\mathbb{P}\frac{1+n_\mathrm{B}(k^0)}{k^3K^2}
        \int_0^\infty dq\,n_\mathrm{F}(q) 
    \bigg[ \nonumber \\
  &- \left(4 q^2+k^2+k_0^2\right)\bigg(
  \Theta(k^2-(2q-k^0)^2)
  - \Theta(k^2-(2q+k^0)^2)\bigg)\nonumber \\
   &+4 q k^0  \bigg(\Theta(k^2-(2q+k^0)^2)
   + \Theta(k^2-(2q-k^0)^2)\bigg)
   \bigg]\nonumber \\
   =&\frac{g^2T_F N_F}{4 \pi ^2 }\int_{K}(2\pi)\mathbb{P}\frac{\frac{1}{2}+n_\mathrm{B}(k^0)}{k^3K^2}
        \int_0^\infty dq\,n_\mathrm{F}(q) 
    \bigg[ \nonumber \\
  &- \Theta(k^2-(2q-k^0)^2)\left((2 q-k^0)^2+k^2\right)
  + \Theta(k^2-(2q+k^0)^2)\left((2 q+k^0)^2+k^2\right)
   \bigg].
    \label{zcfermionqmsq}
\end{align}
For the cut piece, the strategy is as follows: we exploit the frequency
evenness to restrict the frequency integral to twice the positive range.

For the $n_\mathrm{B}(k^0)$-proportional 
part we first perform the $k$ integration analytically, using the PV
prescription for the $k=k^0$ singularities. The $k^0$ and
$q$ integrals can then be done numerically and are free of IR divergences.
The $1/2$-proportional part is carried out by first performing the $k^0$ integration,
to be then followed by the $k$ and $q$ ones. The $k$ integration is divergent and needs 
to be regulated appropriately.
This then yields
\begin{align}
    Z_{g k^{-2}\,F\,\text{cut}}^{(c)}=&\frac{g^2T_F N_FT^2}{96\pi ^2 }
    \bigg\{-\frac{1}{\epsilon^2} -\frac{2\ln\frac{A^{12}\bar\mu}{32\pi T}+1}{\epsilon} 
   -2 \ln^2\frac{A^{12}\bar\mu}{32\pi T}-2 \ln\frac{A^{12}\bar\mu}{32\pi T}   \nonumber \\
   &    -\frac{12 \zeta ''(2)}{\pi ^2}+\frac{72 \zeta'(2)^2}{\pi ^4}+\frac{7 \pi ^2}{12}-6+
   8\ln^22-4 \ln 2 + 1.70118
    \bigg\}.
    \label{zcfermionqmsqfinaltwo}
\end{align}
Here all terms come from the $1/2$-proportional vacuum part with the exception of
the final term, coming from the numerical integrations. 
Summing the cut and pole contribution yields
\begin{align}
    Z_{g\, k^{-2}\,F\,\text{ non-HTL}}^{(c)}=&
    \frac{g^2T_F N_FT^2}{96\pi ^2 }\bigg\{
   \underbrace{4.4713}_{q>k\text{ num}} \quad\underbrace{-\frac{96\ln 2}
   {\pi^2}}_{q>k\text{ subtr}}\quad\underbrace{+3.7933}_{k>q\text{ therm.}} + 1.70118\nonumber \\
   &+\frac{4\ln 2}{\epsilon}
   +8\ln 2\ln\frac{A^{12}\bar\mu }{8\pi T}+\frac{\pi^2}{2}-4+2\ln(2)(1-2\ln 2)
    \bigg\}\label{eq:sumkminus}
\end{align}

The total contribution from the $k^{-2}$-proportional part in
the fermion contribution is, from Eqs.~\eqref{qmsquarenologpiece}, ~\eqref{zcexprhtllogmatsufinal},
 and ~\eqref{eq:sumkminus} therefore
\begin{align}
  Z_{g\, k^{-2}\,F}^{(c)}=& \frac{g^2T_F N_FT^2}{96\pi ^2 }\bigg\{
    -3 \left(4 \ln (2) \ln \left(\frac{2}{A^8}\right)+2\ln (2)\right)+\frac{\pi ^2}{6}-8 \gamma_E  \ln 2\nonumber \\
   &+4.4713 -\frac{96\ln 2}
   {\pi^2}+3.7933 + 1.70118
  \bigg\}.\label{totqmsquarefermion_app}
\end{align}
We can thus see that the entire $k^{-2}$-proportional contribution is 
finite. The result has been confirmed, by numerically performing the sum over the 
Matusbara modes associated with $K$.

\subsection{Fermionic Contribution: \texorpdfstring{$\Pi_{L}-\Pi_T$}{TEXT} Piece}\label{sec:fermion_lmt}

Let us start from the HTL subset of the full contribution
\begin{equation}
    Z_{g\,F\,\text{ L-T HTL}}^{(c)}=i\int_{K}\frac{1+n_\mathrm{B}(k^0)}{(k^--i\varepsilon)^2}\bigg[
 \frac{k_\perp^2(\Pi_L^R(K)-\Pi_T^R(K))_\mathrm{HTL}}{k^2(K^2+i\varepsilon k^0)}-\text{adv}
    \bigg].
    \label{zcexprLTHTL}
\end{equation}
From Eq.~\eqref{retdiffquarksimple} we then have
\begin{align}
   \Pi_L^R(K)_{F\,\mathrm{HTL}}-\Pi_T^R(K)_{F\,\mathrm{HTL}} =&  \frac{g^2 T_F N_F T^2}{12  }
   \bigg[-  \frac{(6 k_0^2-4k^2)}{k^2}  
  +3 \frac{ k^0K^2}{k^3} \ln \frac{k^0+k+i\varepsilon }{k^0-k+i\varepsilon} 
   \bigg].
   \label{retdiffquarksimpleHTL}
\end{align}
The pole contribution is then
\begin{equation}
    Z_{g\,F\,\text{ L-T HTL pole}}^{(c)}=-\frac{g^2 T_F N_F T^2}{12}
    \int_{K}\frac{\frac{1}{2}+n_\mathrm{B}(\vert k^0\vert)}{(k^--i\varepsilon)^2}\frac{(6 k_0^2-4k^2)}{k^2}\bigg[
 \frac{k_\perp^2}{k^2}2\pi\delta(K^2)
    \bigg].
    \label{zcexprLTHTLpole}
\end{equation}
The vacuum part is scale-free, so in DR, we have
\begin{equation}
    Z_{g\,F\,\text{ L-T HTL pole}}^{(c)}=-\frac{g^2 T_F N_F T^2}{6}
    \int_{k}\frac{n_\mathrm{B}(k)}{2k}\bigg[\frac{1}{(k-k^z-i\varepsilon)^2}+\frac{1}{(-k-k^z-i\varepsilon)^2}\bigg] \frac{k_\perp^2}{k^2}.
    \label{zcexprLTHTLpole2}
\end{equation}
We drop the $i\varepsilon$ prescription, because 
DR can make the $x$ integration convergent in the appropriate number of dimensions, i.e.
\begin{align}
    Z_{g\,F\,\text{ L-T HTL pole}}^{(c)}=&-\frac{g^2 T_F N_F T^2}{6} \Big( \frac{\bar{\mu}^{2}e^{\gamma_E}}{4\pi} \Big)^\epsilon \frac{2\pi^{d/2-1/2}}{\Gamma(d/2-1/2)(2\pi)^d}
    \int_0^\infty dk\,k^{d-4}n_\mathrm{B}(k)\nonumber\\
    &\times\int_{-1}^1 dx (1-x^2)^{d/2-3/2}
    \frac{1+x^2}{1-x^2}.\nonumber
    \\=&-\frac{g^2 T_F N_F T^2}{6} \Big( \frac{\bar{\mu}^{2}e^{\gamma_E}}{4\pi} \Big)^\epsilon \frac{2\pi^{d/2-1/2}}{\Gamma(d/2-1/2)(2\pi)^d}
    T^{d-3} \zeta (d-3) \Gamma (d-3)\nonumber\\
    &\frac{\sqrt{\pi } (d-1) \Gamma \left(\frac{d-3}{2}\right)}{2 \Gamma \left(\frac{d}{2}\right)}\nonumber \\
    &\hspace{-2cm}=\frac{g^2 T_F N_F T^2}{48\pi^2}\bigg[\frac{1}{\epsilon^2}
    +\frac{2\ln  \frac{\bar\mu \exp(\gamma_E+1/2) }{4\pi T}  }{\epsilon}
   + 2 \ln^2\frac{\bar\mu \exp(\gamma_E+1/2) }{4\pi T}-4 \gamma _1+\frac{\pi ^2}{4}
    -2 \gamma_E^2+\frac{3}{2}
    \bigg].
    \label{zcexprLTHTLpoleDRfinal}
\end{align}
The cut contribution of the HTL piece is instead
\begin{equation}
    Z_{g\,F\,\text{ L-T HTL cut}}^{(c)}= \frac{g^2 T_F N_F T^2}{4}
    \int_{K}\frac{1+n_\mathrm{B}(k^0)}{(k^--i\varepsilon)^2}
 \frac{k_\perp^2   k^0}{k^5}2\pi \Theta(k^2-k_0^2) .
    \label{zcexprLTHTLcut}
\end{equation}
To simplify things, let us symmetrise the frequency integration
\begin{equation}
    Z_{g\,F\,\text{ L-T HTL cut}}^{(c)}= \frac{g^2 T_F N_F T^2}{4}
    \int_{K} \frac{k_\perp^2   k^0}{k^5}2\pi \Theta(k^2-k_0^2)\frac{1}{2}\bigg[
    \frac{1+n_\mathrm{B}(k^0)}{(k^0-k^z-i\varepsilon)^2}-
    \frac{-n_\mathrm{B}(k^0)}{(-k^0-k^z-i\varepsilon)^2}\bigg] .
    \label{zcexprLTHTLcutsym}
\end{equation}
As this integral is now even, we can restrict over twice the positive range, i.e.
\begin{equation}
    Z_{g\,F\,\text{ L-T HTL cut}}^{(c)}= \frac{g^2 T_F N_F T^2}{4}
    \int_0^\infty \frac{dk^0}{2\pi}\int_{k} \frac{k_\perp^2   k^0}{k^5}2\pi \Theta(k^2-k_0^2)\bigg[
    \frac{1+n_\mathrm{B}(k^0)}{(k^0-k^z-i\varepsilon)^2}+
    \frac{n_\mathrm{B}(k^0)}{(k^0+k^z+i\varepsilon)^2}\bigg] .
    \label{zcexprLTHTLcutsympos}
\end{equation}
The $n_\mathrm{B}$-independent part is now scale-free, while the rest can be done
in a somewhat similar way to the pole piece
\begin{align}
   &Z_{g\,F\,\text{ L-T HTL cut}}^{(c)}=\frac{g^2 T_F N_F T^2}{4}\Big(\frac{\bar{\mu}^{2}e^{\gamma_E}}{4\pi} \Big)^\epsilon \frac{2\pi^{d/2-1/2}}{\Gamma(d/2-1/2)(2\pi)^{d-1}}
    \int_0^\infty \frac{dk^0}{2\pi} k^0 n_\mathrm{B}(k^0)
    \nonumber \\
    &\times\int_{k_0}^\infty dk k^{d-4} \int_{-1}^{1} dx (1-x^2)^{d/2-1/2}
   \bigg[
    \frac{1}{(k^0-k x-i\varepsilon)^2}+
    \frac{1}{(k^0+k x+i\varepsilon)^2}\bigg] .
    \label{zcexprLTHTLcutsymposDR}
\end{align}
Our strategy here is to do the following subtraction: $n_\mathrm{B}(k^0)\to
n_\mathrm{B}(k^0)\mp(T/k^0-\Theta(T-k^0)/2)$. In this way, the subtraction
is finite, and then we re-evaluate the addition in DR. 
For the finite part, we can first perform the angular and momentum integration, i.e.
\begin{align}
    &Z_{g\,F\,\text{ L-T HTL cut finite}}^{(c)}=\frac{g^2 T_F N_F T^2}{4} \frac{1}{2\pi}
    \int_0^{\infty} \frac{dk^0}{2\pi} k^0 \bigg[n_\mathrm{B}(k^0)-\frac{T}{k^0}
    +\frac{\Theta(T-k^0)}{2}\bigg] 
    \nonumber \\
    &\times\int_{k^0}^\infty \frac{dk}{k}\int_{-1}^{1} dx (1-x^2)
   \bigg[
    \frac{1}{(k^0-k x-i\varepsilon)^2}+
    \frac{1}{(k^0+k x+i\varepsilon)^2}\bigg] \nonumber\\
    =&\frac{g^2 T_F N_F T^2}{4} \frac{1}{2\pi}
    \int_0^{\infty} \frac{dk^0}{2\pi} k^0 \bigg[n_\mathrm{B}(k^0)-\frac{T}{k^0}
    +\frac{\Theta(T-k^0)}{2}\bigg]
    \nonumber\\
   &\times  \int_{k^0}^\infty\frac{dk}{k} \bigg[\frac{4 k^0 \tanh ^{-1}\left(\frac{2 k k^0}{k^2+k_0^2}\right)}{k^3}-\frac{8}{k^2}\bigg]\nonumber \\
   =&\frac{2g^2 T_F N_F T^2}{3}(\ln2-1) \frac{1}{(2\pi)^2}
    \int_0^{\infty} \frac{dk^0}{k^0}  \bigg[n_\mathrm{B}(k^0)-\frac{T}{k^0}
    +\frac{\Theta(T-k^0)}{2}\bigg]\nonumber \\
    =&\frac{g^2 T_F N_F T^2}{12 \pi^2}\ln\frac{T}{\omega_{\text{T}}}(\ln2-1),
        \label{zcexprLTHTLcutsymposDRfinite}
\end{align}
where $\omega_{\text{T}}=2\pi T\exp(-\gamma_{E})$.
For the DR part subtracted part we can instead do frequency first and 
forget about the zero-mode subtraction, because it will be scale-free. Then
\begin{align}
    &Z_{g\,F\,\text{ L-T HTL cut div}}^{(c)}=-\frac{g^2 T_F N_F T^2}{4}\Big(\frac{\bar{\mu}^{2}e^{\gamma_E}}{4\pi} \Big)^\epsilon \frac{2\pi^{d/2-1/2}}{\Gamma(d/2-1/2)(2\pi)^{d-1}}
    \int_{0}^\infty dk k^{d-4} 
    \nonumber \\
    &\times\int_0^k \frac{dk^0}{2\pi} k^0
    \frac{\Theta(T-k^0)}{2}\int_{-1}^{1} dx (1-x^2)^{d/2-1/2}
   \bigg[
    \frac{1}{(k^0-k x-i\varepsilon)^2}+
    \frac{1}{(k^0+k x+i\varepsilon)^2}\bigg] .
    \label{zcexprLTHTLcutsymposDRdiv}
\end{align}
We can continue splitting this in the 
$k>T$ and $k<T$ ranges, i.e.
\begin{align}
    &Z_{g\,F\,\text{ L-T HTL cut div}}^{(c)}=-\frac{g^2 T_F N_F T^2}{8}\Big(\frac{\bar{\mu}^{2}e^{\gamma_E}}{4\pi} \Big)^\epsilon \frac{2\pi^{d/2-1/2}}{\Gamma(d/2-1/2)(2\pi)^{d}}
    \nonumber \\
    &\times\int_{0}^\infty dk k^{d-4} \int_{-1}^{1} dx (1-x^2)^{d/2-1/2}
    \bigg\{\Theta(T-k)\bigg[-\mathbb{P}\frac{2}{1-x^2}+\ln\frac{1-x^2}{x^2}\bigg]\nonumber \\
  & +\Theta(k-T)\bigg[\ln\frac{\left\vert T^2-k^2 x^2\right\vert }{k^2 x^2}-\mathbb{P}\frac{2 T^2}{T^2-k^2 x^2}\bigg]
   \bigg\},.
     \label{zcexprLTHTLcutsymposDRdivqzero}
\end{align}
where we have used the fact that the imaginary parts of the frequency integral are odd
in $x$ and thus vanish. The principal value arises 
from the $\varepsilon\to 0$ limit. The $k>T$ slice is finite, since the $k$ integration is
IR divergent, while the $T>k$ part is IR divergent and needs regularisation. We then have
\begin{align}
    &Z_{g\,F\,\text{ L-T HTL cut div}}^{(c)}=-\frac{g^2 T_F N_F T^2}{8}\Big(\frac{\bar{\mu}^{2}e^{\gamma_E}}{4\pi} \Big)^\epsilon \frac{2\pi^{d/2-1/2}}{\Gamma(d/2-1/2)(2\pi)^{d}}
    \nonumber \\
    &\times\bigg\{\int_{-1}^{1} dx (1-x^2)^{d/2-1/2}
    \frac{T^{d-3}}{d-3}\bigg[-\mathbb{P}\frac{2}{1-x^2}+\ln\frac{1-x^2}{x^2}\bigg]\nonumber \\
  & +\int_{-1}^{1} dx (1-x^2)\bigg[\frac{\text{Li}_2\left(x^2\right)}{2}+\ln \left(\frac{x^2}{1-x^2}\right)+\ln ^2\vert x\vert-\frac{\pi ^2}{6}\bigg]
   \bigg\},
     \label{zcexprLTHTLcutsymposDRdivqzerosplit}
\end{align}
which yields
\begin{align}
    &Z_{g\,F\,\text{ L-T HTL cut div}}^{(c)}=-\frac{g^2 T_F N_F T^2}{8}\Big(\frac{\bar{\mu}^{2}e^{\gamma_E}}{4\pi} \Big)^\epsilon \frac{2\pi^{d/2-1/2}}{\Gamma(d/2-1/2)(2\pi)^{d}}
    \nonumber \\
    &\times\bigg\{
    \frac{T^{d-3}}{d-3}
    \frac{\sqrt{\pi } \Gamma \left(\frac{d}{2}+\frac{1}{2}\right) \left(\psi ^{(0)}\left(\frac{d}{2}-\frac{1}{2}\right)+\gamma
   -2+2\ln (2)\right)}{\Gamma \left(\frac{d}{2}+1\right)}\nonumber \\
  & +\frac{1}{9} \left(-8-\pi ^2+8\ln 2)\right)
   \bigg\}\nonumber\\
   =&-\frac{g^2 T_F N_F T^2}{8}\bigg[
   \frac{1-\ln 2}{3\pi^2\epsilon}+\frac{(1-\ln 2)(2\ln\frac{\bar\mu}{2T}+1)
   }{3\pi^2}\bigg]
  ,
     \label{zcexprLTHTLcutsymposDRdivqzerosplitfinal}
\end{align}
so that finally
\begin{equation}
    Z_{g\,F\,\text{ L-T HTL cut}}^{(c)} = 
    \frac{g^2 T_F N_F T^2}{24\pi^2}(\ln(2)-1)\bigg[
   \frac{1}{\epsilon}+2\ln\frac{\bar\mu\exp(\gamma_E+1/2)}{4\pi T}\bigg].
   \label{zcexprLTHTLcutfinal}
\end{equation}
Let us now consider the non-HTL piece, i.e
\begin{align}
   &\delta\Pi_L^R(K)_F-\delta\Pi_T^R(K)_F =  \frac{g^2 T_F N_F K^2}{8 \pi ^2 k^3}
   \int_0^\infty dq \, n_\mathrm{F}(q)  \bigg[24  q k
   \nonumber \\
  &  + \left(12 q^2-k^2+3k_0^2\right) \bigg[\ln\frac{k^0+k-2q+i\varepsilon}{ k^0-k-2q+i\varepsilon}+\ln\frac{k^0-k+2q+i\varepsilon}{ k^0+k+2q+i\varepsilon}\bigg]\nonumber \\
   &+12 q k^0 \bigg[\ln\frac{k^0-k+2q+i\varepsilon}{ k^0+k+2q+i\varepsilon}-\ln\frac{k^0+k-2q+i\varepsilon}{ k^0-k-2q+i\varepsilon}\bigg]
   \bigg],
   \label{retdiffquarksimplenonHTL}
\end{align}
where $\delta\Pi=\Pi-\Pi_\text{HTL}$. 
This then gives
\begin{align}
    &Z_{g\,F\,\text{ L-T non-HTL}} ^{(c)}=\frac{g^2 T_F N_F }{4 \pi  }\int_0^\infty dq \, n_\mathrm{F}(q) \int_{K}\frac{\frac{1}{2}+n_\mathrm{B}(k^0)}{(k^--i\varepsilon)^2}
  \frac{k_\perp^2}{k^5}\nonumber
  \\&\times\bigg[\Theta(k^2-(k^0 - 2 q)^2)\left(3(2q-k^0)^2-k^2\right)
    -\Theta(k^2-(k^0 + 2 q)^2) \left(3(2q+k^0)^2-k^2\right)\bigg].
    \label{zcexprlogs}
\end{align}
Here we used the fact that the 1/2 piece vanishes to restrict to the purely odd 
$1/2+n_\mathrm{B}(k^0)$ component.
Let us again symmetrise and restrict to twice the positive frequency range:
\begin{align}
    &Z_{g\,F\,\text{ L-T non-HTL}} ^{(c)}=\frac{g^2 T_F N_F }{4 \pi  }\int_0^\infty dq \, n_\mathrm{F}(q) \int_0^\infty\frac{dk^0}{2\pi}
    \int_{k}\frac{k_\perp^2}{k^5}\bigg[\frac{\frac{1}{2}+n_\mathrm{B}(k^0)}{(k^0-k^z-i\varepsilon)^2}
    +\frac{\frac{1}{2}+n_\mathrm{B}(k^0)}{(k^0+k^z+i\varepsilon)^2}\bigg]\nonumber\\
    &\times
  \bigg[\Theta(k^2-(k^0 - 2 q)^2)\left(3(2q-k^0)^2-k^2\right) -\Theta(k^2-(k^0 + 2 q)^2) \left (3(2q+k^0)^2-k^2\right)\bigg].
    \label{zcexprlogssym}
\end{align}
This corresponds to
\begin{align}
    &Z_{g\,F\,\text{ L-T non-HTL}} ^{(c)}=\frac{g^2 T_F N_F }{4 \pi  }\int_0^\infty dq \, n_\mathrm{F}(q) \int_0^\infty\frac{dk^0}{2\pi}
    \int_{k}\frac{k_\perp^2}{k^5}\bigg[\frac{1/2+n_\mathrm{B}(k^0)}{(k^0-k^z-i\varepsilon)^2}
    +\frac{1/2+n_\mathrm{B}(k^0)}{(k^0+k^z+i\varepsilon)^2}\bigg]\nonumber\\
    &\times
  \bigg\{\Theta(k-2q)\bigg[\left (3(2q-k^0)^2-k^2\right)\Theta(2q+k-k^0)-
  \left (3(2q+k^0)^2-k^2\right)\Theta(k-2q-k^0)\bigg]\nonumber \\
  &+\Theta(2q-k) \left (3(2q-k^0)^2-k^2\right)\Theta(2q+k-k^0)
  )\Theta(k^0+k-2q)\bigg\}.
    \label{zcexprlogssymthetas}
\end{align}
We start by dealing with the vacuum part
\begin{align}
    &Z_{g\,F\,\text{ L-T non-HTL vac}} ^{(c)}=\frac{g^2 T_F N_F }{8 \pi  }\int_0^\infty dq \, n_\mathrm{F}(q) \int_0^\infty\frac{dk^0}{2\pi}
    \int_{k}\frac{k_\perp^2}{k^5}\bigg[\frac{1}{(k^0-k^z-i\varepsilon)^2}
    +\frac{1}{(k^0+k^z+i\varepsilon)^2}\bigg]\nonumber\\
    &\times
  \bigg\{\Theta(k-2q)\bigg[\left (3(2q-k^0)^2-k^2\right)\Theta(k+2q-k^0)-
  \left (3(2q+k^0)^2-k^2\right)\Theta(k-2q-k^0)\bigg]\nonumber\\
  &+\Theta(2q-k)\left (3(2q-k^0)^2-k^2\right)\Theta(2q+k-k^0)\Theta(k^0-2q+k)
\bigg\},
    \label{zcexprlogssymvac}
\end{align}
which we can simplify as
\begin{align}
    &Z_{g\,F\,\text{ L-T non-HTL vac}} ^{(c)}=\frac{g^2 T_F N_F }{8 \pi  }\int_0^\infty dq \, n_\mathrm{F}(q) \int_0^\infty\frac{dk^0}{2\pi}
    \int_{k}\frac{k_\perp^2}{k^5}\bigg[\frac{1}{(k^0-k^z-i\varepsilon)^2}
    +\frac{1}{(k^0+k^z+i\varepsilon)^2}\bigg]\nonumber\\
    &\times
  \bigg\{\Theta(k-2q)\bigg[\left (3(2q-k^0)^2-k^2\right)\Theta(k+2q-k^0)\Theta(k^0-k+2q)-
  24 q k^0\Theta(k-2q-k^0)\bigg]\nonumber\\
  &+\Theta(2q-k)\left (3(2q-k^0)^2-k^2\right)\Theta(2q+k-k^0)\Theta(k^0-2q+k)
\bigg\},
    \label{zcexprlogssymvacsymp}
\end{align}
The strategy here is the following: we further symmetrise the expression in $k^z$,
to perform the $k^0$ integrations. After that one can take $\varepsilon\to 0$.
Logarithmic pieces will see their argument change to its absolute value, while
possible imaginary parts will cancel out due to $k^z$ symmetrisation (in essence,
the imaginary part is odd in $k^z$ and thus vanishing). Non-logarithmic pieces
present poles on the $k^z$ integration range. If we restrict to twice
the positive $k^z$ range, these occur at $-\vert 2q-k\vert+k^z=0$ and 
are turned into a PV prescription by the $\varepsilon\to 0$ limit. 
We can thus proceed to the $k^z$ integral in the two ranges $2q >k$ and $k>2q$. 
In the latter case, we find two subranges, one for $k_\perp<2q$ and one for 
$k_\perp>2q$. This last one has to be carried out in DR, since it contains a UV divergence. We then find
\begin{align}
    &Z_{g\,F\,\text{ L-T non-HTL vac}}=\frac{g^2 T_F N_F T^2}{48\pi^2}\bigg\{
    \frac{1}{\epsilon^2}+\frac{2}{\epsilon}\ln\frac{\bar\mu A^{12}e^{-1/2}}{16 \pi T}+2 \ln^2\frac{\bar\mu A^{12}e^{-1/2}}{16 \pi T}\nonumber\\
    &+48\ln (A) (\gamma_{E} +\ln (2 \pi )-6\ln(A))
    +\frac{12 \zeta
   ''(2)}{\pi ^2}+\frac{1}{2}
   \text{Li}_2\left(\frac{2}{3}\right)+\frac14\text{Li}_2\left(\frac{3}{4}\right)
   \nonumber\\
    &-\frac12-\frac{11}{2} \ln
   ^2(2)-2 \ln (\pi ) \ln (4 \pi )-2 \gamma_E  \left(\gamma_E +\ln \left(4 \pi
   ^2\right)\right)-\frac12\ln (3) \coth ^{-1}(7)
    \bigg\}.
      \label{zcexprlogssymvacsympfinal}
\end{align}
For the thermal part we have instead
\begin{align}
    &Z_{g\,F\,\text{ L-T non-HTL th}} ^{(c)}=\frac{g^2 T_F N_F }{2 \pi  }\int_0^\infty dq \, n_\mathrm{F}(q) \int_0^\infty\frac{dk^0}{2\pi}
    \int_{0}^\infty \frac{dk}{4\pi^2}\frac{n_\mathrm{B}(k^0)}{k}\nonumber
    \\&\times\bigg[
    \frac{2 k^0 \tanh ^{-1}\left(\frac{2 k k^0}{k^2+k_0^2}\right)}{k^3}-\frac{4}{k^2}\bigg]
    \nonumber\\
    &\times
  \bigg\{\Theta(k-2q)\bigg[\left (3(2q-k^0)^2-k^2\right)\Theta(2q+k-k^0)-
  \left (3(2q+k^0)^2-k^2\right)\Theta(k-2q-k^0)\bigg]\nonumber \\
  &+\Theta(2q-k) \left (3(2q-k^0)^2-k^2\right)\Theta(2q+k-k^0)
  )\Theta(k^0+k-2q)\bigg\}.
    \label{zcexprlogssymthetas3Dthermal}
\end{align}
Here we instead perform $k$ first analytically, followed by $k^0$ and $q$
numerically. We find
\begin{equation}
    Z_{g\,F\,\text{ L-T non-HTL th}} ^{(c)}=\frac{g^2 T_F N_F T^2}{48\pi^2}\bigg[
    \underbrace{-0.438129}_{k<2q}+\underbrace{1.97382}_{k>2q}\bigg].
\end{equation}

Putting everything together we find

\begin{align}
    &Z_{g\,F\,\text{ L-T }} ^{(c)}=\frac{g^2 T_F N_F T^2}{48\pi^2}
    \bigg\{\frac{2}{\epsilon^2}
    +\frac{2\ln  \frac{\bar\mu \exp(\gamma_E+1/2) }{4\pi T}+2(\ln(2)-1) +2\ln\frac{\bar\mu A^{12}e^{-1/2}}{16 \pi T} }{\epsilon}\nonumber \\
   &+ 2 \ln^2\frac{\bar\mu \exp(\gamma_E+1/2) }{4\pi T}-4 \gamma _1+\frac{\pi ^2}{4}
    -2 \gamma_E^2+\frac32
    \nonumber \\
    &+2(\ln(2)-1)\bigg[
   2\ln\frac{\bar\mu\exp(\gamma_E+1/2)}{4\pi T}\bigg]\nonumber \\
   &
    +2 \ln^2\frac{\bar\mu A^{12}e^{-1/2}}{16 \pi T}\nonumber\\
    &+48\ln (A) (\gamma_{E} +\ln (2 \pi )-6\ln(A))
    +\frac{12 \zeta
   ''(2)}{\pi ^2}+\frac{1}{2}
   \text{Li}_2\left(\frac{2}{3}\right)+\frac14\text{Li}_2\left(\frac{3}{4}\right)
   \nonumber\\
    &-\frac{1}{2}-\frac{11}{2} \ln
   ^2(2)-2 \ln (\pi ) \ln (4 \pi )-2 \gamma_E  \left(\gamma_E +\ln \left(4 \pi
   ^2\right)\right)-\frac{1}{2}\ln (3) \coth ^{-1}(7)
    \nonumber \\
    &
    \underbrace{-0.438129}_{k<2q}+\underbrace{1.97382}_{k>2q}
    \bigg\}.
    \label{finalLTfermion_app}
\end{align}
In contrast to the $k^{-2}$ piece, this contribution is 
indeed divergent. Moreover, we hammer home the point that 
these divergences \emph{are not} cancelled by opposing ones coming from the soft 
sector, as EQCD is a theory of gluons only.

\subsection{Gluonic Contribution: \texorpdfstring{$k^{-2}$}{TEXT} Piece}\label{sec:gluon_km}

We persevere with the full contribution to 
diagram $(c)$, starting with the gluonic $k^{-2}$ piece, as we did for 
the fermionic counterpart in Sec.~\ref{sec:fermion_km}.

Our starting point is Eq.~\eqref{gcqminusgmatsum}: we need to be aware 
that there may be divergences, other than the zero-mode contribution, Eq.~\eqref{IRpoleqmsquare}. 
Upon computing the contribution of the non-zero Matsubara modes numerically, we see a (logarithmic) divergence appearing. 
In order to isolate the finite contribution, we perform a subtraction 
on Eq.~\eqref{gcqminusgmatsum} so that we are essentially left with the equivalent fermionic expression (up to the factor of $n_{B}(k)$), which was finite. 
This leaves us with
\begin{align}
    &Z^{(c)}_{g\,k^{-2}\,G\,n\neq 0\,\text{finite}}=-\frac{g^2C_{A}T}{8\pi^2}\sum_{n\neq 0}\int_k\int_{0}^{\infty}dq\frac{1}{k^3}n_{\text{B}}(q)\frac{1}{(-\omega_n^2-k^2)^2}\Bigg(8qk(k^2-\omega_n^2)\nonumber
    \\&-(-\omega_n^2-k^2)(4q^2+k^2-\omega_n^2)\left(\ln\frac{i\omega_n+k-2q}{i\omega_n-k-2q}+\ln\frac{i\omega_n-k+2q}{i\omega_n+k+2q}\right)\nonumber
    \\&-4iq\omega_n(-\omega_n^2-k^2)\left(2\ln\frac{i\omega_n+k}{i\omega_n-k}+\ln\frac{i\omega_n-k+2q}{i\omega_n+k+2q}-\ln\frac{i\omega_n+k-2q}{i\omega_n-k-2q}\right)\Bigg)\nonumber
    \\&=-\frac{g^2C_{A}T^2}{32\pi^2}\times0.00380658
    \label{eq:non_zero_app}
\end{align}
The divergent part then reads
\begin{align}
    Z^{(c)}_{g\,k^{-2}\,G\,n\neq 0\,\text{div}}&=-\frac{g^2C_{A}T}{4\pi^2}\sum_{n\ne 0}\int_k\int_{0}^{\infty}dq\frac{n_{B}(q)}{k(\omega_n^2+k^2)}
    \left(\ln\frac{i\omega_n+k-2q}{i\omega_n-k-2q}+\ln\frac{i\omega_n-k+2q}{i\omega_n+k+2q}\right).
\end{align}
This can be integrated by rephrasing it back in Minkowskian signature
and separating $1/2+n_\mathrm{B}(\vert k^0\vert )$ into a vacuum part, $1/2$,
and a zero-mode subtracted thermal part $n_\mathrm{B}(\vert k^0\vert )-T/|k^0|$,
so as to avoid double-counting the zero mode.
The integration proceeds in much the same way as for the fermion case except here, 
the vacuum part is UV divergent. 
The total reads
\begin{align}
    Z^{(c)}_{g\,k^{-2}\,G\,n\neq 0\,\text{div}}&=
    \frac{g^2 C_A T^2}{48 \pi ^2 } \left(\frac{1}{\epsilon}
    -2 \ln\frac{4\pi  T}{A^{12}\bar \mu} 
    -2.44147 -2.44091\right),
    \label{finaldivpartgqmsquare_app}
\end{align}
where the last two numbers come from the $n_\mathrm{B}(\vert k^0\vert )-T/|k^0|$ contribution 
in the pole and cut channels respectively. Unlike the fermion $k^{-2}$ piece, 
the gluonic one contains a divergence coming from both the zero-mode 
and the non-zero modes. 

Note that the gluonic ``no log'' equivalent (see Eq.~\eqref{qmsquarenologpiece}) is split here between the zero-mode 
contribution, Eq.~\eqref{IRpoleqmsquare} and finite contribution from the 
non-zero modes, Eq.~\eqref{eq:non_zero}\footnote{It can nevertheless easily be obtained analytically
by making the replacement $2N_{F}T_{F}\rightarrow C_{A}$ in Eq.~\eqref{qmsquarenologpiece}.}.

\subsection{Gluonic Contribution: \texorpdfstring{$\Pi_{L}-\Pi_T$}{TEXT} Piece}\label{sec:gluon_lmt}
For the HTL part, the $q$ integrand is given by $q\,n_{\text{B}}(q)$. Therefore, given that
\begin{equation}
    \int_{0}^{\infty}dq q n_{\text{B}}(q)=2\int_{0}^{\infty}dq q n_{\text{F}}(q)=\frac{\pi^2T}{6},
\end{equation}
we may simply make the replacement $2N_{F}T_{F}\rightarrow C_{A}$ in Eq.~\eqref{zcexprhtllogmatsufinal}
to obtain the gluonic equivalent. This yields 
\begin{align}
    Z_{g\,\text{G L-T HTL}}^{(c)}=&
   \frac{g^2 C_A T^2}{48\pi^2}\bigg[\frac{1}{\epsilon^2}
    +\frac{2\ln  \frac{\bar\mu \exp(\gamma_E+1/2) }{4\pi T}  }{\epsilon}
   + 2 \ln^2\frac{\bar\mu \exp(\gamma_E+1/2) }{4\pi T}-4 \gamma _1+\frac{\pi ^2}{4}
    -2 \gamma_E^2+\frac32
    \bigg]\nonumber\\
    &+\frac{g^2 C_A T^2}{24\pi^2}(\ln(2)-1)\bigg[
   \frac{1}{\epsilon}+2\ln\frac{\bar\mu\exp(\gamma_E+1/2)}{4\pi T}\bigg].
    \label{zcexprLTHTLDRfinalg}
\end{align}

For the vacuum part of the non-HTL contribution we have instead
\begin{align}
    &Z_{g\,\text{G L-T non-HTL vac}} ^{(c)}=\frac{g^2 C_A }{16 \pi  }\int_0^\infty dq \, n_\mathrm{B}(q) \int_0^\infty\frac{dk^0}{2\pi}
    \int_{k}\frac{k_\perp^2}{k^5}\bigg[\frac{1}{(k^0-k^z-i\varepsilon)^2}
    +\frac{1}{(k^0+k^z+i\varepsilon)^2}\bigg]\nonumber\\
    &\times
  \bigg\{\Theta(k-2q)\bigg[\left (3(2q-k^0)^2-k^2\right)\Theta(k+2q-k^0)\Theta(k^0-k+2q)-
  24 q k^0\Theta(k-2q-k^0)\bigg]\nonumber\\
  &+\Theta(2q-k)\left (3(2q-k^0)^2-k^2\right)\Theta(2q+k-k^0)\Theta(k^0-2q+k)
\bigg\}.
    \label{zcexprlogssymvacsympg}
\end{align}
Using the methods of Sec.~\ref{sec:fermion_lmt} this gives
\begin{align}
    &Z_{g\,\text{G L-T non-HTL vac}}^{(c)}=\frac{g^2 C_A T^2}{48\pi^2}\bigg\{
    \frac{1}{\epsilon^2}+\frac{2}{\epsilon}\ln\frac{\bar\mu A^{12}e^{-1/2}}{8 \pi T}+2 \ln^2\frac{\bar\mu A^{12}e^{-1/2}}{8 \pi T}\nonumber\\
    &+48\ln (A) (\gamma_{E} +\ln (2 \pi )-6\ln(A))
    +\frac{12 \zeta
   ''(2)}{\pi ^2}+\frac12
   \text{Li}_2\left(\frac{2}{3}\right)+\frac14\text{Li}_2\left(\frac{3}{4}\right)
   \nonumber\\
    &-\frac12-\frac{3}{2} \ln
   ^2(2)-2 \ln (\pi ) \ln (4 \pi )-2 \gamma_E  \left(\gamma_E +\ln \left(4 \pi
   ^2\right)\right)-\frac12\ln (3) \coth ^{-1}(7)
    \bigg\}.
      \label{zcexprlogssymvacsympfinalg}
\end{align}
For the thermal part we have instead to subtract the zero mode, which is evaluated
in Eq.~\eqref{LmTzeromodediv}. No zero-mode subtraction is necessary in the 
HTL part, since the zero mode is scale-free there. This yields
\begin{align}
    &Z_{g\,\text{G L-T non-HTL th}} ^{(c)}=\frac{g^2 C_A }{4 \pi  }\int_0^\infty dq \, n_\mathrm{B}(q) \int_0^\infty\frac{dk^0}{2\pi}
    \int_{0}^\infty \frac{dk }{4\pi^2}\frac{1}{k}\bigg[n_\mathrm{B}(k^0)-\frac{T}{k^0}\bigg]
    \nonumber\\
    &\times\bigg[
        \frac{2 k^0 \tanh ^{-1}\left(\frac{2 k k^0}{k^2+k_0^2}\right)}{k^3}-\frac{4}{k^2}\bigg]
  \bigg\{\Theta(k-2q)\bigg[\left(3(2q-k^0)^2-k^2\right)\Theta(2q+k-k^0)\nonumber \\
  &-
  \left(3(2q+k^0)^2-k^2\right)\Theta(k-2q-k^0)\bigg]\nonumber
  \\&+\Theta(2q-k) \left(3(2q-k^0)^2-k^2\right)\Theta(2q+k-k^0)
  )\Theta(k^0+k-2q)\bigg\}.
    \label{zcexprlogssymthetas3Dthermalg}
\end{align}
Here we instead perform $k$ first analytically, followed by $k^0$ and $q$
numerically. We find
\begin{equation}
    Z_{g\,\text{G L-T non-HTL th}} ^{(c)}=\frac{g^2 C_A T^2}{48\pi^2}\bigg[
    \underbrace{0.145277}_{k<2q}\underbrace{-5.13832}_{k>2q}\bigg].
\end{equation}
The total is then
\begin{align}
    &Z_{g\,\text{G L-T }} ^{(c)}=\frac{g^2 C_A T^2}{48\pi^2}
    \bigg\{\frac{2}{\epsilon^2}
    +\frac{2\ln  \frac{\bar\mu \exp(\gamma_E+1/2) }{4\pi T}+2(\ln(2)-1) +2\ln\frac{\bar\mu A^{12}e^{-1/2}}{8 \pi T} }{\epsilon}\nonumber \\
   &+ 2 \ln^2\frac{\bar\mu \exp(\gamma_E+1/2) }{4\pi T}-4 \gamma _1+\frac{\pi ^2}{4}
    -2 \gamma_E^2+\frac32
    \nonumber \\
    &+2(\ln(2)-1)\bigg[
   2\ln\frac{\bar\mu\exp(\gamma_E+1/2)}{4\pi T}\bigg]\nonumber \\
   &
    +2 \ln^2\frac{\bar\mu A^{12}e^{-1/2}}{8 \pi T}\nonumber\\
    &+48\ln (A) (\gamma_{E} +\ln (2 \pi )-6\ln(A))
    +\frac{12 \zeta
   ''(2)}{\pi ^2}+\frac12
   \text{Li}_2\left(\frac{2}{3}\right)+\frac14\text{Li}_2\left(\frac{3}{4}\right)
   \nonumber\\
    &-\frac12-\frac{3}{2} \ln
   ^2(2)-2 \ln (\pi ) \ln (4 \pi )-2 \gamma_E  \left(\gamma_E +\ln \left(4 \pi
   ^2\right)\right)-\frac12\ln (3) \coth ^{-1}(7)
    \nonumber \\
    &
     +\underbrace{0.145277}_{k<2q}\underbrace{-5.13832}_{k>2q}
    \bigg\}.
    \label{finalLTgluon_app}
\end{align}

\newpage
\bibliographystyle{JHEP}
\bibliography{for_refs.bib}

\providecommand{\href}[2]{#2}\begingroup\raggedright\begin{thebibliography}{100}

\bibitem{Liou:2013qya}
T.~Liou, A.H.~Mueller and B.~Wu, \emph{{Radiative $p_\bot$-broadening of
  high-energy quarks and gluons in QCD matter}},
  \href{https://doi.org/10.1016/j.nuclphysa.2013.08.005}{\emph{Nucl. Phys. A}
  {\bfseries 916} (2013) 102}
  [\href{https://arxiv.org/abs/1304.7677}{{\ttfamily 1304.7677}}].

\bibitem{Blaizot:2013vha}
J.-P.~Blaizot, F.~Dominguez, E.~Iancu and Y.~Mehtar-Tani, \emph{{Probabilistic
  picture for medium-induced jet evolution}},
  \href{https://doi.org/10.1007/JHEP06(2014)075}{\emph{JHEP} {\bfseries 06}
  (2014) 075} [\href{https://arxiv.org/abs/1311.5823}{{\ttfamily 1311.5823}}].

\bibitem{Bjorken:1982tu}
J.D.~Bjorken, \emph{{Energy Loss of Energetic Partons in Quark - Gluon Plasma:
  Possible Extinction of High p(t) Jets in Hadron - Hadron Collisions}}, .

\bibitem{PHENIX:2001hpc}
{\scshape PHENIX} collaboration, \emph{{Suppression of hadrons with large
  transverse momentum in central Au+Au collisions at $\sqrt{s_{NN}}$ =
  130-GeV}}, \href{https://doi.org/10.1103/PhysRevLett.88.022301}{\emph{Phys.
  Rev. Lett.} {\bfseries 88} (2002) 022301}
  [\href{https://arxiv.org/abs/nucl-ex/0109003}{{\ttfamily nucl-ex/0109003}}].

\bibitem{STAR:2002svs}
{\scshape STAR} collaboration, \emph{{Disappearance of back-to-back high
  $p_{T}$ hadron correlations in central Au+Au collisions at $\sqrt{s_{NN}}$ =
  200-GeV}}, \href{https://doi.org/10.1103/PhysRevLett.90.082302}{\emph{Phys.
  Rev. Lett.} {\bfseries 90} (2003) 082302}
  [\href{https://arxiv.org/abs/nucl-ex/0210033}{{\ttfamily nucl-ex/0210033}}].

\bibitem{Mueller:2016gko}
A.H.~Mueller, B.~Wu, B.-W.~Xiao and F.~Yuan, \emph{{Probing Transverse Momentum
  Broadening in Heavy Ion Collisions}},
  \href{https://doi.org/10.1016/j.physletb.2016.10.037}{\emph{Phys. Lett. B}
  {\bfseries 763} (2016) 208}
  [\href{https://arxiv.org/abs/1604.04250}{{\ttfamily 1604.04250}}].

\bibitem{Chen:2016vem}
L.~Chen, G.-Y.~Qin, S.-Y.~Wei, B.-W.~Xiao and H.-Z.~Zhang, \emph{{Probing
  Transverse Momentum Broadening via Dihadron and Hadron-jet Angular
  Correlations in Relativistic Heavy-ion Collisions}},
  \href{https://doi.org/10.1016/j.physletb.2017.09.031}{\emph{Phys. Lett. B}
  {\bfseries 773} (2017) 672}
  [\href{https://arxiv.org/abs/1607.01932}{{\ttfamily 1607.01932}}].

\bibitem{CMS:2017eqd}
{\scshape CMS} collaboration, \emph{{Study of Jet Quenching with $Z+\text{jet}$
  Correlations in Pb-Pb and $pp$ Collisions at ${\sqrt{s}}_{NN}=5.02\text{
  }\text{ }\mathrm{TeV}$}},
  \href{https://doi.org/10.1103/PhysRevLett.119.082301}{\emph{Phys. Rev. Lett.}
  {\bfseries 119} (2017) 082301}
  [\href{https://arxiv.org/abs/1702.01060}{{\ttfamily 1702.01060}}].

\bibitem{Ringer:2019rfk}
F.~Ringer, B.-W.~Xiao and F.~Yuan, \emph{{Can we observe jet $P_T$-broadening
  in heavy-ion collisions at the LHC?}},
  \href{https://doi.org/10.1016/j.physletb.2020.135634}{\emph{Phys. Lett. B}
  {\bfseries 808} (2020) 135634}
  [\href{https://arxiv.org/abs/1907.12541}{{\ttfamily 1907.12541}}].

\bibitem{Chien:2022wiq}
Y.-T.~Chien, R.~Rahn, D.Y.~Shao, W.J.~Waalewijn and B.~Wu, \emph{{Precision
  boson-jet azimuthal decorrelation at hadron colliders}},
  \href{https://doi.org/10.1007/JHEP02(2023)256}{\emph{JHEP} {\bfseries 02}
  (2023) 256} [\href{https://arxiv.org/abs/2205.05104}{{\ttfamily
  2205.05104}}].

\bibitem{CaronHuot:2008ni}
S.~Caron-Huot, \emph{{O(g) plasma effects in jet quenching}},
  \href{https://doi.org/10.1103/PhysRevD.79.065039}{\emph{Phys. Rev. D}
  {\bfseries 79} (2009) 065039}
  [\href{https://arxiv.org/abs/0811.1603}{{\ttfamily 0811.1603}}].

\bibitem{Moore:2019lgw}
G.D.~Moore and N.~Schlusser, \emph{{Transverse momentum broadening from the
  lattice}}, \href{https://doi.org/10.1103/PhysRevD.101.014505}{\emph{Phys.
  Rev. D} {\bfseries 101} (2020) 014505}
  [\href{https://arxiv.org/abs/1911.13127}{{\ttfamily 1911.13127}}].

\bibitem{Moore:2021jwe}
G.D.~Moore, S.~Schlichting, N.~Schlusser and I.~Soudi, \emph{{Non-perturbative
  determination of collisional broadening and medium induced radiation in QCD
  plasmas}}, \href{https://doi.org/10.1007/JHEP10(2021)059}{\emph{JHEP}
  {\bfseries 10} (2021) 059}
  [\href{https://arxiv.org/abs/2105.01679}{{\ttfamily 2105.01679}}].

\bibitem{Schlichting:2021idr}
S.~Schlichting and I.~Soudi, \emph{{Splitting rates in QCD plasmas from a
  nonperturbative determination of the momentum broadening kernel
  C(q\ensuremath{\perp})}},
  \href{https://doi.org/10.1103/PhysRevD.105.076002}{\emph{Phys. Rev. D}
  {\bfseries 105} (2022) 076002}
  [\href{https://arxiv.org/abs/2111.13731}{{\ttfamily 2111.13731}}].

\bibitem{Moore:2020wvy}
G.D.~Moore and N.~Schlusser, \emph{{The nonperturbative contribution to
  asymptotic masses}},
  \href{https://doi.org/10.1103/PhysRevD.102.094512}{\emph{Phys. Rev. D}
  {\bfseries 102} (2020) 094512}
  [\href{https://arxiv.org/abs/2009.06614}{{\ttfamily 2009.06614}}].

\bibitem{Ghiglieri:2021bom}
J.~Ghiglieri, G.D.~Moore, P.~Schicho and N.~Schlusser, \emph{{The
  force-force-correlator in hot QCD perturbatively and from the lattice}},
  \href{https://doi.org/10.1007/JHEP02(2022)058}{\emph{JHEP} {\bfseries 02}
  (2022) 058} [\href{https://arxiv.org/abs/2112.01407}{{\ttfamily
  2112.01407}}].

\bibitem{Caron-Huot:2007zhp}
S.~Caron-Huot, \emph{{Heavy quark energy losses in the quark-gluon plasma :
  beyond leading order}},  Master's thesis, McGill U., 2007.

\bibitem{Ghiglieri:2022gyv}
J.~Ghiglieri and E.~Weitz, \emph{{Classical vs quantum corrections to jet
  broadening in a weakly-coupled Quark-Gluon Plasma}},
  \href{https://doi.org/10.1007/JHEP11(2022)068}{\emph{JHEP} {\bfseries 11}
  (2022) 068} [\href{https://arxiv.org/abs/2207.08842}{{\ttfamily
  2207.08842}}].

\bibitem{Yang:1954ek}
C.-N.~Yang and R.L.~Mills, \emph{{Conservation of Isotopic Spin and Isotopic
  Gauge Invariance}}, \href{https://doi.org/10.1103/PhysRev.96.191}{\emph{Phys.
  Rev.} {\bfseries 96} (1954) 191}.

\bibitem{Fritzsch:1973pi}
H.~Fritzsch, M.~Gell-Mann and H.~Leutwyler, \emph{{Advantages of the Color
  Octet Gluon Picture}},
  \href{https://doi.org/10.1016/0370-2693(73)90625-4}{\emph{Phys. Lett. B}
  {\bfseries 47} (1973) 365}.

\bibitem{Peskin:1995ev}
M.E.~Peskin and D.V.~Schroeder, \emph{{An Introduction to quantum field
  theory}}, Addison-Wesley, Reading, USA (1995).

\bibitem{Srednicki:2007qs}
M.~Srednicki, \emph{{Quantum field theory}}, Cambridge University Press (1,
  2007).

\bibitem{Schwartz:2014sze}
M.D.~Schwartz, \emph{{Quantum Field Theory and the Standard Model}}, Cambridge
  University Press (3, 2014).

\bibitem{Politzer:1973fx}
H.D.~Politzer, \emph{{Reliable Perturbative Results for Strong Interactions?}},
  \href{https://doi.org/10.1103/PhysRevLett.30.1346}{\emph{Phys. Rev. Lett.}
  {\bfseries 30} (1973) 1346}.

\bibitem{Gross:1973id}
D.J.~Gross and F.~Wilczek, \emph{{Ultraviolet Behavior of Nonabelian Gauge
  Theories}}, \href{https://doi.org/10.1103/PhysRevLett.30.1343}{\emph{Phys.
  Rev. Lett.} {\bfseries 30} (1973) 1343}.

\bibitem{Baikov:2016tgj}
P.A.~Baikov, K.G.~Chetyrkin and J.H.~K\"uhn, \emph{{Five-Loop Running of the
  QCD coupling constant}},
  \href{https://doi.org/10.1103/PhysRevLett.118.082002}{\emph{Phys. Rev. Lett.}
  {\bfseries 118} (2017) 082002}
  [\href{https://arxiv.org/abs/1606.08659}{{\ttfamily 1606.08659}}].

\bibitem{Luthe:2016ima}
T.~Luthe, A.~Maier, P.~Marquard and Y.~Schr\"oder, \emph{{Towards the five-loop
  Beta function for a general gauge group}},
  \href{https://doi.org/10.1007/JHEP07(2016)127}{\emph{JHEP} {\bfseries 07}
  (2016) 127} [\href{https://arxiv.org/abs/1606.08662}{{\ttfamily
  1606.08662}}].

\bibitem{Luthe:2017ttg}
T.~Luthe, A.~Maier, P.~Marquard and Y.~Schroder, \emph{{The five-loop Beta
  function for a general gauge group and anomalous dimensions beyond Feynman
  gauge}}, \href{https://doi.org/10.1007/JHEP10(2017)166}{\emph{JHEP}
  {\bfseries 10} (2017) 166}
  [\href{https://arxiv.org/abs/1709.07718}{{\ttfamily 1709.07718}}].

\bibitem{Chetyrkin:2017bjc}
K.G.~Chetyrkin, G.~Falcioni, F.~Herzog and J.A.M.~Vermaseren, \emph{{Five-loop
  renormalisation of QCD in covariant gauges}},
  \href{https://doi.org/10.1007/JHEP10(2017)179}{\emph{JHEP} {\bfseries 10}
  (2017) 179} [\href{https://arxiv.org/abs/1709.08541}{{\ttfamily
  1709.08541}}].

\bibitem{ParticleDataGroup:2018ovx}
{\scshape Particle Data Group} collaboration, \emph{{Review of Particle
  Physics}}, \href{https://doi.org/10.1103/PhysRevD.98.030001}{\emph{Phys. Rev.
  D} {\bfseries 98} (2018) 030001}.

\bibitem{PhysRev.177.2426}
S.L.~Adler, \emph{Axial-vector vertex in spinor electrodynamics},
  \href{https://doi.org/10.1103/PhysRev.177.2426}{\emph{Phys. Rev.} {\bfseries
  177} (1969) 2426}.

\bibitem{Fujikawa:1979ay}
K.~Fujikawa, \emph{{Path Integral Measure for Gauge Invariant Fermion
  Theories}}, \href{https://doi.org/10.1103/PhysRevLett.42.1195}{\emph{Phys.
  Rev. Lett.} {\bfseries 42} (1979) 1195}.

\bibitem{Giusti:1998wy}
L.~Giusti, F.~Rapuano, M.~Talevi and A.~Vladikas, \emph{{The QCD chiral
  condensate from the lattice}},
  \href{https://doi.org/10.1016/S0550-3213(98)00659-2}{\emph{Nucl. Phys. B}
  {\bfseries 538} (1999) 249}
  [\href{https://arxiv.org/abs/hep-lat/9807014}{{\ttfamily hep-lat/9807014}}].

\bibitem{Feynman:1969wa}
R.P.~Feynman, \emph{{The behavior of hadron collisions at extreme energies}},
  {\emph{Conf. Proc. C} {\bfseries 690905} (1969) 237}.

\bibitem{PhysRevD.13.1958}
E.C.~Poggio, H.R.~Quinn and S.~Weinberg, \emph{Smearing method in the quark
  model}, \href{https://doi.org/10.1103/PhysRevD.13.1958}{\emph{Phys. Rev. D}
  {\bfseries 13} (1976) 1958}.

\bibitem{Blaizot:2015lma}
J.-P.~Blaizot and Y.~Mehtar-Tani, \emph{{Jet Structure in Heavy Ion
  Collisions}}, \href{https://doi.org/10.1142/S021830131530012X}{\emph{Int. J.
  Mod. Phys. E} {\bfseries 24} (2015) 1530012}
  [\href{https://arxiv.org/abs/1503.05958}{{\ttfamily 1503.05958}}].

\bibitem{Ellis:1996mzs}
R.K.~Ellis, W.J.~Stirling and B.R.~Webber, \emph{{QCD and collider physics}},
  vol.~8, Cambridge University Press (2, 2011),
  \href{https://doi.org/10.1017/CBO9780511628788}{10.1017/CBO9780511628788}.

\bibitem{Ellis:1976uc}
J.R.~Ellis, M.K.~Gaillard and G.G.~Ross, \emph{{Search for Gluons in e+ e-
  Annihilation}},
  \href{https://doi.org/10.1016/0550-3213(77)90253-X}{\emph{Nucl. Phys. B}
  {\bfseries 111} (1976) 253}.

\bibitem{Wiik:1979cq}
B.H.~Wiik, \emph{{First Results from PETRA}}, {\emph{Conf. Proc. C} {\bfseries
  7906181} (1979) 113}.

\bibitem{Hoche:2014rga}
S.~H\"oche, \emph{{Introduction to parton-shower event generators}},  in
  \emph{{Theoretical Advanced Study Institute in Elementary Particle Physics}:
  {Journeys Through the Precision Frontier: Amplitudes for Colliders}},
  pp.~235--295, 2015, \href{https://doi.org/10.1142/9789814678766_0005}{DOI}
  [\href{https://arxiv.org/abs/1411.4085}{{\ttfamily 1411.4085}}].

\bibitem{Bierlich:2022pfr}
C.~Bierlich et~al., \emph{{A comprehensive guide to the physics and usage of
  PYTHIA 8.3}},  \href{https://arxiv.org/abs/2203.11601}{{\ttfamily
  2203.11601}}.

\bibitem{Bellm:2019zci}
J.~Bellm et~al., \emph{{Herwig 7.2 release note}},
  \href{https://doi.org/10.1140/epjc/s10052-020-8011-x}{\emph{Eur. Phys. J. C}
  {\bfseries 80} (2020) 452}
  [\href{https://arxiv.org/abs/1912.06509}{{\ttfamily 1912.06509}}].

\bibitem{Sherpa:2019gpd}
{\scshape Sherpa} collaboration, \emph{{Event Generation with Sherpa 2.2}},
  \href{https://doi.org/10.21468/SciPostPhys.7.3.034}{\emph{SciPost Phys.}
  {\bfseries 7} (2019) 034} [\href{https://arxiv.org/abs/1905.09127}{{\ttfamily
  1905.09127}}].

\bibitem{Campbell:2022qmc}
J.M.~Campbell et~al., \emph{{Event Generators for High-Energy Physics
  Experiments}},  in \emph{{Snowmass 2021}}, 3, 2022
  [\href{https://arxiv.org/abs/2203.11110}{{\ttfamily 2203.11110}}].

\bibitem{Dokshitzer:1997in}
Y.L.~Dokshitzer, G.D.~Leder, S.~Moretti and B.R.~Webber, \emph{{Better jet
  clustering algorithms}},
  \href{https://doi.org/10.1088/1126-6708/1997/08/001}{\emph{JHEP} {\bfseries
  08} (1997) 001} [\href{https://arxiv.org/abs/hep-ph/9707323}{{\ttfamily
  hep-ph/9707323}}].

\bibitem{Catani:1993hr}
S.~Catani, Y.L.~Dokshitzer, M.H.~Seymour and B.R.~Webber, \emph{{Longitudinally
  invariant $K_t$ clustering algorithms for hadron hadron collisions}},
  \href{https://doi.org/10.1016/0550-3213(93)90166-M}{\emph{Nucl. Phys. B}
  {\bfseries 406} (1993) 187}.

\bibitem{Cacciari:2008gp}
M.~Cacciari, G.P.~Salam and G.~Soyez, \emph{{The anti-$k_t$ jet clustering
  algorithm}}, \href{https://doi.org/10.1088/1126-6708/2008/04/063}{\emph{JHEP}
  {\bfseries 04} (2008) 063} [\href{https://arxiv.org/abs/0802.1189}{{\ttfamily
  0802.1189}}].

\bibitem{Salam:2007xv}
G.P.~Salam and G.~Soyez, \emph{{A Practical Seedless Infrared-Safe Cone jet
  algorithm}}, \href{https://doi.org/10.1088/1126-6708/2007/05/086}{\emph{JHEP}
  {\bfseries 05} (2007) 086} [\href{https://arxiv.org/abs/0704.0292}{{\ttfamily
  0704.0292}}].

\bibitem{Salam:2010nqg}
G.P.~Salam, \emph{{Towards Jetography}},
  \href{https://doi.org/10.1140/epjc/s10052-010-1314-6}{\emph{Eur. Phys. J. C}
  {\bfseries 67} (2010) 637} [\href{https://arxiv.org/abs/0906.1833}{{\ttfamily
  0906.1833}}].

\bibitem{ATLAS:2017bfj}
{\scshape ATLAS} collaboration, \emph{{Search for dark matter and other new
  phenomena in events with an energetic jet and large missing transverse
  momentum using the ATLAS detector}},
  \href{https://doi.org/10.1007/JHEP01(2018)126}{\emph{JHEP} {\bfseries 01}
  (2018) 126} [\href{https://arxiv.org/abs/1711.03301}{{\ttfamily
  1711.03301}}].

\bibitem{CMS:2017zts}
{\scshape CMS} collaboration, \emph{{Search for new physics in final states
  with an energetic jet or a hadronically decaying $W$ or $Z$ boson and
  transverse momentum imbalance at $\sqrt{s}=13\text{ }\text{ }\mathrm{TeV}$}},
  \href{https://doi.org/10.1103/PhysRevD.97.092005}{\emph{Phys. Rev. D}
  {\bfseries 97} (2018) 092005}
  [\href{https://arxiv.org/abs/1712.02345}{{\ttfamily 1712.02345}}].

\bibitem{ATLAS:2018nda}
{\scshape ATLAS} collaboration, \emph{{Search for dark matter in events with a
  hadronically decaying vector boson and missing transverse momentum in $pp$
  collisions at $\sqrt{s} = 13$ TeV with the ATLAS detector}},
  \href{https://doi.org/10.1007/JHEP10(2018)180}{\emph{JHEP} {\bfseries 10}
  (2018) 180} [\href{https://arxiv.org/abs/1807.11471}{{\ttfamily
  1807.11471}}].

\bibitem{CMS:2021snz}
{\scshape CMS} collaboration, \emph{{Search for new particles in events with
  energetic jets and large missing transverse momentum in proton-proton
  collisions at $\sqrt{s}=13~\mathrm{TeV}$}}, .

\bibitem{ATLAS:2023swa}
{\scshape ATLAS} collaboration, \emph{{Search for non-resonant production of
  semi-visible jets using Run\textasciitilde{}2 data in ATLAS}},
  \href{https://arxiv.org/abs/2305.18037}{{\ttfamily 2305.18037}}.

\bibitem{Elfner:2022iae}
H.~Elfner and B.~M\"uller, \emph{{The exploration of hot and dense nuclear
  matter: introduction to relativistic heavy-ion physics}},
  \href{https://doi.org/10.1088/1361-6471/ace824}{\emph{J. Phys. G} {\bfseries
  50} (2023) 103001} [\href{https://arxiv.org/abs/2210.12056}{{\ttfamily
  2210.12056}}].

\bibitem{Gamow1930MassDC}
G.~Gamow, \emph{Mass defect curve and nuclear constitution}, {\emph{Proceedings
  of The Royal Society A: Mathematical, Physical and Engineering Sciences}
  {\bfseries 126} (1930) 632}.

\bibitem{Weizsacker:1935bkz}
C.F.V.~Weizsacker, \emph{{Zur Theorie der Kernmassen}},
  \href{https://doi.org/10.1007/BF01337700}{\emph{Z. Phys.} {\bfseries 96}
  (1935) 431}.

\bibitem{Karnaukhov:2003vp}
V.A.~Karnaukhov et~al., \emph{{Critical temperature for the nuclear liquid gas
  phase transition}},
  \href{https://doi.org/10.1103/PhysRevC.67.011601}{\emph{Phys. Rev. C}
  {\bfseries 67} (2003) 011601}
  [\href{https://arxiv.org/abs/nucl-ex/0302006}{{\ttfamily nucl-ex/0302006}}].

\bibitem{Hagedorn:1965st}
R.~Hagedorn, \emph{{Statistical thermodynamics of strong interactions at
  high-energies}}, {\emph{Nuovo Cim. Suppl.} {\bfseries 3} (1965) 147}.

\bibitem{Cabibbo:1975ig}
N.~Cabibbo and G.~Parisi, \emph{{Exponential Hadronic Spectrum and Quark
  Liberation}}, \href{https://doi.org/10.1016/0370-2693(75)90158-6}{\emph{Phys.
  Lett. B} {\bfseries 59} (1975) 67}.

\bibitem{Fukushima:2013rx}
K.~Fukushima and C.~Sasaki, \emph{{The phase diagram of nuclear and quark
  matter at high baryon density}},
  \href{https://doi.org/10.1016/j.ppnp.2013.05.003}{\emph{Prog. Part. Nucl.
  Phys.} {\bfseries 72} (2013) 99}
  [\href{https://arxiv.org/abs/1301.6377}{{\ttfamily 1301.6377}}].

\bibitem{Fischer:2018sdj}
C.S.~Fischer, \emph{{QCD at finite temperature and chemical potential from
  Dyson\textendash{}Schwinger equations}},
  \href{https://doi.org/10.1016/j.ppnp.2019.01.002}{\emph{Prog. Part. Nucl.
  Phys.} {\bfseries 105} (2019) 1}
  [\href{https://arxiv.org/abs/1810.12938}{{\ttfamily 1810.12938}}].

\bibitem{Gelis:2021zmx}
F.~Gelis, \emph{{Some Aspects of the Theory of Heavy Ion Collisions}},
  \href{https://doi.org/10.1088/1361-6633/abec2e}{\emph{Rept. Prog. Phys.}
  {\bfseries 84} (2021) 056301}
  [\href{https://arxiv.org/abs/2102.07604}{{\ttfamily 2102.07604}}].

\bibitem{Polyakov:1975rs}
A.M.~Polyakov, \emph{{Compact Gauge Fields and the Infrared Catastrophe}},
  \href{https://doi.org/10.1016/0370-2693(75)90162-8}{\emph{Phys. Lett. B}
  {\bfseries 59} (1975) 82}.

\bibitem{Kuti:1980gh}
J.~Kuti, J.~Polonyi and K.~Szlachanyi, \emph{{Monte Carlo Study of SU(2) Gauge
  Theory at Finite Temperature}},
  \href{https://doi.org/10.1016/0370-2693(81)90987-4}{\emph{Phys. Lett. B}
  {\bfseries 98} (1981) 199}.

\bibitem{McLerran:1981pb}
L.D.~McLerran and B.~Svetitsky, \emph{{Quark Liberation at High Temperature: A
  Monte Carlo Study of SU(2) Gauge Theory}},
  \href{https://doi.org/10.1103/PhysRevD.24.450}{\emph{Phys. Rev. D} {\bfseries
  24} (1981) 450}.

\bibitem{Borsanyi:2010bp}
{\scshape Wuppertal-Budapest} collaboration, \emph{{Is there still any $T_c$
  mystery in lattice QCD? Results with physical masses in the continuum limit
  III}}, \href{https://doi.org/10.1007/JHEP09(2010)073}{\emph{JHEP} {\bfseries
  09} (2010) 073} [\href{https://arxiv.org/abs/1005.3508}{{\ttfamily
  1005.3508}}].

\bibitem{PhysRevLett.65.2491}
F.R.~Brown, F.P.~Butler, H.~Chen, N.H.~Christ, Z.~Dong, W.~Schaffer et~al.,
  \emph{On the existence of a phase transition for qcd with three light
  quarks}, \href{https://doi.org/10.1103/PhysRevLett.65.2491}{\emph{Phys. Rev.
  Lett.} {\bfseries 65} (1990) 2491}.

\bibitem{HotQCD:2018pds}
{\scshape HotQCD} collaboration, \emph{{Chiral crossover in QCD at zero and
  non-zero chemical potentials}},
  \href{https://doi.org/10.1016/j.physletb.2019.05.013}{\emph{Phys. Lett. B}
  {\bfseries 795} (2019) 15}
  [\href{https://arxiv.org/abs/1812.08235}{{\ttfamily 1812.08235}}].

\bibitem{Nagata:2021ugx}
K.~Nagata, \emph{{Finite-density lattice QCD and sign problem: Current status
  and open problems}},
  \href{https://doi.org/10.1016/j.ppnp.2022.103991}{\emph{Prog. Part. Nucl.
  Phys.} {\bfseries 127} (2022) 103991}
  [\href{https://arxiv.org/abs/2108.12423}{{\ttfamily 2108.12423}}].

\bibitem{Hasenfratz:1983ba}
P.~Hasenfratz and F.~Karsch, \emph{{Chemical Potential on the Lattice}},
  \href{https://doi.org/10.1016/0370-2693(83)91290-X}{\emph{Phys. Lett. B}
  {\bfseries 125} (1983) 308}.

\bibitem{Alford:1998sd}
M.G.~Alford, A.~Kapustin and F.~Wilczek, \emph{{Imaginary chemical potential
  and finite fermion density on the lattice}},
  \href{https://doi.org/10.1103/PhysRevD.59.054502}{\emph{Phys. Rev. D}
  {\bfseries 59} (1999) 054502}
  [\href{https://arxiv.org/abs/hep-lat/9807039}{{\ttfamily hep-lat/9807039}}].

\bibitem{Hands:1999md}
S.~Hands, J.B.~Kogut, M.-P.~Lombardo and S.E.~Morrison, \emph{{Symmetries and
  spectrum of SU(2) lattice gauge theory at finite chemical potential}},
  \href{https://doi.org/10.1016/S0550-3213(99)00364-8}{\emph{Nucl. Phys. B}
  {\bfseries 558} (1999) 327}
  [\href{https://arxiv.org/abs/hep-lat/9902034}{{\ttfamily hep-lat/9902034}}].

\bibitem{Fodor:2001au}
Z.~Fodor and S.D.~Katz, \emph{{A New method to study lattice QCD at finite
  temperature and chemical potential}},
  \href{https://doi.org/10.1016/S0370-2693(02)01583-6}{\emph{Phys. Lett. B}
  {\bfseries 534} (2002) 87}
  [\href{https://arxiv.org/abs/hep-lat/0104001}{{\ttfamily hep-lat/0104001}}].

\bibitem{Gupta:2004pk}
S.~Gupta, \emph{{Lattice QCD with chemical potential: Evading the fermion-sign
  problem}}, \href{https://doi.org/10.1007/BF02704891}{\emph{Pramana}
  {\bfseries 63} (2004) 1211}.

\bibitem{Fukushima:2003fw}
K.~Fukushima, \emph{{Chiral effective model with the Polyakov loop}},
  \href{https://doi.org/10.1016/j.physletb.2004.04.027}{\emph{Phys. Lett. B}
  {\bfseries 591} (2004) 277}
  [\href{https://arxiv.org/abs/hep-ph/0310121}{{\ttfamily hep-ph/0310121}}].

\bibitem{Ratti:2005jh}
C.~Ratti, M.A.~Thaler and W.~Weise, \emph{{Phases of QCD: Lattice
  thermodynamics and a field theoretical model}},
  \href{https://doi.org/10.1103/PhysRevD.73.014019}{\emph{Phys. Rev. D}
  {\bfseries 73} (2006) 014019}
  [\href{https://arxiv.org/abs/hep-ph/0506234}{{\ttfamily hep-ph/0506234}}].

\bibitem{Borsanyi:2013bia}
S.~Borsanyi, Z.~Fodor, C.~Hoelbling, S.D.~Katz, S.~Krieg and K.K.~Szabo,
  \emph{{Full result for the QCD equation of state with 2+1 flavors}},
  \href{https://doi.org/10.1016/j.physletb.2014.01.007}{\emph{Phys. Lett. B}
  {\bfseries 730} (2014) 99} [\href{https://arxiv.org/abs/1309.5258}{{\ttfamily
  1309.5258}}].

\bibitem{Alford:1997zt}
M.G.~Alford, K.~Rajagopal and F.~Wilczek, \emph{{QCD at finite baryon density:
  Nucleon droplets and color superconductivity}},
  \href{https://doi.org/10.1016/S0370-2693(98)00051-3}{\emph{Phys. Lett. B}
  {\bfseries 422} (1998) 247}
  [\href{https://arxiv.org/abs/hep-ph/9711395}{{\ttfamily hep-ph/9711395}}].

\bibitem{Berges:1998rc}
J.~Berges and K.~Rajagopal, \emph{{Color superconductivity and chiral symmetry
  restoration at nonzero baryon density and temperature}},
  \href{https://doi.org/10.1016/S0550-3213(98)00620-8}{\emph{Nucl. Phys. B}
  {\bfseries 538} (1999) 215}
  [\href{https://arxiv.org/abs/hep-ph/9804233}{{\ttfamily hep-ph/9804233}}].

\bibitem{Son:1998uk}
D.T.~Son, \emph{{Superconductivity by long range color magnetic interaction in
  high density quark matter}},
  \href{https://doi.org/10.1103/PhysRevD.59.094019}{\emph{Phys. Rev. D}
  {\bfseries 59} (1999) 094019}
  [\href{https://arxiv.org/abs/hep-ph/9812287}{{\ttfamily hep-ph/9812287}}].

\bibitem{Alford:2007xm}
M.G.~Alford, A.~Schmitt, K.~Rajagopal and T.~Sch\"afer, \emph{{Color
  superconductivity in dense quark matter}},
  \href{https://doi.org/10.1103/RevModPhys.80.1455}{\emph{Rev. Mod. Phys.}
  {\bfseries 80} (2008) 1455}
  [\href{https://arxiv.org/abs/0709.4635}{{\ttfamily 0709.4635}}].

\bibitem{Annala:2019puf}
E.~Annala, T.~Gorda, A.~Kurkela, J.~N\"attil\"a and A.~Vuorinen,
  \emph{{Evidence for quark-matter cores in massive neutron stars}},
  \href{https://doi.org/10.1038/s41567-020-0914-9}{\emph{Nature Phys.}
  {\bfseries 16} (2020) 907}
  [\href{https://arxiv.org/abs/1903.09121}{{\ttfamily 1903.09121}}].

\bibitem{PhysRevLett.119.161101}
{\scshape LIGO Scientific Collaboration and Virgo Collaboration} collaboration,
  \emph{Gw170817: Observation of gravitational waves from a binary neutron star
  inspiral}, \href{https://doi.org/10.1103/PhysRevLett.119.161101}{\emph{Phys.
  Rev. Lett.} {\bfseries 119} (2017) 161101}.

\bibitem{PhysRevLett.121.161101}
{\scshape The LIGO Scientific Collaboration and the Virgo Collaboration}
  collaboration, \emph{Gw170817: Measurements of neutron star radii and
  equation of state},
  \href{https://doi.org/10.1103/PhysRevLett.121.161101}{\emph{Phys. Rev. Lett.}
  {\bfseries 121} (2018) 161101}.

\bibitem{Shuryak:1980tp}
E.V.~Shuryak, \emph{{Quantum Chromodynamics and the Theory of Superdense
  Matter}}, \href{https://doi.org/10.1016/0370-1573(80)90105-2}{\emph{Phys.
  Rept.} {\bfseries 61} (1980) 71}.

\bibitem{Gardim:2019xjs}
F.G.~Gardim, G.~Giacalone, M.~Luzum and J.-Y.~Ollitrault, \emph{{Thermodynamics
  of hot strong-interaction matter from ultrarelativistic nuclear collisions}},
  \href{https://doi.org/10.1038/s41567-020-0846-4}{\emph{Nature Phys.}
  {\bfseries 16} (2020) 615}
  [\href{https://arxiv.org/abs/1908.09728}{{\ttfamily 1908.09728}}].

\bibitem{Iancu:2012xa}
E.~Iancu, \emph{{QCD in heavy ion collisions}},  in \emph{{2011 European School
  of High-Energy Physics}}, pp.~197--266, 2014,
  \href{https://doi.org/10.5170/CERN-2014-003.197}{DOI}
  [\href{https://arxiv.org/abs/1205.0579}{{\ttfamily 1205.0579}}].

\bibitem{Busza:2018rrf}
W.~Busza, K.~Rajagopal and W.~van~der Schee, \emph{{Heavy Ion Collisions: The
  Big Picture, and the Big Questions}},
  \href{https://doi.org/10.1146/annurev-nucl-101917-020852}{\emph{Ann. Rev.
  Nucl. Part. Sci.} {\bfseries 68} (2018) 339}
  [\href{https://arxiv.org/abs/1802.04801}{{\ttfamily 1802.04801}}].

\bibitem{Bjorken:1982qr}
J.D.~Bjorken, \emph{{Highly Relativistic Nucleus-Nucleus Collisions: The
  Central Rapidity Region}},
  \href{https://doi.org/10.1103/PhysRevD.27.140}{\emph{Phys. Rev. D} {\bfseries
  27} (1983) 140}.

\bibitem{McLerran:1993ni}
L.D.~McLerran and R.~Venugopalan, \emph{{Computing quark and gluon distribution
  functions for very large nuclei}},
  \href{https://doi.org/10.1103/PhysRevD.49.2233}{\emph{Phys. Rev. D}
  {\bfseries 49} (1994) 2233}
  [\href{https://arxiv.org/abs/hep-ph/9309289}{{\ttfamily hep-ph/9309289}}].

\bibitem{McLerran:1993ka}
L.D.~McLerran and R.~Venugopalan, \emph{{Gluon distribution functions for very
  large nuclei at small transverse momentum}},
  \href{https://doi.org/10.1103/PhysRevD.49.3352}{\emph{Phys. Rev. D}
  {\bfseries 49} (1994) 3352}
  [\href{https://arxiv.org/abs/hep-ph/9311205}{{\ttfamily hep-ph/9311205}}].

\bibitem{Gelis:2010nm}
F.~Gelis, E.~Iancu, J.~Jalilian-Marian and R.~Venugopalan, \emph{{The Color
  Glass Condensate}},
  \href{https://doi.org/10.1146/annurev.nucl.010909.083629}{\emph{Ann. Rev.
  Nucl. Part. Sci.} {\bfseries 60} (2010) 463}
  [\href{https://arxiv.org/abs/1002.0333}{{\ttfamily 1002.0333}}].

\bibitem{Baier:2000sb}
R.~Baier, A.H.~Mueller, D.~Schiff and D.T.~Son, \emph{{'Bottom up'
  thermalization in heavy ion collisions}},
  \href{https://doi.org/10.1016/S0370-2693(01)00191-5}{\emph{Phys. Lett. B}
  {\bfseries 502} (2001) 51}
  [\href{https://arxiv.org/abs/hep-ph/0009237}{{\ttfamily hep-ph/0009237}}].

\bibitem{Kurkela:2018vqr}
A.~Kurkela, A.~Mazeliauskas, J.-F.~Paquet, S.~Schlichting and D.~Teaney,
  \emph{{Effective kinetic description of event-by-event pre-equilibrium
  dynamics in high-energy heavy-ion collisions}},
  \href{https://doi.org/10.1103/PhysRevC.99.034910}{\emph{Phys. Rev. C}
  {\bfseries 99} (2019) 034910}
  [\href{https://arxiv.org/abs/1805.00961}{{\ttfamily 1805.00961}}].

\bibitem{ATLAS:2010isq}
{\scshape ATLAS} collaboration, \emph{{Observation of a Centrality-Dependent
  Dijet Asymmetry in Lead-Lead Collisions at $\sqrt{s_{NN}}=2.77$ TeV with the
  ATLAS Detector at the LHC}},
  \href{https://doi.org/10.1103/PhysRevLett.105.252303}{\emph{Phys. Rev. Lett.}
  {\bfseries 105} (2010) 252303}
  [\href{https://arxiv.org/abs/1011.6182}{{\ttfamily 1011.6182}}].

\bibitem{CMS:2011iwn}
{\scshape CMS} collaboration, \emph{{Observation and studies of jet quenching
  in PbPb collisions at nucleon-nucleon center-of-mass energy = 2.76 TeV}},
  \href{https://doi.org/10.1103/PhysRevC.84.024906}{\emph{Phys. Rev. C}
  {\bfseries 84} (2011) 024906}
  [\href{https://arxiv.org/abs/1102.1957}{{\ttfamily 1102.1957}}].

\bibitem{Wang:1998bha}
X.-N.~Wang, \emph{{Effect of jet quenching on high $p_{T}$ hadron spectra in
  high-energy nuclear collisions}},
  \href{https://doi.org/10.1103/PhysRevC.58.2321}{\emph{Phys. Rev. C}
  {\bfseries 58} (1998) 2321}
  [\href{https://arxiv.org/abs/hep-ph/9804357}{{\ttfamily hep-ph/9804357}}].

\bibitem{Shukla:2001mb}
P.~Shukla, \emph{{Glauber model for heavy ion collisions from low-energies to
  high-energies}},  \href{https://arxiv.org/abs/nucl-th/0112039}{{\ttfamily
  nucl-th/0112039}}.

\bibitem{Stock:2020blh}
R.~Stock, \emph{{ROY GLAUBER: In Memoriam the Glauber Model in High Energy
  Nucleus-Nucleus Collisions}},
  \href{https://doi.org/10.1007/978-3-030-53448-6_1}{\emph{Springer Proc.
  Phys.} {\bfseries 250} (2020) 3}.

\bibitem{Cunqueiro:2021wls}
L.~Cunqueiro and A.M.~Sickles, \emph{{Studying the QGP with Jets at the LHC and
  RHIC}}, \href{https://doi.org/10.1016/j.ppnp.2022.103940}{\emph{Prog. Part.
  Nucl. Phys.} {\bfseries 124} (2022) 103940}
  [\href{https://arxiv.org/abs/2110.14490}{{\ttfamily 2110.14490}}].

\bibitem{Apolinario:2022vzg}
L.~Apolin\'ario, Y.-J.~Lee and M.~Winn, \emph{{Heavy quarks and jets as probes
  of the QGP}}, \href{https://doi.org/10.1016/j.ppnp.2022.103990}{\emph{Prog.
  Part. Nucl. Phys.} {\bfseries 127} (2022) 103990}
  [\href{https://arxiv.org/abs/2203.16352}{{\ttfamily 2203.16352}}].

\bibitem{Matsui:1986dk}
T.~Matsui and H.~Satz, \emph{{$J/\psi$ Suppression by Quark-Gluon Plasma
  Formation}}, \href{https://doi.org/10.1016/0370-2693(86)91404-8}{\emph{Phys.
  Lett. B} {\bfseries 178} (1986) 416}.

\bibitem{Tang:2020ame}
Z.~Tang, Z.-B.~Tang, W.~Zha, W.-M.~Zha, Y.~Zhang and Y.-F.~Zhang, \emph{{An
  experimental review of open heavy flavor and quarkonium production at RHIC}},
  \href{https://doi.org/10.1007/s41365-020-00785-8}{\emph{Nucl. Sci. Tech.}
  {\bfseries 31} (2020) 81} [\href{https://arxiv.org/abs/2105.11656}{{\ttfamily
  2105.11656}}].

\bibitem{ATLAS:2018gwx}
{\scshape ATLAS} collaboration, \emph{{Measurement of the nuclear modification
  factor for inclusive jets in Pb+Pb collisions at $\sqrt{s_\mathrm{NN}}=5.02$
  TeV with the ATLAS detector}},
  \href{https://doi.org/10.1016/j.physletb.2018.10.076}{\emph{Phys. Lett. B}
  {\bfseries 790} (2019) 108}
  [\href{https://arxiv.org/abs/1805.05635}{{\ttfamily 1805.05635}}].

\bibitem{ALICE:2015xmh}
{\scshape ALICE} collaboration, \emph{{Direct photon production in Pb-Pb
  collisions at $\sqrt{s_{NN}} =$ 2.76 TeV}},
  \href{https://doi.org/10.1016/j.physletb.2016.01.020}{\emph{Phys. Lett. B}
  {\bfseries 754} (2016) 235}
  [\href{https://arxiv.org/abs/1509.07324}{{\ttfamily 1509.07324}}].

\bibitem{Gale:2009gc}
C.~Gale, \emph{{Photon Production in Hot and Dense Strongly Interacting
  Matter}},
  \href{https://doi.org/10.1007/978-3-642-01539-7_15}{\emph{Landolt-Bornstein}
  {\bfseries 23} (2010) 445} [\href{https://arxiv.org/abs/0904.2184}{{\ttfamily
  0904.2184}}].

\bibitem{Wiedemann:2021bwz}
U.A.~Wiedemann, \emph{{HIP and HEP}},
  \href{https://doi.org/10.22323/1.390.0046}{\emph{PoS} {\bfseries ICHEP2020}
  (2021) 046} [\href{https://arxiv.org/abs/2101.01971}{{\ttfamily
  2101.01971}}].

\bibitem{ATLAS:2012at}
{\scshape ATLAS} collaboration, \emph{{Measurement of the azimuthal anisotropy
  for charged particle production in $\sqrt{s_{NN}}=2.76$ TeV lead-lead
  collisions with the ATLAS detector}},
  \href{https://doi.org/10.1103/PhysRevC.86.014907}{\emph{Phys. Rev. C}
  {\bfseries 86} (2012) 014907}
  [\href{https://arxiv.org/abs/1203.3087}{{\ttfamily 1203.3087}}].

\bibitem{Ollitrault:1992bk}
J.-Y.~Ollitrault, \emph{{Anisotropy as a signature of transverse collective
  flow}}, \href{https://doi.org/10.1103/PhysRevD.46.229}{\emph{Phys. Rev. D}
  {\bfseries 46} (1992) 229}.

\bibitem{Dusling:2007gi}
K.~Dusling and D.~Teaney, \emph{{Simulating elliptic flow with viscous
  hydrodynamics}},
  \href{https://doi.org/10.1103/PhysRevC.77.034905}{\emph{Phys. Rev. C}
  {\bfseries 77} (2008) 034905}
  [\href{https://arxiv.org/abs/0710.5932}{{\ttfamily 0710.5932}}].

\bibitem{Song:2007ux}
H.~Song and U.W.~Heinz, \emph{{Causal viscous hydrodynamics in 2+1 dimensions
  for relativistic heavy-ion collisions}},
  \href{https://doi.org/10.1103/PhysRevC.77.064901}{\emph{Phys. Rev. C}
  {\bfseries 77} (2008) 064901}
  [\href{https://arxiv.org/abs/0712.3715}{{\ttfamily 0712.3715}}].

\bibitem{Alver:2010dn}
B.H.~Alver, C.~Gombeaud, M.~Luzum and J.-Y.~Ollitrault, \emph{{Triangular flow
  in hydrodynamics and transport theory}},
  \href{https://doi.org/10.1103/PhysRevC.82.034913}{\emph{Phys. Rev. C}
  {\bfseries 82} (2010) 034913}
  [\href{https://arxiv.org/abs/1007.5469}{{\ttfamily 1007.5469}}].

\bibitem{phenix:2003qra}
{\scshape PHENIX} collaboration, \emph{{Elliptic flow of identified hadrons in
  Au+Au collisions at s(NN)**(1/2) = 200-GeV}},
  \href{https://doi.org/10.1103/PhysRevLett.91.182301}{\emph{Phys. Rev. Lett.}
  {\bfseries 91} (2003) 182301}
  [\href{https://arxiv.org/abs/nucl-ex/0305013}{{\ttfamily nucl-ex/0305013}}].

\bibitem{star:2004jwm}
{\scshape STAR} collaboration, \emph{{Azimuthal anisotropy in Au+Au collisions
  at s(NN)**(1/2) = 200-GeV}},
  \href{https://doi.org/10.1103/PhysRevC.72.014904}{\emph{Phys. Rev. C}
  {\bfseries 72} (2005) 014904}
  [\href{https://arxiv.org/abs/nucl-ex/0409033}{{\ttfamily nucl-ex/0409033}}].

\bibitem{ALICE:2011ab}
{\scshape ALICE} collaboration, \emph{{Higher harmonic anisotropic flow
  measurements of charged particles in Pb-Pb collisions at $\sqrt{s_{NN}}$=2.76
  TeV}}, \href{https://doi.org/10.1103/PhysRevLett.107.032301}{\emph{Phys. Rev.
  Lett.} {\bfseries 107} (2011) 032301}
  [\href{https://arxiv.org/abs/1105.3865}{{\ttfamily 1105.3865}}].

\bibitem{cms:2013jlh}
{\scshape CMS} collaboration, \emph{{Multiplicity and Transverse Momentum
  Dependence of Two- and Four-Particle Correlations in pPb and PbPb
  Collisions}},
  \href{https://doi.org/10.1016/j.physletb.2013.06.028}{\emph{Phys. Lett. B}
  {\bfseries 724} (2013) 213}
  [\href{https://arxiv.org/abs/1305.0609}{{\ttfamily 1305.0609}}].

\bibitem{ALICE:2016kpq}
{\scshape ALICE} collaboration, \emph{{Correlated event-by-event fluctuations
  of flow harmonics in Pb-Pb collisions at $\sqrt{s_{_{\rm NN}}}=2.76$ TeV}},
  \href{https://doi.org/10.1103/PhysRevLett.117.182301}{\emph{Phys. Rev. Lett.}
  {\bfseries 117} (2016) 182301}
  [\href{https://arxiv.org/abs/1604.07663}{{\ttfamily 1604.07663}}].

\bibitem{Policastro:2001yc}
G.~Policastro, D.T.~Son and A.O.~Starinets, \emph{{The Shear viscosity of
  strongly coupled N=4 supersymmetric Yang-Mills plasma}},
  \href{https://doi.org/10.1103/PhysRevLett.87.081601}{\emph{Phys. Rev. Lett.}
  {\bfseries 87} (2001) 081601}
  [\href{https://arxiv.org/abs/hep-th/0104066}{{\ttfamily hep-th/0104066}}].

\bibitem{Kovtun:2003wp}
P.~Kovtun, D.T.~Son and A.O.~Starinets, \emph{{Holography and hydrodynamics:
  Diffusion on stretched horizons}},
  \href{https://doi.org/10.1088/1126-6708/2003/10/064}{\emph{JHEP} {\bfseries
  10} (2003) 064} [\href{https://arxiv.org/abs/hep-th/0309213}{{\ttfamily
  hep-th/0309213}}].

\bibitem{Kovtun:2004de}
P.~Kovtun, D.T.~Son and A.O.~Starinets, \emph{{Viscosity in strongly
  interacting quantum field theories from black hole physics}},
  \href{https://doi.org/10.1103/PhysRevLett.94.111601}{\emph{Phys. Rev. Lett.}
  {\bfseries 94} (2005) 111601}
  [\href{https://arxiv.org/abs/hep-th/0405231}{{\ttfamily hep-th/0405231}}].

\bibitem{Maldacena:1997re}
J.M.~Maldacena, \emph{{The Large N limit of superconformal field theories and
  supergravity}}, \href{https://doi.org/10.4310/ATMP.1998.v2.n2.a1}{\emph{Adv.
  Theor. Math. Phys.} {\bfseries 2} (1998) 231}
  [\href{https://arxiv.org/abs/hep-th/9711200}{{\ttfamily hep-th/9711200}}].

\bibitem{Arnold:2000dr}
P.B.~Arnold, G.D.~Moore and L.G.~Yaffe, \emph{{Transport coefficients in high
  temperature gauge theories. 1. Leading log results}},
  \href{https://doi.org/10.1088/1126-6708/2000/11/001}{\emph{JHEP} {\bfseries
  11} (2000) 001} [\href{https://arxiv.org/abs/hep-ph/0010177}{{\ttfamily
  hep-ph/0010177}}].

\bibitem{Arnold:2003zc}
P.B.~Arnold, G.D.~Moore and L.G.~Yaffe, \emph{{Transport coefficients in high
  temperature gauge theories. 2. Beyond leading log}},
  \href{https://doi.org/10.1088/1126-6708/2003/05/051}{\emph{JHEP} {\bfseries
  05} (2003) 051} [\href{https://arxiv.org/abs/hep-ph/0302165}{{\ttfamily
  hep-ph/0302165}}].

\bibitem{Ghiglieri:2018dib}
J.~Ghiglieri, G.D.~Moore and D.~Teaney, \emph{{QCD Shear Viscosity at (almost)
  NLO}}, \href{https://doi.org/10.1007/JHEP03(2018)179}{\emph{JHEP} {\bfseries
  03} (2018) 179} [\href{https://arxiv.org/abs/1802.09535}{{\ttfamily
  1802.09535}}].

\bibitem{CMS:2010ifv}
{\scshape CMS} collaboration, \emph{{Observation of Long-Range Near-Side
  Angular Correlations in Proton-Proton Collisions at the LHC}},
  \href{https://doi.org/10.1007/JHEP09(2010)091}{\emph{JHEP} {\bfseries 09}
  (2010) 091} [\href{https://arxiv.org/abs/1009.4122}{{\ttfamily 1009.4122}}].

\bibitem{Nagle:2018nvi}
J.L.~Nagle and W.A.~Zajc, \emph{{Small System Collectivity in Relativistic
  Hadronic and Nuclear Collisions}},
  \href{https://doi.org/10.1146/annurev-nucl-101916-123209}{\emph{Ann. Rev.
  Nucl. Part. Sci.} {\bfseries 68} (2018) 211}
  [\href{https://arxiv.org/abs/1801.03477}{{\ttfamily 1801.03477}}].

\bibitem{Polchinski:2001tt}
J.~Polchinski and M.J.~Strassler, \emph{{Hard scattering and gauge / string
  duality}}, \href{https://doi.org/10.1103/PhysRevLett.88.031601}{\emph{Phys.
  Rev. Lett.} {\bfseries 88} (2002) 031601}
  [\href{https://arxiv.org/abs/hep-th/0109174}{{\ttfamily hep-th/0109174}}].

\bibitem{Erlich:2005qh}
J.~Erlich, E.~Katz, D.T.~Son and M.A.~Stephanov, \emph{{QCD and a holographic
  model of hadrons}},
  \href{https://doi.org/10.1103/PhysRevLett.95.261602}{\emph{Phys. Rev. Lett.}
  {\bfseries 95} (2005) 261602}
  [\href{https://arxiv.org/abs/hep-ph/0501128}{{\ttfamily hep-ph/0501128}}].

\bibitem{DaRold:2005mxj}
L.~Da~Rold and A.~Pomarol, \emph{{Chiral symmetry breaking from five
  dimensional spaces}},
  \href{https://doi.org/10.1016/j.nuclphysb.2005.05.009}{\emph{Nucl. Phys. B}
  {\bfseries 721} (2005) 79}
  [\href{https://arxiv.org/abs/hep-ph/0501218}{{\ttfamily hep-ph/0501218}}].

\bibitem{Casero:2007ae}
R.~Casero, E.~Kiritsis and A.~Paredes, \emph{{Chiral symmetry breaking as open
  string tachyon condensation}},
  \href{https://doi.org/10.1016/j.nuclphysb.2007.07.009}{\emph{Nucl. Phys. B}
  {\bfseries 787} (2007) 98}
  [\href{https://arxiv.org/abs/hep-th/0702155}{{\ttfamily hep-th/0702155}}].

\bibitem{Gursoy:2007cb}
U.~Gursoy and E.~Kiritsis, \emph{{Exploring improved holographic theories for
  QCD: Part I}},
  \href{https://doi.org/10.1088/1126-6708/2008/02/032}{\emph{JHEP} {\bfseries
  02} (2008) 032} [\href{https://arxiv.org/abs/0707.1324}{{\ttfamily
  0707.1324}}].

\bibitem{Gursoy:2007er}
U.~Gursoy, E.~Kiritsis and F.~Nitti, \emph{{Exploring improved holographic
  theories for QCD: Part II}},
  \href{https://doi.org/10.1088/1126-6708/2008/02/019}{\emph{JHEP} {\bfseries
  02} (2008) 019} [\href{https://arxiv.org/abs/0707.1349}{{\ttfamily
  0707.1349}}].

\bibitem{Jarvinen:2011qe}
M.~Jarvinen and E.~Kiritsis, \emph{{Holographic Models for QCD in the Veneziano
  Limit}}, \href{https://doi.org/10.1007/JHEP03(2012)002}{\emph{JHEP}
  {\bfseries 03} (2012) 002} [\href{https://arxiv.org/abs/1112.1261}{{\ttfamily
  1112.1261}}].

\bibitem{Alho:2012mh}
T.~Alho, M.~J\"arvinen, K.~Kajantie, E.~Kiritsis and K.~Tuominen, \emph{{On
  finite-temperature holographic QCD in the Veneziano limit}},
  \href{https://doi.org/10.1007/JHEP01(2013)093}{\emph{JHEP} {\bfseries 01}
  (2013) 093} [\href{https://arxiv.org/abs/1210.4516}{{\ttfamily 1210.4516}}].

\bibitem{Connors:2017ptx}
M.~Connors, C.~Nattrass, R.~Reed and S.~Salur, \emph{{Jet measurements in heavy
  ion physics}}, \href{https://doi.org/10.1103/RevModPhys.90.025005}{\emph{Rev.
  Mod. Phys.} {\bfseries 90} (2018) 025005}
  [\href{https://arxiv.org/abs/1705.01974}{{\ttfamily 1705.01974}}].

\bibitem{Wang:2001ifa}
X.-N.~Wang and X.-f.~Guo, \emph{{Multiple parton scattering in nuclei: Parton
  energy loss}},
  \href{https://doi.org/10.1016/S0375-9474(01)01130-7}{\emph{Nucl. Phys. A}
  {\bfseries 696} (2001) 788}
  [\href{https://arxiv.org/abs/hep-ph/0102230}{{\ttfamily hep-ph/0102230}}].

\bibitem{Majumder:2009zu}
A.~Majumder, \emph{{The In-medium scale evolution in jet modification}},
  \href{https://arxiv.org/abs/0901.4516}{{\ttfamily 0901.4516}}.

\bibitem{Caucal:2018dla}
P.~Caucal, E.~Iancu, A.H.~Mueller and G.~Soyez, \emph{{Vacuum-like jet
  fragmentation in a dense QCD medium}},
  \href{https://doi.org/10.1103/PhysRevLett.120.232001}{\emph{Phys. Rev. Lett.}
  {\bfseries 120} (2018) 232001}
  [\href{https://arxiv.org/abs/1801.09703}{{\ttfamily 1801.09703}}].

\bibitem{Caucal:2019uvr}
P.~Caucal, E.~Iancu and G.~Soyez, \emph{{Deciphering the $z_g$ distribution in
  ultrarelativistic heavy ion collisions}},
  \href{https://doi.org/10.1007/JHEP10(2019)273}{\emph{JHEP} {\bfseries 10}
  (2019) 273} [\href{https://arxiv.org/abs/1907.04866}{{\ttfamily
  1907.04866}}].

\bibitem{Caucal:2020xad}
P.~Caucal, E.~Iancu, A.H.~Mueller and G.~Soyez, \emph{{Nuclear modification
  factors for jet fragmentation}},
  \href{https://doi.org/10.1007/JHEP10(2020)204}{\emph{JHEP} {\bfseries 10}
  (2020) 204} [\href{https://arxiv.org/abs/2005.05852}{{\ttfamily
  2005.05852}}].

\bibitem{Thoma:1990fm}
M.H.~Thoma and M.~Gyulassy, \emph{{Quark Damping and Energy Loss in the High
  Temperature {QCD}}},
  \href{https://doi.org/10.1016/S0550-3213(05)80031-8}{\emph{Nucl. Phys. B}
  {\bfseries 351} (1991) 491}.

\bibitem{Mrowczynski:1991da}
S.~Mrowczynski, \emph{{Energy loss of a high-energy parton in the quark - gluon
  plasma}}, \href{https://doi.org/10.1016/0370-2693(91)90188-V}{\emph{Phys.
  Lett. B} {\bfseries 269} (1991) 383}.

\bibitem{Djordjevic:2006tw}
M.~Djordjevic, \emph{{Collisional energy loss in a finite size QCD matter}},
  \href{https://doi.org/10.1103/PhysRevC.74.064907}{\emph{Phys. Rev. C}
  {\bfseries 74} (2006) 064907}
  [\href{https://arxiv.org/abs/nucl-th/0603066}{{\ttfamily nucl-th/0603066}}].

\bibitem{Braaten:1991jj}
E.~Braaten and M.H.~Thoma, \emph{{Energy loss of a heavy fermion in a hot
  plasma}}, \href{https://doi.org/10.1103/PhysRevD.44.1298}{\emph{Phys. Rev. D}
  {\bfseries 44} (1991) 1298}.

\bibitem{Peigne:2007sd}
S.~Peigne and A.~Peshier, \emph{{Collisional Energy Loss of a Fast Muon in a
  Hot QED Plasma}},
  \href{https://doi.org/10.1103/PhysRevD.77.014015}{\emph{Phys. Rev. D}
  {\bfseries 77} (2008) 014015}
  [\href{https://arxiv.org/abs/0710.1266}{{\ttfamily 0710.1266}}].

\bibitem{Peigne:2008wu}
S.~Peigne and A.V.~Smilga, \emph{{Energy losses in a hot plasma revisited}},
  \href{https://doi.org/10.3367/UFNe.0179.200907a.0697}{\emph{Phys. Usp.}
  {\bfseries 52} (2009) 659} [\href{https://arxiv.org/abs/0810.5702}{{\ttfamily
  0810.5702}}].

\bibitem{Baier:2000mf}
R.~Baier, D.~Schiff and B.G.~Zakharov, \emph{{Energy loss in perturbative
  QCD}}, \href{https://doi.org/10.1146/annurev.nucl.50.1.37}{\emph{Ann. Rev.
  Nucl. Part. Sci.} {\bfseries 50} (2000) 37}
  [\href{https://arxiv.org/abs/hep-ph/0002198}{{\ttfamily hep-ph/0002198}}].

\bibitem{Zakharov:2007pj}
B.G.~Zakharov, \emph{{Parton energy loss in an expanding quark-gluon plasma:
  Radiative versus collisional}},
  \href{https://doi.org/10.1134/S0021364007190034}{\emph{JETP Lett.} {\bfseries
  86} (2007) 444} [\href{https://arxiv.org/abs/0708.0816}{{\ttfamily
  0708.0816}}].

\bibitem{Arnold:2001ba}
P.B.~Arnold, G.D.~Moore and L.G.~Yaffe, \emph{{Photon emission from
  ultrarelativistic plasmas}},
  \href{https://doi.org/10.1088/1126-6708/2001/11/057}{\emph{JHEP} {\bfseries
  11} (2001) 057} [\href{https://arxiv.org/abs/hep-ph/0109064}{{\ttfamily
  hep-ph/0109064}}].

\bibitem{Arnold:2001ms}
P.B.~Arnold, G.D.~Moore and L.G.~Yaffe, \emph{{Photon emission from quark gluon
  plasma: Complete leading order results}},
  \href{https://doi.org/10.1088/1126-6708/2001/12/009}{\emph{JHEP} {\bfseries
  12} (2001) 009} [\href{https://arxiv.org/abs/hep-ph/0111107}{{\ttfamily
  hep-ph/0111107}}].

\bibitem{Arnold:2002ja}
P.B.~Arnold, G.D.~Moore and L.G.~Yaffe, \emph{{Photon and gluon emission in
  relativistic plasmas}},
  \href{https://doi.org/10.1088/1126-6708/2002/06/030}{\emph{JHEP} {\bfseries
  06} (2002) 030} [\href{https://arxiv.org/abs/hep-ph/0204343}{{\ttfamily
  hep-ph/0204343}}].

\bibitem{Arnold:2002zm}
P.B.~Arnold, G.D.~Moore and L.G.~Yaffe, \emph{{Effective kinetic theory for
  high temperature gauge theories}},
  \href{https://doi.org/10.1088/1126-6708/2003/01/030}{\emph{JHEP} {\bfseries
  01} (2003) 030} [\href{https://arxiv.org/abs/hep-ph/0209353}{{\ttfamily
  hep-ph/0209353}}].

\bibitem{Jeon:2003gi}
S.~Jeon and G.D.~Moore, \emph{{Energy loss of leading partons in a thermal QCD
  medium}}, \href{https://doi.org/10.1103/PhysRevC.71.034901}{\emph{Phys. Rev.
  C} {\bfseries 71} (2005) 034901}
  [\href{https://arxiv.org/abs/hep-ph/0309332}{{\ttfamily hep-ph/0309332}}].

\bibitem{Aurenche:1998nw}
P.~Aurenche, F.~Gelis, R.~Kobes and H.~Zaraket, \emph{{Bremsstrahlung and
  photon production in thermal QCD}},
  \href{https://doi.org/10.1103/PhysRevD.58.085003}{\emph{Phys. Rev. D}
  {\bfseries 58} (1998) 085003}
  [\href{https://arxiv.org/abs/hep-ph/9804224}{{\ttfamily hep-ph/9804224}}].

\bibitem{Landau:1953um}
L.D.~Landau and I.~Pomeranchuk, \emph{{Limits of applicability of the theory of
  bremsstrahlung electrons and pair production at high-energies}}, {\emph{Dokl.
  Akad. Nauk Ser. Fiz.} {\bfseries 92} (1953) 535}.

\bibitem{Landau:1953gr}
L.D.~Landau and I.~Pomeranchuk, \emph{{Electron cascade process at very
  high-energies}}, {\emph{Dokl. Akad. Nauk Ser. Fiz.} {\bfseries 92} (1953)
  735}.

\bibitem{Migdal:1956tc}
A.B.~Migdal, \emph{{Bremsstrahlung and pair production in condensed media at
  high-energies}}, \href{https://doi.org/10.1103/PhysRev.103.1811}{\emph{Phys.
  Rev.} {\bfseries 103} (1956) 1811}.

\bibitem{CaronHuot:2010bp}
S.~Caron-Huot and C.~Gale, \emph{{Finite-size effects on the radiative energy
  loss of a fast parton in hot and dense strongly interacting matter}},
  \href{https://doi.org/10.1103/PhysRevC.82.064902}{\emph{Phys. Rev. C}
  {\bfseries 82} (2010) 064902}
  [\href{https://arxiv.org/abs/1006.2379}{{\ttfamily 1006.2379}}].

\bibitem{Altarelli:1977zs}
G.~Altarelli and G.~Parisi, \emph{{Asymptotic Freedom in Parton Language}},
  \href{https://doi.org/10.1016/0550-3213(77)90384-4}{\emph{Nucl. Phys. B}
  {\bfseries 126} (1977) 298}.

\bibitem{Arnold:2008zu}
P.B.~Arnold and C.~Dogan, \emph{{QCD Splitting/Joining Functions at Finite
  Temperature in the Deep LPM Regime}},
  \href{https://doi.org/10.1103/PhysRevD.78.065008}{\emph{Phys. Rev. D}
  {\bfseries 78} (2008) 065008}
  [\href{https://arxiv.org/abs/0804.3359}{{\ttfamily 0804.3359}}].

\bibitem{Apolinario:2014csa}
L.~Apolin\'ario, N.~Armesto, J.G.~Milhano and C.A.~Salgado,
  \emph{{Medium-induced gluon radiation and colour decoherence beyond the soft
  approximation}}, \href{https://doi.org/10.1007/JHEP02(2015)119}{\emph{JHEP}
  {\bfseries 02} (2015) 119} [\href{https://arxiv.org/abs/1407.0599}{{\ttfamily
  1407.0599}}].

\bibitem{Isaksen:2023nlr}
J.H.~Isaksen and K.~Tywoniuk, \emph{{Precise description of medium-induced
  emissions}}, \href{https://doi.org/10.1007/JHEP09(2023)049}{\emph{JHEP}
  {\bfseries 09} (2023) 049}
  [\href{https://arxiv.org/abs/2303.12119}{{\ttfamily 2303.12119}}].

\bibitem{Arnold:2008iy}
P.B.~Arnold, \emph{{Simple Formula for High-Energy Gluon Bremsstrahlung in a
  Finite, Expanding Medium}},
  \href{https://doi.org/10.1103/PhysRevD.79.065025}{\emph{Phys. Rev. D}
  {\bfseries 79} (2009) 065025}
  [\href{https://arxiv.org/abs/0808.2767}{{\ttfamily 0808.2767}}].

\bibitem{Arnold:2015qya}
P.~Arnold and S.~Iqbal, \emph{{The LPM effect in sequential bremsstrahlung}},
  \href{https://doi.org/10.1007/JHEP09(2016)072}{\emph{JHEP} {\bfseries 04}
  (2015) 070} [\href{https://arxiv.org/abs/1501.04964}{{\ttfamily
  1501.04964}}].

\bibitem{Ghiglieri:2020dpq}
J.~Ghiglieri, A.~Kurkela, M.~Strickland and A.~Vuorinen, \emph{{Perturbative
  Thermal QCD: Formalism and Applications}},
  \href{https://doi.org/10.1016/j.physrep.2020.07.004}{\emph{Phys. Rept.}
  {\bfseries 880} (2020) 1} [\href{https://arxiv.org/abs/2002.10188}{{\ttfamily
  2002.10188}}].

\bibitem{Peshier:2008zz}
A.~Peshier, \emph{{QCD running coupling and collisional jet quenching}},
  \href{https://doi.org/10.1088/0954-3899/35/4/044028}{\emph{J. Phys. G}
  {\bfseries 35} (2008) 044028}.

\bibitem{Aurenche:2002pd}
P.~Aurenche, F.~Gelis and H.~Zaraket, \emph{{A Simple sum rule for the thermal
  gluon spectral function and applications}},
  \href{https://doi.org/10.1088/1126-6708/2002/05/043}{\emph{JHEP} {\bfseries
  05} (2002) 043} [\href{https://arxiv.org/abs/hep-ph/0204146}{{\ttfamily
  hep-ph/0204146}}].

\bibitem{Moliere:1947zza}
G.~Moliere, \emph{{Theorie der Streuung schneller geladener Teilchen I.
  Einzelstreuung am abgeschirmten Coulomb-Feld}}, {\emph{Z. Naturforsch. A}
  {\bfseries 2} (1947) 133}.

\bibitem{Moliere:1948zz}
G.~Moliere, \emph{{Theory of the scattering of fast charged particles. 2.
  Repeated and multiple scattering}}, {\emph{Z. Naturforsch. A} {\bfseries 3}
  (1948) 78}.

\bibitem{Qin:2007rn}
G.-Y.~Qin, J.~Ruppert, C.~Gale, S.~Jeon, G.D.~Moore and M.G.~Mustafa,
  \emph{{Radiative and collisional jet energy loss in the quark-gluon plasma at
  RHIC}}, \href{https://doi.org/10.1103/PhysRevLett.100.072301}{\emph{Phys.
  Rev. Lett.} {\bfseries 100} (2008) 072301}
  [\href{https://arxiv.org/abs/0710.0605}{{\ttfamily 0710.0605}}].

\bibitem{Schenke:2009gb}
B.~Schenke, C.~Gale and S.~Jeon, \emph{{MARTINI: An Event generator for
  relativistic heavy-ion collisions}},
  \href{https://doi.org/10.1103/PhysRevC.80.054913}{\emph{Phys. Rev. C}
  {\bfseries 80} (2009) 054913}
  [\href{https://arxiv.org/abs/0909.2037}{{\ttfamily 0909.2037}}].

\bibitem{Gyulassy:1993hr}
M.~Gyulassy and X.-n.~Wang, \emph{{Multiple collisions and induced gluon
  Bremsstrahlung in QCD}},
  \href{https://doi.org/10.1016/0550-3213(94)90079-5}{\emph{Nucl. Phys. B}
  {\bfseries 420} (1994) 583}
  [\href{https://arxiv.org/abs/nucl-th/9306003}{{\ttfamily nucl-th/9306003}}].

\bibitem{Arnold:2008vd}
P.B.~Arnold and W.~Xiao, \emph{{High-energy jet quenching in weakly-coupled
  quark-gluon plasmas}},
  \href{https://doi.org/10.1103/PhysRevD.78.125008}{\emph{Phys. Rev. D}
  {\bfseries 78} (2008) 125008}
  [\href{https://arxiv.org/abs/0810.1026}{{\ttfamily 0810.1026}}].

\bibitem{Arnold:2007pg}
P.B.~Arnold, \emph{{Quark-Gluon Plasmas and Thermalization}},
  \href{https://doi.org/10.1142/S021830130700832X}{\emph{Int. J. Mod. Phys. E}
  {\bfseries 16} (2007) 2555}
  [\href{https://arxiv.org/abs/0708.0812}{{\ttfamily 0708.0812}}].

\bibitem{Armesto:2011ht}
N.~Armesto et~al., \emph{{Comparison of Jet Quenching Formalisms for a
  Quark-Gluon Plasma 'Brick'}},
  \href{https://doi.org/10.1103/PhysRevC.86.064904}{\emph{Phys. Rev. C}
  {\bfseries 86} (2012) 064904}
  [\href{https://arxiv.org/abs/1106.1106}{{\ttfamily 1106.1106}}].

\bibitem{Baier:1994bd}
R.~Baier, Y.L.~Dokshitzer, S.~Peigne and D.~Schiff, \emph{{Induced gluon
  radiation in a QCD medium}},
  \href{https://doi.org/10.1016/0370-2693(94)01617-L}{\emph{Phys. Lett. B}
  {\bfseries 345} (1995) 277}
  [\href{https://arxiv.org/abs/hep-ph/9411409}{{\ttfamily hep-ph/9411409}}].

\bibitem{Baier:1996vi}
R.~Baier, Y.L.~Dokshitzer, A.H.~Mueller, S.~Peigne and D.~Schiff, \emph{{The
  Landau-Pomeranchuk-Migdal effect in QED}},
  \href{https://doi.org/10.1016/0550-3213(96)00426-9}{\emph{Nucl. Phys. B}
  {\bfseries 478} (1996) 577}
  [\href{https://arxiv.org/abs/hep-ph/9604327}{{\ttfamily hep-ph/9604327}}].

\bibitem{Baier:1996kr}
R.~Baier, Y.L.~Dokshitzer, A.H.~Mueller, S.~Peigne and D.~Schiff,
  \emph{{Radiative energy loss of high-energy quarks and gluons in a finite
  volume quark - gluon plasma}},
  \href{https://doi.org/10.1016/S0550-3213(96)00553-6}{\emph{Nucl. Phys. B}
  {\bfseries 483} (1997) 291}
  [\href{https://arxiv.org/abs/hep-ph/9607355}{{\ttfamily hep-ph/9607355}}].

\bibitem{Baier:1996sk}
R.~Baier, Y.L.~Dokshitzer, A.H.~Mueller, S.~Peigne and D.~Schiff,
  \emph{{Radiative energy loss and p(T) broadening of high-energy partons in
  nuclei}}, \href{https://doi.org/10.1016/S0550-3213(96)00581-0}{\emph{Nucl.
  Phys. B} {\bfseries 484} (1997) 265}
  [\href{https://arxiv.org/abs/hep-ph/9608322}{{\ttfamily hep-ph/9608322}}].

\bibitem{Zakharov:1996fv}
B.G.~Zakharov, \emph{{Fully quantum treatment of the Landau-Pomeranchuk-Migdal
  effect in QED and QCD}}, \href{https://doi.org/10.1134/1.567126}{\emph{JETP
  Lett.} {\bfseries 63} (1996) 952}
  [\href{https://arxiv.org/abs/hep-ph/9607440}{{\ttfamily hep-ph/9607440}}].

\bibitem{Zakharov:1997uu}
B.G.~Zakharov, \emph{{Radiative energy loss of high-energy quarks in finite
  size nuclear matter and quark - gluon plasma}},
  \href{https://doi.org/10.1134/1.567389}{\emph{JETP Lett.} {\bfseries 65}
  (1997) 615} [\href{https://arxiv.org/abs/hep-ph/9704255}{{\ttfamily
  hep-ph/9704255}}].

\bibitem{Zakharov:1998sv}
B.G.~Zakharov, \emph{{Light cone path integral approach to the
  Landau-Pomeranchuk-Migdal effect}}, {\emph{Phys. Atom. Nucl.} {\bfseries 61}
  (1998) 838} [\href{https://arxiv.org/abs/hep-ph/9807540}{{\ttfamily
  hep-ph/9807540}}].

\bibitem{Baier:1998kq}
R.~Baier, Y.L.~Dokshitzer, A.H.~Mueller and D.~Schiff, \emph{{Medium induced
  radiative energy loss: Equivalence between the BDMPS and Zakharov
  formalisms}},
  \href{https://doi.org/10.1016/S0550-3213(98)00546-X}{\emph{Nucl. Phys. B}
  {\bfseries 531} (1998) 403}
  [\href{https://arxiv.org/abs/hep-ph/9804212}{{\ttfamily hep-ph/9804212}}].

\bibitem{Salgado:2003gb}
C.A.~Salgado and U.A.~Wiedemann, \emph{{Calculating quenching weights}},
  \href{https://doi.org/10.1103/PhysRevD.68.014008}{\emph{Phys. Rev. D}
  {\bfseries 68} (2003) 014008}
  [\href{https://arxiv.org/abs/hep-ph/0302184}{{\ttfamily hep-ph/0302184}}].

\bibitem{Armesto:2003jh}
N.~Armesto, C.A.~Salgado and U.A.~Wiedemann, \emph{{Medium induced gluon
  radiation off massive quarks fills the dead cone}},
  \href{https://doi.org/10.1103/PhysRevD.69.114003}{\emph{Phys. Rev. D}
  {\bfseries 69} (2004) 114003}
  [\href{https://arxiv.org/abs/hep-ph/0312106}{{\ttfamily hep-ph/0312106}}].

\bibitem{Gyulassy:2000er}
M.~Gyulassy, P.~Levai and I.~Vitev, \emph{{Reaction operator approach to
  nonAbelian energy loss}},
  \href{https://doi.org/10.1016/S0550-3213(00)00652-0}{\emph{Nucl. Phys. B}
  {\bfseries 594} (2001) 371}
  [\href{https://arxiv.org/abs/nucl-th/0006010}{{\ttfamily nucl-th/0006010}}].

\bibitem{Gyulassy:2001nm}
M.~Gyulassy, P.~Levai and I.~Vitev, \emph{{Jet tomography of Au+Au reactions
  including multigluon fluctuations}},
  \href{https://doi.org/10.1016/S0370-2693(02)01990-1}{\emph{Phys. Lett. B}
  {\bfseries 538} (2002) 282}
  [\href{https://arxiv.org/abs/nucl-th/0112071}{{\ttfamily nucl-th/0112071}}].

\bibitem{Wiedemann:2000za}
U.A.~Wiedemann, \emph{{Gluon radiation off hard quarks in a nuclear
  environment: Opacity expansion}},
  \href{https://doi.org/10.1016/S0550-3213(00)00457-0}{\emph{Nucl. Phys. B}
  {\bfseries 588} (2000) 303}
  [\href{https://arxiv.org/abs/hep-ph/0005129}{{\ttfamily hep-ph/0005129}}].

\bibitem{Mehtar-Tani:2019tvy}
Y.~Mehtar-Tani, \emph{{Gluon bremsstrahlung in finite media beyond multiple
  soft scattering approximation}},
  \href{https://doi.org/10.1007/JHEP07(2019)057}{\emph{JHEP} {\bfseries 07}
  (2019) 057} [\href{https://arxiv.org/abs/1903.00506}{{\ttfamily
  1903.00506}}].

\bibitem{Ovanesyan:2011kn}
G.~Ovanesyan and I.~Vitev, \emph{{Medium-induced parton splitting kernels from
  Soft Collinear Effective Theory with Glauber gluons}},
  \href{https://doi.org/10.1016/j.physletb.2011.11.040}{\emph{Phys. Lett. B}
  {\bfseries 706} (2012) 371}
  [\href{https://arxiv.org/abs/1109.5619}{{\ttfamily 1109.5619}}].

\bibitem{Sievert:2018imd}
M.D.~Sievert and I.~Vitev, \emph{{Quark branching in QCD matter to any order in
  opacity beyond the soft gluon emission limit}},
  \href{https://doi.org/10.1103/PhysRevD.98.094010}{\emph{Phys. Rev. D}
  {\bfseries 98} (2018) 094010}
  [\href{https://arxiv.org/abs/1807.03799}{{\ttfamily 1807.03799}}].

\bibitem{Djordjevic:2008iz}
M.~Djordjevic and U.W.~Heinz, \emph{{Radiative energy loss in a finite
  dynamical QCD medium}},
  \href{https://doi.org/10.1103/PhysRevLett.101.022302}{\emph{Phys. Rev. Lett.}
  {\bfseries 101} (2008) 022302}
  [\href{https://arxiv.org/abs/0802.1230}{{\ttfamily 0802.1230}}].

\bibitem{Isaksen:2022pkj}
J.H.~Isaksen, A.~Takacs and K.~Tywoniuk, \emph{{A unified picture of
  medium-induced radiation}},
  \href{https://doi.org/10.1007/JHEP02(2023)156}{\emph{JHEP} {\bfseries 02}
  (2023) 156} [\href{https://arxiv.org/abs/2206.02811}{{\ttfamily
  2206.02811}}].

\bibitem{Mehtar-Tani:2019ygg}
Y.~Mehtar-Tani and K.~Tywoniuk, \emph{{Improved opacity expansion for
  medium-induced parton splitting}},
  \href{https://doi.org/10.1007/JHEP06(2020)187}{\emph{JHEP} {\bfseries 06}
  (2020) 187} [\href{https://arxiv.org/abs/1910.02032}{{\ttfamily
  1910.02032}}].

\bibitem{Barata:2020rdn}
J.a.~Barata, Y.~Mehtar-Tani, A.~Soto-Ontoso and K.~Tywoniuk, \emph{{Revisiting
  transverse momentum broadening in dense QCD media}},
  \href{https://doi.org/10.1103/PhysRevD.104.054047}{\emph{Phys. Rev. D}
  {\bfseries 104} (2021) 054047}
  [\href{https://arxiv.org/abs/2009.13667}{{\ttfamily 2009.13667}}].

\bibitem{Barata:2021wuf}
J.a.~Barata, Y.~Mehtar-Tani, A.~Soto-Ontoso and K.~Tywoniuk,
  \emph{{Medium-induced radiative kernel with the Improved Opacity Expansion}},
  \href{https://doi.org/10.1007/JHEP09(2021)153}{\emph{JHEP} {\bfseries 09}
  (2021) 153} [\href{https://arxiv.org/abs/2106.07402}{{\ttfamily
  2106.07402}}].

\bibitem{Andres:2020vxs}
C.~Andres, L.~Apolin\'ario and F.~Dominguez, \emph{{Medium-induced gluon
  radiation with full resummation of multiple scatterings for realistic
  parton-medium interactions}},
  \href{https://doi.org/10.1007/JHEP07(2020)114}{\emph{JHEP} {\bfseries 07}
  (2020) 114} [\href{https://arxiv.org/abs/2002.01517}{{\ttfamily
  2002.01517}}].

\bibitem{Mehtar-Tani:2010ebp}
Y.~Mehtar-Tani, C.A.~Salgado and K.~Tywoniuk, \emph{{Anti-angular ordering of
  gluon radiation in QCD media}},
  \href{https://doi.org/10.1103/PhysRevLett.106.122002}{\emph{Phys. Rev. Lett.}
  {\bfseries 106} (2011) 122002}
  [\href{https://arxiv.org/abs/1009.2965}{{\ttfamily 1009.2965}}].

\bibitem{Mehtar-Tani:2011hma}
Y.~Mehtar-Tani, C.A.~Salgado and K.~Tywoniuk, \emph{{Jets in QCD Media: From
  Color Coherence to Decoherence}},
  \href{https://doi.org/10.1016/j.physletb.2011.12.042}{\emph{Phys. Lett. B}
  {\bfseries 707} (2012) 156}
  [\href{https://arxiv.org/abs/1102.4317}{{\ttfamily 1102.4317}}].

\bibitem{Mehtar-Tani:2011vlz}
Y.~Mehtar-Tani and K.~Tywoniuk, \emph{{Jet coherence in QCD media: the antenna
  radiation spectrum}},
  \href{https://doi.org/10.1007/JHEP01(2013)031}{\emph{JHEP} {\bfseries 01}
  (2013) 031} [\href{https://arxiv.org/abs/1105.1346}{{\ttfamily 1105.1346}}].

\bibitem{Blaizot:2012fh}
J.-P.~Blaizot, F.~Dominguez, E.~Iancu and Y.~Mehtar-Tani, \emph{{Medium-induced
  gluon branching}}, \href{https://doi.org/10.1007/JHEP01(2013)143}{\emph{JHEP}
  {\bfseries 01} (2013) 143} [\href{https://arxiv.org/abs/1209.4585}{{\ttfamily
  1209.4585}}].

\bibitem{Blaizot:2013hx}
J.-P.~Blaizot, E.~Iancu and Y.~Mehtar-Tani, \emph{{Medium-induced QCD cascade:
  democratic branching and wave turbulence}},
  \href{https://doi.org/10.1103/PhysRevLett.111.052001}{\emph{Phys. Rev. Lett.}
  {\bfseries 111} (2013) 052001}
  [\href{https://arxiv.org/abs/1301.6102}{{\ttfamily 1301.6102}}].

\bibitem{Blanco:2020uzy}
E.~Blanco, K.~Kutak, W.~P\l{}aczek, M.~Rohrmoser and R.~Straka, \emph{{Medium
  induced QCD cascades: broadening and rescattering during branching}},
  \href{https://doi.org/10.1007/JHEP04(2021)014}{\emph{JHEP} {\bfseries 04}
  (2021) 014} [\href{https://arxiv.org/abs/2009.03876}{{\ttfamily
  2009.03876}}].

\bibitem{Blanco:2021usa}
E.~Blanco, K.~Kutak, W.~Placzek, M.~Rohrmoser and K.~Tywoniuk, \emph{{System of
  evolution equations for quark and gluon jet quenching with broadening}},
  \href{https://doi.org/10.1140/epjc/s10052-022-10311-2}{\emph{Eur. Phys. J. C}
  {\bfseries 82} (2022) 355}
  [\href{https://arxiv.org/abs/2109.05918}{{\ttfamily 2109.05918}}].

\bibitem{Adhya:2022tcn}
S.P.~Adhya, K.~Kutak, W.~P\l{}aczek, M.~Rohrmoser and K.~Tywoniuk,
  \emph{{Transverse momentum broadening of medium-induced cascades in expanding
  media}}, \href{https://doi.org/10.1140/epjc/s10052-023-11555-2}{\emph{Eur.
  Phys. J. C} {\bfseries 83} (2023) 512}
  [\href{https://arxiv.org/abs/2211.15803}{{\ttfamily 2211.15803}}].

\bibitem{Arnold:2016kek}
P.~Arnold, H.-C.~Chang and S.~Iqbal, \emph{{The LPM effect in sequential
  bremsstrahlung 2: factorization}},
  \href{https://doi.org/10.1007/JHEP09(2016)078}{\emph{JHEP} {\bfseries 09}
  (2016) 078} [\href{https://arxiv.org/abs/1605.07624}{{\ttfamily
  1605.07624}}].

\bibitem{Arnold:2016mth}
P.~Arnold, H.-C.~Chang and S.~Iqbal, \emph{{The LPM effect in sequential
  bremsstrahlung: dimensional regularization}},
  \href{https://doi.org/10.1007/JHEP10(2016)100}{\emph{JHEP} {\bfseries 10}
  (2016) 100} [\href{https://arxiv.org/abs/1606.08853}{{\ttfamily
  1606.08853}}].

\bibitem{Arnold:2018yjd}
P.~Arnold and S.~Iqbal, \emph{{In-medium loop corrections and longitudinally
  polarized gauge bosons in high-energy showers}},
  \href{https://doi.org/10.1007/JHEP12(2018)120}{\emph{JHEP} {\bfseries 12}
  (2018) 120} [\href{https://arxiv.org/abs/1806.08796}{{\ttfamily
  1806.08796}}].

\bibitem{Arnold:2018fjr}
P.~Arnold, S.~Iqbal and T.~Rase, \emph{{Strong- vs. weak-coupling pictures of
  jet quenching: a dry run using QED}},
  \href{https://doi.org/10.1007/JHEP05(2019)004}{\emph{JHEP} {\bfseries 05}
  (2019) 004} [\href{https://arxiv.org/abs/1810.06578}{{\ttfamily
  1810.06578}}].

\bibitem{Arnold:2019qqc}
P.~Arnold, \emph{{Landau-Pomeranchuk-Migdal effect in sequential
  bremsstrahlung: From large-$N$ QCD to $N$=3 via the SU($N$) analog of Wigner
  6-$j$ symbols}},
  \href{https://doi.org/10.1103/PhysRevD.100.034030}{\emph{Phys. Rev. D}
  {\bfseries 100} (2019) 034030}
  [\href{https://arxiv.org/abs/1904.04264}{{\ttfamily 1904.04264}}].

\bibitem{Arnold:2020uzm}
P.~Arnold, T.~Gorda and S.~Iqbal, \emph{{The LPM effect in sequential
  bremsstrahlung: nearly complete results for QCD}},
  \href{https://doi.org/10.1007/JHEP11(2020)053}{\emph{JHEP} {\bfseries 11}
  (2020) 053} [\href{https://arxiv.org/abs/2007.15018}{{\ttfamily
  2007.15018}}].

\bibitem{Arnold:2021pin}
P.~Arnold, T.~Gorda and S.~Iqbal, \emph{{The LPM effect in sequential
  bremsstrahlung: analytic results for sub-leading (single) logarithms}},
  \href{https://doi.org/10.1007/JHEP04(2022)085}{\emph{JHEP} {\bfseries 04}
  (2022) 085} [\href{https://arxiv.org/abs/2112.05161}{{\ttfamily
  2112.05161}}].

\bibitem{Arnold:2022epx}
P.~Arnold and O.~Elgedawy, \emph{{The LPM Effect in sequential bremsstrahlung:
  $1/N_c^2$ corrections}},  \href{https://arxiv.org/abs/2202.04662}{{\ttfamily
  2202.04662}}.

\bibitem{Arnold:2022fku}
P.~Arnold, T.~Gorda and S.~Iqbal, \emph{{The LPM effect in sequential
  bremsstrahlung: incorporation of
  \textquotedblleft{}instantaneous\textquotedblright{} interactions for QCD}},
  \href{https://doi.org/10.1007/JHEP11(2022)130}{\emph{JHEP} {\bfseries 11}
  (2022) 130} [\href{https://arxiv.org/abs/2209.03971}{{\ttfamily
  2209.03971}}].

\bibitem{Arnold:2022mby}
P.~Arnold, O.~Elgedawy and S.~Iqbal, \emph{{Are gluon showers inside a
  quark-gluon plasma strongly coupled? a theorist's test}},
  \href{https://arxiv.org/abs/2212.08086}{{\ttfamily 2212.08086}}.

\bibitem{Arnold:2023qwi}
P.~Arnold, O.~Elgedawy and S.~Iqbal, \emph{{The LPM effect in sequential
  bremsstrahlung: gluon shower development}},
  \href{https://arxiv.org/abs/2302.10215}{{\ttfamily 2302.10215}}.

\bibitem{Wu:2011kc}
B.~Wu, \emph{{On $p_\bot$-broadening of high energy partons associated with the
  LPM effect in a finite-volume QCD medium}},
  \href{https://doi.org/10.1007/JHEP10(2011)029}{\emph{JHEP} {\bfseries 10}
  (2011) 029} [\href{https://arxiv.org/abs/1102.0388}{{\ttfamily 1102.0388}}].

\bibitem{Blaizot:2019muz}
J.-P.~Blaizot and F.~Dominguez, \emph{{Radiative corrections to the jet
  quenching parameter in dilute and dense media}},
  \href{https://doi.org/10.1103/PhysRevD.99.054005}{\emph{Phys. Rev. D}
  {\bfseries 99} (2019) 054005}
  [\href{https://arxiv.org/abs/1901.01448}{{\ttfamily 1901.01448}}].

\bibitem{Iancu:2014kga}
E.~Iancu, \emph{{The non-linear evolution of jet quenching}},
  \href{https://doi.org/10.1007/JHEP10(2014)095}{\emph{JHEP} {\bfseries 10}
  (2014) 095} [\href{https://arxiv.org/abs/1403.1996}{{\ttfamily 1403.1996}}].

\bibitem{Iancu:2014sha}
E.~Iancu and D.N.~Triantafyllopoulos, \emph{{Running coupling effects in the
  evolution of jet quenching}},
  \href{https://doi.org/10.1103/PhysRevD.90.074002}{\emph{Phys. Rev. D}
  {\bfseries 90} (2014) 074002}
  [\href{https://arxiv.org/abs/1405.3525}{{\ttfamily 1405.3525}}].

\bibitem{Caucal:2021lgf}
P.~Caucal and Y.~Mehtar-Tani, \emph{{Anomalous diffusion in QCD matter}},
  \href{https://doi.org/10.1103/PhysRevD.106.L051501}{\emph{Phys. Rev. D}
  {\bfseries 106} (2022) L051501}
  [\href{https://arxiv.org/abs/2109.12041}{{\ttfamily 2109.12041}}].

\bibitem{Caucal:2022fhc}
P.~Caucal and Y.~Mehtar-Tani, \emph{{Universality aspects of quantum
  corrections to transverse momentum broadening in QCD media}},
  \href{https://doi.org/10.1007/JHEP09(2022)023}{\emph{JHEP} {\bfseries 09}
  (2022) 023} [\href{https://arxiv.org/abs/2203.09407}{{\ttfamily
  2203.09407}}].

\bibitem{Wu:2014nca}
B.~Wu, \emph{{Radiative energy loss and radiative $p_{\bot}$-broadening of
  high-energy partons in QCD matter}},
  \href{https://doi.org/10.1007/JHEP12(2014)081}{\emph{JHEP} {\bfseries 12}
  (2014) 081} [\href{https://arxiv.org/abs/1408.5459}{{\ttfamily 1408.5459}}].

\bibitem{Blaizot:2014bha}
J.-P.~Blaizot and Y.~Mehtar-Tani, \emph{{Renormalization of the jet-quenching
  parameter}},
  \href{https://doi.org/10.1016/j.nuclphysa.2014.05.018}{\emph{Nucl. Phys. A}
  {\bfseries 929} (2014) 202}
  [\href{https://arxiv.org/abs/1403.2323}{{\ttfamily 1403.2323}}].

\bibitem{Pich:1998xt}
A.~Pich, \emph{{Effective field theory: Course}},  in \emph{{Les Houches Summer
  School in Theoretical Physics, Session 68: Probing the Standard Model of
  Particle Interactions}}, pp.~949--1049, 6, 1998
  [\href{https://arxiv.org/abs/hep-ph/9806303}{{\ttfamily hep-ph/9806303}}].

\bibitem{Stewart:2014}
I.~W.Stewart, \emph{Effective field theory lecture notes}, .

\bibitem{Manohar:2018aog}
A.V.~Manohar, \emph{{Introduction to Effective Field Theories}},
  \href{https://arxiv.org/abs/1804.05863}{{\ttfamily 1804.05863}}.

\bibitem{Caswell:1985ui}
W.E.~Caswell and G.P.~Lepage, \emph{{Effective Lagrangians for Bound State
  Problems in QED, QCD, and Other Field Theories}},
  \href{https://doi.org/10.1016/0370-2693(86)91297-9}{\emph{Phys. Lett. B}
  {\bfseries 167} (1986) 437}.

\bibitem{Brambilla:1999xf}
N.~Brambilla, A.~Pineda, J.~Soto and A.~Vairo, \emph{{Potential NRQCD: An
  Effective theory for heavy quarkonium}},
  \href{https://doi.org/10.1016/S0550-3213(99)00693-8}{\emph{Nucl. Phys. B}
  {\bfseries 566} (2000) 275}
  [\href{https://arxiv.org/abs/hep-ph/9907240}{{\ttfamily hep-ph/9907240}}].

\bibitem{Bauer:2000yr}
C.W.~Bauer, S.~Fleming, D.~Pirjol and I.W.~Stewart, \emph{{An Effective field
  theory for collinear and soft gluons: Heavy to light decays}},
  \href{https://doi.org/10.1103/PhysRevD.63.114020}{\emph{Phys. Rev. D}
  {\bfseries 63} (2001) 114020}
  [\href{https://arxiv.org/abs/hep-ph/0011336}{{\ttfamily hep-ph/0011336}}].

\bibitem{Bauer:2001ct}
C.W.~Bauer and I.W.~Stewart, \emph{{Invariant operators in collinear effective
  theory}}, \href{https://doi.org/10.1016/S0370-2693(01)00902-9}{\emph{Phys.
  Lett. B} {\bfseries 516} (2001) 134}
  [\href{https://arxiv.org/abs/hep-ph/0107001}{{\ttfamily hep-ph/0107001}}].

\bibitem{Bauer:2001yt}
C.W.~Bauer, D.~Pirjol and I.W.~Stewart, \emph{{Soft collinear factorization in
  effective field theory}},
  \href{https://doi.org/10.1103/PhysRevD.65.054022}{\emph{Phys. Rev. D}
  {\bfseries 65} (2002) 054022}
  [\href{https://arxiv.org/abs/hep-ph/0109045}{{\ttfamily hep-ph/0109045}}].

\bibitem{Bauer:2002nz}
C.W.~Bauer, S.~Fleming, D.~Pirjol, I.Z.~Rothstein and I.W.~Stewart, \emph{{Hard
  scattering factorization from effective field theory}},
  \href{https://doi.org/10.1103/PhysRevD.66.014017}{\emph{Phys. Rev. D}
  {\bfseries 66} (2002) 014017}
  [\href{https://arxiv.org/abs/hep-ph/0202088}{{\ttfamily hep-ph/0202088}}].

\bibitem{Beneke:2002ph}
M.~Beneke, A.P.~Chapovsky, M.~Diehl and T.~Feldmann, \emph{{Soft collinear
  effective theory and heavy to light currents beyond leading power}},
  \href{https://doi.org/10.1016/S0550-3213(02)00687-9}{\emph{Nucl. Phys. B}
  {\bfseries 643} (2002) 431}
  [\href{https://arxiv.org/abs/hep-ph/0206152}{{\ttfamily hep-ph/0206152}}].

\bibitem{Weinberg:1978kz}
S.~Weinberg, \emph{{Phenomenological Lagrangians}},
  \href{https://doi.org/10.1016/0378-4371(79)90223-1}{\emph{Physica A}
  {\bfseries 96} (1979) 327}.

\bibitem{Gasser:1983yg}
J.~Gasser and H.~Leutwyler, \emph{{Chiral Perturbation Theory to One Loop}},
  \href{https://doi.org/10.1016/0003-4916(84)90242-2}{\emph{Annals Phys.}
  {\bfseries 158} (1984) 142}.

\bibitem{Buchmuller:1985jz}
W.~Buchmuller and D.~Wyler, \emph{{Effective Lagrangian Analysis of New
  Interactions and Flavor Conservation}},
  \href{https://doi.org/10.1016/0550-3213(86)90262-2}{\emph{Nucl. Phys. B}
  {\bfseries 268} (1986) 621}.

\bibitem{Grzadkowski:2010es}
B.~Grzadkowski, M.~Iskrzynski, M.~Misiak and J.~Rosiek, \emph{{Dimension-Six
  Terms in the Standard Model Lagrangian}},
  \href{https://doi.org/10.1007/JHEP10(2010)085}{\emph{JHEP} {\bfseries 10}
  (2010) 085} [\href{https://arxiv.org/abs/1008.4884}{{\ttfamily 1008.4884}}].

\bibitem{Beneke:1997zp}
M.~Beneke and V.A.~Smirnov, \emph{{Asymptotic expansion of Feynman integrals
  near threshold}},
  \href{https://doi.org/10.1016/S0550-3213(98)00138-2}{\emph{Nucl. Phys. B}
  {\bfseries 522} (1998) 321}
  [\href{https://arxiv.org/abs/hep-ph/9711391}{{\ttfamily hep-ph/9711391}}].

\bibitem{Smirnov:2002pj}
V.A.~Smirnov, \emph{{Applied asymptotic expansions in momenta and masses}},
  {\emph{Springer Tracts Mod. Phys.} {\bfseries 177} (2002) 1}.

\bibitem{Bellac:2011kqa}
M.L.~Bellac, \emph{{Thermal Field Theory}}, Cambridge Monographs on
  Mathematical Physics, Cambridge University Press (3, 2011),
  \href{https://doi.org/10.1017/CBO9780511721700}{10.1017/CBO9780511721700}.

\bibitem{Feynman:100771}
R.P.~Feynman and A.R.~Hibbs, \emph{{Quantum mechanics and path integrals}},
  International series in pure and applied physics, McGraw-Hill, New York, NY
  (1965).

\bibitem{Matsubara:1955ws}
T.~Matsubara, \emph{{A New approach to quantum statistical mechanics}},
  \href{https://doi.org/10.1143/PTP.14.351}{\emph{Prog. Theor. Phys.}
  {\bfseries 14} (1955) 351}.

\bibitem{Kapusta:1989tk}
J.I.~Kapusta, \emph{{Finite Temperature Field Theory}}, Cambridge Monographs on
  Mathematical Physics, Cambridge University Press, Cambridge (1989).

\bibitem{Linde:1980ts}
A.D.~Linde, \emph{{Infrared Problem in Thermodynamics of the Yang-Mills Gas}},
  \href{https://doi.org/10.1016/0370-2693(80)90769-8}{\emph{Phys. Lett. B}
  {\bfseries 96} (1980) 289}.

\bibitem{Braaten:1995cm}
E.~Braaten and A.~Nieto, \emph{{Effective field theory approach to high
  temperature thermodynamics}},
  \href{https://doi.org/10.1103/PhysRevD.51.6990}{\emph{Phys. Rev. D}
  {\bfseries 51} (1995) 6990}
  [\href{https://arxiv.org/abs/hep-ph/9501375}{{\ttfamily hep-ph/9501375}}].

\bibitem{Braaten:1995jr}
E.~Braaten and A.~Nieto, \emph{{Free energy of QCD at high temperature}},
  \href{https://doi.org/10.1103/PhysRevD.53.3421}{\emph{Phys. Rev. D}
  {\bfseries 53} (1996) 3421}
  [\href{https://arxiv.org/abs/hep-ph/9510408}{{\ttfamily hep-ph/9510408}}].

\bibitem{Kajantie:1995dw}
K.~Kajantie, M.~Laine, K.~Rummukainen and M.E.~Shaposhnikov, \emph{{Generic
  rules for high temperature dimensional reduction and their application to the
  standard model}},
  \href{https://doi.org/10.1016/0550-3213(95)00549-8}{\emph{Nucl. Phys. B}
  {\bfseries 458} (1996) 90}
  [\href{https://arxiv.org/abs/hep-ph/9508379}{{\ttfamily hep-ph/9508379}}].

\bibitem{Kajantie:1997tt}
K.~Kajantie, M.~Laine, K.~Rummukainen and M.E.~Shaposhnikov, \emph{{3-D SU(N) +
  adjoint Higgs theory and finite temperature QCD}},
  \href{https://doi.org/10.1016/S0550-3213(97)00425-2}{\emph{Nucl. Phys. B}
  {\bfseries 503} (1997) 357}
  [\href{https://arxiv.org/abs/hep-ph/9704416}{{\ttfamily hep-ph/9704416}}].

\bibitem{Laine:2005ai}
M.~Laine and Y.~Schroder, \emph{{Two-loop QCD gauge coupling at high
  temperatures}},
  \href{https://doi.org/10.1088/1126-6708/2005/03/067}{\emph{JHEP} {\bfseries
  03} (2005) 067} [\href{https://arxiv.org/abs/hep-ph/0503061}{{\ttfamily
  hep-ph/0503061}}].

\bibitem{Ghisoiu:2015uza}
I.~Ghisoiu, J.~Moller and Y.~Schroder, \emph{{Debye screening mass of hot
  Yang-Mills theory to three-loop order}},
  \href{https://doi.org/10.1007/JHEP11(2015)121}{\emph{JHEP} {\bfseries 11}
  (2015) 121} [\href{https://arxiv.org/abs/1509.08727}{{\ttfamily
  1509.08727}}].

\bibitem{Kajantie:2002wa}
K.~Kajantie, M.~Laine, K.~Rummukainen and Y.~Schroder, \emph{{The Pressure of
  hot QCD up to g6 ln(1/g)}},
  \href{https://doi.org/10.1103/PhysRevD.67.105008}{\emph{Phys. Rev. D}
  {\bfseries 67} (2003) 105008}
  [\href{https://arxiv.org/abs/hep-ph/0211321}{{\ttfamily hep-ph/0211321}}].

\bibitem{Vuorinen:2003fs}
A.~Vuorinen, \emph{{The Pressure of QCD at finite temperatures and chemical
  potentials}}, \href{https://doi.org/10.1103/PhysRevD.68.054017}{\emph{Phys.
  Rev. D} {\bfseries 68} (2003) 054017}
  [\href{https://arxiv.org/abs/hep-ph/0305183}{{\ttfamily hep-ph/0305183}}].

\bibitem{Kajantie:2003ax}
K.~Kajantie, M.~Laine, K.~Rummukainen and Y.~Schroder, \emph{{Four loop vacuum
  energy density of the SU(N(c)) + adjoint Higgs theory}},
  \href{https://doi.org/10.1088/1126-6708/2003/04/036}{\emph{JHEP} {\bfseries
  04} (2003) 036} [\href{https://arxiv.org/abs/hep-ph/0304048}{{\ttfamily
  hep-ph/0304048}}].

\bibitem{Hietanen:2004ew}
A.~Hietanen, K.~Kajantie, M.~Laine, K.~Rummukainen and Y.~Schroder,
  \emph{{Plaquette expectation value and gluon condensate in three
  dimensions}},
  \href{https://doi.org/10.1088/1126-6708/2005/01/013}{\emph{JHEP} {\bfseries
  01} (2005) 013} [\href{https://arxiv.org/abs/hep-lat/0412008}{{\ttfamily
  hep-lat/0412008}}].

\bibitem{DiRenzo:2006nh}
F.~Di~Renzo, M.~Laine, V.~Miccio, Y.~Schroder and C.~Torrero, \emph{{The
  Leading non-perturbative coefficient in the weak-coupling expansion of hot
  QCD pressure}},
  \href{https://doi.org/10.1088/1126-6708/2006/07/026}{\emph{JHEP} {\bfseries
  07} (2006) 026} [\href{https://arxiv.org/abs/hep-ph/0605042}{{\ttfamily
  hep-ph/0605042}}].

\bibitem{Dolan:1973qd}
L.~Dolan and R.~Jackiw, \emph{{Symmetry Behavior at Finite Temperature}},
  \href{https://doi.org/10.1103/PhysRevD.9.3320}{\emph{Phys. Rev. D} {\bfseries
  9} (1974) 3320}.

\bibitem{Schwinger:1960qe}
J.S.~Schwinger, \emph{{Brownian motion of a quantum oscillator}},
  \href{https://doi.org/10.1063/1.1703727}{\emph{J. Math. Phys.} {\bfseries 2}
  (1961) 407}.

\bibitem{Keldysh:1964ud}
L.V.~Keldysh, \emph{{Diagram technique for nonequilibrium processes}},
  {\emph{Zh. Eksp. Teor. Fiz.} {\bfseries 47} (1964) 1515}.

\bibitem{NIEMI1984105}
A.~Niemi and G.~Semenoff, \emph{Finite-temperature quantum field theory in
  minkowski space},
  \href{https://doi.org/https://doi.org/10.1016/0003-4916(84)90082-4}{\emph{Annals
  of Physics} {\bfseries 152} (1984) 105}.

\bibitem{Niemi:1983ea}
A.J.~Niemi and G.W.~Semenoff, \emph{{Thermodynamic Calculations in Relativistic
  Finite Temperature Quantum Field Theories}},
  \href{https://doi.org/10.1016/0550-3213(84)90123-8}{\emph{Nucl. Phys. B}
  {\bfseries 230} (1984) 181}.

\bibitem{Kubo:1957mj}
R.~Kubo, \emph{{Statistical mechanical theory of irreversible processes. 1.
  General theory and simple applications in magnetic and conduction problems}},
  \href{https://doi.org/10.1143/JPSJ.12.570}{\emph{J. Phys. Soc. Jap.}
  {\bfseries 12} (1957) 570}.

\bibitem{PhysRev.115.1342}
P.C.~Martin and J.~Schwinger, \emph{Theory of many-particle systems. i},
  \href{https://doi.org/10.1103/PhysRev.115.1342}{\emph{Phys. Rev.} {\bfseries
  115} (1959) 1342}.

\bibitem{Chou:1984es}
K.-c.~Chou, Z.-b.~Su, B.-l.~Hao and L.~Yu, \emph{{Equilibrium and
  Nonequilibrium Formalisms Made Unified}},
  \href{https://doi.org/10.1016/0370-1573(85)90136-X}{\emph{Phys. Rept.}
  {\bfseries 118} (1985) 1}.

\bibitem{Braaten:1989mz}
E.~Braaten and R.D.~Pisarski, \emph{{Soft Amplitudes in Hot Gauge Theories: A
  General Analysis}},
  \href{https://doi.org/10.1016/0550-3213(90)90508-B}{\emph{Nucl. Phys. B}
  {\bfseries 337} (1990) 569}.

\bibitem{Braaten:1991gm}
E.~Braaten and R.D.~Pisarski, \emph{{Simple effective Lagrangian for hard
  thermal loops}}, \href{https://doi.org/10.1103/PhysRevD.45.R1827}{\emph{Phys.
  Rev. D} {\bfseries 45} (1992) R1827}.

\bibitem{Frenkel:1989br}
J.~Frenkel and J.C.~Taylor, \emph{{High Temperature Limit of Thermal QCD}},
  \href{https://doi.org/10.1016/0550-3213(90)90661-V}{\emph{Nucl. Phys. B}
  {\bfseries 334} (1990) 199}.

\bibitem{Frenkel:1991ts}
J.~Frenkel and J.C.~Taylor, \emph{{Hard thermal QCD, forward scattering and
  effective actions}},
  \href{https://doi.org/10.1016/0550-3213(92)90480-Y}{\emph{Nucl. Phys. B}
  {\bfseries 374} (1992) 156}.

\bibitem{Taylor:1990ia}
J.C.~Taylor and S.M.H.~Wong, \emph{{The Effective Action of Hard Thermal Loops
  in {QCD}}}, \href{https://doi.org/10.1016/0550-3213(90)90240-E}{\emph{Nucl.
  Phys. B} {\bfseries 346} (1990) 115}.

\bibitem{CaronHuot:2007nw}
S.~Caron-Huot, \emph{{Hard thermal loops in the real-time formalism}},
  \href{https://doi.org/10.1088/1126-6708/2009/04/004}{\emph{JHEP} {\bfseries
  04} (2009) 004} [\href{https://arxiv.org/abs/0710.5726}{{\ttfamily
  0710.5726}}].

\bibitem{Karsch:1998tx}
F.~Karsch, M.~Oevers and P.~Petreczky, \emph{{Screening masses of hot SU(2)
  gauge theory from the 3-d adjoint Higgs model}},
  \href{https://doi.org/10.1016/S0370-2693(98)01248-9}{\emph{Phys. Lett. B}
  {\bfseries 442} (1998) 291}
  [\href{https://arxiv.org/abs/hep-lat/9807035}{{\ttfamily hep-lat/9807035}}].

\bibitem{Cucchieri:2000cy}
A.~Cucchieri, F.~Karsch and P.~Petreczky, \emph{{Magnetic screening in hot
  nonAbelian gauge theory}},
  \href{https://doi.org/10.1016/S0370-2693(00)01331-9}{\emph{Phys. Lett. B}
  {\bfseries 497} (2001) 80}
  [\href{https://arxiv.org/abs/hep-lat/0004027}{{\ttfamily hep-lat/0004027}}].

\bibitem{Weldon:1982aq}
H.A.~Weldon, \emph{{Covariant Calculations at Finite Temperature: The
  Relativistic Plasma}},
  \href{https://doi.org/10.1103/PhysRevD.26.1394}{\emph{Phys. Rev. D}
  {\bfseries 26} (1982) 1394}.

\bibitem{Baym:1990uj}
G.~Baym, H.~Monien, C.J.~Pethick and D.G.~Ravenhall, \emph{{Transverse
  Interactions and Transport in Relativistic Quark - Gluon and Electromagnetic
  Plasmas}}, \href{https://doi.org/10.1103/PhysRevLett.64.1867}{\emph{Phys.
  Rev. Lett.} {\bfseries 64} (1990) 1867}.

\bibitem{Pisarski:1990ds}
R.D.~Pisarski, \emph{{Resummation and the gluon damping rate in hot QCD}},
  \href{https://doi.org/10.1016/0375-9474(91)90325-Z}{\emph{Nucl. Phys. A}
  {\bfseries 525} (1991) 175}.

\bibitem{Braaten:1990it}
E.~Braaten and R.D.~Pisarski, \emph{{Calculation of the gluon damping rate in
  hot QCD}}, \href{https://doi.org/10.1103/PhysRevD.42.2156}{\emph{Phys. Rev.
  D} {\bfseries 42} (1990) 2156}.

\bibitem{Pisarski:1989wb}
R.D.~Pisarski, \emph{{Renormalized Fermion Propagator in Hot Gauge Theories}},
  \href{https://doi.org/10.1016/0375-9474(89)90620-9}{\emph{Nucl. Phys. A}
  {\bfseries 498} (1989) 423C}.

\bibitem{Kalashnikov:1979cy}
O.K.~Kalashnikov and V.V.~Klimov, \emph{{Polarization Tensor in QCD for Finite
  Temperature and Density}}, {\emph{Sov. J. Nucl. Phys.} {\bfseries 31} (1980)
  699}.

\bibitem{CaronHuot:2008uw}
S.~Caron-Huot, \emph{{On supersymmetry at finite temperature}},
  \href{https://doi.org/10.1103/PhysRevD.79.125002}{\emph{Phys. Rev. D}
  {\bfseries 79} (2009) 125002}
  [\href{https://arxiv.org/abs/0808.0155}{{\ttfamily 0808.0155}}].

\bibitem{Kapusta:2006pm}
J.I.~Kapusta and C.~Gale, \emph{{Finite-temperature field theory: Principles
  and applications}}, Cambridge Monographs on Mathematical Physics, Cambridge
  University Press (2011),
  \href{https://doi.org/10.1017/CBO9780511535130}{10.1017/CBO9780511535130}.

\bibitem{simonguy}
S.~Caron-Huot and G.D.~Moore, \emph{{Heavy quark diffusion in QCD and N=4 SYM
  at next-to-leading order}},
  \href{https://doi.org/10.1088/1126-6708/2008/02/081}{\emph{JHEP} {\bfseries
  0802} (2008) 081} [\href{https://arxiv.org/abs/0801.2173}{{\ttfamily
  0801.2173}}].

\bibitem{Ghiglieri:2013gia}
J.~Ghiglieri, J.~Hong, A.~Kurkela, E.~Lu, G.D.~Moore and D.~Teaney,
  \emph{{Next-to-leading order thermal photon production in a weakly coupled
  quark-gluon plasma}},
  \href{https://doi.org/10.1007/JHEP05(2013)010}{\emph{JHEP} {\bfseries 05}
  (2013) 010} [\href{https://arxiv.org/abs/1302.5970}{{\ttfamily 1302.5970}}].

\bibitem{Ghiglieri:2015zma}
J.~Ghiglieri and D.~Teaney, \emph{{Parton energy loss and momentum broadening
  at NLO in high temperature QCD plasmas}},
  \href{https://doi.org/10.1142/S0218301315300131}{\emph{Int. J. Mod. Phys. E}
  {\bfseries 24} (2015) 1530013}
  [\href{https://arxiv.org/abs/1502.03730}{{\ttfamily 1502.03730}}].

\bibitem{CasalderreySolana:2007qw}
J.~Casalderrey-Solana and D.~Teaney, \emph{{Transverse Momentum Broadening of a
  Fast Quark in a N=4 Yang Mills Plasma}},
  \href{https://doi.org/10.1088/1126-6708/2007/04/039}{\emph{JHEP} {\bfseries
  04} (2007) 039} [\href{https://arxiv.org/abs/hep-th/0701123}{{\ttfamily
  hep-th/0701123}}].

\bibitem{DEramo:2012uzl}
F.~D'Eramo, M.~Lekaveckas, H.~Liu and K.~Rajagopal, \emph{{Momentum Broadening
  in Weakly Coupled Quark-Gluon Plasma (with a view to finding the
  quasiparticles within liquid quark-gluon plasma)}},
  \href{https://doi.org/10.1007/JHEP05(2013)031}{\emph{JHEP} {\bfseries 05}
  (2013) 031} [\href{https://arxiv.org/abs/1211.1922}{{\ttfamily 1211.1922}}].

\bibitem{Benzke:2012sz}
M.~Benzke, N.~Brambilla, M.A.~Escobedo and A.~Vairo, \emph{{Gauge invariant
  definition of the jet quenching parameter}},
  \href{https://doi.org/10.1007/JHEP02(2013)129}{\emph{JHEP} {\bfseries 02}
  (2013) 129} [\href{https://arxiv.org/abs/1208.4253}{{\ttfamily 1208.4253}}].

\bibitem{Panero:2013pla}
M.~Panero, K.~Rummukainen and A.~Sch\"afer, \emph{{Lattice Study of the Jet
  Quenching Parameter}},
  \href{https://doi.org/10.1103/PhysRevLett.112.162001}{\emph{Phys. Rev. Lett.}
  {\bfseries 112} (2014) 162001}
  [\href{https://arxiv.org/abs/1307.5850}{{\ttfamily 1307.5850}}].

\bibitem{Laine:2013lia}
M.~Laine and A.~Rothkopf, \emph{{Light-cone Wilson loop in classical lattice
  gauge theory}}, \href{https://doi.org/10.1007/JHEP07(2013)082}{\emph{JHEP}
  {\bfseries 07} (2013) 082} [\href{https://arxiv.org/abs/1304.4443}{{\ttfamily
  1304.4443}}].

\bibitem{Becher:2015hka}
T.~Becher, M.~Neubert, L.~Rothen and D.Y.~Shao, \emph{{Effective Field Theory
  for Jet Processes}},
  \href{https://doi.org/10.1103/PhysRevLett.116.192001}{\emph{Phys. Rev. Lett.}
  {\bfseries 116} (2016) 192001}
  [\href{https://arxiv.org/abs/1508.06645}{{\ttfamily 1508.06645}}].

\bibitem{Peter:1997me}
M.~Peter, \emph{{The Static potential in QCD: A Full two loop calculation}},
  \href{https://doi.org/10.1016/S0550-3213(97)00373-8}{\emph{Nucl. Phys. B}
  {\bfseries 501} (1997) 471}
  [\href{https://arxiv.org/abs/hep-ph/9702245}{{\ttfamily hep-ph/9702245}}].

\bibitem{Berwein:2012mw}
M.~Berwein, N.~Brambilla, J.~Ghiglieri and A.~Vairo, \emph{{Renormalization of
  the cyclic Wilson loop}},
  \href{https://doi.org/10.1007/JHEP03(2013)069}{\emph{JHEP} {\bfseries 03}
  (2013) 069} [\href{https://arxiv.org/abs/1212.4413}{{\ttfamily 1212.4413}}].

\bibitem{Kovchegov:2012mbw}
Y.V.~Kovchegov and E.~Levin, \emph{{Quantum Chromodynamics at High Energy}},
  vol.~33, Oxford University Press (2013),
  \href{https://doi.org/10.1017/9781009291446}{10.1017/9781009291446}.

\bibitem{Ghiglieri:2015ala}
J.~Ghiglieri, G.D.~Moore and D.~Teaney, \emph{{Jet-Medium Interactions at NLO
  in a Weakly-Coupled Quark-Gluon Plasma}},
  \href{https://doi.org/10.1007/JHEP03(2016)095}{\emph{JHEP} {\bfseries 03}
  (2016) 095} [\href{https://arxiv.org/abs/1509.07773}{{\ttfamily
  1509.07773}}].

\bibitem{Heinz:1986kz}
U.W.~Heinz, K.~Kajantie and T.~Toimela, \emph{{Gauge Covariant Linear Response
  Analysis of QCD Plasma Oscillations}},
  \href{https://doi.org/10.1016/0003-4916(87)90002-9}{\emph{Annals Phys.}
  {\bfseries 176} (1987) 218}.

\bibitem{Manuel:2016wqs}
C.~Manuel, J.~Soto and S.~Stetina, \emph{{On-shell effective field theory: A
  systematic tool to compute power corrections to the hard thermal loops}},
  \href{https://doi.org/10.1103/PhysRevD.94.025017}{\emph{Phys. Rev. D}
  {\bfseries 94} (2016) 025017}
  [\href{https://arxiv.org/abs/1603.05514}{{\ttfamily 1603.05514}}].

\bibitem{Carignano:2017ovz}
S.~Carignano, C.~Manuel and J.~Soto, \emph{{Power corrections to the HTL
  effective Lagrangian of QED}},
  \href{https://doi.org/10.1016/j.physletb.2018.03.012}{\emph{Phys. Lett. B}
  {\bfseries 780} (2018) 308}
  [\href{https://arxiv.org/abs/1712.07949}{{\ttfamily 1712.07949}}].

\bibitem{Arnold:2009mr}
P.B.~Arnold, \emph{{High-energy gluon bremsstrahlung in a finite medium:
  harmonic oscillator versus single scattering approximation}},
  \href{https://doi.org/10.1103/PhysRevD.80.025004}{\emph{Phys. Rev. D}
  {\bfseries 80} (2009) 025004}
  [\href{https://arxiv.org/abs/0903.1081}{{\ttfamily 0903.1081}}].

\bibitem{Andres:2020kfg}
C.~Andres, F.~Dominguez and M.~Gonzalez~Martinez, \emph{{From soft to hard
  radiation: the role of multiple scatterings in medium-induced gluon
  emissions}}, \href{https://doi.org/10.1007/JHEP03(2021)102}{\emph{JHEP}
  {\bfseries 03} (2021) 102}
  [\href{https://arxiv.org/abs/2011.06522}{{\ttfamily 2011.06522}}].

\bibitem{Ghiglieri:2023cyw}
J.~Ghiglieri, G.D.~Moore, P.~Schicho, N.~Schlusser and E.~Weitz, \emph{{Hard
  parton dispersion in the quark-gluon plasma, non-perturbatively}},  in
  \emph{{11th International Conference on Hard and Electromagnetic Probes of
  High-Energy Nuclear Collisions}: {Hard Probes 2023}}, 7, 2023
  [\href{https://arxiv.org/abs/2307.09297}{{\ttfamily 2307.09297}}].

\bibitem{Yazdi:2022bru}
R.M.~Yazdi, S.~Shi, C.~Gale and S.~Jeon, \emph{{Leading order, next-to-leading
  order, and nonperturbative parton collision kernels: Effects in static and
  evolving media}},
  \href{https://doi.org/10.1103/PhysRevC.106.064902}{\emph{Phys. Rev. C}
  {\bfseries 106} (2022) 064902}
  [\href{https://arxiv.org/abs/2206.05855}{{\ttfamily 2206.05855}}].

\bibitem{Shi:2022rja}
S.~Shi, R.~Modarresi~Yazdi, C.~Gale and S.~Jeon, \emph{{Comparing the martini
  and cujet models for jet quenching: Medium modification of jets and jet
  substructure}},
  \href{https://doi.org/10.1103/PhysRevC.107.034908}{\emph{Phys. Rev. C}
  {\bfseries 107} (2023) 034908}
  [\href{https://arxiv.org/abs/2212.05944}{{\ttfamily 2212.05944}}].

\end{thebibliography}\endgroup
\newpage
\thispagestyle{empty}\null\newpage
\newpage 
\includepdf[pages={2}]{Cover.pdf}
\end{document}